\documentclass[aps,pra,reprint,onecolumn,superscriptaddress]{revtex4}

\usepackage{amsmath,amsthm,amssymb,graphicx,xcolor,float,makecell,multirow}

\usepackage{esint}
\usepackage[unicode=true]{hyperref}
\hypersetup{
     colorlinks=true,       		
     linkcolor=blue,          	
     citecolor=red,            
     urlcolor=magenta,           	
 }

\usepackage{mathrsfs,mathtools,mathdots,dsfont}
\usepackage{listings}
\usepackage{subfigure}

\newtheorem*{theorem*}{Theorem}

\newtheorem*{corollary*}{Corollary}

\newtheorem*{lemma*}{Lemma}

\newtheorem*{proposition*}{Example*}

\newtheorem*{conjecture*}{Conjecture}
\theoremstyle{definition}

\newtheorem*{definition*}{Definition}
\theoremstyle{remark}
\newtheorem{remark}{Remark}
\newtheorem*{remark*}{Remark}

\begin{document}

\title{Exact Solutions of the SU(2) Yang-Mills Equations from a Static Ansatz}

\author{Yu-Xuan Zhang}
\affiliation{School of Physics, Nankai University, Tianjin 300071, People's Republic of China}

\author{Jing-Ling Chen}
\email{chenjl@nankai.edu.cn}
\affiliation{Theoretical Physics Division, Chern Institute of Mathematics, Nankai University, Tianjin 300071, People's Republic of
	 	China}

\date{\today}

\begin{abstract}
We present a systematic study of static solutions to the source-free SU(2) Yang-Mills equations, in which the gauge potential explicitly depends on spin operators. By employing the \emph{vector potential extraction approach} --- which requires the total angular momentum operator (orbital plus spin) to satisfy the standard angular momentum algebra --- we derive the most general form of the spin vector potential $\vec{A}$. This leads to the static ansatz $\{ \vec{A} = [k_1(\hat{r}\times\vec{\Gamma}) + k_2\vec{\Gamma} + k_3(\vec{\Gamma}\cdot\hat{r})\hat{r}]/r, \varphi = f_1(r)\,(\vec{\Gamma}\cdot\hat{r}) + f_2(r)\}$, parametrized by three constants $\{k_1, k_2, k_3\}$ and two radial functions $\{f_1(r), f_2(r)\}$. After substituting this static ansatz into the Yang-Mills equations we obtain a set of consistency equations. Solving these equations provides a complete classification of the exact static solutions, including both real and complex families. The known simple SU(2) static solution $\{\vec{A}=\tilde{k} (\hat{r}\times\vec{\Gamma})/r, \varphi=\kappa/r \}$ is recovered as a special case. Our classification reveals new static configurations that could be valuable for non-perturbative studies and for models where the internal spin couples to non-Abelian gauge fields.
\end{abstract}

\maketitle


\section{Introduction}

The Yang-Mills theory (YM) \cite{Yang1954} is the foundation of modern gauge theories of fundamental interactions \cite{Weinberg1995,Zee2010}. Despite its success, the YM equations are highly nonlinear coupled partial differential equations, making the search for exact classical solutions a challenging yet crucial task. Classical solutions play a crucial role in revealing the non-perturbative dynamics of gauge theories, such as vacuum tunneling and confinement mechanisms \cite{Callan1978}.

Historically, the quest for exact or topologically non-trivial configurations has driven major theoretical breakthroughs. Early attempts to find point-like singularities yielded the pure SU(2) Wu-Yang monopole \cite{WuYang1969}, which was later regularized by coupling to a Higgs field, resulting in the finite-energy 't Hooft-Polyakov monopole \cite{tHooft1974,Prasad1975}. Subsequent discoveries further unveiled the rich topological structure of the vacuum, including instantons in Euclidean spacetime \cite{BPST1975,ADHM1978,Nahm1982} and dyons \cite{JuliaZee1975}. Traditionally, discovering these exact solutions has relied heavily on symmetry-based reductions \cite{Witten1977, Forgacs1980} and specific algebraic ansatz techniques \cite{Corrigan1977,Actor1979}. In particular, solutions that couple the gauge potential to internal spin degrees of freedom are of interest, as they may shed light on the role of angular momentum and color-spin vector decompositions in non-Abelian gauge theories \cite{Faddeev1999}.

 Recently, a new heuristic approach has emerged. A simple SU(2) static solution was proposed in \cite{Zhou2025}, featuring  a spin vector potential $\vec{A} \propto (\vec{r}\times\vec{S})/{r^2}$ and a Coulomb-type scalar potential $\varphi \propto {1}/{r}$. Remarkably, this solution satisfies the YM equations only when a specific quantization condition on the coupling constants holds. The form of this vector potential closely resembles that obtained by a \emph{vector potential extraction approach} (VPEA) introduced for the Abelian case in \cite{Chen2025}. The VPEA starts from an angular momentum operator $\vec{L}=\vec{\ell}+\vec{G}$ and imposes the standard angular momentum algebra $\vec{L}\times\vec{L}={\rm i} \hbar\vec{L}$ (with $\vec{\ell}=\vec{r}\times \vec{p}$ being the orbital angular momentum operator). It has successfully reproduced several well-known U(1) gauge potentials, including those for the Aharonov-Bohm effect \cite{2005QParadox}, uniform magnetic fields \cite{1999Jackson},
 and the Dirac monopole \cite{Dirac1931,WuYang1975,WuYang1976}.

In the non-Abelian case, the VPEA naturally leads to a spin-dependent gauge potential, but the question arises: What is its most general form? In this work, we generalize the VPEA to its most general form by allowing an arbitrary linear combination of the three independent spin-dependent vectors that can appear in the operator $\vec{G}$:
\begin{eqnarray}
&& \vec{S}, \;\;\;\; (\vec{S}\cdot\hat{r})\,\hat{r}, \;\;\;\; \hat{r}\times\vec{S},
\end{eqnarray}
with $\hat{r}=\vec{r}/r$. From this, we extract the corresponding gauge potential and systematically solve the SU(2) YM equations for static fields.

Our main result is a complete classification of static solutions based on the following static ansatz:
\begin{eqnarray}\label{eq:ansatz}
&&\vec{A} = \frac{1}{r}\Bigl[k_1(\hat{r}\times\vec{\Gamma}) + k_2\vec{\Gamma} + k_3(\vec{\Gamma}\cdot\hat{r})\hat{r}\Bigr], \nonumber\\
&& \varphi = f_1(r)\,(\vec{\Gamma}\cdot\hat{r}) + f_2(r),
\end{eqnarray}
where \(\vec{\Gamma}\) are the Pauli matrices (for spin-1/2 generators) and $\vec{S}=(\hbar/2) \vec{\Gamma}$. The parameters $k_1, k_2, k_3$ and the radial functions $f_1(r), f_2(r)$ are to be determined. Substituting this ansatz into the YM equations reduces the problem to a set of ordinary differential equations and algebraic constraints, which we solve case by case.
The exact solutions we obtained are remarkably rich. Both real and complex solutions are presented, and we show how the previously known simple solution of Ref. \cite{Zhou2025} emerges as a special subclass. The inclusion of complex solutions is motivated by modern analytical methods, such as resurgence theory and the complexification of path integrals \cite{Witten2010,Dunne2012,Basar2013,Aniceto2019}, as well as early explorations of complex gauge fields \cite{Oh1990PLB}. Furthermore, the exact static solutions found here provide a natural starting point for future investigations into color radiation \cite{Oh1986PRD} and non-Abelian progressive waves \cite{Oh1985JMP,Rabinowitch2026}.

The paper is organized as follows. In Sec. \ref{sec-4} we recall the YM equations and a known simple SU(2) static solution. In Sec. \ref{sec-5} we review the VPEA for the Abelian case and extends it to the non-Abelian SU(2) case, leading to the general form of spin vector potential. In Sec. \ref{sec-6} we present the general static ansatz, calculate the electric-like and magnetic-like fields, derive the constraint equations, solve them, and classify all static solutions. Finally, Sec. \ref{sec-7} offers our conclusions and discussion.

\section{The Yang-Mills Equations and A Simple SU(2) Static Solution }\label{sec-4}

In this work, we use the Gaussian units. The metric tensor of space-time is taken as
\begin{eqnarray}
\eta_{\mu\nu} = \text{diag}(1, -1, -1, -1), \;\;\;\;\;\;\; (\mu, \nu=0, 1, 2, 3).
\end{eqnarray}
The relation between the four-potential and the scalar-vector potential is given by
\begin{eqnarray}
A_{\mu} = (\varphi, -\vec{A}), \;\;\;\;\;\; (\mu=0, 1, 2, 3),
\end{eqnarray}
with $A_0=\varphi$, $\vec{A}=(A_1, A_2, A_3)$ or $\vec{A}=(A_x, A_y, A_z)$. Here \emph{the scalar potential} $\varphi$ and \emph{the vector potential} $\vec{A}$ have the same physics dimension.

\subsection{The Yang-Mills Equations}

The Yang-Mills equations in vacuum without sources (i.e., the 4-vector current ${J^\nu}=0$) are given by \cite{Yang1954}
      \begin{subequations}
            \begin{align}
                  & D_\mu {F}^{\mu\nu} \equiv\partial_\mu {F}^{\mu\nu}
                        + {\frac{{\rm i} g}{c\hbar}}\Bigl[{A}_\mu ,\ {F}^{\mu\nu}\Bigr]=0,
                        \label{eq:YM1a} \\
                  & D_\mu {F}_{\nu\gamma} +D_\nu {F}_{\gamma\mu}
                        +D_\gamma {F}_{\mu\nu} =0. \label{eq:YM1b}
            \end{align}
      \end{subequations}
Here the covariant derivative $D_{\mu}$ is defined as
\begin{eqnarray}
D_{\mu}=\partial_{\mu}+ \frac{{\rm i} g}{c\hbar} {A}_{\mu},  \quad\;\;\;\;  (\mu=0, 1, 2, 3),
\end{eqnarray}
and the field strength tensor is
      \begin{equation}\label{eq:nonabelian}
            {F}_{\mu\nu} =\partial_\mu {A}_{\nu}
                  -\partial_\nu {A}_{\mu}
                  + \frac{{\rm i} g}{c\hbar} [{A}_{\mu},\ {A}_{\nu}], \;\;\;\;\; (\mu, \nu=0, 1, 2, 3),
      \end{equation}
with $g$ being the gauge coupling parameter, $c$ the speed of light in vacuum and $\hbar$ the Planck constant. The matrix form of field tensor $F_{\mu\nu}$ is given by \cite{1999Jackson}
   \begin{equation}
       F_{\mu\nu}=\begin{bmatrix}
            0 & {E}_x & {E}_y & {E}_z \\
            -{E}_x & 0 & -{B}_z & {B}_y \\
            -{E}_y & {B}_z & 0 & -{B}_x \\
            -{E}_z & -{B}_y & {B}_x & 0
      \end{bmatrix}.
     \end{equation}
      Based on which, we can have the vector forms of ``magnetic'' field and ``electric'' field as
      \begin{subequations}
            \begin{eqnarray}
                  \vec{{B}} &=& \vec{\nabla}\times\vec{{A}}
                        - \frac{{\rm i} g}{c\hbar}\left(\vec{{A}}\times\vec{{A}}\right),
                      \label{eq:mag-2}  \\
                  \vec{{E}} &=& -\dfrac{1}{c}
                              \dfrac{\partial\,\vec{{A}}}{\partial\,t}
                        -\vec{\nabla}\varphi- \frac{{\rm i} g}{c\hbar} \left[\varphi,\ \vec{{A}}
                        \right]. \label{eq:ele-2}
            \end{eqnarray}
      \end{subequations}
Accordingly, we can write down the YM equations in terms of $\{\vec{B}, \vec{E}, \vec{A}, \varphi\}$ as
      \begin{subequations}
            \begin{eqnarray}
                  && \vec{\nabla}\cdot\vec{E}\ {-}\boxed{{\rm i}\,\frac{g}{\hbar c}\Bigl(
                        \vec{A}\cdot\vec{E}
                        -\vec{E}\cdot\vec{A}\Bigr)}=0, \label{eq:DivEYM-a} \\
                  && -\dfrac{1}{c} \dfrac{\partial}{\partial\,t} \vec{B}
                        -\vec{\nabla}\times\vec{E} {-}\boxed{{\rm i}\,\frac{g}{\hbar c}\biggl(\Bigl[\varphi,\
                                    \vec{B}\Bigr]
                              -\vec{A}\times\vec{E}
                              -\vec{E}\times\vec{A}\biggr)}=0, \label{eq:CurlEYM} \\
                  && \vec{\nabla}\cdot\vec{B}\ {-}\boxed{{\rm i}\,\frac{g}{\hbar c}\Bigl(
                        \vec{A}\cdot\vec{B}
                        -\vec{B}\cdot\vec{A}\Bigr)}=0, \label{eq:DivBYM} \\
                  && -\dfrac{1}{c} \dfrac{\partial}{\partial\,t} \vec{E}
                        +\vec{\nabla}\times\vec{B}  {-}\boxed{{\rm i}\,\frac{g}{\hbar c}\biggl(\Bigl[\varphi,\
                                    \vec{E}\Bigr]
                              +\vec{A}\times\vec{B}
                              +\vec{B}\times\vec{A}\biggr)}=0. \label{eq:CurlBYM-a}
            \end{eqnarray}
      \end{subequations}
Here the YM equation (\ref{eq:YM1a}) yields Eq. (\ref{eq:DivEYM-a}) and Eq. (\ref{eq:CurlBYM-a}), the YM equation (\ref{eq:YM1b}) is just the Bianchi identity, which yields Eq. (\ref{eq:CurlEYM}) and Eq. (\ref{eq:DivBYM}).
If one neglects the ``boxed'' terms (or just let $g=0$), then the YM equations reduce to the usual forms of Maxwell's equations (in vacuum without sources). Eq. (\ref{eq:CurlEYM}) and Eq. (\ref{eq:DivBYM}) are originated from the Bianchi identity (\ref{eq:YM1b}), which is satisfied automatically. Therefore, to find the static solutions of the YM equations, we only need to focus on Eq. (\ref{eq:DivEYM-a}) and Eq. (\ref{eq:CurlBYM-a}), namely,
      \begin{subequations}
            \begin{eqnarray}
                  && \vec{\nabla}\cdot\vec{E}\ {-} {{\rm i}\,\frac{g}{\hbar c}\Bigl(
                        \vec{A}\cdot\vec{E}
                        -\vec{E}\cdot\vec{A}\Bigr)}=0, \label{eq:DivEYM-b} \\
                  && -\dfrac{1}{c} \dfrac{\partial}{\partial\,t} \vec{E}
                        +\vec{\nabla}\times\vec{B}  {-} {{\rm i}\,\frac{g}{\hbar c}\biggl(\Bigl[\varphi,\
                                    \vec{E}\Bigr]
                              +\vec{A}\times\vec{B}
                              +\vec{B}\times\vec{A}\biggr)}=0. \label{eq:CurlBYM-b}
            \end{eqnarray}
      \end{subequations}
In comparison to Maxwell's equations, one can call Eq. (\ref{eq:DivEYM-b}) and Eq. (\ref{eq:CurlBYM-b}) as the generalized Gauss's law and Amp\`ere's law, respectively.

\subsection{A Simple SU(2) Static Solution of the Yang-Mills Equations }

A simple SU(2) static solution of the YM equations has been presented in Ref. \cite{Zhou2025}. Here we make a brief review. For simplicity, let us denote a new parameter
\begin{eqnarray}
\tilde{g}=\frac{g}{c\hbar},
\end{eqnarray}
and consider further the potentials and the fields are static, then Eqs. (\ref{eq:DivEYM-b}) and (\ref{eq:CurlBYM-b}) can be simplified as
      \begin{subequations}
            \begin{eqnarray}
                  && \vec{\nabla}\cdot\vec{E}\ {-}{{\rm i}\,\tilde{g}\Bigl(
                        \vec{A}\cdot\vec{E}
                        -\vec{E}\cdot\vec{A}\Bigr)}=0, \label{eq:DivEYM-c} \\
                  && \vec{\nabla}\times\vec{B}  {-}{{\rm i}\,\tilde{g}\biggl(\Bigl[\varphi,\
                                    \vec{E}\Bigr]
                              +\vec{A}\times\vec{B}
                              +\vec{B}\times\vec{A}\biggr)}=0, \label{eq:CurlBYM-c}
            \end{eqnarray}
\end{subequations}
with the ``magnetic'' and ``electric'' fields being
\begin{eqnarray} \label{eq:magele}
\vec{{B}} &=& \vec{\nabla}\times\vec{{A}}
                        - {\rm i} \tilde{g} \left(\vec{{A}}\times\vec{{A}}\right), \nonumber\\
\vec{{E}} &=&
-\vec{\nabla}\varphi- {\rm i} \tilde{g} \left[\varphi,\ \vec{{A}}\right].
\end{eqnarray}
The simple SU(2) static solution is given by $\mathcal{S}_{\rm YM}= \{\vec{A}, \, \varphi\}$, where the gauge potentials are given by \cite{Zhou2025}
\begin{eqnarray}\label{eq:Aphi-1}
\vec{A}=\tilde{k}\frac{\vec{r}\times\vec{S}}{r^2}, \quad\;\;\;\;  \varphi=\frac{\kappa}{r},
\end{eqnarray}
or
\begin{eqnarray}
\vec{A}=k\frac{\hat{r}\times\vec{\Gamma}}{r}, \quad\;\;\;\;  \varphi=\frac{\kappa}{r},
\end{eqnarray}
under the following constraint condition
\begin{eqnarray}\label{eq:constr}
\tilde{g} \hbar \tilde{k}=1, \;\;\; {\rm or} \;\;\; 2.
 \end{eqnarray}
Here $\kappa$ and $\tilde{k}$ are real constant numbers, $\varphi$ is the Coulomb-type scalar potential, $\vec{r}$ is position operator with $r=|\vec{r}|$ and $\hat{r}=\vec{r}/r$,  $\vec{S}=(S_x, S_y, S_z)$ is the spin-$s$ angular momentum operator (with the spin value $s=0, 1/2, 1, ...$), which satisfies the following
commutation relations
\begin{eqnarray}\label{angularm1}
&& [S_x, S_y]= \mathrm{i} \hbar\; S_z, \;\;\; [S_y, S_z]= \mathrm{i} \hbar\; S_x,\;\;\;  [S_z, S_x]= \mathrm{i} \hbar\; S_y.
\end{eqnarray}
or in the vector form  as
$ \vec{S} \times \vec{S} = \mathrm{i} \hbar\; \vec{S}$.
The square of $\vec{S}$ satisfies
$ \vec{S}^{\,2} = s(s+1) \hbar^2\, \mathbb{I}$,
with $\mathbb{I}$ being the identity matrix. The operator $\Gamma$ is defined through
\begin{equation}
      \vec{S} = \frac{\hbar}{2} \, \vec{\Gamma},
\end{equation}
hence $\vec{\Gamma}$ is dimensionless, and the parameter $\tilde{k}=2k/{\hbar}$.
Specially, when $s=1/2$, one has $\vec{S}^2 =(3/4) \hbar^2\, \openone$, $\openone$ is the $2\times 2$ identity matrix, and the operator $\vec{\Gamma}$ can be realized by $\vec{\Gamma} = \vec{\sigma}$, with $\vec{\sigma}=(\sigma_x, \sigma_y, \sigma_z)$ being the vector of Pauli matrices.
In this work, we restrict our study to $\vec{\Gamma} = \vec{\sigma}$ and view $\vec{\sigma}$ as the generators of the SU(2) gauge group. In Supplemental Material (SM) \cite{SM}, one can verify  Eq. (\ref{eq:Aphi-1}) is indeed a simple solution of Eqs. (\ref{eq:DivEYM-c}) and (\ref{eq:CurlBYM-c}).

\begin{remark}
When $k=0$ (or $\tilde{k}=0$), the static solution
\begin{eqnarray}
\mathcal{S}_{\rm YM}= \left\{\vec{A}=\tilde{k}\frac{\vec{r}\times\vec{S}}{r^2}, \, \varphi=\frac{\kappa}{r}\right\}
\end{eqnarray}
reduces to the static solution of Maxwell's equations, i.e.,
\begin{eqnarray}
\mathcal{S}_{\rm Maxwell}= \left\{\vec{A}=0, \, \varphi=\frac{\kappa}{r}\right\}
\end{eqnarray}
which represents the Coulomb-type scalar potential.  $\blacksquare$
\end{remark}

\begin{remark}
After substituting the constraint condition (\ref{eq:constr}) into the ``electric'' and ``magnetic'' fields, we have
\begin{eqnarray}
\vec{E}=\dfrac{\kappa}{r^3} \vec{r}, \;\;\;\quad \vec{B} = -\frac{k (\vec{r} \cdot \vec{\Gamma}) }{r^4}  \vec{r},
\;\;\;\quad  {\rm for} \;\;\;\; \tilde{g} \hbar \tilde{k}=1,
\end{eqnarray}
and
\begin{eqnarray}
\vec{E}=\dfrac{\kappa}{r^3} \vec{r}, \;\;\;\quad \vec{B} = 0, \;\;\;\quad {\rm for} \;\;\;\; \tilde{g} \hbar \tilde{k}=2.
\end{eqnarray}
It can be noticed that for the constraint condition $ \tilde{g} \hbar \tilde{k}=2$, the ``magnetic'' field is zero.
$\blacksquare$
\end{remark}

\section{An Intuitive Approach to Extract Vector Potential}\label{sec-5}

The readers might be curious why the spin vector potential (\ref{eq:Aphi-1}) takes the following form
\begin{eqnarray}\label{eq:Aphi-2}
\vec{A}=\tilde{k}\frac{\vec{r}\times\vec{S}}{r^2},
\end{eqnarray}
 or where does the spin vector potential come from. Actually, such a simple form appeared in Ref. \cite{Chen2025} as a physical hypothesis by considering a particle with a spin operator $\vec{S}$. In \cite{Chen2025}, an intuitive approach to extract the vector potential from the orbital angular momentum operator has been proposed. For simplicity, we would like to call such an approach as the \emph{vector potential extraction approach} (VPEA). In this section, let us briefly review and advance study the VPEA, which is crucial to establish the general static solutions for the SU(2) YM theory.

\subsection{Brief Review of the VPEA for the Abelian Case}

Let us make a brief review of the \emph{vector potential extraction approach}. The starting point is the orbital angular momentum
\begin{eqnarray}
\vec{\ell}&=&(\ell_x, \ell_y, \ell_z )=\vec{r}\times\vec{p},
\end{eqnarray}
where $\vec{p}$ is just the linear momentum operator. Essentially, the VPEA is a kind of approach mapping the linear momentum $\vec{p}$ to the canonical momentum $\vec{\Pi}$, i.e.,
\begin{eqnarray}
\vec{p} \mapsto \vec{\Pi}= \vec{p}-\frac{q}{c} \vec{A}.
\end{eqnarray}
Now let us illustrate the procedure. Mathematically, $\vec{\ell}$ satisfies the commutation relations of angular momentum operator, i.e.,
\begin{eqnarray}\label{angularm1a}
&& [\ell_x, \ell_y]= \mathrm{i} \hbar\; \ell_z, \;\;\;\;\; [\ell_y, \ell_z]= \mathrm{i} \hbar\; \ell_x, \;\;\;\;\;
 [\ell_z, \ell_x]= \mathrm{i} \hbar\; \ell_y,
\end{eqnarray}
or in the vector form as
\begin{eqnarray}\label{angularm3}
&& \vec{\ell} \times \vec{\ell} = \mathrm{i} \hbar\; \vec{\ell}.
\end{eqnarray}
Now we perform a ``displacement'' for the orbital angular momentum $\vec{\ell}$ with an operator $\vec{G}$, i.e.,
\begin{eqnarray}\label{G-2m}
\vec{L}&=&\vec{\ell}+q \vec{G}=\vec{r}\times\vec{p}+q \vec{G},
\end{eqnarray}
where $q$ is a real number (Note: Alternatively one may also absorb the parameter $q$ to operator $G$, such that the displacement simply becomes $\vec{L}=\vec{\ell}+\vec{G}$). Furthermore, we require that the resultant operator $\vec{L}$ is \emph{still} an angular momentum operator, i.e., $\vec{L}$ satisfies
\begin{eqnarray}\label{M-4}
&& \vec{L} \times \vec{L} = \mathrm{i} \hbar\; \vec{L}.
\end{eqnarray}
We then recast Eq. (\ref{G-2m}) to the following equivalent form
\begin{eqnarray}\label{G-2n}
\vec{L}&=&\vec{r}\times\vec{p}+q \vec{G}=\vec{r}\times\vec{p}+q \left(G_r \hat{e}_r+G_\theta \hat{e}_\theta+G_\phi \hat{e}_\phi\right)\nonumber\\
&=&\vec{r}\times\vec{p}+q \left[G_r \hat{e}_r+G_\theta (-\hat{e}_r\times \hat{e}_\phi)+G_\phi (\hat{e}_r\times \hat{e}_\theta)\right]\nonumber\\
&=&\vec{r}\times\left[\vec{p}+\frac{q}{r}\left(-G_\theta \hat{e}_\phi+G_\phi  \hat{e}_\theta\right)\right]+q G_r \hat{e}_r,
\end{eqnarray}
and view the operator
\begin{eqnarray}\label{G-2p}
&& \vec{\Pi}:=\vec{p}+\frac{q}{r}\left(-G_\theta \hat{e}_\phi+G_\phi  \hat{e}_\theta\right)=\vec{p}-\frac{q}{c}\vec{A}
\end{eqnarray}
as the canonical momentum, here $\{\hat{e}_r, \hat{e}_\theta, \hat{e}_\phi\}$ is the spherical coordinate system and $\hat{e}_r=\vec{r}/r$. Thus, from Eq. (\ref{G-2p}) one can extract a vector potential $\vec{A}$ as
\begin{eqnarray}\label{G-2r}
 \vec{A}&=&\frac{c}{r} \left(G_\theta \hat{e}_\phi - G_\phi  \hat{e}_\theta\right)\nonumber\\
&=& \dfrac{c}{r}\left|\begin{array}{ccc}
                  \hat{e}_r & \hat{e}_\theta & \hat{e}_\phi \\
                  1 & 0 & 0 \\
                  G_r & G_\theta & G_\phi
            \end{array}\right| = c \dfrac{\hat{e}_r\times\vec{G}}{r} \nonumber\\
            &=& c \dfrac{\vec{r}\times\vec{G}}{r^2}.
\end{eqnarray}
For the Abelian case, the three components of $\vec{G}=(G_x, G_y, G_z)$ are commutative operators. Based on this approach, we can successfully derive four kinds of vector potentials: (i) the magnetic vector potential $\vec{A}_{\rm M}$, {$(\rho>r_0)$}, used in the magnetic AB effect; (ii) the Wu-Yang U(1) monopole vector potential ($\vec{A}_a$ and $\vec{A}_b$); (iii) the magnetic vector potential $\vec{A}$ corresponding to a uniform magnetic field along the $z$-axis (i.e., $\vec{B}=B \hat{e}_z$); (iv) the vector potential corresponding to a toroidal solenoid. One may refer the detailed derivations in SM \cite{SM}.

\subsection{Brief Review of the VPEA for the Non-Abelian Case}

For the non-Abelian case, the approach is similar. The difference is that now the components of $\vec{G}$ are non-abelian operators. For a simple example, let
\begin{eqnarray}
q\vec{G}=\vec{S},
\end{eqnarray}
where $\vec{S}$ is a spin-$s$ angular momentum operator $(s=0, 1/2, 1, 3/2, \cdots)$. In this situation, it is easy to have that the resultant vector
\begin{eqnarray}\label{eq:ls-1}
\vec{L}&=&\vec{r}\times\vec{p}+q \vec{G}=\vec{r}\times\vec{p}+\vec{S}
\end{eqnarray}
is also an angular momentum operator. Then due to the VPEA, we can derive the previous spin vector potential
\begin{eqnarray}\label{eq:svp-1}
 \vec{A}&=& c \dfrac{\vec{r}\times\vec{G}}{r^2}=\frac{c}{q} \dfrac{\vec{r}\times\vec{S}}{r^2}.
\end{eqnarray}
Then we explain the form of spin vector potential in Eq. (\ref{eq:Aphi-2}) with
\begin{eqnarray}
\tilde{k}=\frac{c}{q}.
\end{eqnarray}

\subsection{Determining the General Form of Spin Vector Potential Based on VPEA}

From previous section, we have known that the spin vector potential takes the following form
\begin{eqnarray}\label{eq:svp-2}
 \vec{A} & \propto&  \dfrac{\vec{r}\times\vec{S}}{r^2},
\end{eqnarray}
which comes from the angular momentum operator $\vec{L}= \vec{r}\times\vec{p}+\vec{S}$
by the VPEA. It gives rise to natural question: Is Eq. (\ref{eq:svp-2}) the most general form of the spin vector potential $\vec{A}$? Mathematically, from the viewpoint of VPEA, it is equivalent to ask: Given the orbital angular momentum operator $\vec{\ell}=\vec{r}\times \vec{p}$ and the spin angular momentum operator $\vec{S}$, and we define
\begin{eqnarray}
\vec{L}=F(\vec{\ell}, \vec{S})
\end{eqnarray}
as an operator function of $\vec{\ell}$ and $\vec{S}$, then what is the general form of $\vec{L}$, if it is still required to be an angular momentum operator? In the next subsection, let us study the general form of $\vec{L}=F(\vec{\ell}, \vec{S})$.

\subsubsection{The General Form of $\vec{L}=F(\vec{\ell}, \vec{S})$}

Frankly speaking, the expression  of (\ref{eq:ls-1}) is obtained by a simple ``displacement'' (i.e., $\vec{L} = \vec{\ell}+\vec{S}$), which is merely a simple realization of $\vec{L}=F(\vec{\ell}, \vec{S})$. Based on two facts: (i) $\vec{\ell}$ and $\vec{L}$ are angular momentum operator, and (ii) the operator $\vec{L}$ is generated from the operator $\vec{\ell}$, thus the other way to construct $\vec{L}$ is to introduce the unitary transformation $\mathcal{U}$ (i.e., a ``rotation''). Explicitly, one has
\begin{eqnarray}\label{eq:ls-2}
\vec{L}&=& \mathcal{U} \,\vec{\ell}\,  \mathcal{U}^\dagger.
\end{eqnarray}
Such a unitary transformation acting on $\vec{\ell}$ automatically guarantees that $\vec{L}$ is an angular momentum operator.

Since we expect that the spin vector potential depends only on the position operator $\vec{r}$ and the spin operator $\vec{S}$ (i.e., it does not depend on the momentum operator $\vec{p}$), thus the unitary matrix $\mathcal{U}$ is a function of $\vec{r}$ and $\vec{S}$. In this work, for simplicity, let us restrict to spin-1/2 case, then we may write
\begin{eqnarray}\label{eq:U-1}
\mathcal{U}= {\rm e}^{{\rm i} \theta \vec{\Gamma}\cdot \hat{r}},
\end{eqnarray}
with
\begin{eqnarray}
\vec{\Gamma}=\vec{\sigma}, \;\;\;\; \hat{r} = \frac{\vec{r}}{r}, \;\;\; (\vec{\Gamma} \cdot \hat{r})^2=\openone.
\end{eqnarray}
Due to
\begin{eqnarray}
&& \mathcal{U}= {\rm e}^{{\rm i} \theta \vec{\Gamma}\cdot \hat{r}} = \cos\theta \openone+ {\rm i}  \sin\theta (\vec{\Gamma} \cdot \hat{r}), \nonumber\\
&& \mathcal{U}^\dagger= {\rm e}^{-{\rm i} \theta \vec{\Gamma}\cdot \hat{r}}= \cos \theta \openone- {\rm i} \sin \theta (\vec{\Gamma} \cdot \hat{r}),
\end{eqnarray}
we can have
\begin{eqnarray}
 \vec{L}&=&(\cos\theta \openone+ {\rm i} \vec{\Gamma} \cdot \hat{r} \sin\theta) \, \vec{\ell}\, (\cos\theta \openone - {\rm i} \vec{\Gamma} \cdot \hat{r}\sin\theta)\nonumber\\
&=& \vec{\ell} \cos^2\theta + (\vec{\Gamma} \cdot \hat{r}) \,\vec{\ell} \, (\vec{\Gamma} \cdot \hat{r}) \sin^2\theta + {\rm i} [ \vec{\Gamma} \cdot\hat{r}, \; \vec{\ell} ] \sin\theta \cos\theta.
\end{eqnarray}
Because
\begin{eqnarray}
 [ \vec{\ell}, \; \vec{\Gamma} \cdot \hat{r} ]= -{\rm i} \hbar ( \hat{r}\times \vec{\Gamma}),
\end{eqnarray}
\begin{eqnarray}
 (\vec{\Gamma} \cdot \hat{r})\; \vec{\ell}\; (\vec{\Gamma} \cdot \hat{r})
  &=& \vec{\ell} + \hbar \left[\vec{\Gamma} - (\vec{\Gamma}\cdot \hat{r}) \hat{r}\right],
\end{eqnarray}
we have
\begin{eqnarray}\label{eq:L-1}
\vec{L}
&=& \vec{\ell}- \hbar (  \hat{r}\times\vec{\Gamma})\sin\theta \cos\theta+ \hbar\left[\vec{\Gamma}-(\vec{\Gamma} \cdot \hat{r})\hat{r}\right]\sin^2\theta,
\end{eqnarray}
or in terms of $\vec{S}$ and $\vec{r}$ as
\begin{eqnarray}\label{eq:disp-1}
\vec{L}&=& \vec{\ell}- 2 \sin\theta \cos\theta\frac{\vec{r}\times\vec{S}}{r} + 2 \sin^2\theta \left[\vec{S}-\frac{(\vec{S} \cdot \vec{r})\vec{r}}{r^2}\right].
\end{eqnarray}

\begin{remark} From the viewpoint of displacement, Eq. (\ref{eq:disp-1}) can be written as
\begin{eqnarray}\label{eq:disp-2}
\vec{L}&=& \vec{\ell} +q \vec{G},
\end{eqnarray}
with
\begin{eqnarray}\label{eq:disp-3}
q \vec{G}&=&  - 2 \sin\theta \cos\theta\frac{\vec{r}\times\vec{S}}{r} + 2 \sin^2\theta \left[\vec{S}-\frac{(\vec{S} \cdot \vec{r})\vec{r}}{r^2}\right].
\end{eqnarray}

(i) When $\theta=\pi/4$, one has $2 \sin\theta \cos\theta=1$, $2 \sin^2\theta=1$, and
\begin{eqnarray}
\vec{L}&=& \vec{\ell}- 2 \sin\theta \cos\theta\frac{\vec{r}\times\vec{S}}{r} + 2 \sin^2\theta \left[\vec{S}-\frac{(\vec{S} \cdot \vec{r})\vec{r}}{r^2}\right]\nonumber\\
&=& \vec{\ell} + \vec{S}-  \left[\frac{\vec{r}\times\vec{S}}{r}+\frac{(\vec{S} \cdot \vec{r})\vec{r}}{r^2}\right],
\end{eqnarray}
which is different from $\vec{L}= \vec{\ell} + \vec{S}$ by an additional term $-  \left[\frac{\vec{r}\times\vec{S}}{r}+\frac{(\vec{S} \cdot \vec{r})\vec{r}}{r^2}\right]$. This also means that the ``rotation'' $\mathcal{U} \,\vec{\ell}\,  \mathcal{U}^\dagger$ cannot generate the ``displacement'' $\vec{L}= \vec{\ell} + \vec{S}$.

(ii) When $\theta=\pi/2$, one has $2 \sin\theta \cos\theta=0$, $2 \sin^2\theta=2$, and
\begin{eqnarray}
\vec{L}&=& \vec{\ell}- 2 \sin\theta \cos\theta\frac{\vec{r}\times\vec{S}}{r} + 2 \sin^2\theta \left[\vec{S}-\frac{(\vec{S} \cdot \vec{r})\vec{r}}{r^2}\right]\nonumber\\
&=& \vec{\ell}+2\vec{S}-2\frac{(\vec{S} \cdot \vec{r})\vec{r}}{r^2}=\vec{\ell}+2\vec{S}-2(\vec{S} \cdot \hat{r})\hat{r},
\end{eqnarray}
which is also an angular momentum operator, but is different from $\vec{L}= \vec{\ell} + \vec{S}$. $\blacksquare$
\end{remark}

\begin{remark}Based on (i) the simple ``displacement'' $\vec{L}= \vec{\ell} + \vec{S}$, and (ii) the result (\ref{eq:disp-1}) obtained from the ``rotation'', we can find that the most general displacement is of the following form
\begin{eqnarray}\label{eq:disp-4}
\vec{L}&=& \vec{\ell}+ a_1 \vec{S} + a_2 (\vec{S} \cdot \hat{r})\hat{r} + a_3 (\hat{r}\times\vec{S}),
\end{eqnarray}
i.e., $\vec{L} = \vec{\ell} +q \vec{G}$
with
\begin{eqnarray}\label{eq:disp-5}
q \vec{G}&=&  a_1 \vec{S} + a_2 (\vec{S} \cdot \hat{r})\hat{r} + a_3 (\hat{r}\times\vec{S}),
\end{eqnarray}
namely, the operator $\vec{G}$ is generally a linear combination of $\vec{S}$, $(\vec{S} \cdot \hat{r})\hat{r}$, and $(\hat{r}\times\vec{S})$, here $a_1$, $a_2$, $a_3$ are {some coefficients}, which will be determined by the following definition of angular momentum operator
\begin{eqnarray}
\vec{L} \times \vec{L}={\rm i} \hbar \vec{L}.
\end{eqnarray}
For the symmetrical reason, it is sufficient to check its $z$-component. In SM \cite{SM}, we have calculated $(\vec{L} \times \vec{L})_z=[L_x, L_y]$ as
\begin{eqnarray}[L_x, L_y]
&=& {\rm i}\hbar \left\{ \ell_z + (a_1^2 + a_1 a_2 - a_2) S_z + (3a_2 + a_3^2 - a_1 a_2) (\vec{S} \cdot \hat{r})\frac{z}{r} + (a_3 + a_1 a_3 + a_2 a_3) (\hat{r}\times\vec{S})_z \right\}.
\end{eqnarray}
What we need is $[L_x, L_y] = {\rm i}\hbar L_z$, i.e.,
\begin{eqnarray}
[L_x, L_y] &=& {\rm i}\hbar \left[ \ell_z + a_1 S_z + a_2 (\vec{S} \cdot \hat{r})\frac{z}{r} + a_3 (\hat{r}\times\vec{S})_z \right].
\end{eqnarray}
After comparing the coefficients of the corresponding terms, we obtain the following equations for $a_1, a_2, a_3$:
\begin{eqnarray}\label{eq:a123}
a_1^2 + a_1 a_2 - a_2 &=& a_1, \nonumber\\
3a_2 + a_3^2 - a_1 a_2 &=& a_2, \nonumber\\
a_3 + a_1 a_3 + a_2 a_3 &=& a_3.
\end{eqnarray}

\emph{Analysis 1.---}From the third of Eq. (\ref{eq:a123}), we have
\begin{eqnarray}
a_3(a_1 + a_2) = 0.
\end{eqnarray}
If $a_1 + a_2\neq0$, we must have $a_3 = 0$, the remaining equations give $a_1 = 1$ and $a_2 = 0$, which corresponds to the simple ``displacement'' result $\vec{L} = \vec{\ell} + \vec{S}$. If $a_1 +a_2=0$, the  equations yield $a_2=-a_1$, and
\begin{eqnarray}
-2a_1 + a_3^2 + a_1^2 &=& 0,
\end{eqnarray}
i.e.,
\begin{eqnarray}\label{eq:a2a3}
(a_1 - 1)^2 + a_3^2 = 1.
\end{eqnarray}
If the coefficients $a_2$ and $a_3$ are real numbers, then Eq. (\ref{eq:a2a3})  describes a continuous family of solutions parameterized by $\theta$. One can parameterize them as
\begin{eqnarray}
&& a_1 - 1 = -\cos 2\theta, \nonumber\\
&& a_3 = -\sin 2\theta.
\end{eqnarray}
We find that
\begin{eqnarray}
a_1 &=& 2\sin^2\theta, \nonumber\\
a_2 &=& -2\sin^2\theta, \nonumber\\
a_3 &=& - 2\sin\theta\cos\theta,
\end{eqnarray}
which just comes from the ``rotation'' solution
\begin{eqnarray}
\vec{L}&=& \vec{\ell}- 2 \sin\theta \cos\theta\frac{\vec{r}\times\vec{S}}{r} + 2 \sin^2\theta \left[\vec{S}-\frac{(\vec{S} \cdot \vec{r})\vec{r}}{r^2}\right].
\end{eqnarray}

\emph{Analysis 2.---} Note that here the derivation only depends on the commutation relations of the angular momentum operator (i.e., $\vec{S} \times \vec{S} ={\rm i} \hbar \vec{S}$), without using the special spin-1/2 condition (i.e., $(\vec{\Gamma} \cdot \hat{r})^2=\openone$), thus this result is also valid for any spin-$s$ angular momentum operator.
In summary, for arbitrary spin-$s$ angular momentum, the solution of the function $\vec{L}=F(\vec{\ell}, \vec{S})$ is given by
\begin{equation}
    \vec{L}=F(\vec{\ell}, \vec{S}) = \begin{cases}
        \vec{\ell}+\vec{S}, \\
        \vec{\ell}- 2 \sin\theta \cos\theta\dfrac{\vec{r}\times\vec{S}}{r} + 2 \sin^2\theta \left[\vec{S}-\dfrac{(\vec{S} \cdot \vec{r})\vec{r}}{r^2}\right].
    \end{cases}
\end{equation}
However, for simplicity, in this work, we restrict our study on the spin-1/2 case.

\end{remark}

\subsubsection{Extracting General Spin Vector Potential from $\vec{L}=F(\vec{\ell}, \vec{S})$ }

From previous section, we have known that the general solutions of the angular momentum operator $\vec{L}=F(\vec{\ell}, \vec{S})$ are given by
\begin{equation}
    \vec{L}=F(\vec{\ell}, \vec{S}) =  \vec{\ell} + a_1 \vec{S} + a_2 (\vec{S} \cdot \hat{r})\hat{r} + a_3 (\hat{r}\times\vec{S}),
\end{equation}
with \emph{solution 1} as $\{a_1=1, a_2=0, a_3=0\}$ and \emph{solution 2} as $\{a_2=-a_1, (a_1 - 1)^2 + a_3^2 = 1\}$. Due to
$\vec{L}= \vec{\ell} +q \vec{G}$ and
\begin{eqnarray}\label{eq:disp-5a}
q \vec{G}&=&  a_1 \vec{S} + a_2 (\vec{S} \cdot \hat{r})\hat{r} + a_3 (\hat{r}\times\vec{S}).
\end{eqnarray}
Then based on the VPEA, we can extract the spin vector potential as
\begin{eqnarray}
 \vec{A}&=& c \dfrac{\vec{r}\times\vec{G}}{r^2}=\frac{c}{q} \dfrac{\hat{r}\times  \left[a_1 \vec{S} + a_2 (\vec{S} \cdot \hat{r})\hat{r} + a_3 (\hat{r}\times\vec{S})\right]}{r}\nonumber\\
&=&\frac{c}{q} \dfrac{\hat{r}\times  \left[a_1 \vec{S} +  a_3 (\hat{r}\times\vec{S})\right]}{r}= a_1 \frac{c}{q} \dfrac{\hat{r}\times \vec{S}}{r}+
a_3\frac{c}{q} \dfrac{\hat{r}\times (\hat{r}\times\vec{S})}{r}\nonumber\\
&=&  a_1 \frac{c}{q} \dfrac{\hat{r}\times \vec{S}}{r}+
a_3\frac{c}{q} \dfrac{(\vec{S} \cdot \hat{r})\hat{r} - \vec{S}}{r}\nonumber\\
&=& \frac{1}{r} \frac{c}{q}  \left[a_1 (\hat{r}\times \vec{S})-a_3 \vec{S} +
a_3 (\vec{S} \cdot \hat{r})\hat{r}\right].
\end{eqnarray}
Specially, for $a_1=1$ and $a_3=0$, we can recover the previous spin vector potential in Eq. (\ref{eq:svp-1}).

\begin{remark}
In the spherical coordinate system, the spin vector potential can be written as
\begin{eqnarray}
\vec{A} &=& A_r \hat{e}_r+   A_\theta  \hat{e}_\theta+  A_\phi \hat{e}_\phi.
\end{eqnarray}
One can check that $A_r=0$. Explicitly, one has
\begin{eqnarray}
 A_r &=& \vec{A}\cdot \hat{e}_r = \frac{1}{r} \frac{c}{q}  \left[a_1 (\hat{r}\times \vec{S})-a_3 \vec{S} +
a_3 (\vec{S} \cdot \hat{r})\hat{r}\right]\cdot \hat{e}_r \nonumber\\
&=& \frac{1}{r} \frac{c}{q}  \left[a_1 (\hat{r}\times \vec{S})-a_3 \vec{S} +
a_3 (\vec{S} \cdot \hat{r})\hat{r}\right]\cdot \hat{r} =
\frac{1}{r} \frac{c}{q}  \left[-a_3 \vec{S} +
a_3 (\vec{S} \cdot \hat{r})\hat{r}\right]\cdot \hat{r}\nonumber\\
& =&
\frac{1}{r} \frac{c}{q} a_3 \left[-\vec{S}\cdot \hat{r} +
 (\vec{S} \cdot \hat{r})\right]=0.
\end{eqnarray}
$\blacksquare$
\end{remark}

\section{General Static Solutions of the SU(2) Yang-Mills Theory}\label{sec-6}

\subsection{The Static Ansatz}

In previous section, we have determined the general angular momentum operator $\vec{L}=F(\vec{\ell}, \vec{S})$ as
\begin{equation}
    \vec{L}=F(\vec{\ell}, \vec{S}) =  \vec{\ell} + a_1 \vec{S} + a_2 (\vec{S} \cdot \hat{r})\hat{r} + a_3 (\hat{r}\times\vec{S}),
\end{equation}
where \emph{solution 1} is $\{a_1=1, a_2=0, a_3=0\}$, which comes from the simple ``displacement'', and \emph{solution 2} is $\{a_2=-a_1, (a_1 - 1)^2 + a_3^2 = 1\}$, which comes from the ``rotation''. Accordingly, based on the VPEA, one can extract the spin vector potential as
\begin{eqnarray}
 \vec{A}&=& \frac{1}{r} \frac{c}{q}  \left[a_1 (\hat{r}\times \vec{S})-a_3 \vec{S} +
a_3 (\vec{S} \cdot \hat{r})\hat{r}\right].
\end{eqnarray}
Note that the term $a_2 (\vec{S} \cdot \hat{r})\hat{r}$ in operator $\vec{L}$ has no contribution to the vector potential $\vec{A}$, thus $\{a_1=1, a_3=0\}$ corresponds to the vector potential of \emph{solution 1}, and $\{a_2=-a_1, (a_1 - 1)^2 + a_3^2 = 1\}$ corresponds to vector potential of \emph{the solution 2}.

One may observe that the spin vector potentials of \emph{solution 1} and \emph{solution 2} share the following form
\begin{eqnarray}
\vec{A} = \frac{1}{r} \left[ k_1 (\hat{r} \times \vec{\Gamma}) + k_2 \vec{\Gamma} + k_3 (\vec{\Gamma} \cdot \hat{r})\hat{r} \right].
\end{eqnarray}
This is why we have chosen the general static ansatz as in Eq.~(\ref{eq:ansatz}). Moreover, the scalar potential takes the form $\varphi = f_1(r)\,(\vec{\Gamma}\cdot\hat{r}) + f_2(r)$ because $(\vec{\Gamma} \cdot \hat{r})^2 = \openone$, which prevents the appearance of higher-order terms $(\vec{\Gamma} \cdot \hat{r})^n$ with $n \ge 2$.
Based on the static ansatz, we shall solve the general static solutions for the SU(2) YM equations.

\subsection{The Constraint Equations Obtained from the Yang-Mills Equations}

Based on the static ansatz, we come to exactly solve the SU(2) YM equations. Our task is to solve the generalized Gauss's law and Amp\`ere's law as follows
      \begin{subequations}
            \begin{eqnarray}
                  && \vec{\nabla}\cdot\vec{E}\ {-} {{\rm i}\,\tilde{g}\Bigl(
                        \vec{A}\cdot\vec{E}
                        -\vec{E}\cdot\vec{A}\Bigr)}=0, \label{eq:DivEYM} \\
                  && \vec{\nabla}\times\vec{B}  {-} {{\rm i}\,\tilde{g}\biggl(\Bigl[\varphi,\
                                    \vec{E}\Bigr]
                              +\vec{A}\times\vec{B}
                              +\vec{B}\times\vec{A}\biggr)}=0, \label{eq:CurlBYM}
            \end{eqnarray}
      \end{subequations}
with the following quantities
\begin{eqnarray}
\vec{A} = \frac{1}{r} \left[ k_1 (\hat{r} \times \vec{\Gamma}) + k_2 \vec{\Gamma} + k_3 \hat{r} (\vec{\Gamma} \cdot \hat{r}) \right],
 \end{eqnarray}
\begin{eqnarray}
\varphi=f_1(r)(\vec{\Gamma}\cdot\hat{r})+f_2(r),
 \end{eqnarray}
\begin{eqnarray}\label{eq:B-2}
\vec{B}=\frac{1}{r^2} \left\{ \left[ 2k_1 \left( \tilde{g} k_1 - 1 \right) - 2\tilde{g} k_2 k_3 \right] \left( \vec{\Gamma} \cdot \hat{r} \right) \hat{r}+ \left( k_2 + k_3 \right) \left( 2\tilde{g} k_1 - 1 \right) \left( \hat{r} \times \vec{\Gamma} \right) + 2\tilde{g} k_2 \left( k_2 + k_3 \right) \vec{\Gamma} \right\},
\end{eqnarray}
\begin{eqnarray}
\vec{E}=-  \frac{\partial f_2(r)}{\partial r} \hat{r}- \frac{f_1(r)}{r} \left( 1 - 2\tilde{g} k_1 \right) \vec{\Gamma} - \hat{r} (\vec{\Gamma} \cdot \hat{r}) \left[ \frac{\partial f_1(r)}{\partial r} - \frac{f_1(r)}{r} \left( 1 - 2\tilde{g} k_1 \right) \right] - 2\tilde{g} k_2 \frac{f_1(r)}{r}\left(\hat{r} \times \vec{\Gamma}\right).
\end{eqnarray}
Here $\vec{B}$ and $\vec{E}$ can be calculated directly based on the static ansatz. For the detailed calculations, one may refer to SM \cite{SM}.

\subsubsection{Checking the Generalized Amp\`ere's Law}

Let us first check (\ref{eq:CurlBYM}). For the vector identity, due to symmetry, we only need to verify the $z$-component. The detailed calculations can be found in SM \cite{SM}. We can have the individual terms as follows:

\begin{eqnarray}\label{eq:AB}
&& \left(\vec{A}\times\vec{B}+\vec{B}\times\vec{A}\right)_z=
\left[ A_x, B_y \right] - \left[ A_y, B_x \right] \nonumber\\
&=& \frac{2{\rm i}}{r^3} \left\{ \left[ k_1 \left( 2k_1 \left( \tilde{g} k_1 - 1 \right) - 2\tilde{g} k_2 k_3 \right) + \left( k_2 + k_3 \right)^2 \left( 2\tilde{g} k_1 - 1 \right) + 2\tilde{g} k_1 k_2 \left( k_2 + k_3 \right) \right] \left( \hat{r}_x \Gamma_y - \hat{r}_y \Gamma_x \right) \right. \nonumber\\
&& + \left[ k_2 \left( 2k_1 \left( \tilde{g} k_1 - 1 \right) - 2\tilde{g} k_2 k_3 \right) + 2\tilde{g} k_2 \left( k_2 + k_3 \right) \left( 2k_2 + k_3 \right) \right] \Gamma_z \nonumber\\
&& \left. + \left[ -k_2 \left( 2k_1 \left( \tilde{g} k_1 - 1 \right) - 2\tilde{g} k_2 k_3 \right) + 2k_1 \left( k_2 + k_3 \right) \left( 2\tilde{g} k_1 - 1 \right) - 2\tilde{g} k_2 k_3 \left( k_2 + k_3 \right) \right] \hat{r}_z \left( \vec{\Gamma} \cdot \hat{r} \right) \right\}.
\end{eqnarray}

\begin{eqnarray}
\partial_x B_y - \partial_y B_x
&=& \frac{k_2+k_3}{r^3} \left( 2\tilde{g} k_1 - 1 \right) \left[ \Gamma_z - 3\hat{r}_z \left( \vec{\Gamma} \cdot \hat{r} \right) \right] - \frac{2}{r^3} \left[ \tilde{g} \left( {k_1^2 + 2k_2^2 + k_2 k_3} \right) - k_1 \right] \left( \hat{r}_x \Gamma_y - \hat{r}_y \Gamma_x \right).
\end{eqnarray}

\begin{eqnarray}
[\varphi, E_z] &=& 2{\rm i}f_1(r) \biggl\{\frac{f_1(r)}{r} \left( 1 - 2\tilde{g} k_1 \right)(\hat{r} \times \vec{\Gamma})_z+2\tilde{g} k_2 \frac{f_1(r)}{r} [ \hat{r}_z (\vec{\Gamma}\cdot\hat{r}) - \Gamma_z ] \biggr\}.
\end{eqnarray}
After substituting them into the $z$-component of Eq. (\ref{eq:CurlBYM}), i.e.,
\begin{eqnarray}
 \partial_x B_y - \partial_y B_x  {-}{{\rm i}\,\tilde{g}\biggl(\Bigl[\varphi,E_z\Bigr]+[A_x, B_y] - [A_y, B_x]\biggr)}=0,
\end{eqnarray}
we can get three constraint equations from the coefficients matching of $(\hat{r} \times \vec{\Gamma})_z $, $\Gamma_z$ and $\hat{r}_z (\vec{\Gamma}\cdot\hat{r})$, respectively. Explicitly, we have

(a) The coefficient of $(\hat{r} \times \vec{\Gamma})_z$:
\begin{eqnarray}
&& -\frac{2}{r^3} \left( \tilde{g} \left( {k_1^2 + 2k_2^2 + k_2 k_3} \right) - k_1 \right) + \frac{2\tilde{g} f_1^2 \left( r \right)}{r} \left( 1 - 2\tilde{g} k_1 \right) \nonumber \\
&& + \frac{2\tilde{g}}{r^3} \left[ k_1 \left( 2k_1 \left( \tilde{g} k_1 - 1 \right) - 2\tilde{g} k_2 k_3 \right) + \left( k_2 + k_3 \right)^2 \left( 2\tilde{g} k_1 - 1 \right) + 2\tilde{g} k_1 k_2 \left( k_2 + k_3 \right) \right] = 0.
\end{eqnarray}

(b) The coefficient of $\Gamma_z$:
\begin{eqnarray}
&& \frac{k_2 + k_3}{r^3} \left( 2\tilde{g} k_1 - 1 \right) - \frac{4\tilde{g}^2 k_2 f_1^2 \left( r \right)}{r} \nonumber \\
&& + \frac{2\tilde{g}}{r^3} \left[ k_2 \left( 2k_1 \left( \tilde{g} k_1 - 1 \right) - 2\tilde{g} k_2 k_3 \right) + 2\tilde{g} k_2 \left( k_2 + k_3 \right) \left( 2k_2 + k_3 \right) \right] = 0.
\end{eqnarray}

(c) The coefficient of $\hat{r}_z (\vec{\Gamma} \cdot \hat{r})$:
\begin{eqnarray}
&& -\frac{3 \left( k_2 + k_3 \right)}{r^3} \left( 2\tilde{g} k_1 - 1 \right) + \frac{4\tilde{g}^2 k_2 f_1^2 \left( r \right)}{r} \nonumber \\
&& + \frac{2\tilde{g}}{r^3} \left[ -k_2 \left( 2k_1 \left( \tilde{g} k_1 - 1 \right) - 2\tilde{g} k_2 k_3 \right) + 2k_1 \left( k_2 + k_3 \right) \left( 2\tilde{g} k_1 - 1 \right) - 2\tilde{g} k_2 k_3 \left( k_2 + k_3 \right) \right] = 0.
\end{eqnarray}

\subsubsection{Checking the Generalized Gauss's Law}

We now verify Eq. (\ref{eq:DivEYM}). For convenience, we have denote $f_j'(r)=\dfrac{\partial f_j(r)}{\partial r}$, $(j=1, 2)$. We can have \cite{SM}
\begin{eqnarray}
\nabla \cdot \vec{E}
 & = & - \left[ f_2''\left(r\right) + \frac{2}{r} f_2'\left(r\right) \right] - \left(\vec{\Gamma} \cdot \hat{r}\right) \left[ f_1''\left(r\right) + \frac{2}{r} f_1'\left(r\right) - \frac{2 f_1\left(r\right)}{r^2} \left( 1 - 2\tilde{g} k_1 \right) \right].
\end{eqnarray}

\begin{eqnarray}
[A_x, E_x] +[A_y, E_y]+[A_z, E_z]=-4i\frac{1}{r^2}\left(k_1-2\tilde{g} k_1^2-2\tilde{g} k_2^2\right)f_1(r)(\vec{\Gamma} \cdot \hat{r}).
\end{eqnarray}
From Eq. (\ref{eq:DivEYM}) we have
\begin{eqnarray}
 \vec{\nabla}\cdot\vec{E}\ {-}{{\rm i}\,\tilde{g}\Bigl([A_x, E_x] +[A_y, E_y]+[A_z, E_z]\Bigr)}=0,
\end{eqnarray}
we can get two constraint equations from matching coefficients. Explicitly, we have

(a) The coefficient of constant term:
\begin{eqnarray}
 f_2''(r) + \frac{2}{r} f_2'(r)=0
\end{eqnarray}

(b) The coefficient of $\vec{\Gamma} \cdot \hat{r}$:
\begin{eqnarray}
4\frac{1}{r^2}\tilde{g}\left(k_1-2\tilde{g} k_1^2-2\tilde{g} k_2^2\right)f_1(r) +  \left[f_1''(r) + \frac{2}{r} f_1'(r) - \frac{2 f_1(r)}{r^2} \left(1 - 2\tilde{g} k_1\right)\right]=0
\end{eqnarray}
In fact, the \(k_3\) term does not affect the generalized Gauss's law.

In summary, we now obtain five constraint equations for two functions $f_1(r)$, $f_2(r)$, the parameters $\tilde{g}$ and $k_j$'s $(j=1, 2, 3)$,  which are as follows
\begin{eqnarray}\label{eq:f2}
 f_2''(r) + \frac{2}{r} f_2'(r)=0,
\end{eqnarray}
\begin{eqnarray}\label{eq:e1}
4\frac{\tilde{g}}{r^2} \left( k_1 - 2\tilde{g} k_1^2 - 2\tilde{g} k_2^2 \right) f_1(r) + f_1''(r) + \frac{2}{r} f_1'(r) - \frac{2f_1(r)}{r^2} \left( 1 - 2\tilde{g} k_1 \right) = 0,
\end{eqnarray}
\begin{eqnarray}\label{eq:e2}
& & -\frac{2}{r^3} \left[ \tilde{g} \left( k_1^2 + 2k_2^2 + k_2 k_3 \right) - k_1 \right] + \frac{2\tilde{g} f_1^2 \left( r \right)}{r} \left( 1 - 2\tilde{g} k_1 \right) \nonumber \\
& & + \frac{2\tilde{g}}{r^3} \left[ k_1 \left( 2k_1 \left( \tilde{g} k_1 - 1 \right) - 2\tilde{g} k_2 k_3 \right) + \left( k_2 + k_3 \right)^2 \left( 2\tilde{g} k_1 - 1 \right) + 2\tilde{g} k_1 k_2 \left( k_2 + k_3 \right) \right] = 0,
\end{eqnarray}
\begin{eqnarray}\label{eq:e3}
& & \frac{k_2 + k_3}{r^3} \left( 2\tilde{g} k_1 - 1 \right) - \frac{4\tilde{g}^2 k_2 f_1^2 \left( r \right)}{r} \nonumber \\
& & + \frac{2\tilde{g}}{r^3} \left[ k_2 \left( 2k_1 \left( \tilde{g} k_1 - 1 \right) - 2\tilde{g} k_2 k_3 \right) + 2\tilde{g} k_2 \left( k_2 + k_3 \right) \left( 2k_2 + k_3 \right) \right] = 0,
\end{eqnarray}
\begin{eqnarray}\label{eq:e4}
& & -\frac{3 \left( k_2 + k_3 \right)}{r^3} \left( 2\tilde{g} k_1 - 1 \right) + \frac{4\tilde{g}^2 k_2 f_1^2 \left( r \right)}{r} \nonumber \\
& & + \frac{2\tilde{g}}{r^3} \left[ -k_2 \left( 2k_1 \left( \tilde{g} k_1 - 1 \right) - 2\tilde{g} k_2 k_3 \right) + 2k_1 \left( k_2 + k_3 \right) \left( 2\tilde{g} k_1 - 1 \right) - 2\tilde{g} k_2 k_3 \left( k_2 + k_3 \right) \right] = 0.
\end{eqnarray}

Note that Eq. (\ref{eq:f2}) only involves the function $f_2(r)$, whose solution is
\begin{eqnarray}
f_2(r)=\frac{C_1}{r} + C_0,
\end{eqnarray}
where $C_0$ is trivial, one may set $C_0=0$. Eqs. (\ref{eq:e1})-(\ref{eq:e4}) involves the function $f_1(r)$ and the parameters $\tilde{g}$ and $k_j$'s.

\subsection{Solving the Constraint Equations}

We now divide the study into four cases: (1) $g=0$ and $k_1=0$;  (2) $g=0$ and $k_1\neq 0$; (3) $g\neq 0$ and $k_1=0$; (4) $g\neq 0$ and $k_1\neq 0$. And we shall solve them case by case. For simplicity, here we focus on the real static solutions, such that $\vec{A}$ and $\varphi$ are hermitian. For the complex static solutions, one may find in SM \cite{SM}.

\subsubsection{The Case of $\{g=0, k_1=0\}$}

In this case, from Eqs. (\ref{eq:e1})-(\ref{eq:e4}) we have
\begin{eqnarray}\label{eq:e1-A}
 f_1''(r) + \frac{2}{r} f_1'(r) - \frac{2f_1(r)}{r^2}= 0,
\end{eqnarray}
\begin{eqnarray}
 k_2+k_3=0.
\end{eqnarray}
Then we have
\begin{eqnarray}
 f_1(r)=\frac{C_2}{r^2}+C_3 r,
\end{eqnarray}
with $k_2=-k_3$. Here $k_2$, $k_3$, $C_2$ and $C_3$ are real numbers.

\subsubsection{The Case of $\{g=0, k_1\neq 0\}$}

In this case, from Eqs. (\ref{eq:e1})-(\ref{eq:e4}) we have
\begin{eqnarray}
 f_1''(r) + \frac{2}{r} f_1'(r) - \frac{2f_1(r)}{r^2}= 0,
\end{eqnarray}
\begin{eqnarray}\label{eq:e2-AA}
& & \frac{2}{r^3} k_1=0,
\end{eqnarray}
\begin{eqnarray}
 k_2+k_3=0.
\end{eqnarray}
However, since $k_1\neq 0$, then Eq. (\ref{eq:e2-AA}) cannot be valid. Therefore, no real solutions for this case.

\subsubsection{The Case of $\{g\neq 0, k_1=0\}$}

In this case, from Eqs. (\ref{eq:e1})-(\ref{eq:e4}) we have
\begin{eqnarray}\label{eq:e1-a}
4\frac{\tilde{g}}{r^2} \left( - 2\tilde{g} k_2^2 \right) f_1(r) + f_1''(r) + \frac{2}{r} f_1'(r) - \frac{2f_1(r)}{r^2} = 0,
\end{eqnarray}

\begin{eqnarray}\label{eq:e2-a}
& & -\frac{2}{r^3} \left[ \tilde{g} \left(2k_2^2 + k_2 k_3 \right) \right] + \frac{2\tilde{g} f_1^2 \left( r \right)}{r} + \frac{2\tilde{g}}{r^3} \left[- \left( k_2 + k_3 \right)^2 \right] = 0,
\end{eqnarray}

\begin{eqnarray}\label{eq:e3-a}
& & -\frac{k_2 + k_3}{r^3} - \frac{4\tilde{g}^2 k_2 f_1^2 \left( r \right)}{r}  + \frac{2\tilde{g}}{r^3} \left[ k_2 \left(- 2\tilde{g} k_2 k_3 \right) + 2\tilde{g} k_2 \left( k_2 + k_3 \right) \left( 2k_2 + k_3 \right) \right] = 0,
\end{eqnarray}

\begin{eqnarray}\label{eq:e4-a}
& & \frac{3 \left( k_2 + k_3 \right)}{r^3} + \frac{4\tilde{g}^2 k_2 f_1^2 \left( r \right)}{r} + \frac{2\tilde{g}}{r^3} \left[ -k_2 \left( - 2\tilde{g} k_2 k_3 \right)  - 2\tilde{g} k_2 k_3 \left( k_2 + k_3 \right) \right] = 0.
\end{eqnarray}
By adding Eq. (\ref{eq:e3-a}) and Eq. (\ref{eq:e4-a}), one has
\begin{eqnarray}
& & \frac{2 \left( k_2 + k_3 \right)}{r^3}  + \frac{2\tilde{g}}{r^3} \left[ 2\tilde{g} k_2 \left( k_2 + k_3 \right) \left( 2k_2 + k_3 \right) - 2\tilde{g} k_2 k_3 \left( k_2 + k_3 \right) \right] = 0,
\end{eqnarray}
i.e.,
\begin{eqnarray}
& & \left( k_2 + k_3 \right) (1+{4}\tilde{g}^2 k_2^2)= 0,
\end{eqnarray}
i.e.,
\begin{eqnarray}
& & k_2 + k_3= 0,
\end{eqnarray}
i.e.,
\begin{eqnarray}\label{eq:k2k3-1}
& & k_2=-k_3.
\end{eqnarray}
Based on Eq. (\ref{eq:k2k3-1}), from Eq. (\ref{eq:e3-a}) and Eq. (\ref{eq:e4-a}) we have
\begin{eqnarray}
& & \frac{4 \tilde{g}^2 k_2 f_1^2 \left( r \right)}{r} + \frac{2 \tilde{g} }{r^3}k_2 \left(2\tilde{g} k_2 k_3 \right) = 0,
\end{eqnarray}
i.e.,
\begin{eqnarray}
& &  k_3 f_1^2 (r)= \frac{1}{r^2}  k_3^3.
\end{eqnarray}
There are two possibilities: (i) $k_3=0$; (ii) $k_3\neq 0$ and $f_1^2(r)=\dfrac{k_3^2}{r^2}$.

\emph{ (i) The Case of $k_3=0$.---} In this case, one has $k_2=k_3=0$. Then Eq. (\ref{eq:e1-a}) and Eq. (\ref{eq:e2-a}) become
\begin{eqnarray}\label{eq:e1-aa}
 f_1''(r) + \frac{2}{r} f_1'(r) - \frac{2f_1(r)}{r^2} = 0,
\end{eqnarray}
\begin{eqnarray}\label{eq:e2-aa}
& & \frac{2\tilde{g} f_1^2 \left( r \right)}{r} = 0,
\end{eqnarray}
which yield
\begin{eqnarray}
 f_1(r) = 0.
\end{eqnarray}

\emph{(ii) The Case of $k_3\neq 0$ and $f_1^2(r)=\dfrac{k_3^2}{r^2}$.---} In this case, one has $k_2=-k_3$. Then Eq. (\ref{eq:e2-a}) becomes
\begin{eqnarray}
& & -\frac{2}{r^3} \left[ \tilde{g} \left(k_3^2 \right) \right] + \frac{2 \tilde{g} f_1^2 \left( r \right)}{r} = 0,
\end{eqnarray}
i.e.,
\begin{eqnarray}
& &  f_1^2(r)= \frac{k_3^2}{r^2},
\end{eqnarray}
which is automatically valid. Similarly, Eq. (\ref{eq:e1-a}) becomes
\begin{eqnarray}
-\frac{8 \tilde{g}^2 k_3^2}{r^2} f_1(r) + f_1''(r) + \frac{2}{r} f_1'(r) - \frac{2f_1(r)}{r^2} = 0.
\end{eqnarray}
Let $f_1(r)=\dfrac{C}{r}$, with $C=\pm \sqrt{k_3^2}$, one has
\begin{eqnarray}
-\frac{8 \tilde{g}^2 k_3^2}{r^2} \left(\dfrac{C}{r}\right) + \left(\dfrac{C}{r}\right)'' + \frac{2}{r} \left(\dfrac{C}{r}\right)' - \frac{2}{r^2}\left(\dfrac{C}{r}\right) = 0,
\end{eqnarray}
i.e.,
\begin{eqnarray}
4 \tilde{g}^2 k_2^2+1= 0,
\end{eqnarray}
which cannot be valid when $k_2$ is a real number. Thus in this case, there is no real solution.

\subsubsection{The Case of $\{g\neq 0, k_1\neq 0\}$}

In this case, from Eqs. (\ref{eq:e1})-(\ref{eq:e4}) we have
\begin{eqnarray}\label{eq:e1-b}
4\frac{\tilde{g}}{r^2} \left( k_1 - 2\tilde{g} k_1^2 - 2\tilde{g} k_2^2 \right) f_1(r) + f_1''(r) + \frac{2}{r} f_1'(r) - \frac{2f_1(r)}{r^2} \left( 1 - 2\tilde{g} k_1 \right) = 0,
\end{eqnarray}
\begin{eqnarray}\label{eq:e2-b}
& & -\frac{2}{r^3} \left[ \tilde{g} \left( k_1^2 + 2k_2^2 + k_2 k_3 \right) - k_1 \right] + \frac{2\tilde{g} f_1^2 \left( r \right)}{r} \left( 1 - 2\tilde{g} k_1 \right) \nonumber \\
& & + \frac{2\tilde{g}}{r^3} \left[ k_1 \left( 2k_1 \left( \tilde{g} k_1 - 1 \right) - 2\tilde{g} k_2 k_3 \right) + \left( k_2 + k_3 \right)^2 \left( 2\tilde{g} k_1 - 1 \right) + 2\tilde{g} k_1 k_2 \left( k_2 + k_3 \right) \right] = 0,
\end{eqnarray}
\begin{eqnarray}\label{eq:e3-b}
& & \frac{k_2 + k_3}{r^3} \left( 2\tilde{g} k_1 - 1 \right) - \frac{4\tilde{g}^2 k_2 f_1^2 \left( r \right)}{r} \nonumber \\
& & + \frac{2\tilde{g}}{r^3} \left[ k_2 \left( 2k_1 \left( \tilde{g} k_1 - 1 \right) - 2\tilde{g} k_2 k_3 \right)
+ 2\tilde{g} k_2 \left( k_2 + k_3 \right) \left( 2k_2 + k_3 \right) \right] = 0,
\end{eqnarray}
\begin{eqnarray}\label{eq:e4-b}
& & -\frac{3 \left( k_2 + k_3 \right)}{r^3} \left( 2\tilde{g} k_1 - 1 \right) + \frac{4\tilde{g}^2 k_2 f_1^2 \left( r \right)}{r} \nonumber \\
& & + \frac{2\tilde{g}}{r^3} \left[ -k_2 \left( 2k_1 \left( \tilde{g} k_1 - 1 \right) - 2\tilde{g} k_2 k_3 \right) + 2k_1 \left( k_2 + k_3 \right) \left( 2\tilde{g} k_1 - 1 \right) - 2\tilde{g} k_2 k_3 \left( k_2 + k_3 \right) \right] = 0.
\end{eqnarray}
After adding Eq. (\ref{eq:e3-b}) and Eq. (\ref{eq:e4-b}), we have
\begin{eqnarray}
  -\frac{2 \left( k_2 + k_3 \right)}{r^3} \left( 2\tilde{g} k_1 - 1 \right)  + \frac{2\tilde{g}}{r^3} \left[  2\tilde{g} k_2 \left( k_2 + k_3 \right) \left( 2k_2 + k_3 \right)+ 2k_1 \left( k_2 + k_3 \right) \left( 2\tilde{g} k_1 - 1 \right) - 2\tilde{g} k_2 k_3 \left( k_2 + k_3 \right) \right] = 0,
\end{eqnarray}
i.e.,
\begin{eqnarray}
  \left( k_2 + k_3 \right) \left[ 1
  + 4\tilde{g} \left(   \tilde{g} k_2^2 + \tilde{g} k_1^2- k_1\right)\right] = 0.
\end{eqnarray}
Then one has {three} possibilities: (i) $k_2=k_3=0$; (ii) $k_2 =-k_3$ and $k_3\neq 0$; (iii) $k_2 + k_3\neq 0$ and $\left[ 1  + 4\tilde{g} \left(   \tilde{g} k_2^2 + \tilde{g} k_1^2- k_1\right)\right]=0$.

\vspace{4mm}

\centerline{\emph{4.1. The Case of $k_2=k_3=0$}}

\vspace{4mm}

In this case, Eq. (\ref{eq:e3-b}) and Eq. (\ref{eq:e4-b}) are automatically satisfied. Eq. (\ref{eq:e2-b}) become
\begin{eqnarray}
& & -\frac{2}{r^3} \left[ \tilde{g}  k_1^2  - k_1 \right] + \frac{2\tilde{g} f_1^2 \left( r \right)}{r} \left( 1 - 2\tilde{g} k_1 \right)+ \frac{2\tilde{g}}{r^3} \left[ 2 k_1^2  \left( \tilde{g} k_1 - 1 \right)  \right] = 0,
\end{eqnarray}
i.e.,
\begin{eqnarray}
& & \tilde{g} f_1^2(r)  \left( 1 - 2\tilde{g} k_1 \right)=\frac{k_1\left( \tilde{g}  k_1  - 1 \right)}{r^2} \left(1- 2 \tilde{g} k_1 \right),
\end{eqnarray}
which leads to two possibilities: (i) $2\tilde{g} k_1=1$; (ii) $2\tilde{g} k_1-1\neq 0$ and $f_1^2(r)=\dfrac{k_1\left( \tilde{g}  k_1  - 1 \right)/\tilde{g}}{r^2}$.

(i) The case of $2\tilde{g} k_1=1$. {This situation also means $\left[ 1  + 4\tilde{g} \left(   \tilde{g} k_2^2 + \tilde{g} k_1^2- k_1\right)\right]=0$ with $k_2=0$}. In this case, Eq. (\ref{eq:e1-b}) becomes
\begin{eqnarray}
 f_1''(r) + \frac{2}{r} f_1'(r)= 0,
\end{eqnarray}
which yields
\begin{eqnarray}
 f_1(r)=\frac{C_2}{r} + C_3.
\end{eqnarray}

(ii) The case of $2\tilde{g} k_1-1\neq 0$ and $f_1^2(r)=\dfrac{k_1\left( \tilde{g}  k_1  - 1 \right)/\tilde{g}}{r^2}$. For simplicity, we let
{\begin{eqnarray}\label{eq:kappa-1}
 f_1(r)=\frac{C}{r}, \;\;\;\;\; C= \pm \sqrt{k_1\left( \tilde{g}  k_1  - 1 \right)/\tilde{g}}.
\end{eqnarray}
}

(ii-1) If $C=0$, then we have
\begin{eqnarray}
 \tilde{g}  k_1  = 1,
\end{eqnarray}
thus $f_1(r)=0$. In this situation, one may check that Eq. (\ref{eq:e1-b}) is automatically satisfied.

(ii-2)  If $C\neq 0$ or $\tilde{g}  k_1  \neq  1$, after substituting Eq. (\ref{eq:kappa-1}) into Eq. (\ref{eq:e1-b}), we have
\begin{eqnarray}
4\frac{\tilde{g}}{r^2} \left( k_1 - 2\tilde{g} k_1^2 \right) \left(\dfrac{C}{r}\right) + \left(\dfrac{C}{r}\right)'' + \frac{2}{r} \left(\dfrac{C}{r}\right)' - \frac{2}{r^2} \left( 1 - 2\tilde{g} k_1 \right) \left(\dfrac{C}{r}\right)= 0,
\end{eqnarray}
i.e.,
\begin{eqnarray}
-(2\tilde{g} k_1 -1)^2= 0,
\end{eqnarray}
which cannot be valid for $2\tilde{g} k_1-1\neq 0$. Thus there is no real solution for this case.

\vspace{4mm}

\centerline{\emph{4.2. $k_2 =-k_3$ and $k_3\neq 0$}}

\vspace{4mm}

In this case, Eq. (\ref{eq:e3-b}) and Eq. (\ref{eq:e4-b}) reduces to
\begin{eqnarray}
& & \frac{4\tilde{g}^2 k_2 f_1^2 \left( r \right)}{r}  + \frac{2\tilde{g}}{r^3} \left[ -k_2 \left( 2k_1 \left( \tilde{g} k_1 - 1 \right) - 2\tilde{g} k_2 k_3 \right)\right] = 0,
\end{eqnarray}
i.e.,
\begin{eqnarray}
& & 4\tilde{g}^2 k_2 f_1^2(r) = \frac{4\tilde{g} k_2}{r^2} \left[ k_1 \left( \tilde{g} k_1 - 1 \right) - \tilde{g} k_2 k_3 \right],
\end{eqnarray}
i.e.,
\begin{eqnarray}
& & \tilde{g} f_1^2(r) = \frac{1}{r^2} \left[ k_1 \left( \tilde{g} k_1 - 1 \right) +\tilde{g} k_3^2 \right],
\end{eqnarray}
i.e.,
\begin{eqnarray}\label{eq:ff-1}
& & f_1^2(r) = \frac{\left( \tilde{g} k_1^2+\tilde{g} k_3^2 -k_1\right)/\tilde{g}}{r^2}.
\end{eqnarray}
After substituting Eq. (\ref{eq:ff-1}) into Eq. (\ref{eq:e2-b}), we find that Eq. (\ref{eq:e2-b}) is satisfied automatically.

Now, the function $f_1(r)$ has the structure
\begin{eqnarray}
 f_1(r)=\frac{C}{r}, \;\;\;\;\; C= \pm \sqrt{\left( \tilde{g} k_1^2+\tilde{g} k_3^2 -k_1\right)/\tilde{g}}.
\end{eqnarray}

(i) If $C=0$, we have $f_1(r)=0$, one may check that Eq. (\ref{eq:e1-b}) is automatically satisfied.
In this situation, we have
\begin{eqnarray}\label{eq:ka}
 \tilde{g} k_1^2+\tilde{g} k_3^2 -k_1=0.
\end{eqnarray}
By considering
\begin{eqnarray}
&& k_1= \frac{c\hbar}{2g} a_1, \;\;\;\; k_2=-\frac{c\hbar}{2g}a_3, \;\;\;\; k_3=\frac{c\hbar}{2g}a_3, \;\;\;\;\tilde{g}=\frac{g}{c\hbar},
\end{eqnarray}
Eq. (\ref{eq:ka}) becomes
\begin{eqnarray}
 a_1^2+a_3^2 - 2 a_1=0,
\end{eqnarray}
which is just the previous constraint
\begin{eqnarray}
(a_1 - 1)^2 + a_3^2 = 1.
\end{eqnarray}
For the real static solution, one can have the following parameterization
\begin{eqnarray}
&& a_1 = 2\sin^2\theta, \;\;\;\; a_3 = - 2\sin\theta\cos\theta,
\end{eqnarray}
thus
\begin{eqnarray}
&& k_1= \frac{c\hbar}{2g} a_1 = \frac{c\hbar}{g} \sin^2\theta = \frac{1}{\tilde{g}} \sin^2\theta , \;\;\;\; k_3=-\frac{1}{\tilde{g}}\sin\theta\cos\theta,
\end{eqnarray}
with $k_3 \neq 0$, i.e., $\theta \neq \pi/2, 3\pi/2$.

(ii) If $C\neq 0$, i.e., $\tilde{g} k_1^2+\tilde{g} k_3^2 -k_1 \neq 0$,
then from Eq. (\ref{eq:e1-b}) we have
\begin{eqnarray}
4\frac{\tilde{g}}{r^2} \left( k_1 - 2\tilde{g} k_1^2 - 2\tilde{g} k_2^2 \right) \left(\dfrac{C}{r}\right) + \left(\dfrac{C}{r}\right)'' + \frac{2}{r} \left(\dfrac{C}{r}\right)' - \frac{2}{r^2} \left( 1 - 2\tilde{g} k_1 \right) \left(\dfrac{C}{r}\right)= 0,
\end{eqnarray}
i.e.,
\begin{eqnarray}
\tilde{g} k_1^2 + \tilde{g} k_3^2 -k_1 =-\frac{1}{4\tilde{g}}.
\end{eqnarray}
Based on which, we have
\begin{eqnarray}
 C= \pm \sqrt{\left( \tilde{g} k_1^2+\tilde{g} k_3^2 -k_1\right)/\tilde{g}}=\pm \sqrt{-\frac{1}{4\tilde{g}^2}}=\pm \frac{{\rm i}}{2|\tilde{g}|},
\end{eqnarray}
 which is a complex number. Thus in this case, there is no real solution.

\vspace{4mm}

\centerline{\emph{4.3. The Case of $k_2 + k_3\neq 0$ and $\left[ 1  + 4\tilde{g} \left(   \tilde{g} k_2^2 + \tilde{g} k_1^2- k_1\right)\right]=0$}}

\vspace{4mm}

In this case, we have
\begin{eqnarray}
1  + 4\tilde{g} \left(   \tilde{g} k_2^2 + \tilde{g} k_1^2- k_1\right) =0,
\end{eqnarray}
i.e.,
\begin{eqnarray}
(2\tilde{g} k_2)^2  + (2\tilde{g} k_1 -1)^2 =0.
\end{eqnarray}
Based on which we have $k_2=0$ and
\begin{eqnarray}
2\tilde{g} k_1=1.
\end{eqnarray}
From Eq. (\ref{eq:e3-b}) we have
\begin{eqnarray}
& & \frac{k_3}{r^3} \left( 2\tilde{g} k_1 - 1 \right)= 0,
\end{eqnarray}
which is automatically satisfied. Based on $2\tilde{g} k_1=1$, one similarly verifies that Eq. (\ref{eq:e2-b}) is also automatically satisfied.

From Eq. (\ref{eq:e1-b}) we have
\begin{eqnarray}
f_1''(r) + \frac{2}{r} f_1'(r) = 0,
\end{eqnarray}
which yields
\begin{eqnarray}
 f_1(r)=\frac{C_2}{r} + C_3.
\end{eqnarray}
with $C_2$ and $C_3$ are real numbers.

In summary, we would like to list all the real static solutions of the Yang-Mills equations in
Table \ref{Tab:real}.

\begin{table}[H]
\centering
\caption{All the real static solutions of the Yang-Mills equations. The spin vector potential is given by $\vec{A} = \frac{1}{r} \left[ k_1 (\hat{r} \times \vec{\Gamma}) + k_2 \vec{\Gamma} + k_3 (\vec{\Gamma} \cdot \hat{r})\hat{r} \right]$, and the scalar potential is $ \varphi = f_1(r) ( \vec{\Gamma}\cdot \hat{r}) +  f_2(r)$. Here ``static'' means the solutions $\{\vec{A}, \varphi\}$ are time-independent, ``real'' means three parameters $k_1$, $k_2$, $k_3$ and two functions $f_1(r)$, $f_2(r)$ are all real. The parameter $\tilde{g}=g/c\hbar$, with $\hbar$ being Planck's constant, $c$ is the speed of light in vacuum, $g$ is the gauge coupling parameter, and they are all real numbers. $g$ can be greater than, equal to, or less than 0. The solution of $f_2(r)$ is given by $f_2(r)=\frac{C_1}{r}$, and $C_1$ is real. }
\begin{tabular}{|c|c|c|c|c|}
\hline
\(f_1(r)\) & \(k_1\) & \(k_2\) & \(k_3\) & Remarks \\
\hline
\multicolumn{5}{|c|}{\textbf{Case 1: $\{g=0, k_1=0\}$}} \\
\hline
\(\dfrac{C_2}{r^2}+C_3 r\) & \(0\) & $k_2 \in \mathbb{R}$ & \(-k_2\) & $C_2$ and $C_3$ are real \\
\hline
\multicolumn{5}{|c|}{\textbf{Case 2: $\{g=0, k_1\neq 0\}$}} \\
\hline
\(--\) & \(--\) & \(--\) & \(--\) & No Solutions\\
\hline
\multicolumn{5}{|c|}{\textbf{Case 3: $\{g\neq 0, k_1= 0\}$}} \\
\hline
\(0\) & \(0\) & \(0\) & \(0\) & $\vec{A}=0$, and $\varphi$ is Coulomb-type potential  \\
\hline
\multicolumn{5}{|c|}{\textbf{Case 4: $\{g\neq 0, k_1\neq  0\}$}} \\
\hline
\(\dfrac{C_2}{r} + C_3\) & \(\dfrac{1}{2\tilde{g}}\) & \(0\) & \(0\) & The constraint condition $2\tilde{g} k_1=1$ \\
\hline
\(0\) & \(\dfrac{1}{\tilde{g}}\) & \(0\) & \(0\) & The constraint condition $\tilde{g} k_1=1$ \\
\hline
\(0\) & \(\dfrac{1}{\tilde{g}} \sin^2\theta\) & \(-k_3\) & \(-\dfrac{1}{\tilde{g}}\sin\theta\cos\theta \neq 0\) & Constraint $\tilde{g} k_1^2+\tilde{g} k_3^2 -k_1=0$ \\
\hline
\(\dfrac{C_2}{r} + C_3\) & \(\dfrac{1}{2\tilde{g}}\) & \(0\) & \(k_3\neq 0\) & The constraint condition $2\tilde{g} k_1=1$ \\
\hline
\end{tabular}\label{Tab:real}
\end{table}

In addition, we would like to list the complex static solutions of the Yang-Mills equation in
Table \ref{Tab:complex}. For the detailed calculation, one may find in SM \cite{SM}.

\begin{table}[H]
\centering
\footnotesize
\caption{All the complex static solutions of the Yang-Mills equations. The spin vector potential is given by $\vec{A} = \frac{1}{r} \left[ k_1 (\hat{r} \times \vec{\Gamma}) + k_2 \vec{\Gamma} + k_3 (\vec{\Gamma} \cdot \hat{r})\hat{r} \right]$, and the scalar potential is $ \varphi = f_1(r) ( \vec{\Gamma}\cdot \hat{r}) +  f_2(r)$. Here ``static'' means the solutions $\{\vec{A}, \varphi\}$ are time-independent, ``complex'' means at least one of three parameters $k_1$, $k_2$, $k_3$ or two functions $f_1(r)$, $f_2(r)$ is complex. The parameter $\tilde{g}=g/c\hbar$, with $\hbar$ being Planck's constant, $c$ is the speed of light in vacuum, $g$ is the gauge coupling parameter, and they are all real numbers. $g$ can be greater than, equal to, or less than 0. The solution of $f_2(r)$ is given by $f_2(r)=\frac{C_1}{r}$. }
\begin{tabular}{|c|c|c|c|c|}
\hline
\(f_1(r)\) & \(k_1\) & \(k_2\) & \(k_3\) & Remarks \\
\hline
\multicolumn{5}{|c|}{\textbf{Case 1: $\{g=0, k_1=0\}$}} \\
\hline
\(\dfrac{C_2}{r^2}+C_3 r\) & \(0\) & $k_2 \in \mathbb{C}$ & \(-k_2\) & At least one of $\{C_1, C_2, C_3, k_2\}$ is complex \\
\hline
\multicolumn{5}{|c|}{\textbf{Case 2: $\{g=0, k_1\neq 0\}$}} \\
\hline
\(--\) & \(--\) & \(--\) & \(--\) & No Solutions\\
\hline
\multicolumn{5}{|c|}{\textbf{Case 3: $\{g\neq 0, k_1= 0\}$}} \\
\hline
\(0\) & \(0\) & \(0\) & \(0\) & $C_1$ is complex  \\
\hline
\(\dfrac{\pm \frac{{\rm i}}{2|\tilde{g}|}}{r}\) & \(0\) & \(\pm\frac{{\rm i}}{2|\tilde{g}|}\) & \(\mp \frac{{\rm i}}{2|\tilde{g}|}\) & At least one of $\{C_1, C, k_2, k_3\}$ is complex \\
\hline
\(0\) & \(0\) & \(\pm \frac{{\rm i}}{2|\tilde{g}|}\) & \(\frac{-3\pm {\rm i} \sqrt{3}}{2} k_2\) & $C_1$ is a real or complex number \\
\hline
\(\dfrac{\pm \sqrt{3 k_2^2+ 3k_2 k_3 +k_3^2}}{r}\) & \(0\) & \(\pm \frac{{\rm i}}{2|\tilde{g}|}\) & \(k_3\neq -k_2\) & $C_1$ is a real or complex number \\
\hline
\multicolumn{5}{|c|}{\textbf{Case 4: $\{g\neq 0, k_1\neq  0\}$}} \\
\hline
\(\dfrac{C_2}{r} + C_3\) & \(\dfrac{1}{2\tilde{g}}\) & \(0\) & \(0\) & At least one of $\{C_1, C_2, C_3\}$ is complex \\
\hline
\(0\) & \(\dfrac{1}{\tilde{g}}\) & \(0\) & \(0\) & $C_1$ is a complex number \\
\hline
\(0\) & \(\dfrac{1}{\tilde{g}} \sin^2\theta\) & \(-k_3\) & \(-\dfrac{1}{\tilde{g}}\sin\theta\cos\theta \neq 0\) & $C_1$ complex, constraint $\tilde{g} k_1^2+\tilde{g} k_3^2 -k_1=0$ \\
\hline
\(0\) & \(\dfrac{1}{2\tilde{g}} (1 \pm \cosh \vartheta)\) & \(-k_3\) & \(\dfrac{{\rm i}}{2\tilde{g}} \sinh \vartheta \neq 0\) & Constraint $\tilde{g} k_1^2+\tilde{g} k_3^2 -k_1=0$ \\
\hline
\(0\) & \(\dfrac{1}{2\tilde{g}} (1+ {\rm i}\sinh^2 \vartheta)\) & \(-k_3\) & \(\pm \cosh \vartheta \neq 0\) & Constraint $\tilde{g} k_1^2+\tilde{g} k_3^2 -k_1=0$ \\
\hline
\(\dfrac{\pm \frac{{\rm i}}{2|\tilde{g}|}}{r}\) & Satisfies constraint & \(-k_3\) & \(k_3\neq 0\) & Constraint $\tilde{g} k_1^2+\tilde{g} k_3^2 -k_1=-\frac{1}{4\tilde{g}}$ \\
\hline
\(\dfrac{C_2}{r} + C_3\) & \(\dfrac{1}{2\tilde{g}}\) & \(0\) & \(k_3\neq 0\) & Constraint $2\tilde{g} k_1=1$, at least one of $\{C_1, C_2, C_3\}$ is complex \\
\hline
\(0\) & \(k_1\neq  \frac{1}{2\tilde{g}}\) & \(\pm {\rm i} \left|\frac{2\tilde{g} k_1 -1}{2\tilde{g}}\right|\) & Satisfies constraint & $(k_2 + k_3)( 2\tilde{g} k_1 - 1 ) + k_2[ -1+4\tilde{g}^2(k_2+k_3)^2]=0$ \\
\hline
\(\dfrac{C}{r}\) & \(k_1\neq  \frac{1}{2\tilde{g}}\) & \(\pm {\rm i} \left|\frac{2\tilde{g} k_1 -1}{2\tilde{g}}\right|\) & Arbitrary & $C^2= \dfrac{(k_2 + k_3)( 2\tilde{g} k_1 - 1 ) + k_2[ -1+4\tilde{g}^2(k_2+k_3)^2]}{4\tilde{g}^2 k_2}$ \\
\hline
\end{tabular}\label{Tab:complex}
\end{table}

\section{Conclusion and Discussion}\label{sec-7}

In this work, we have achieved a complete classification of static solutions to the vacuum SU(2) YM equations that carry an explicit dependence on spin degrees of freedom. By generalizing the vector potential extraction approach (VPEA) from the Abelian to the non-Abelian case, we derived the  general form of the spin vector potential compatible with the angular momentum algebra. This led to the static ansatz (\ref{eq:ansatz}), whose substitution into the YM equations yielded a closed set of constraint equations. Solving these constraints case by case produced the rich families of real and complex static solutions summarized in Tables~\ref{Tab:real} and~\ref{Tab:complex}.

\emph{Summary of Key Results.---}

\begin{itemize}
    \item \textbf{Recovery of known solutions}: Our classification successfully reproduces the simple SU(2) static solution of Ref.~\cite{Zhou2025} as the special case \(k_2=k_3=0\) with \(2\tilde{g} k_1=1\) or \(2\) (i.e., $\tilde{g} \hbar \tilde{k}=1$ or 2). This serves as a consistency check for our general framework.

    \item \textbf{New real static solutions}: Beyond the known solution, we discovered several families of real solutions with non-zero \(k_2, k_3\) and non-trivial radial functions \(f_1(r)\). These configurations demonstrate that the internal spin structure allows for significantly richer vacuum configurations than those captured by standard point-source models~\cite{Faddeev1999}. Notably, many of these new solutions satisfy the VPEA-derived constraint \(k_2+k_3=0\) and exhibit a purely radial, Coulomb-like electric field with a vanishing magnetic field.

    \item \textbf{Complex static solutions}: By relaxing the reality condition, we uncovered a rich set of complex solutions. While their direct physical interpretation remains open, such complex classical configurations are increasingly recognized as important in modern theoretical frameworks, including the analytic continuation of gauge theories~\cite{Witten2010} and resurgence theory~
        \cite{Dunne2012, Basar2013, Aniceto2019}. They also echo earlier explorations of complex solutions in topologically massive gauge theories~\cite{Oh1990PLB}.

    \item \textbf{Role of the VPEA constraints}: The angular-momentum constraints derived from the VPEA, namely \((a_1-1)^2+a_3^2=1\) (equivalently \(\tilde{g}(k_1^2+k_3^2)-k_1=0\) and \(k_2+k_3=0\)), emerge as a special subclass of our general solutions. When these constraints are imposed, the magnetic field vanishes identically (\(\vec{B}=0\)), and the solution reduces to a pure gauge configuration (up to the scalar potential). This provides a clear physical interpretation of the VPEA conditions.
\end{itemize}

\emph{Physical Implications.---} The existence of these spin-dependent static solutions suggests a novel phenomenon in non-Abelian gauge theory: The gauge field can support ``Coulomb-like'' interactions that couple directly to the internal spin of a test particle. This is reminiscent of the classical color-electric fields generated by non-Abelian point sources~\cite{Oh1986PRD}, but with a crucial difference: here, the interaction acquires a directional dependence through the \(\vec{\Gamma}\cdot\hat{r}\) and \(\hat{r}\times\vec{\Gamma}\) structures.

A natural question arises: {Can these solutions be interpreted as describing a ``spin-charged'' object?} If so, how would such an object differ from the well-studied Wu-Yang~\cite{WuYang1969} or 't Hooft-Polyakov~\cite{tHooft1974} monopoles? Unlike those topologically stable solitons, our solutions are not necessarily finite-energy or topologically protected. Nevertheless, they may represent important saddle-point configurations in the path integral, contributing to non-perturbative phenomena such as vacuum tunneling or confinement. The fact that some solutions exhibit a vanishing magnetic field while others do not suggests a possible phase structure depending on the parameters \(k_i\).

\emph{Limitations and Future Directions.---} Several natural extensions of this work present themselves:

\begin{enumerate}
    \item \textbf{Higher spins}: Our analysis was restricted to spin-\(1/2\) for simplicity. Extending the VPEA to higher spins (using the appropriate spin matrices) is straightforward in principle, though the algebraic constraints will become more complex. Such an extension could reveal spin-dependent effects that scale with the spin quantum number.

    \item \textbf{Time-dependent generalizations}: We have considered only static fields. Relaxing this assumption could uncover non-Abelian progressive waves~\cite{Oh1985JMP} or other wave-like solutions recently discussed in the literature~\cite{Rabinowitch2026,ZhangChen2026}. The VPEA framework might offer a new perspective on constructing such dynamical solutions.

    \item \textbf{Physical properties of the new solutions}: It would be worthwhile to compute the classical energy density, total energy, and possible topological charges of the new real solutions. Are any of them normalizable? Could they serve as backgrounds for semi-classical quantization~\cite{Callan1978, Prasad1975}?

    \item \textbf{Interpretation of complex solutions}: The complex solutions we found require further physical interpretation. They might be relevant in the context of Euclidean instantons~\cite{BPST1975}, dyon configurations~\cite{JuliaZee1975}, or non-Hermitian extensions of Yang-Mills theory. Their role in resurgence theory, where complex ``ghost'' instantons contribute to transseries expansions, is particularly intriguing~\cite{Dunne2012, Basar2013, Aniceto2019}.
\end{enumerate}

 Finally, we note that the vector potential extraction approach has proven to be a powerful heuristic tool for generating gauge potentials that encode angular momentum information. Its success in the Abelian case---reproducing the AB effect~\cite{2005QParadox}, uniform magnetic field~\cite{1999Jackson}, Dirac monopole~\cite{Dirac1931}, and Wu-Yang monopole~\cite{WuYang1975,WuYang1976}---and now in the non-Abelian SU(2) case, suggests that the method may have even wider applicability. One could, for example, apply the same idea to higher-dimensional gauge theories, to other gauge groups such as SU(3), or explore its potential connections to the algebraic matrix constructions underlying the ADHM~\cite{ADHM1978} and Nahm~\cite{Nahm1982} methods. In summary, this work provides a {complete and systematic classification of static SU(2) Yang-Mills solutions with a spin-dependent vector potential} derived from the VPEA. The resulting families---both real and complex---significantly expand the known landscape of exact solutions and open up new avenues for exploring the role of spin in non-Abelian gauge theories, both at the classical and, potentially, the quantum level.

\emph{Acknowledgments.---}This work is supported by the Quantum Science and Technology-National Science and Technology Major Project (Grant No. 2024ZD0301000), and the National Natural Science Foundation of China (Grant No. 12275136).


\end{document}


\title{Supplemental Material for\\``Exact Solutions of the SU(2) Yang-Mills Equations from a Static Ansatz''}

\author{Yu-Xuan Zhang}
\affiliation{School of Physics, Nankai University, Tianjin 300071, People's Republic of China}

\author{Jing-Ling Chen}
\email{chenjl@nankai.edu.cn}
\affiliation{Theoretical Physics Division, Chern Institute of Mathematics, Nankai University, Tianjin 300071, People's Republic of
	 	China}

\date{\today}


\maketitle\tableofcontents


\section{Explanation of Notation}\label{sec-2}

In this work, we use the Gaussian units. The metric tensor of space-time is taken as
\begin{eqnarray}
\eta_{\mu\nu} = \text{diag}(1, -1, -1, -1), \;\;\;\;\;\;\; (\mu, \nu=0, 1, 2, 3).
\end{eqnarray}
The relation between the four-potential and the scalar-vector potential is given by
\begin{eqnarray}
A_{\mu} = (\varphi, -\vec{A}), \;\;\;\;\;\; (\mu=0, 1, 2, 3),
\end{eqnarray}
with $A_0=\varphi$, $\vec{A}=(A_1, A_2, A_3)$ or $\vec{A}=(A_x, A_y, A_z)$. It is clear that the scalar potential $\varphi$ and the vector potential $\vec{A}$ have the same physics dimension. In this work, the static ansatz is given by
\begin{eqnarray}\label{eq:ansatz}
&&\vec{A} = \frac{1}{r}\Bigl[k_1(\hat{r}\times\vec{\Gamma}) + k_2\vec{\Gamma} + k_3(\vec{\Gamma}\cdot\hat{r})\hat{r}\Bigr], \nonumber\\
&& \varphi = f_1(r)\,(\vec{\Gamma}\cdot\hat{r}) + f_2(r),
\end{eqnarray}
where \(\vec{\Gamma}\) are the Pauli matrices (for spin-1/2 generators), and $\vec{S}=(\hbar/2)\vec{\Gamma}$.

\newpage

\section{Gauge Covariant Dirac Equation and Gauge Invariance of the Yang-Mills Equations} \label{sec-3}

\subsection{From the Ordinary Dirac Equation to the Gauge Covariant Dirac Equation}

Let us consider the following ordinary Dirac equation (i.e., Dirac's equation for a free particle without the vector and scalar potentials)
\begin{eqnarray}\label{eq:Dirac}
{\rm i}\hbar\frac{\partial \psi}{\partial t} = \left[c\vec{\alpha} \cdot \vec{p} + \beta m c^2 \right] \psi,
\end{eqnarray}
where $\psi$ is the wavefunction, $m$ is mass, $c$ is the speed of light in vacuum, $\hbar$ is Planck's constant, $\vec{p}=(p_1, p_2, p_3)$ or $\vec{p}=(p_x, p_y, p_z)$ is the linear momentum, $\{\vec{\alpha}, \beta\}$ are Dirac's matrices with
\begin{equation}
            \vec{\alpha}=
            \left(\begin{array}{cc}
                    0 & \vec{\sigma} \\
                  \vec{\sigma} & 0 \\
                  \end{array}
                \right)=\sigma_x\otimes\vec{\sigma},\;\;\;\;\;
             \beta=
             \left(\begin{array}{cc}
                  \openone & 0 \\
                  0 & -\openone \\
                  \end{array}
                \right)
           =\sigma_z\otimes\openone,
      \end{equation}
$\openone$ is the $2\times 2$ identity matrix, and $\vec{\sigma}=(\sigma_x, \sigma_y, \sigma_z)$ is the vector of Pauli matrices with
\begin{eqnarray}
 \label{S-3}
  && \sigma_x=
  \left(
    \begin{array}{cc}
      0 & 1 \\
      1 & 0 \\
    \end{array}
  \right),\;\;\;\sigma_y=
  \left(
    \begin{array}{cc}
      0 & -{\rm i} \\
      {\rm i} & 0 \\
    \end{array}
  \right),\;\;\;  \sigma_z=
  \left(
    \begin{array}{cc}
      1 & 0 \\
      0 & -1 \\
    \end{array}
  \right).
   \end{eqnarray}
Due to
\begin{eqnarray}
p_i=-{\rm i} \hbar\partial_i = -{\rm i}\hbar \frac{\partial }{\partial x_i}, \quad (i=1, 2, 3),
\end{eqnarray}
or
\begin{eqnarray}
\vec{p}= -{\rm i} \hbar \left(\frac{\partial }{\partial x}, \frac{\partial }{\partial y},
\frac{\partial }{\partial z} \right)=-{\rm i} \hbar \vec{\nabla} ,
\end{eqnarray}
Eq. (\ref{eq:Dirac}) can be written as
\begin{eqnarray}
{\rm i}\hbar\frac{\partial \psi}{\partial t} = \left[-{\rm i} c \hbar \vec{\alpha} \cdot \vec{\nabla} + \beta m c^2 \right] \psi.
\end{eqnarray}

Now, we apply a gauge transformation $U$ on the wave function $\psi$, namely,
\begin{eqnarray}
 \psi \mapsto \psi'=U  \psi, \quad U={\rm e}^{{\rm i} g \theta^a T^a},
\end{eqnarray}
where $T^a$'s are Lie algebra generators, $\theta^a$'s are transformation parameters, and $g$ is the gauge coupling parameter.
After such a gauge transformation, the Dirac equation becomes
\begin{eqnarray}\label{eq:Dirac-1}
 {\rm i}\hbar\frac{\partial (U\psi)}{\partial t}= \left[-{\rm i} c \hbar \vec{\alpha} \cdot \vec{\nabla} + \beta m c^2 \right] (U\psi).
\end{eqnarray}

\begin{remark}If the gauge transformation is \emph{global}, i.e., the transformation matrix $U$ (or the parameter $\theta^a$) does not depend on
the space-time coordinates $x_\mu$ (with $x_\mu=(ct, -x, -y, -z)$), then
\begin{eqnarray}
 && \frac{\partial (U\psi)}{\partial t}= U  \frac{\partial \psi}{\partial t}, \nonumber\\
 && \vec{\nabla} (U\psi)= U (\vec{\nabla} \psi).
\end{eqnarray}
From Eq. (\ref{eq:Dirac-1}), we have
\begin{eqnarray}
U {\rm i}\hbar\frac{\partial \psi}{\partial t}= U\, \left[-{\rm i} c \hbar \vec{\alpha} \cdot \vec{\nabla} + \beta m c^2 \right] \psi.
\end{eqnarray}
i.e.,
\begin{eqnarray}
U^{-1} U \, {\rm i}\hbar\frac{\partial \psi}{\partial t}=U^{-1}  U \, \left[-{\rm i} c \hbar \vec{\alpha} \cdot \vec{\nabla} + \beta m c^2 \right] \psi.
\end{eqnarray}
i.e.,
\begin{eqnarray}
 {\rm i}\hbar\frac{\partial \psi}{\partial t}= \left[-{\rm i} c \hbar \vec{\alpha} \cdot \vec{\nabla} + \beta m c^2 \right] \psi,
\end{eqnarray}
which is just Eq. (\ref{eq:Dirac}). Thus the ordinary Dirac equation is invariant under the global gauge transformation. $\blacksquare$
\end{remark}

\begin{remark}If the gauge transformation is \emph{local}, i.e., the transformation matrix $U$ (or the parameter $\theta^a \equiv\theta^a (x_\mu)$) depends on the space-time coordinates $x_\mu$, then generally one has
\begin{eqnarray}
&&\vec{\alpha} \cdot \vec{\nabla}(U\psi)\neq U(\vec{\alpha} \cdot \vec{\nabla}\psi),\nonumber\\
&&\frac{\partial (U\psi)}{\partial t}\neq U \left(\frac{\partial \psi}{\partial t}\right).
\end{eqnarray}
It is obvious that the solution of the Dirac equation (i.e., $U\psi$) no longer satisfies the ordinary Dirac equation (\ref{eq:Dirac}) after the gauge transformation $U$. This fact tells us that the ordinary Dirac equation is not gauge invariant under the local gauge transformation.

%

Now, we would like to generalize the Dirac equation to gauge covariant form. A direct idea is to replace the ordinary derivative \(\partial_\mu\) (\(\mu = 0, 1, 2, 3\)) with a new ``derivative'', called the \emph{gauge covariant derivative} $D_\mu$ $(\mu = 0, 1, 2, 3)$. The key property of gauge covariance requires that when \(\psi\) transforms as \(\psi \to \psi' = U\psi\), the covariant derivative transforms covariantly, i.e.,
\begin{eqnarray}
 D_\mu \mapsto D'_\mu = U D_\mu U^{-1},
 \end{eqnarray}
such that $(D_\mu \psi)' = D'_\mu \psi' = U(D_\mu \psi)$. This ensures that an equation written in terms of $D_\mu$ maintains its form under a local gauge transformation.

Different to the ordinary derivative, the gauge covariant derivative satisfies
\begin{eqnarray}
 UD_{\mu}(\psi)=D_{\mu}'(\psi').
\end{eqnarray}
Now, let us discuss the form and the gauge transformation of the gauge covariant derivative. At first, we suppose its form is given by
\begin{eqnarray}
 D_{\mu}=\partial_{\mu}+\frac{{\rm i} g}{c\hbar}A_{\mu}, \quad (\mu=0, 1, 2, 3),
\end{eqnarray}
where $A_\mu$ is called the gauge potential. Denote
\begin{eqnarray}
D_\mu'=\partial_{\mu}+\frac{{\rm i} g}{c\hbar}A_{\mu}'.
\end{eqnarray}
Because
\begin{eqnarray}
 UD_{\mu}(\psi)=D_{\mu}'(\psi')=D_{\mu}'(U\psi),
\end{eqnarray}
and
\begin{eqnarray}
 D_{\mu}'(U\psi) &=& \partial_\mu(U\psi)+\frac{{\rm i} g}{c\hbar}A_{\mu}'U\psi = (\partial_{\mu}U)\psi+U(\partial_{\mu}\psi)+\frac{{\rm i} g}{c\hbar}A_{\mu}'U\psi, \\
 UD_{\mu}(\psi) &=& U \left(\partial_{\mu}\psi+\frac{{\rm i} g}{c\hbar}A_{\mu}\psi\right),
\end{eqnarray}
we can have
\begin{eqnarray}
U \left(\partial_{\mu}\psi+\frac{{\rm i} g}{c\hbar}A_{\mu}\psi\right)=(\partial_{\mu}U)\psi+U(\partial_{\mu}\psi)+\frac{{\rm i} g}{c\hbar}A_{\mu}'U\psi,
\end{eqnarray}
i.e.,
\begin{eqnarray}
(\partial_{\mu}U)\psi+\frac{{\rm i} g}{c\hbar}A_{\mu}'U\psi = U\frac{{\rm i} g}{c\hbar}A_{\mu}\psi,
\end{eqnarray}
i.e.,
\begin{eqnarray}
A_{\mu}' = UA_{\mu}U^{-1} + {\rm i} \frac{c\hbar}{g}(\partial_{\mu}U)U^{-1}.
\end{eqnarray}

Based on the above discussion, we can make a summary as follows. If we take
\begin{eqnarray}
\partial_\mu\mapsto D_{\mu}=\partial_{\mu}+\frac{{\rm i} g}{c\hbar}A_{\mu}, \quad \mu=0,1,2,3,
\end{eqnarray}
and define the gauge transformation of the potential as
\begin{eqnarray}
A_{\mu}' = UA_{\mu}U^{-1} + {\rm i} \frac{c\hbar}{g}(\partial_{\mu}U)U^{-1},
\end{eqnarray}
then we can have
\begin{eqnarray}
 UD_{\mu}(\psi)=D_{\mu}'(\psi').
\end{eqnarray}
Based on that, the gauge covariant Dirac Equation is given by
\begin{eqnarray}
 {\rm i}\hbar c D_0\psi= \left[-{\rm i} c \hbar \vec{\alpha} \cdot (D_1,D_2,D_3) +\beta m c^2 \right] \psi.
\end{eqnarray}
Now we perform the gauge transformation
\begin{eqnarray}
\psi\mapsto\psi', \quad \quad D_\mu\mapsto D'_\mu,
\end{eqnarray}
we can have
\begin{eqnarray}
 {\rm i}\hbar c D'_0\psi'= \left[-{\rm i} c \hbar \vec{\alpha} \cdot (D'_1,D'_2,D'_3) +\beta m c^2 \right] \psi'.
\end{eqnarray}
Because
\begin{eqnarray}
 UD_{\mu}(\psi)=D_{\mu}'(\psi'),
\end{eqnarray}
we can have
\begin{eqnarray}
 {\rm i}\hbar c UD_0\psi= U\left[-{\rm i} c \hbar \vec{\alpha} \cdot (D_1,D_2,D_3) +\beta m c^2 \right] \psi,
\end{eqnarray}
i.e.,
\begin{eqnarray}
 U^{-1}{\rm i}\hbar c UD_0\psi=  U^{-1} U\left[-{\rm i} c \hbar \vec{\alpha} \cdot (D_1,D_2,D_3) +\beta m c^2 \right] \psi.
\end{eqnarray}
Due to $U^{-1}U=1$, we can have
\begin{eqnarray}
 {\rm i}\hbar c D_0\psi= \left[-{\rm i} c \hbar \vec{\alpha} \cdot (D_1,D_2,D_3) +\beta m c^2 \right] \psi,
\end{eqnarray}
which is the same as the original form. Therefore, the gauge covariant Dirac equation is invariant under the local gauge transformation. $\blacksquare$
\end{remark}

\begin{remark}
Now, let us write down the gauge covariant Dirac equation explicitly. Because
\begin{eqnarray}
A_\mu = (\varphi, -\vec{A}) = (\varphi, -A_i), \quad (\mu=0, 1, 2, 3), \quad (i=1, 2, 3),
\end{eqnarray}
and
\begin{eqnarray}
D_\mu= \partial_\mu + \frac{{\rm i} g}{c\hbar}A_\mu,\quad\quad (\mu=0, 1, 2, 3),
\end{eqnarray}
we can have
\begin{eqnarray}
D_0 &=& \frac{1}{c}\partial_t + \frac{{\rm i} g}{c\hbar}\varphi, \\
D_i &=&   \partial_i - \frac{{\rm i} g}{c\hbar}A_i, \quad (i=1, 2, 3).
\end{eqnarray}
Recall the form of the  gauge covariant Dirac equation
\begin{eqnarray}
 {\rm i}\hbar c D_0\psi= \left[- {\rm i} c \hbar \vec{\alpha} \cdot (D_1,D_2,D_3) +\beta m c^2 \right] \psi,
\end{eqnarray}
we can rewrite it as
\begin{eqnarray}
{\rm i}\hbar\frac{\partial \psi}{\partial t} + {\rm i} \hbar\frac{ {\rm i} g}{\hbar}\varphi\psi = \left[-{\rm i} c \hbar \vec{\alpha} \cdot \left( \vec{\nabla} - \frac{{\rm i} g}{c\hbar}\vec{A}\right) + \beta m c^2 \right] \psi,
\end{eqnarray}
i.e.,
\begin{eqnarray}\label{eq:Dirac-3}
 {\rm i} \hbar\frac{\partial \psi}{\partial t} = \left[c\vec{\alpha} \cdot\left( \vec{p} - \frac{g}{c}\vec{A}\right) + \beta m c^2 + g\varphi\right] \psi.
\end{eqnarray}
Based on Eq. (\ref{eq:Dirac}), one knows that the Dirac Hamiltonian for a free particle is given by
\begin{eqnarray}\label{eq:Dirac-H0}
H_{\rm Dirac}=c\vec{\alpha} \cdot \vec{p} + \beta m c^2.
\end{eqnarray}
However, based on Eq. (\ref{eq:Dirac-3}), one knows that the Dirac Hamiltonian with gauge potential is given by
\begin{eqnarray}\label{eq:Dirac-Hg}
H_{\rm Dirac}^g= c\vec{\alpha} \cdot\left( \vec{p} - \frac{g}{c}\vec{A}\right) + \beta m c^2 + g\varphi.
\end{eqnarray}
When $g=0$ or without gauge potentials, $H_{\rm Dirac}^g$ naturally reduces to the free Dirac Hamiltonian $H_{\rm Dirac}$.
$\blacksquare$
\end{remark}

\begin{remark}
\textcolor{blue}{The Explanation of Physical Dimension.} Let us consider the exponential structure of gauge transformation
\begin{eqnarray}
U = e^{{\rm i} g \theta^a T^a},
\end{eqnarray}
where the Lie algebra generators $T^a$'s are dimensionless. (i) Since the quantity in the exponent must be dimensionless,
one must have
\begin{eqnarray}
[g \theta] = 1.
\end{eqnarray}
(ii) Furthermore, since
\begin{eqnarray}
A_{\mu}' = UA_{\mu}U^{-1} + {\rm i} \frac{c\hbar}{g}(\partial_{\mu}U)U^{-1},
\end{eqnarray}
we require the following dimensional equality
\begin{eqnarray}
\left[\frac{c\hbar}{g}\partial_{\mu}\right] = [A_{\mu}].
\end{eqnarray}
(iii) Finally, based on Eq. (\ref{eq:Dirac-Hg}) we know that $gA_{\mu}$ has the physical dimension of energy $E$, i.e.,
\begin{eqnarray}
[gA_{\mu}] = [E]
\end{eqnarray}
Based on the above three points, we find that in our definition of the covariant derivative
\begin{eqnarray}
\partial_{\mu} \mapsto D_{\mu} = \partial_{\mu} + \frac{{\rm i} g}{c\hbar}A_{\mu}, \quad \mu=0, 1, 2, 3,
\end{eqnarray}
the terms $\partial_{\mu}$ and $\dfrac{{\rm i} g}{c\hbar}A_{\mu}$ have the same dimension. This aligns with physical intuition. $\blacksquare$
\end{remark}

\subsection{Gauge Covariance of $F_{\mu\nu}$ and Gauge Invariance of the Yang-Mills Equations}

Now we start to discuss the gauge covariance of the field strength tensor. Keep the same notation with previous sections, we denote the gauge covariant derivative $D_{\mu}$ as
\begin{eqnarray}\label{eq:cd-1}
D_{\mu} = \partial_{\mu} + \frac{{\rm i} g}{c\hbar}A_{\mu}, \quad (\mu=0, 1, 2, 3).
\end{eqnarray}
Based on which, we define the field strength tensor as the commutator of the gauge covariant derivative
\begin{eqnarray}
\frac{{\rm i} g}{c\hbar} F_{\mu\nu} = [D_\mu, D_\nu], \quad (\mu, \nu=0, 1, 2, 3).
\end{eqnarray}
After some calculation, we can have the field strength tensor as
\begin{eqnarray}\label{eq:Fmn-1}
 F_{\mu\nu} &=& \frac{c\hbar}{{\rm i} g} [D_\mu, D_\nu] \nonumber\\
 &=& \frac{c\hbar}{{\rm i} g} \left[\partial_{\mu} + \frac{{\rm i} g}{c\hbar}A_{\mu}, \partial_{\nu} + \frac{{\rm i} g}{c\hbar}A_{\nu}\right] \nonumber \\
 &=& \partial_\mu A_\nu - \partial_\nu A_\mu + \frac{{\rm i} g}{\hbar c} [A_\mu, A_\nu], \quad\quad\quad (\mu, \nu=0, 1, 2, 3).
\end{eqnarray}

\begin{remark}
Let us discuss the gauge transformation of the gauge covariant derivative. Because
\begin{eqnarray}
 D_{\mu}'(\psi') = U D_{\mu}(\psi),
\end{eqnarray}
and consider that $\psi = U^{-1} \psi'$, we can have
\begin{eqnarray}
 D_{\mu}' \psi' = U D_{\mu} U^{-1} \psi'.
\end{eqnarray}
Since $\psi'$ is arbitrary, we finally have
\begin{eqnarray}
 D_{\mu}' = U D_{\mu} U^{-1}.
\end{eqnarray}
$\blacksquare$
\end{remark}

\begin{remark}
Based on the former remark, we can have
\begin{equation}
F'_{\mu\nu} = [D_\mu', D_\nu'] = [U D_\mu U^{-1}, U D_\nu U^{-1}] = U [D_\mu, D_\nu] U^{-1} = U F_{\mu\nu} U^{-1}.
\end{equation}
Then the gauge transformation of the field strength tensor is given by
\begin{equation}
F_{\mu\nu}' = U F_{\mu\nu} U^{-1}.
\end{equation}
Thus, the field strength tensor $F_{\mu\nu}$ is gauge covariant.
$\blacksquare$
\end{remark}

\begin{remark}
Now we denote the adjoint gauge covariant derivative (because it acts on objects in the adjoint representation, ensuring the covariance of operations on the field strength tensor. We still use the notation $D_\mu$) as
\begin{eqnarray}
D_\mu=\partial_\mu+\frac{{\rm i}g}{\hbar c} \left[A_\mu,\; \right].
\end{eqnarray}
Based on that, we can check that the adjoint gauge covariant derivative acting on the field strength tensor transforms covariantly. We can have
\begin{eqnarray}
D'_\mu=\partial_\mu+\frac{{\rm i}g}{\hbar c} \left[A'_\mu,\; \right],
\end{eqnarray}
then
\begin{eqnarray}
D_\mu'F'^{\mu\nu}=\partial_\mu F'^{\mu\nu}+\frac{{\rm i} g}{\hbar c} \left[A'_\mu, F'^{\mu\nu}\right],
\end{eqnarray}
i.e.,
\begin{eqnarray}
D_\mu'F'^{\mu\nu}=\partial_\mu \left(U F^{\mu\nu}U^{-1}\right)+\frac{{\rm i}g}{\hbar c}\left[UA_{\mu}U^{-1}
+{\rm i}\frac{c\hbar}{g}(\partial_{\mu}U)U^{-1}, U F^{\mu\nu}U^{-1}\right].
\end{eqnarray}
Notice that
\begin{eqnarray}
\frac{{\rm i} g}{\hbar c}\left[{\rm i}\frac{c\hbar}{g}(\partial_{\mu}U)U^{-1}, U F^{\mu\nu}U^{-1}\right]=-\left[(\partial_{\mu}U)U^{-1}, U F^{\mu\nu}U^{-1}\right],
\end{eqnarray}
\begin{eqnarray}
\partial_\mu \left(UF^{\mu\nu}U^{-1}\right)=(\partial_\mu U)F^{\mu\nu}U^{-1}+U\partial_\mu (F^{\mu\nu})U^{-1}+ U F^{\mu\nu}(\partial_\mu U^{-1}).
\end{eqnarray}
and
\begin{eqnarray}
\partial_\mu(UU^{-1})=0=\partial_\mu(U)U^{-1}+U\partial_\mu(U^{-1}),
\end{eqnarray}
we can find
\begin{eqnarray}
U^{-1}\partial_\mu(U)U^{-1}+\partial_\mu(U^{-1})=0.
\end{eqnarray}
Then we can have
\begin{eqnarray}
U F^{\mu\nu}U^{-1}(\partial_{\mu}U)U^{-1}+ U F^{\mu\nu}(\partial_\mu U^{-1})=0.
\end{eqnarray}
As a result, we can have
\begin{eqnarray}
D_\mu'F'^{\mu\nu} &=& \partial_\mu \left(U F^{\mu\nu}U^{-1}\right)+\frac{{\rm i}g}{\hbar c}\left[UA_{\mu}U^{-1}
+{\rm i}\frac{c\hbar}{g}(\partial_{\mu}U)U^{-1}, UF^{\mu\nu}U^{-1}\right] \nonumber\\
&=&\partial_\mu \left(U F^{\mu\nu}U^{-1}\right)+\frac{{\rm i}g}{\hbar c} \left[UA_{\mu}U^{-1}, UF^{\mu\nu}U^{-1}\right]-\left[(\partial_{\mu}U)U^{-1}, UF^{\mu\nu}U^{-1}\right]\nonumber\\
&=&\partial_\mu \left(U F^{\mu\nu}U^{-1}\right)-\left[(\partial_{\mu}U)U^{-1}, UF^{\mu\nu}U^{-1}\right]
+\frac{{\rm i}g}{\hbar c} \left[UA_{\mu}U^{-1}, UF^{\mu\nu}U^{-1}\right]\nonumber\\
&=&[(\partial_\mu U)F^{\mu\nu}U^{-1}+U\partial_\mu (F^{\mu\nu})U^{-1}+ UF^{\mu\nu}(\partial_\mu U^{-1})]\nonumber\\
&& -[(\partial_{\mu}U)U^{-1}UF^{\mu\nu}U^{-1}-UF^{\mu\nu}U^{-1}(\partial_{\mu}U)U^{-1}]+
\frac{{\rm i}g}{\hbar c} \left[UA_{\mu}U^{-1}, UF^{\mu\nu}U^{-1}\right]\nonumber\\
&=&[(\partial_\mu U)F^{\mu\nu}U^{-1}+U\partial_\mu (F^{\mu\nu})U^{-1}+ UF^{\mu\nu}(\partial_\mu U^{-1})]\nonumber\\
&& -[(\partial_{\mu}U)F^{\mu\nu}U^{-1}-UF^{\mu\nu}U^{-1}(\partial_{\mu}U)U^{-1}]+\frac{{\rm i}g}{\hbar c} \left[UA_{\mu}U^{-1}, UF^{\mu\nu}U^{-1}\right]\nonumber\\
&=&[U\partial_\mu (F^{\mu\nu})U^{-1}+ UF^{\mu\nu}(\partial_\mu U^{-1})]+UF^{\mu\nu}U^{-1}(\partial_{\mu}U)U^{-1}
+\frac{{\rm i}g}{\hbar c} \left[UA_{\mu}U^{-1}, UF^{\mu\nu}U^{-1}\right]\nonumber\\
&=&U\partial_\mu (F^{\mu\nu})U^{-1}+\frac{{\rm i}g}{\hbar c} \left[UA_{\mu}U^{-1}, UF^{\mu\nu}U^{-1}\right]\nonumber\\
&=&U \left\{\partial_\mu (F^{\mu\nu})+\frac{{\rm i}g}{\hbar c} \left[A_{\mu}, F^{\mu\nu}\right]\right\} U^{-1},
\end{eqnarray}
i.e.,
\begin{eqnarray}
{D'_\mu F'^{\mu\nu}}=U D_\mu F^{\mu\nu}U^{-1}.
\end{eqnarray}

The Yang-Mills equations without source are given by
\begin{eqnarray}
D_\mu F^{\mu\nu}=0.
\end{eqnarray}
Based on our past discussion, we can have
\begin{eqnarray}
D_\mu F^{\mu\nu}=0 \Rightarrow D_\mu'F'^{\mu\nu}=0.
\end{eqnarray}
Then it is clear that the Yang-Mills equations are gauge invariant. Additionally, by observing the indices of the equations, we can find that it is also Lorentz covariant.
$\blacksquare$
\end{remark}

\newpage

\section{The Yang-Mills Equations and A Simple SU(2) Static Solution }\label{sec-4}

\subsection{The Yang-Mills Equations}

The Yang-Mills equations in vacuum without sources (i.e., the 4-vector current ${J^\nu}=0$) are given by \cite{Yang1954}
      \begin{subequations}
            \begin{align}
                  & D_\mu {F}^{\mu\nu} \equiv\partial_\mu {F}^{\mu\nu}
                        + {\frac{{\rm i} g}{c\hbar}}\Bigl[{A}_\mu ,\ {F}^{\mu\nu}\Bigr]=0,
                        \label{eq:YM1a} \\
                  & D_\mu {F}_{\nu\gamma} +D_\nu {F}_{\gamma\mu}
                        +D_\gamma {F}_{\mu\nu} =0. \label{eq:YM1b}
            \end{align}
      \end{subequations}
Here the covariant derivative $D_{\mu}$ is defined as (i.e., Eq. (\ref{eq:cd-1}))
\begin{eqnarray}
D_{\mu}=\partial_{\mu}+ \frac{{\rm i} g}{c\hbar} {A}_{\mu},  \quad\;\;\;\;  (\mu=0, 1, 2, 3),
\end{eqnarray}
and the field strength tensor is (i.e., Eq. (\ref{eq:Fmn-1}))
      \begin{equation}\label{eq:nonabelian}
            {F}_{\mu\nu} =\partial_\mu {A}_{\nu}
                  -\partial_\nu {A}_{\mu}
                  + \frac{{\rm i} g}{c\hbar} [{A}_{\mu},\ {A}_{\nu}], \;\;\;\;\; (\mu, \nu=0, 1, 2, 3),
      \end{equation}
where $g$ is the gauge coupling parameter, ${A}_\mu=(\varphi, -\vec{A})$, $\varphi$ is the \emph{scalar potential}, and $\vec{A}=(A_x, A_y, A_z)$ is the \emph{vector potential}.

The matrix form of field tensor $F_{\mu\nu}$ is given by \cite{1999Jackson}
   \begin{equation}
       F_{\mu\nu}=\begin{bmatrix}
            0 & {E}_x & {E}_y & {E}_z \\
            -{E}_x & 0 & -{B}_z & {B}_y \\
            -{E}_y & {B}_z & 0 & -{B}_x \\
            -{E}_z & -{B}_y & {B}_x & 0
      \end{bmatrix},
   \end{equation}
   or
     \begin{equation}
                        {F}^{\mu\nu} =\begin{bmatrix}
                              0 & -{E}_x & -{E}_y & -{E}_z \\
                              {E}_x & 0 & -{B}_z & {B}_y \\
                              {E}_y & {B}_z & 0 & -{B}_x \\
                              {E}_z & -{B}_y & {B}_x & 0
                        \end{bmatrix}.
                  \end{equation}
Based on which, we can have the vector forms of ``magnetic'' field and ``electric'' field as
      \begin{subequations}
            \begin{eqnarray}
                  \vec{{B}} &=& \vec{\nabla}\times\vec{{A}}
                        - \frac{{\rm i} g}{c\hbar}\left(\vec{{A}}\times\vec{{A}}\right),
                      \label{eq:mag-2}  \\
                  \vec{{E}} &=& -\dfrac{1}{c}
                              \dfrac{\partial\,\vec{{A}}}{\partial\,t}
                        -\vec{\nabla}\varphi- \frac{{\rm i} g}{c\hbar} \left[\varphi,\ \vec{{A}}
                        \right], \label{eq:ele-2}
            \end{eqnarray}
      \end{subequations}
Accordingly, we can write down the Yang-Mills equations in terms of $\{\vec{B}, \vec{E}, \vec{A}, \varphi\}$ as
      \begin{subequations}
            \begin{eqnarray}
                  && \vec{\nabla}\cdot\vec{E}\ {-}\boxed{{\rm i}\,\frac{g}{\hbar c}\Bigl(
                        \vec{A}\cdot\vec{E}
                        -\vec{E}\cdot\vec{A}\Bigr)}=0, \label{eq:DivEYM-a} \\
                  && -\dfrac{1}{c} \dfrac{\partial}{\partial\,t} \vec{B}
                        -\vec{\nabla}\times\vec{E} {-}\boxed{{\rm i}\,\frac{g}{\hbar c}\biggl(\Bigl[\varphi,\
                                    \vec{B}\Bigr]
                              -\vec{A}\times\vec{E}
                              -\vec{E}\times\vec{A}\biggr)}=0, \label{eq:CurlEYM} \\
                  && \vec{\nabla}\cdot\vec{B}\ {-}\boxed{{\rm i}\,\frac{g}{\hbar c}\Bigl(
                        \vec{A}\cdot\vec{B}
                        -\vec{B}\cdot\vec{A}\Bigr)}=0, \label{eq:DivBYM} \\
                  && -\dfrac{1}{c} \dfrac{\partial}{\partial\,t} \vec{E}
                        +\vec{\nabla}\times\vec{B}  {-}\boxed{{\rm i}\,\frac{g}{\hbar c}\biggl(\Bigl[\varphi,\
                                    \vec{E}\Bigr]
                              +\vec{A}\times\vec{B}
                              +\vec{B}\times\vec{A}\biggr)}=0. \label{eq:CurlBYM-a}
            \end{eqnarray}
      \end{subequations}
Here the Yang-Mills equation (\ref{eq:YM1a}) yields Eq. (\ref{eq:DivEYM-a}) and Eq. (\ref{eq:CurlBYM-a}), the Yang-Mills equation (\ref{eq:YM1b}) is just the Bianchi identity, which yields Eq. (\ref{eq:CurlEYM}) and Eq. (\ref{eq:DivBYM}).

\begin{remark} \textcolor{blue}{The Reduction to Maxwell-Type Equations.}
If one neglects the ``boxed'' terms (or just let $g=0$), then Eqs. (\ref{eq:DivEYM})-(\ref{eq:CurlBYM}) reduce to the usual forms of Maxwell's equations (in vacuum without sources). Namely,
\begin{subequations}
            \begin{eqnarray}
                  && \vec{\nabla}\cdot\vec{{E}}=0, \label{eq:DivEYM-M} \\
                  && -\dfrac{1}{c} \dfrac{\partial \vec{{B}}}{\partial\,t}
                        -\vec{\nabla}\times\vec{{E}}=0, \label{eq:CurlEYM-M} \\
                  && \vec{\nabla}\cdot\vec{{B}}=0, \label{eq:DivBYM-M} \\
                  && -\dfrac{1}{c} \dfrac{\partial \textcolor{blue}{\vec{E}}}{\partial\,t}
                        +\vec{\nabla}\times\vec{{B}}=0. \label{eq:CurlBYM-M}
            \end{eqnarray}
\end{subequations}
$\blacksquare$
\end{remark}

\begin{remark} \textcolor{blue}{Eq. (\ref{eq:CurlEYM}) and Eq. (\ref{eq:DivBYM}) are satisfied automatically.} Eq. (\ref{eq:CurlEYM}) and Eq. (\ref{eq:DivBYM}) are originated from the Bianchi identity (\ref{eq:YM1b}). Now we check that the Bianchi identity is automatically satisfied.
We need to prove
\begin{eqnarray}
D_\mu {F}_{\nu\gamma} + D_\nu {F}_{\gamma\mu} + D_\gamma {F}_{\mu\nu}
&=& 0,
\end{eqnarray}
Due to
\begin{eqnarray}
&& D_\mu {F}_{\nu\gamma} = [D_\mu, {F}_{\nu\gamma}], \nonumber\\
&& F_{\mu\nu} = \frac{c\hbar}{{\rm i} g} [D_\mu, D_\nu],
\end{eqnarray}
we need to prove
\begin{eqnarray}
D_\mu {F}_{\nu\gamma} + D_\nu {F}_{\gamma\mu} + D_\gamma {F}_{\mu\nu}
&=& \left[D_\mu, \left(\frac{c\hbar}{{\rm i} g} [D_\nu, D_\gamma]\right)\right]+
\left[D_\nu, \left(\frac{c\hbar}{{\rm i} g} [D_\gamma, D_\mu]\right)\right]+
\left[D_\gamma, \left(\frac{c\hbar}{{\rm i} g} [D_\mu, D_\nu]\right)\right]=0,
\end{eqnarray}
i.e.,
\begin{eqnarray}\label{jacobi}
[D_\mu, [D_\nu, D_\gamma]]+ [D_\nu, [D_\gamma, D_\mu]]+ [D_\gamma, [D_\mu, D_\nu]]=0.
\end{eqnarray}
Eq. (\ref{jacobi}) is just the Jacobi identity. This ends the proof. $\blacksquare$
\end{remark}

\begin{remark} Therefore, to find the solution of the Yang-Mills equations, we only need to focus on Eq. (\ref{eq:DivEYM-a}) and Eq. (\ref{eq:CurlBYM-a}), namely,
      \begin{subequations}
            \begin{eqnarray}
                  && \vec{\nabla}\cdot\vec{E}\ {-}\boxed{{\rm i}\,\frac{g}{\hbar c}\Bigl(
                        \vec{A}\cdot\vec{E}
                        -\vec{E}\cdot\vec{A}\Bigr)}=0, \label{eq:DivEYM-b} \\
                  && -\dfrac{1}{c} \dfrac{\partial}{\partial\,t} \vec{E}
                        +\vec{\nabla}\times\vec{B}  {-}\boxed{{\rm i}\,\frac{g}{\hbar c}\biggl(\Bigl[\varphi,\
                                    \vec{E}\Bigr]
                              +\vec{A}\times\vec{B}
                              +\vec{B}\times\vec{A}\biggr)}=0. \label{eq:CurlBYM-b}
            \end{eqnarray}
      \end{subequations}
$\blacksquare$
\end{remark}

\subsection{A Simple SU(2) Static Solution of the Yang-Mills Equations }

In \cite{Zhou2025}, the researchers \textcolor{blue}{have} presented a simple SU(2) static solution of the Yang-Mills equations. In this section, we would like to make a brief review and also present a direct verification. Since the Bianchi identity (\ref{eq:YM1b}) is automatically \textcolor{blue}{satisfied}, we only need to solve the Yang-Mills equation (\ref{eq:YM1a}), i.e., Eqs. (\ref{eq:DivEYM-b}) and (\ref{eq:CurlBYM-b}).

For simplicity, let us denote a new parameter
\begin{eqnarray}
\tilde{g}=\frac{g}{c\hbar},
\end{eqnarray}
then Eqs. (\ref{eq:DivEYM-b}) and (\ref{eq:CurlBYM-b}) can be simplified as
      \begin{subequations}
            \begin{eqnarray}
                  && \vec{\nabla}\cdot\vec{E}\ {-}{{\rm i}\,\tilde{g}\Bigl(
                        \vec{A}\cdot\vec{E}
                        -\vec{E}\cdot\vec{A}\Bigr)}=0, \label{eq:DivEYM-c} \\
                  && -\dfrac{1}{c} \dfrac{\partial}{\partial\,t} \vec{E}
                        +\vec{\nabla}\times\vec{B}  {-}{{\rm i}\,\tilde{g}\biggl(\Bigl[\varphi,\
                                    \vec{E}\Bigr]
                              +\vec{A}\times\vec{B}
                              +\vec{B}\times\vec{A}\biggr)}=0, \label{eq:CurlBYM-c}
            \end{eqnarray}
\end{subequations}
with the ``magnetic'' and ``electric'' fields as
\begin{eqnarray} \label{eq:magele}
\vec{{B}} &=& \vec{\nabla}\times\vec{{A}}
                        - {\rm i} \tilde{g} \left(\vec{{A}}\times\vec{{A}}\right), \nonumber\\
\vec{{E}} &=& -\dfrac{1}{c} \dfrac{\partial\,\vec{{A}}}{\partial\,t}
-\vec{\nabla}\varphi- {\rm i} \tilde{g} \left[\varphi,\ \vec{{A}}\right].
\end{eqnarray}
The simple SU(2) static solution is given by $\mathcal{S}_{\rm YM}= \{\vec{A}, \, \varphi\}$ \cite{Zhou2025}, where the gauge potentials are given by
\begin{eqnarray}\label{eq:Aphi-1}
\vec{A}=\tilde{k}\frac{\vec{r}\times\vec{S}}{r^2}, \quad\;\;\;\;  \varphi=\frac{\kappa}{r},
\end{eqnarray}
or
\begin{eqnarray}
\vec{A}=k\frac{\vec{r}\times\vec{\Gamma}}{r^2}, \quad\;\;\;\;  \varphi=\frac{\kappa}{r},
\end{eqnarray}
under the following constraint condition
\begin{eqnarray}\label{eq:constr}
\tilde{g} \hbar \tilde{k}=1, \;\;\; {\rm or} \;\;\; 2.
 \end{eqnarray}
Here $\kappa$ and $\tilde{k}$ are real constant numbers, $\varphi$ is the Coulomb-type potential, $\vec{r}$ is position operator with $r=|\vec{r}|$,  $\vec{S}=(S_x, S_y, S_z)$ is the spin-$s$ angular momentum operator (with the spin value $s=0, 1/2, 1, ...$), which satisfies the following
commutation relations
\begin{eqnarray}\label{angularm1}
&& [S_x, S_y]= \mathrm{i} \hbar\; S_z, \;\;\; [S_y, S_z]= \mathrm{i} \hbar\; S_x,\;\;\;  [S_z, S_x]= \mathrm{i} \hbar\; S_y.
\end{eqnarray}
or in the vector form  as
\begin{eqnarray}
&& \vec{S} \times \vec{S} = \mathrm{i} \hbar\; \vec{S}.
\end{eqnarray}
The square of $\vec{S}$ satisfies
\begin{eqnarray}
&& \vec{S}^{\,2} = s(s+1) \hbar^2\, \mathbb{I},
\end{eqnarray}
with $\mathbb{I}$ being the identity matrix. The operator $\Gamma$ is defined through
\begin{equation}
      \vec{S} = \frac{\hbar}{2} \, \vec{\Gamma},
\end{equation}
hence $\vec{\Gamma}$ is dimensionless, and the parameter
\begin{eqnarray}
&& \tilde{k}=k\frac{2}{\hbar}.
\end{eqnarray}
Specially, when $s=1/2$, one has $\vec{S}^2 =(3/4) \hbar^2\, \openone$, $\openone$ is the $2\times 2$ identity matrix, and the operator $\vec{\Gamma}$ can be realized by
\begin{equation}
      \vec{\Gamma} = \vec{\sigma},
\end{equation}
where $\vec{\sigma}=(\sigma_x, \sigma_y, \sigma_z)$ is the vector of Pauli matrices.

In the following, let us verify that Eq. (\ref{eq:Aphi-1}) is a simple solution of Eqs. (\ref{eq:DivEYM-c}) and (\ref{eq:CurlBYM-c}).
\begin{proof}
(i) \textcolor{blue}{Verification of Eq. (\ref{eq:DivEYM-c}).} Because
\begin{eqnarray}
-\dfrac{1}{c} \dfrac{\partial\,\vec{{A}}}{\partial\,t}&=&0, \nonumber\\
-\vec{\nabla}\varphi &=&
- \dfrac{\partial }{\partial x}\left(\dfrac{\kappa}{r}\right) \hat{e}_x
-\dfrac{\partial }{\partial y}\left(\dfrac{\kappa}{r}\right) \hat{e}_y
-\dfrac{\partial }{\partial z}\left(\dfrac{\kappa}{r}\right) \hat{e}_z\nonumber\\
&=& - \dfrac{\partial }{\partial r}\left(\dfrac{\kappa}{r}\right) \dfrac{x}{r}\hat{e}_x
-\dfrac{\partial }{\partial r}\left(\dfrac{\kappa}{r}\right) \dfrac{y}{r} \hat{e}_y
-\dfrac{\partial }{\partial r}\left(\dfrac{\kappa}{r}\right) \dfrac{z}{r} \hat{e}_z=\dfrac{\kappa}{r^3} \vec{r},
\nonumber\\
\left[\varphi,\ \vec{{A}}\right] &=& 0,
\end{eqnarray}
then we have the ``electric'' field as
\begin{eqnarray}
\vec{{E}} &=& -\dfrac{1}{c} \dfrac{\partial\,\vec{{A}}}{\partial\,t}
-\vec{\nabla}\varphi- {\rm i} \tilde{g} \left[\varphi,\ \vec{{A}}\right]=\dfrac{\kappa}{r^3} \vec{r},
\end{eqnarray}
which yields
\begin{eqnarray}\label{eq:E1}
 \vec{\nabla}\cdot\vec{E}&=& \vec{\nabla}\cdot \left(\dfrac{\kappa}{r^2} \hat{e}_r\right)
 =\kappa\, \vec{\nabla}\cdot \left(\dfrac{x}{r^3} \hat{e}_x+ \dfrac{y}{r^3} \hat{e}_y+\dfrac{z}{r^3} \hat{e}_z\right)
 = \kappa\,\left[\dfrac{\partial }{\partial x}\left(\dfrac{x}{r^3} \right)+ \dfrac{\partial }{\partial y}\left(\dfrac{y}{r^3} \right)
 +\dfrac{\partial }{\partial z}\left(\dfrac{z}{r^3} \right) \right]\nonumber\\
 &=& \kappa \left[\left(\dfrac{1}{r^3}-3\dfrac{x^2}{r^5} \right)+ \left(\dfrac{1}{r^3}-3\dfrac{y^2}{r^5} \right)+\left(\dfrac{1}{r^3}-3\dfrac{z^2}{r^5} \right)\right]=\kappa \left[\dfrac{3}{r^3}-3\dfrac{x^2+y^2+z^2}{r^5} \right]\nonumber\\
 &=& \kappa \left[\dfrac{3}{r^3}-3\dfrac{1}{r^3} \right]=0.
\end{eqnarray}

Because
\begin{eqnarray}
\vec{r} \cdot \vec{A}= \vec{A} \cdot \vec{r}=0,
\end{eqnarray}
then we have
\begin{eqnarray}\label{eq:E2}
&& \vec{A}\cdot\vec{E}-\vec{E}\cdot\vec{A}=0.
\end{eqnarray}
Due to Eq. (\ref{eq:E1}) and Eq. (\ref{eq:E2}), Eq. (\ref{eq:DivEYM-c}) is verified.

(ii) \textcolor{blue}{Verification of Eq. (\ref{eq:CurlBYM-c}).} Firstly let us calculate the ``magnetic'' field. Because
\begin{eqnarray}
(\vec{\nabla} \times \vec{{A}})_z &=&\tilde{k} \left(\vec{\nabla}\times\frac{\vec{r}\times\vec{S}}{r^2}\right)_z
= \tilde{k}\left[\partial_x\left(\frac{(\vec{r}\times\vec{S})_y}{r^2}\right)-\partial_y\left(\frac{(\vec{r}\times\vec{S})_x}{r^2}\right)\right]\nonumber\\
&=&\tilde{k}\left[\partial_x\left(\frac{z S_x-xS_z}{r^2}\right)-\partial_y\left(\frac{yS_z-zS_y}{r^2}\right)\right]\nonumber\\
&=&\tilde{k}\left[ \frac{x^2 S_z - y^2 S_z - z^2 S_z - 2xz S_x}{r^4}-\frac{x^2 S_z - y^2 S_z + z^2 S_z + 2yz S_y}{r^4}\right]\nonumber\\
&=&\tilde{k}\frac{ -2z^2 S_z - 2xz S_x - 2yz S_y}{r^4}=-2\tilde{k}\frac{ x S_x +y S_y+ z S_z}{r^4}z\nonumber\\
&=&-2\tilde{k} \frac{\vec{r}\cdot \vec{S}}{r^4}z,
\end{eqnarray}
then we have
\begin{eqnarray}
(\vec{\nabla} \times \vec{{A}}) &=&-2k \frac{\vec{r}\cdot \vec{S}}{r^4} {\vec{r}}.
\end{eqnarray}
Because
\begin{eqnarray}
( \vec{{A}} \times \vec{{A}})_z&=& [A_x, A_y]=\tilde{k}^2\frac{1}{r^4}[yS_z-zS_y,z S_x-xS_z]\nonumber\\
&=& {\rm i} \hbar \tilde{k}^2\frac{1}{r^4}(yzS_y+z^2S_z+zxS_x)= {\rm i} \hbar \tilde{k}^2\frac{1}{r^4}(xS_x+yS_y+zS_z)z\nonumber\\
&=& {\rm i} \hbar \; \tilde{k}^2 \; \frac{\vec{r}\cdot\vec{S}}{r^4} z,
\end{eqnarray}
then we have
\begin{eqnarray}
&& \vec{{A}} \times \vec{{A}}= {\rm i} \hbar \; \tilde{k}^2 \; \frac{\vec{r}\cdot\vec{S}}{r^4} \; \vec{r}.
\end{eqnarray}
Consequently, we obtain the ``magnetic'' field as
\begin{eqnarray}
\vec{{B}} &=& \vec{\nabla}\times\vec{{A}} -{\rm i}\,\tilde{g}\bigl(\vec{{A}}\times\vec{{A}}\bigr) = -2 \tilde{k} \frac{\vec{r}\cdot \vec{S}}{r^4}\vec{r} - {\rm i}\,\tilde{g} \left[{\rm i} \hbar \; \tilde{k}^2 \; \frac{\vec{r}\cdot\vec{S}}{r^4} \; \vec{r}\right]\nonumber\\
&=& -2\tilde{k} \frac{\vec{r}\cdot \vec{S}}{r^4}\vec{r} + \tilde{g}  \hbar \; \tilde{k}^2 \; \frac{\vec{r}\cdot\vec{S}}{r^4} \; \vec{r}=(-2+\tilde{g} \hbar \tilde{k})\tilde{k} \frac{\vec{r}\cdot \vec{S}}{r^4}\vec{r}\nonumber\\
&=&-(2-\tilde{g} \hbar \tilde{k})\tilde{k} \frac{\vec{r}\cdot \vec{S}}{r^4}\vec{r}.
\end{eqnarray}

Since $\vec{{E}}$ is time-independent and $[\varphi, \vec{{E}}]=0$, then Eq. (\ref{eq:CurlBYM-c}) becomes
\begin{eqnarray}
 && \vec{\nabla}\times\vec{{B}} {-} {\rm i}\, \tilde{g}\biggl(\vec{{A}}\times\vec{{B}}
                              +\vec{{B}}\times\vec{{A}}\biggr)=0. \label{eq:CurlBYM-1}
\end{eqnarray}
Because
 \begin{eqnarray}
  &&\left(\vec{\nabla}\times\vec{{B}}\right)_z =  \dfrac{\partial {B}_y}{\partial x} - \dfrac{\partial {B}_x}{\partial y}\nonumber\\
  &=&  \dfrac{\partial }{\partial x}\left[-(2-\tilde{g} \hbar \tilde{k})\tilde{k} \frac{\vec{r}\cdot \vec{S}}{r^4}y\right]
  - \dfrac{\partial }{\partial y}\left[-(2-\tilde{g} \hbar \tilde{k})\tilde{k} \frac{\vec{r}\cdot \vec{S}}{r^4}x\right]\nonumber\\
    &=&  -(2-\tilde{g} \hbar \tilde{k})\tilde{k}\dfrac{\partial }{\partial x}\left[ \frac{\vec{r}\cdot \vec{S}}{r^4}\right]y
    +(2-\tilde{g} \hbar \tilde{k})\tilde{k} \dfrac{\partial }{\partial y}\left[\frac{\vec{r}\cdot \vec{S}}{r^4}\right]x\nonumber\\
   &=&  -(2-\tilde{g} \hbar \tilde{k})\tilde{k} \left\{\dfrac{\partial }{\partial x}\left[ \frac{1}{r^4}\right] \left(\vec{r}\cdot \vec{S}\right)y
   +\dfrac{\partial \left(\vec{r}\cdot \vec{S}\right)}{\partial x}\frac{1}{r^4} y \right\}
   +(2-\tilde{g} \hbar \tilde{k})\tilde{k} \left\{\dfrac{\partial }{\partial y}\left[ \frac{1}{r^4}\right] \left(\vec{r}\cdot \vec{S}\right)x
   +\dfrac{\partial \left(\vec{r}\cdot \vec{S}\right)}{\partial y}\frac{1}{r^4} x \right\}\nonumber\\
   &=&  -(2-\tilde{g} \hbar \tilde{k})\tilde{k} \left\{\dfrac{\partial }{\partial r}\left[ \frac{1}{r^4}\right] \dfrac{x}{r}\left(\vec{r}\cdot \vec{S}\right)y
   +S_x\frac{1}{r^4} y \right\}+(2-\tilde{g} \hbar \tilde{k})\tilde{k} \left\{\dfrac{\partial }{\partial r}\left[ \frac{1}{r^4}\right] \dfrac{y}{r}\left(\vec{r}\cdot \vec{S}\right)x
   +S_y\frac{1}{r^4} x \right\}\nonumber\\
     &=& (2-\tilde{g} \hbar \tilde{k})\tilde{k} \frac{1}{r^4} \left( x S_y-y S_x\right)\nonumber\\
   &=& (2-\tilde{g} \hbar \tilde{k})\tilde{k} \frac{1}{r^4} \left(\vec{r}\times \vec{S}\right)_z,
 \end{eqnarray}
 then we have
 \begin{eqnarray}
  \vec{\nabla}\times\vec{{B}}
   &=& (2-\tilde{g} \hbar \tilde{k})\tilde{k} \frac{1}{r^4} \left(\vec{r}\times \vec{S}\right),
 \end{eqnarray}

Because
\begin{eqnarray}
   \vec{{A}}\times\vec{{B}} &=& \left(\tilde{k} \frac{\vec{r} \times \vec{S}}{r^2}\right)\times \left[-(2-\tilde{g} \hbar \tilde{k})\tilde{k} \frac{\vec{r}\cdot \vec{S}}{r^4}\vec{r}\right]\nonumber\\
   &=& -(2-\tilde{g} \hbar \tilde{k})\tilde{k}^2 \dfrac{1}{r^6}\left[\left( \vec{r} \times \vec{S}\right)\times\vec{r}\right] \left(\vec{r}\cdot \vec{S}\right)\nonumber\\
  &=& -(2-\tilde{g} \hbar \tilde{k})\tilde{k}^2 \dfrac{1}{r^6} \left[(\vec{r}\cdot\vec{r})\vec{S}-\vec{r}(\vec{S}\cdot\vec{r})\right] \left(\vec{r}\cdot \vec{S}\right)\nonumber\\
  &=& -(2-\tilde{g} \hbar \tilde{k})\tilde{k}^2  \left[\dfrac{1}{r^4}\vec{S}\left(\vec{r}\cdot \vec{S}\right)-\dfrac{1}{r^6}\vec{r}(\vec{r}\cdot\vec{S})^2\right],\nonumber\\
  \vec{{B}}\times\vec{{A}} &=& - \left(\vec{{A}}\times\vec{{B}}\right)^\dag
   = (2-\tilde{g} \hbar \tilde{k})\tilde{k}^2  \left[\dfrac{1}{r^4}\left(\vec{r}\cdot \vec{S}\right) \vec{S}-\dfrac{1}{r^6}\vec{r}(\vec{r}\cdot\vec{S})^2\right],
\end{eqnarray}
\begin{eqnarray}
   \left[\vec{r} \cdot \vec{S},  {S}_z\right]&=&\left[x S_x+ y S_y+ z S_z,  {S}_z\right]=  (-{\rm i} \hbar) x S_y+ ({\rm i} \hbar) y S_x= -{\rm i} \hbar\left(\vec{r}\times \vec{S}\right)_z,\nonumber\\
\left[\vec{r} \cdot \vec{S},  \vec{S}\right]&=&\left[x S_x+ y S_y+ z S_z,  {S}_z\right]=  (-{\rm i} \hbar) x S_y+ ({\rm i} \hbar) y S_x= -{\rm i} \hbar\left(\vec{r}\times \vec{S}\right),
\end{eqnarray}
thus
\begin{eqnarray}
 && \vec{{A}}\times\vec{{B}} +\vec{{B}}\times\vec{{A}}
  =(2-\tilde{g} \hbar \tilde{k})\tilde{k}^2  \dfrac{1}{r^4} \left[\vec{r}\cdot \vec{S}, \vec{S}\right]=
  {-}({\rm i} \hbar)(2-\tilde{g} \hbar \tilde{k})\tilde{k}^2  \dfrac{1}{r^4}\left(\vec{r}\times \vec{S}\right).
 \end{eqnarray}
 Consequently, one has
\begin{eqnarray}
 && \vec{\nabla}\times\vec{{B}} {-} {\rm i}\, \tilde{g}\biggl(\vec{{A}}\times\vec{{B}}
                              +\vec{{B}}\times\vec{{A}}\biggr)\nonumber\\
 &=&(2-\tilde{g} \hbar \tilde{k})\tilde{k} \frac{1}{r^4} \left(\vec{r}\times \vec{S}\right)
 {-} {\rm i}\, \tilde{g} \left[{-}({\rm i} \hbar)(2-\tilde{g} \hbar \tilde{k})\tilde{k}^2  \dfrac{1}{r^4}\left(\vec{r}\times \vec{S}\right)\right]\nonumber\\
 &=&(2-\tilde{g} \hbar \tilde{k})\tilde{k} \frac{1}{r^4} \left(\vec{r}\times \vec{S}\right)
 - \tilde{g} \hbar\tilde{k} \left[(2-\tilde{g} \hbar \tilde{k})\tilde{k}  \dfrac{1}{r^4}\left(\vec{r}\times \vec{S}\right)\right]\nonumber\\
 &=&(1-\tilde{g} \hbar \tilde{k})(2-\tilde{g} \hbar \tilde{k})\tilde{k} \frac{1}{r^4} \left(\vec{r}\times \vec{S}\right),
 \end{eqnarray}
which is zero under the constraint condition (\ref{eq:constr}), i.e., $\tilde{g} \hbar \tilde{k}=1$, or $2$. This ends the proof.
\end{proof}

\begin{remark}
When $k=0$ (or $\tilde{k}=0$), the static solution
\begin{eqnarray}
\mathcal{S}_{\rm YM}= \left\{\vec{A}=\tilde{k}\frac{\vec{r}\times\vec{S}}{r^2}, \, \varphi=\frac{\kappa}{r}\right\}
\end{eqnarray}
reduces to the static solution of Maxwell's equations, i.e.,
\begin{eqnarray}
\mathcal{S}_{\rm Maxwell}= \left\{\vec{A}=0, \, \varphi=\frac{\kappa}{r}\right\}
\end{eqnarray}
which represents the Coulomb-type potential. Thus, in some sense, Maxwell's equations can predict the existence of the usual Coulomb interaction in Nature, while the Yang-Mills equations can similarly predict a kind of spin-dependent Coulomb-type interaction \cite{Zhou2025}. $\blacksquare$
\end{remark}

\begin{remark}
Due to
\begin{eqnarray}
\tilde{g}=\frac{g}{c\hbar}, \; \quad \tilde{k}=k\frac{2}{\hbar},
\end{eqnarray}
and the constraint condition
\begin{eqnarray}
 \tilde{g} \hbar \tilde{k}=1, \;\; {\rm or} \;\;\; 2,
\end{eqnarray}
 we can have
\begin{eqnarray}\label{eq:constr-1}
 \tilde{g} \hbar \tilde{k}=\frac{g}{c\hbar} \hbar k\frac{2}{\hbar}=\frac{2gk}{\hbar c}=1, \;\; {\rm or} \;\;\; 2,
\end{eqnarray}
i.e.,
\begin{eqnarray}\label{eq:gk-1}
\frac{gk}{\hbar c}=1, \;\; {\rm or} \;\;\;  \frac{1}{2}.
\end{eqnarray}
After substituting this constraint condition into the ``electric'' and ``magnetic'' fields, we have
\begin{eqnarray}
\vec{E}=\dfrac{\kappa}{r^3} \vec{r}, \;\;\;\quad \vec{B} = -\frac{k (\vec{r} \cdot \vec{\Gamma}) }{r^4}  \vec{r},
\;\;\;\quad  {\rm for} \;\;\;\; \tilde{g} \hbar \tilde{k}=1,
\end{eqnarray}
and
\begin{eqnarray}
\vec{E}=\dfrac{\kappa}{r^3} \vec{r}, \;\;\;\quad \vec{B} = 0, \;\;\;\quad {\rm for} \;\;\;\; \tilde{g} \hbar \tilde{k}=2.
\end{eqnarray}
It can be noticed that for the constraint condition $ \tilde{g} \hbar \tilde{k}=2$, the ``magnetic'' field is zero.
$\blacksquare$
\end{remark}

\begin{remark}
Here we discuss some dimensional issues. Based on Eq. (\ref{eq:gk-1}), one has known that the quantity $\frac{gk}{\hbar c}$ is dimensionless, i.e.,
\begin{eqnarray}
\left[\frac{gk}{\hbar c}\right]=1.
\end{eqnarray}
Recalling the fine structure constant
\begin{eqnarray}\label{eq:fs-1}
\alpha_{\rm e} = \dfrac{e^2}{\hbar c} \approx \dfrac{1}{137.0359} \approx \dfrac{1}{137},
\end{eqnarray}
which is a mysterious dimensionless number, we know easily that
\begin{eqnarray}
[\hbar c]=[Q]^2,
\end{eqnarray}
where $[Q]$ has the dimension of electric charge. Therefore, we have
\begin{eqnarray}
[g][k]=[Q]^2.
\end{eqnarray}
If we impose the constraint
\begin{eqnarray}
[g]=[Q],
\end{eqnarray}
then it follows that
\begin{eqnarray}
[k]=[Q].
\end{eqnarray}
That is, in this case both $g$ and $k$ have the dimension of electric charge.
$\blacksquare$
\end{remark}

\newpage

\section{An Intuitive Approach to Extract Vector Potential}\label{sec-5}

The readers might be curious why the spin vector potential (\ref{eq:Aphi-1}) takes the following form
\begin{eqnarray}\label{eq:Aphi-2}
\vec{A}=\tilde{k}\frac{\vec{r}\times\vec{S}}{r^2},
\end{eqnarray}
and where does the spin vector potential come from. Actually, such a simple form appeared in Ref. \cite{Chen2025} as a physical hypothesis by considering a particle with a spin operator $\vec{S}$. In \cite{Chen2025}, an intuitive approach to extract the vector potential from orbital angular momentum has been proposed. For simplicity, we would like to call such an approach as the \emph{vector potential extraction approach} (VPEA).
In this section, let us briefly review and advance study the VPEA, which is crucial to establish the general static solutions for the SU(2) Yang-Mills theory.

\subsection{Brief Review of the Vector Potential Extraction Approach for the Abelian Case}

Let us make a brief review of the \emph{vector potential extraction approach}. The starting point is the orbital angular momentum
\begin{eqnarray}
\vec{\ell}&=&(\ell_x, \ell_y, \ell_z )=\vec{r}\times\vec{p},
\end{eqnarray}
where $\vec{p}$ is just the linear momentum operator. Essentially, the VPEA is a kind of approach mapping the linear momentum $\vec{p}$ to the canonical momentum $\vec{\Pi}$, i.e.,
\begin{eqnarray}
\vec{p} \mapsto \vec{\Pi}= \vec{p}-\frac{q}{c} \vec{A}.
\end{eqnarray}
Now let us illustrate the procedure. Mathematically, $\vec{\ell}$ satisfies the commutation relations of angular momentum operator, i.e.,
\begin{eqnarray}\label{angularm1a}
&& [\ell_x, \ell_y]= \mathrm{i} \hbar\; \ell_z, \;\;\;\;\; [\ell_y, \ell_z]= \mathrm{i} \hbar\; \ell_x, \;\;\;\;\;
 [\ell_z, \ell_x]= \mathrm{i} \hbar\; \ell_y,
\end{eqnarray}
or in the vector form as
\begin{eqnarray}\label{angularm3}
&& \vec{\ell} \times \vec{\ell} = \mathrm{i} \hbar\; \vec{\ell}.
\end{eqnarray}
Now we perform a displacement for the orbital angular momentum $\vec{\ell}$ with an operator $\vec{G}$, i.e.,
\begin{eqnarray}\label{G-2m}
\vec{L}&=&\vec{\ell}+q \vec{G}=\vec{r}\times\vec{p}+q \vec{G},
\end{eqnarray}
where $q$ is a real number (Note: Alternatively one may also absorb the parameter $q$ to operator $G$, such that the displacement simply becomes $\vec{L}=\vec{\ell}+\vec{G}$). Furthermore, we require that the resultant operator $\vec{L}$ is \emph{still} an angular momentum operator, i.e., $\vec{L}$ satisfies
\begin{eqnarray}\label{M-4}
&& \vec{L} \times \vec{L} = \mathrm{i} \hbar\; \vec{L}.
\end{eqnarray}
We then recast Eq. (\ref{G-2m}) to the following equivalent form
\begin{eqnarray}\label{G-2n}
\vec{L}&=&\vec{r}\times\vec{p}+q \vec{G}=\vec{r}\times\vec{p}+q \left(G_r \hat{e}_r+G_\theta \hat{e}_\theta+G_\phi \hat{e}_\phi\right)\nonumber\\
&=&\vec{r}\times\vec{p}+q \left[G_r \hat{e}_r+G_\theta (-\hat{e}_r\times \hat{e}_\phi)+G_\phi (\hat{e}_r\times \hat{e}_\theta)\right]\nonumber\\
&=&\vec{r}\times\left[\vec{p}+\frac{q}{r}\left(-G_\theta \hat{e}_\phi+G_\phi  \hat{e}_\theta\right)\right]+q G_r \hat{e}_r,
\end{eqnarray}
and view the operator
\begin{eqnarray}\label{G-2p}
&& \vec{\Pi}:=\vec{p}+\frac{q}{r}\left(-G_\theta \hat{e}_\phi+G_\phi  \hat{e}_\theta\right)=\vec{p}-\frac{q}{c}\vec{A}
\end{eqnarray}
as the canonical momentum, here $\{\hat{e}_r, \hat{e}_\theta, \hat{e}_\phi\}$ is the spherical coordinate system and $\hat{e}_r=\vec{r}/r$. Thus, from Eq. (\ref{G-2p}) one can extract a vector potential $\vec{A}$ as
\begin{eqnarray}\label{G-2r}
 \vec{A}&=&\frac{c}{r} \left(G_\theta \hat{e}_\phi - G_\phi  \hat{e}_\theta\right)\nonumber\\
&=& \dfrac{c}{r}\left|\begin{array}{ccc}
                  \hat{e}_r & \hat{e}_\theta & \hat{e}_\phi \\
                  1 & 0 & 0 \\
                  G_r & G_\theta & G_\phi
            \end{array}\right| = c \dfrac{\hat{e}_r\times\vec{G}}{r} \nonumber\\
            &=& c \dfrac{\vec{r}\times\vec{G}}{r^2}.
\end{eqnarray}
Based on this approach, we can derive four kinds of vector potentials: (i) the magnetic vector potential $\vec{A}_{\rm M}$, {$(\rho>r_0)$}, used in the magnetic AB effect; (ii) the Wu-Yang U(1) monopole vector potential ($\vec{A}_a$ and $\vec{A}_b$); (iii) the magnetic vector potential $\vec{A}$ corresponding to a uniform magnetic field along the $z$-axis (i.e., $\vec{B}=B \hat{e}_z$); (iv) the vector potential corresponding to a toroidal solenoid. (See these four examples in next section).


\begin{remark}
For the Abelian case (i.e., the U(1) case), the operator $\vec{G}$ only depends on the coordinates $x, y, z$ (or $r, \theta, \phi$), then one simply has
\begin{eqnarray}
 \label{M-5}
 \vec{G}\times \vec{G}=0,
 \end{eqnarray}
or
\begin{eqnarray}
&& [G_i, G_j]=0, \;\;\;\;\;(i, j=x, y, z).
 \end{eqnarray}
Then one has
\begin{eqnarray}
 \label{M-6}
&& \vec{L}\times \vec{L}=(\vec{\ell}+q \vec{G})\times (\vec{\ell}+q \vec{G})=\vec{\ell}\times \vec{\ell}+ q(\vec{\ell} \times \vec{G}+\vec{G}\times \vec{\ell}) +q^2\vec{G}\times\vec{G},
 \end{eqnarray}
i.e.,
\begin{eqnarray}
&& \vec{L}\times \vec{L}=(\vec{\ell}+q \vec{G})\times (\vec{\ell}+q \vec{G})=\vec{\ell}\times \vec{\ell}+ q(\vec{\ell} \times \vec{G}+\vec{G}\times \vec{\ell}),
 \end{eqnarray}
i.e.,
\begin{eqnarray}
 \label{M-6b}
&& \vec{L}\times \vec{L}=\mathrm{i} \hbar \;\vec{\ell}+ q(\vec{\ell} \times \vec{G}+\vec{G}\times \vec{\ell}).
 \end{eqnarray}
By comparing Eq. (\ref{M-4}) and Eq. (\ref{M-6b}), we must have
\begin{eqnarray}
 \label{M-7}
&& \vec{\ell} \times \vec{G}+\vec{G}\times \vec{\ell} =\mathrm{i} \hbar \;\vec{G}.
 \end{eqnarray}

Because
\begin{eqnarray}
 \label{M-8a}
 (\vec{\ell} \times \vec{G})_z&=&\left[(\vec{r}\times \vec{p})\times \vec{G}\right]_z =\vec{r} \cdot (p_z \vec{G})-z (\vec{p}\cdot\vec{G})\nonumber\\
&=& x (p_z G_x)+y (p_z G_y)+z (p_z G_z)-z (p_x G_x)-z (p_y G_y)-z (p_z G_z)\nonumber\\
&=& p_z (x G_x+y G_y)-z (p_x G_x+p_y G_y),
 \end{eqnarray}
\begin{eqnarray}
 \label{M-8b}
 (\vec{G}\times \vec{\ell} )_z &=& \left[\vec{G}\times(\vec{r}\times \vec{p})\right]_z =\vec{G} \cdot(z \;\vec{p})-(\vec{G}\cdot\vec{r})p_z\nonumber\\
&=& G_x (z \;p_x)+G_y (z \;p_y)+G_z (z \;p_z) -(G_x x)p_z-(G_y y)p_z-(G_z z)p_z\nonumber\\
&=&z (G_x p_x+G_y p_y)-(G_x x+ G_y y)p_z\nonumber\\
&=& z (G_x p_x+G_y p_y)-(x G_x + y G_y )p_z,
\end{eqnarray}
then we have
\begin{eqnarray}
 \label{M-9}
 (\vec{\ell} \times \vec{G})_z+(\vec{G}\times \vec{\ell} )_z &=&p_z (x G_x+y G_y)-z (p_x G_x+p_y G_y)+z (G_x p_x+G_y p_y)-(x G_x + y G_y )p_z\nonumber\\
&=&p_z (x G_x+y G_y+z G_z)-z (p_x G_x+p_y G_y+p_z G_z)+z (G_x p_x+G_y p_y+G_z p_z)\nonumber\\
&&  -(x G_x + y G_y +z G_z)p_z-p_z z G_z+ zp_z G_z-z G_zp_z+z G_zp_z\nonumber\\
&=& [p_z, \vec{r}\cdot \vec{G}]-z \;\left([p_x, G_x]+[p_y, G_y]+[p_z, G_z]\right)-[p_z, z] G_z\nonumber\\
&=&- \mathrm{i} \hbar \frac{\partial  (\vec{r}\cdot \vec{G})}{\partial z}+ z \;\mathrm{i} \hbar \left(\frac{\partial G_x}{\partial x}+\frac{\partial G_y}{\partial y}+\frac{\partial G_z}{\partial z}\right)+\mathrm{i} \hbar G_z\nonumber\\
&=& - \mathrm{i} \hbar \frac{\partial  (\vec{r}\cdot \vec{G})}{\partial z}+ z \;\mathrm{i} \hbar\left( \vec{\nabla}\cdot\vec{G}\right)+\mathrm{i} \hbar G_z.
 \end{eqnarray}
Namely, we have
 \begin{eqnarray}
 \label{M-11}
 \vec{\ell} \times \vec{G}+\vec{G}\times \vec{\ell} =- \mathrm{i} \hbar \left[\vec{\nabla} (\vec{r}\cdot \vec{G})\right]+ \mathrm{i} \hbar \vec{r}  \; \left(\vec{\nabla}\cdot\vec{G}\right)+\mathrm{i} \hbar \vec{G},
 \end{eqnarray}
with the gradient operator
\begin{eqnarray}
 \label{M-12}
&&  \vec{\nabla} f={\rm grad} f= \hat{e}_x \frac{\partial f}{\partial x} + \hat{e}_y\frac{\partial f}{\partial y} +\hat{e}_z\frac{\partial f}{\partial z} .
 \end{eqnarray}
By comparing Eq. (\ref{M-7}) and Eq. (\ref{M-11}), we have the condition for $\vec{G}$ as
\begin{eqnarray}
 \label{M-13}
&&  \vec{\nabla} (\vec{r}\cdot \vec{G})= \vec{r} \left(\vec{\nabla}\cdot\vec{G}\right).
 \end{eqnarray}
In the spherical coordinate system, one has
\begin{eqnarray}\label{G-3}
\vec{G}=G_r \hat{e}_r+ G_\theta \hat{e}_\theta+G_\phi \hat{e}_\phi,
\end{eqnarray}
and
\begin{eqnarray}\label{N-1b}
\vec{\nabla}=\hat{e}_r \frac{\partial }{\partial r}+\hat{e}_\theta \frac{1}{r}\frac{\partial }{\partial \theta}+\hat{e}_\phi \frac{1}{r\sin\theta}\frac{\partial }{\partial \phi}.
\end{eqnarray}
One can have the divergence as
\begin{eqnarray}
 \label{G-12}
&&  \vec{\nabla}\cdot\vec{G}=\frac{1}{r^2}\biggr[ \frac{\partial (r^2 G_r)}{\partial r}\biggr]+\frac{1}{r\sin\theta}\biggr[ \frac{\partial (\sin\theta G_\theta)}{\partial \theta}\biggr]+\frac{1}{r\sin\theta}\biggr[ \frac{\partial G_\phi}{\partial \phi}\biggr],
 \end{eqnarray}
and the curl as
\begin{eqnarray}
\vec{\nabla} \times \vec{G} = \frac{1}{r^2 \sin \theta}
\begin{vmatrix}
\hat{e}_r & r \hat{e}_\theta & r \sin \theta \, \hat{e}_\phi \\
\dfrac{\partial}{\partial r} & \dfrac{\partial}{\partial \theta} & \dfrac{\partial}{\partial \phi} \\
G_r & r G_\theta & r \sin \theta \, G_\phi
\end{vmatrix}.
 \end{eqnarray}
or
\begin{eqnarray}
\vec{\nabla} \times \vec{G} = \frac{1}{r \sin \theta} \left[ \frac{\partial}{\partial \theta} (\sin \theta \, G_\phi) - \frac{\partial G_\theta}{\partial \phi} \right] \hat{e}_r
+ \frac{1}{r} \left[ \frac{1}{\sin \theta} \frac{\partial G_r}{\partial \phi} - \frac{\partial}{\partial r} (r G_\phi) \right] \hat{e}_\theta
+ \frac{1}{r} \left[ \frac{\partial}{\partial r} (r G_\theta) - \frac{\partial G_r}{\partial \theta} \right] \hat{e}_\phi.
 \end{eqnarray}
$\blacksquare$
\end{remark}

\begin{remark} Let us consider the case of $G_r=0$. Since $\vec{r}\cdot \vec{G}=0$, then the condition (\ref{M-13}) for $\vec{G}$ becomes
\begin{eqnarray}
  \vec{\nabla}\cdot\vec{G}&=&\frac{1}{r^2}\biggr[ \frac{\partial (r^2 G_r)}{\partial r}\biggr]+\frac{1}{r\sin\theta}\biggr[ \frac{\partial (\sin\theta G_\theta)}{\partial \theta}\biggr]+\frac{1}{r\sin\theta}\biggr[ \frac{\partial G_\phi}{\partial \phi}\biggr]\nonumber\\
&=& \frac{1}{r\sin\theta}\biggr[ \frac{\partial (\sin\theta G_\theta)}{\partial \theta}\biggr]+\frac{1}{r\sin\theta}\biggr[ \frac{\partial G_\phi}{\partial \phi}\biggr]=0,
 \end{eqnarray}
 i.e.,
\begin{eqnarray}
 \frac{\partial (\sin\theta G_\theta)}{\partial \theta}+ \frac{\partial G_\phi}{\partial \phi}=0.
 \end{eqnarray}
For a simple situation, one may let
\begin{eqnarray}
 \label{G-13b}
&&   \frac{\partial (\sin\theta G_\theta)}{\partial \theta} =0, \;\;\;  \frac{\partial G_\phi}{\partial \phi}=0.
 \end{eqnarray}
Thus, it is easy to have the solution of $\vec{G}$ as
\begin{eqnarray}
\label{solution-G-1}
 &&\vec{G}= \frac{W_1(r, \phi)}{\sin\theta} \hat{e}_\theta+  W_2(r, \theta) \hat{e}_\phi,
  \end{eqnarray}
where $W_1(r,\phi)$ is function depending only on $r$ and $\phi$, and $W_2(r,\theta)$ is function depending only on $r$ and $\theta$. In this case, one can express the vector potential as
\begin{eqnarray}\label{eq:AG-1}
 \vec{A}&=&c \dfrac{\vec{r}\times\vec{G}}{r^2}=\frac{c}{r} \left(G_\theta \hat{e}_\phi - G_\phi  \hat{e}_\theta\right)\nonumber\\
  &=& \frac{c}{r} \left( \frac{W_1(r, \phi)}{\sin\theta} \hat{e}_\phi - W_2(r, \theta)  \hat{e}_\theta\right).
\end{eqnarray}
$\blacksquare$
\end{remark}

\subsection{Four Examples for the Abelian Case}

\subsubsection{Example 1}

In the magnetic Aharonov-Bohm (AB) effect, the magnetic vector potential is given by \cite{2005QParadox}
\begin{equation}\label{eq:v-10}
   \vec{A}_{\rm M}= \begin{cases}
        & \dfrac{B \sqrt{x^2+y^2}}{2} \hat{e}_\phi,  \; \;\; (\rho< r_0)\\
        & \\
        &  \dfrac{\Phi_{\rm M}}{2\pi\sqrt{x^2+y^2}} \hat{e}_\phi,\;\; (\rho> r_0)
    \end{cases}
  \end{equation}
where $\hat{e}_\phi= (-\sin\phi, \cos\phi, 0)$, $\rho=r\sin\theta= \sqrt{x^2+y^2}$, $r_0$ is the radius of the solenoid, $B$ is a real number, and
\begin{eqnarray}
 \label{M-3d-a}
\Phi_{\rm M}=B\pi r_0^2
\end{eqnarray}
is the magnetic flux. Here $\rho> r_0$ represents the region outside of the solenoid, and $\rho< r_0$ represents the region inside of the solenoid.

In cylindrical coordinate system, the vector potential is written as
\begin{eqnarray}
\vec{A} &=& A_\rho \hat{e}_\rho+   A_\phi  \hat{e}_\phi+  A_z \hat{e}_z,
\end{eqnarray}
with
\begin{eqnarray}
 A_\rho=0, \;\;\;   \;\;\; A_z =0.
\end{eqnarray}
Based on Eq. (\ref{eq:v-10}), for the region of $\rho> r_0$, the vector potential is given by
\begin{eqnarray}
\vec{A}_{\rm M}=\dfrac{\Phi_{\rm M}}{2\pi\sqrt{x^2+y^2}}=\dfrac{\Phi_{\rm M}}{2\pi \rho},
\end{eqnarray}
one can have the magnetic field as
\begin{eqnarray}
\vec{{B}} &=& \vec{\nabla}\times\vec{{A}}\nonumber\\
&=& \left(\frac{1}{\rho}\frac{\partial A_z}{\partial \phi}-\frac{\partial A_\phi}{\partial z}\right)\hat{e}_\rho+
\left(\frac{\partial A_\rho}{\partial z}-\frac{\partial A_z}{\partial \rho}\right)\hat{e}_\phi+
 \left(\frac{1}{\rho}\frac{\partial (\rho A_\phi)}{\partial \rho}-\frac{1}{\rho}\frac{\partial A_\rho}{\partial \phi}\right)\hat{e}_z\nonumber\\
&=& \left(-\frac{\partial A_\phi}{\partial z}\right)\hat{e}_\rho+
 \left(\frac{1}{\rho}\frac{\partial (\rho A_\phi)}{\partial \rho}\right)\hat{e}_z=0,
\end{eqnarray}
namely, there is actually no magnetic field outside the solenoid.
For Eq. (\ref{eq:AG-1}), by letting
\begin{eqnarray}
 \vec{A} &=& \frac{c}{r} \left( \frac{W_1(r, \phi)}{\sin\theta} \hat{e}_\phi - W_2(r, \theta)  \hat{e}_\theta\right)=\dfrac{\Phi_{\rm M}}{2\pi\sqrt{x^2+y^2}} \hat{e}_\phi,
\end{eqnarray}
we can have
\begin{eqnarray}
   c \, \frac{W_1(r, \phi)}{\rho} \hat{e}_\phi -\frac{c}{r}  W_2(r, \theta)  \hat{e}_\theta =\dfrac{\Phi_{\rm M}}{2\pi \rho} \hat{e}_\phi.
\end{eqnarray}
In other words, by selecting
\begin{eqnarray}
 \label{M-3c}
W_1(r, \phi)= \frac{\Phi_{\rm M}}{2\pi c},\;\;\;\;\;  W_2(r, \theta)=0,
\end{eqnarray}
one obtains
\begin{eqnarray}\label{eq:v-10b}
  \vec{A}= \vec{A}_{\rm M},\;\;\;\;\; {\rm for}\;\;\;  (\rho > r_0).
  \end{eqnarray}
This fact implies that the magnetic vector potential $\vec{A}_{\rm M}$, $(\rho > r_0)$, can be derived strictly from the \emph{vector potential extraction approach}.

\subsubsection{Example 2}

\begin{remark}
Let us consider the case with $G_r\neq 0, G_\theta\neq 0, G_\phi=0$. Namely, we have
\begin{eqnarray}
\label{Gr-1aa}
 &&\vec{G}= G_r \hat{e}_r +G_\theta \hat{e}_\theta.
  \end{eqnarray}
By substituting Eq. (\ref{Gr-1aa}) into Eq. (\ref{M-13}), we have
 \begin{eqnarray}
 \label{Gr-2a}
\hat{e}_r \frac{\partial (r G_r)}{\partial r}+\hat{e}_\theta \frac{1}{r}\frac{\partial (r G_r)}{\partial \theta}+\hat{e}_\phi \frac{1}{r\sin\theta}\frac{\partial (r G_r)}{\partial \phi} = \hat{e}_r \biggr(\frac{1}{r}\biggr[ \frac{\partial (r^2 G_r)}{\partial r}\biggr]+\frac{1}{\sin\theta}\biggr[ \frac{\partial (\sin\theta G_\theta)}{\partial \theta}\biggr]\biggr),
 \end{eqnarray}
which leads to
 \begin{eqnarray}
 \label{Gr-3a}
&& \frac{\partial (r G_r)}{\partial r}= \frac{1}{r}\biggr[ \frac{\partial (r^2 G_r)}{\partial r}\biggr]+\frac{1}{\sin\theta}\biggr[ \frac{\partial (\sin\theta G_\theta)}{\partial \theta}\biggr], \;\;\;\;\;\nonumber\\
&& \frac{\partial (r G_r)}{\partial \theta}=0,\;\;\;\;\;\nonumber\\
&& \frac{\partial (r G_r)}{\partial \phi}=0.
 \end{eqnarray}
The last two equations of (\ref{Gr-3a}) implies that $G_r$ is a function depending only on $r$. We have from the first equation of (\ref{Gr-3a}) that
  \begin{eqnarray}
 \label{Gr-4}
&& \frac{\partial (r G_r)}{\partial r}= \frac{1}{r}\biggr[ \frac{\partial (r^2 G_r)}{\partial r}\biggr]+\frac{1}{\sin\theta}\biggr[ \frac{\partial (\sin\theta G_\theta)}{\partial \theta}\biggr],\nonumber\\
&&\Rightarrow\;\;\;G_r+r \frac{\partial  G_r}{\partial r}=2G_r+r\frac{\partial  G_r}{\partial r}+\frac{1}{\sin\theta}\biggr[ \frac{\partial (\sin\theta G_\theta)}{\partial \theta}\biggr],\nonumber\\
 &&\Rightarrow\;\;\;G_r+\frac{1}{\sin\theta}\biggr[ \frac{\partial (\sin\theta G_\theta)}{\partial \theta}\biggr]=0,\nonumber\\
  &&\Rightarrow\;\;\; \frac{\partial (\sin\theta G_\theta)}{\partial \theta}=-G_r \sin\theta,\nonumber\\
  &&\Rightarrow\;\;\; \int \frac{\partial (\sin\theta G_\theta)}{\partial \theta} d\theta=-\int G_r \sin\theta d\theta,\nonumber\\
  &&\Rightarrow\;\;\; \sin\theta G_\theta=-G_r\int \sin\theta d\theta,\nonumber\\
  &&\Rightarrow\;\;\; \sin\theta G_\theta=G_r (\cos\theta+C), \nonumber\\
  &&\Rightarrow\;\;\;  G_\theta=G_r \frac{\cos\theta+C}{\sin\theta},
\end{eqnarray}
where $C$ is real constant number.

By substituting Eq. (\ref{Gr-4}) into Eq. (\ref{Gr-1aa}), we have
\begin{eqnarray}
\label{Gr-1ab}
 \vec{G}&=& G_r \hat{e}_r +G_\theta \hat{e}_\theta=G_r \hat{e}_r +G_\theta (-\hat{e}_r \times \hat{e}_\phi)\nonumber\\
 &=&G_r \hat{e}_r -G_r \frac{\cos\theta+C}{\sin\theta}(\hat{e}_r \times \hat{e}_\phi) \nonumber\\
 &= & G_r \frac{\vec{r}}{r} -G_r \frac{\cos\theta+C}{r\sin\theta}(\vec{r} \times \hat{e}_\phi),
  \end{eqnarray}
therefore
\begin{eqnarray}
 \label{M-3d}
 \vec{L}&=&\vec{\ell}+q \vec{G}=\vec{r}\times \vec{p}+q\vec{r}\times\left(-G_r \frac{\cos\theta+C}{r\sin\theta} \hat{e}_\phi\right)+q G_r \frac{\vec{r}}{r}\nonumber\\
 &=&\vec{r}\times \left(\vec{p}-q G_r \frac{\cos\theta+C}{r\sin\theta} \hat{e}_\phi\right)+q G_r \frac{\vec{r}}{r}.
\end{eqnarray}
$\blacksquare$
\end{remark}
Let us compare Eq. (\ref{M-3d}) with the Wu-Yang angular momentum operator
\begin{equation}
 \label{mo-eq3aa}
 \vec{L}=\vec{r}\times \left(\vec{p}-Ze \vec{A}\right)-q\frac{\vec{r}}{r},
\end{equation}
we easily find that for
\begin{eqnarray}
 \label{M-3e}
G_r=-1, \;\;\; q=Zeg, \;\;\; C=-1,
\end{eqnarray}
one immediately obtains the Wu-Yang monopole vector potential in the region $a$ as
\begin{eqnarray}
\label{commu-6-a}
 \vec{A}=\vec{A}_a &=& \frac{g}{r} \frac{1-\cos\theta}{\sin\theta}\; \hat{e}_\phi.
 \end{eqnarray}
Similarly, for
\begin{eqnarray}
 \label{M-3f}
G_r=-1, \;\;\; q=Zeg, \;\;\; C=1,
\end{eqnarray}
one immediately obtains the Wu-Yang monopole vector potential in the region $b$ as
\begin{eqnarray}
\label{commu-6-b}
 \vec{A}= \vec{A}_b &=& -\frac{g}{r} \frac{1+\cos\theta}{\sin\theta}\; \hat{e}_\phi.
\end{eqnarray}
Thus, we have strictly derived the Wu-Yang U(1) monopole vector potential based on the \emph{vector potential extraction approach}.

\begin{remark}Let us calculate the corresponding magnetic field. For $\vec{A}= \vec{A}_a$, we have
\begin{eqnarray}
\vec{{B}} &=& \vec{\nabla}\times\vec{{A}}\nonumber\\
&=&  \frac{1}{r \sin \theta} \left[ \frac{\partial}{\partial \theta} (\sin \theta \, A_\phi) - \frac{\partial A_\theta}{\partial \phi} \right] \hat{e}_r
+ \frac{1}{r} \left[ \frac{1}{\sin \theta} \frac{\partial A_r}{\partial \phi} - \frac{\partial}{\partial r} (r A_\phi) \right] \hat{e}_\theta
+ \frac{1}{r} \left[ \frac{\partial}{\partial r} (r A_\theta) - \frac{\partial A_r}{\partial \theta} \right] \hat{e}_\phi \nonumber\\
&=&  \frac{1}{r \sin \theta} \left[ \frac{\partial}{\partial \theta} (\sin \theta \, A_\phi) \right] \hat{e}_r
+ \frac{1}{r} \left[  - \frac{\partial}{\partial r} (r A_\phi) \right] \hat{e}_\theta=\frac{1}{r \sin \theta} \left[ \frac{\partial}{\partial \theta} (\sin \theta \, A_\phi) \right] \hat{e}_r \nonumber\\
&=&   \frac{1}{r \sin \theta} \left[ \frac{\partial}{\partial \theta} \left(\sin \theta \, \frac{g}{r} \frac{1-\cos\theta}{\sin\theta}\right) \right] \hat{e}_r
 \nonumber\\
&=&  \frac{g}{r^2 \sin \theta} \left[ \frac{\partial}{\partial \theta} (1-\cos\theta) \right] \hat{e}_r\nonumber\\
&=&  \frac{g}{r^2 \sin \theta} \sin\theta \hat{e}_r
 \nonumber\\
&=&  g\frac{\vec{r}}{r^3}.
\end{eqnarray}
For $\vec{A}= \vec{A}_b$, we have
\begin{eqnarray}
\vec{{B}} &=& \vec{\nabla}\times\vec{{A}}\nonumber\\
&=&  \frac{1}{r \sin \theta} \left[ \frac{\partial}{\partial \theta} (\sin \theta \, A_\phi) \right] \hat{e}_r
= \frac{1}{r \sin \theta} \left[ \frac{\partial}{\partial \theta} \left(\sin \theta \, \frac{-g}{r} \frac{1+\cos\theta}{\sin\theta}\right) \right] \hat{e}_r
 \nonumber\\
&=&  \frac{-g}{r^2 \sin \theta} \left[ \frac{\partial}{\partial \theta} (1+\cos\theta) \right] \hat{e}_r
=  \frac{g}{r^2 \sin \theta} \sin\theta \hat{e}_r
 \nonumber\\
&=&  g\frac{\vec{r}}{r^3}.
\end{eqnarray}
One finds that the direction of the magnetic field is along the $\hat{e}_r$. $\blacksquare$
\end{remark}

\subsubsection{Example 3}


Let us look at Eq. (\ref{eq:v-10}), for the region of $\rho< r_0$, in cylindrical coordinate system the vector potential is given by
\begin{eqnarray}
\vec{A}=\dfrac{B \sqrt{x^2+y^2}}{2} \hat{e}_\phi= \dfrac{B r \sin\theta}{2} \hat{e}_\phi = \dfrac{B \rho}{2} \hat{e}_\phi,
\end{eqnarray}
or, in Cartesian coordinate system the vector potential is given by
\begin{eqnarray}
\vec{A}&=&\dfrac{B r \sin\theta}{2} \hat{e}_\phi = \dfrac{B r \sin\theta}{2} \left(-\sin\phi \hat{e}_x+\cos\phi \hat{e}_y\right)\nonumber\\
&=&\dfrac{B}{2} \left(-r \sin\theta \sin\phi \hat{e}_x+r \sin\theta\cos\phi \hat{e}_y\right)=
\dfrac{B}{2} \left(-y \hat{e}_x+ x \hat{e}_y\right)\nonumber\\
&=&-\dfrac{B}{2} \left(y \hat{e}_x- x \hat{e}_y\right)= -\frac{B}{2} (\vec{r}\times \hat{z}).
\end{eqnarray}
In cylindrical coordinate system, one can have the magnetic field as
\begin{eqnarray}
\vec{{B}} &=& \vec{\nabla}\times\vec{{A}}\nonumber\\
&=& \left(\frac{1}{\rho}\frac{\partial A_z}{\partial \phi}-\frac{\partial A_\phi}{\partial z}\right)\hat{e}_\rho+
\left(\frac{\partial A_\rho}{\partial z}-\frac{\partial A_z}{\partial \rho}\right)\hat{e}_\phi+
 \left(\frac{1}{\rho}\frac{\partial (\rho A_\phi)}{\partial \rho}-\frac{1}{\rho}\frac{\partial A_\rho}{\partial \phi}\right)\hat{e}_z\nonumber\\
&=& \left(-\frac{\partial A_\phi}{\partial z}\right)\hat{e}_\rho+
 \left(\frac{1}{\rho}\frac{\partial (\rho A_\phi)}{\partial \rho}\right)\hat{e}_z=
 \left(\frac{1}{\rho}\frac{\partial (\rho A_\phi)}{\partial \rho}\right)\hat{e}_z\nonumber\\
&=& \left(\frac{1}{\rho}\frac{\partial (B\rho^2/2)}{\partial \rho}\right)\hat{e}_z=B \hat{e}_z.
\end{eqnarray}
Or, in Cartesian coordinate system, one can have the magnetic field as
\begin{eqnarray}
\vec{{B}} &=& \vec{\nabla}\times\vec{{A}}=\dfrac{B}{2} \vec{\nabla}\times \left(-y \hat{e}_x+x \hat{e}_y\right)\nonumber\\
&=& \dfrac{B}{2}\left|\begin{array}{ccc}
                  \hat{e}_x & \hat{e}_y & \hat{e}_z \\
                  \dfrac{\partial }{\partial x} & \dfrac{\partial }{\partial y} & \dfrac{\partial }{\partial z} \\
                  -y & x & 0
            \end{array}\right| =B \hat{e}_z.
\end{eqnarray}
namely, the magnetic field is a uniform field along the $z$-direction.

Now, it gives rise to natural question: \emph{Can one derive the vector potential corresponding to the uniform magnetic field along the $z$-direction (without considering the constraint of $\rho< r_0$)?} We find that, perhaps due to \emph{some topological reasons}, one is not able to strictly derive the corresponding vector potential, however, one can derive it by taking some reasonable limits from the  Wu-Yang U(1) monopole vector potential.

\begin{remark} \textcolor{blue}{The Difficulty of Strict Derivation.} Let us come to show the difficulty of strict derivation.
Based on the following relation
\begin{eqnarray}
 \vec{A}&=&\frac{c}{r} \left(G_\theta \hat{e}_\phi - G_\phi  \hat{e}_\theta\right)=\dfrac{B \sqrt{x^2+y^2}}{2} \hat{e}_\phi,
\end{eqnarray}
one can have
\begin{eqnarray}
 \frac{c}{r} G_\theta = \dfrac{B \sqrt{x^2+y^2}}{2}, \;\;\;\;\;\;  G_\phi=0,
\end{eqnarray}
i.e.,
\begin{eqnarray}
 \frac{c}{r} G_\theta = \dfrac{B r \sin\theta}{2}, \;\;\;\;\;\;  G_\phi=0,
\end{eqnarray}
i.e.,
\begin{eqnarray}
 G_\theta = \dfrac{B r^2 \sin\theta}{2c}, \;\;\;\;\;\;  G_\phi=0.
\end{eqnarray}

Let us check the condition (\ref{M-13}) requiring by the angular momentum, i.e.,
\begin{eqnarray}\label{eq:cond-1}
&&  \vec{\nabla} (\vec{r}\cdot \vec{G})= \vec{r} \left(\vec{\nabla}\cdot\vec{G}\right).
 \end{eqnarray}
In spherical coordinate system, we can have
\begin{eqnarray}
&&  \vec{\nabla} (r G_r)= \vec{r} \left(\vec{\nabla}\cdot\vec{G}\right),
 \end{eqnarray}
i.e.,
\begin{eqnarray}
&&  \vec{\nabla} (r G_r)= \vec{r} \left(\frac{1}{r^2}\biggr[ \frac{\partial (r^2 G_r)}{\partial r}\biggr]+\frac{1}{r\sin\theta}\biggr[ \frac{\partial (\sin\theta G_\theta)}{\partial \theta}\biggr]+\frac{1}{r\sin\theta}\biggr[ \frac{\partial G_\phi}{\partial \phi}\biggr]\right),
 \end{eqnarray}
i.e.,
\begin{eqnarray}\label{eq:Gr-2}
&&  \vec{\nabla} (r G_r)= \left(\frac{1}{r^2}\biggr[ \frac{\partial (r^2 G_r)}{\partial r}\biggr]+\frac{1}{r\sin\theta}\biggr[ \frac{\partial (\sin\theta G_\theta)}{\partial \theta}\biggr]+\frac{1}{r\sin\theta}\biggr[ \frac{\partial G_\phi}{\partial \phi}\biggr]\right) r \hat{e}_r.
 \end{eqnarray}
Due to
\begin{eqnarray}
&&\vec{\nabla} f = \frac{\partial f}{\partial r} \hat{e}_r + \frac{1}{r} \frac{\partial f}{\partial \theta} \hat{e}_\theta + \frac{1}{r \sin\theta} \frac{\partial f}{\partial \phi} \hat{e}_\phi,
 \end{eqnarray}
from Eq. (\ref{eq:Gr-2}) we have
 \begin{eqnarray}
 \label{Gr-3ab}
&& \frac{\partial (r G_r)}{\partial \theta}=0,\;\;\;\;\;\frac{\partial (r G_r)}{\partial \phi}=0,
 \end{eqnarray}
which implies that ``\emph{$G_r$ must depend only on $r$}''. We then have
\begin{eqnarray}
&& \frac{\partial (r G_r)}{\partial r}= \left(\frac{1}{r^2}\biggr[ \frac{\partial (r^2 G_r)}{\partial r}\biggr]+\frac{1}{r\sin\theta}\biggr[ \frac{\partial (\sin\theta G_\theta)}{\partial \theta}\biggr]+\frac{1}{r\sin\theta}\biggr[ \frac{\partial G_\phi}{\partial \phi}\biggr]\right) r,
 \end{eqnarray}
i.e.,
\begin{eqnarray}
&& \frac{\partial (r G_r)}{\partial r}= \left(\frac{1}{r^2}\biggr[ \frac{\partial (r^2 G_r)}{\partial r}\biggr]+\frac{1}{r\sin\theta}\biggr[ \frac{\partial (\sin\theta \frac{B r^2 \sin\theta}{2c})}{\partial \theta}\biggr]\right) r,
 \end{eqnarray}
i.e.,
\begin{eqnarray}
&& \frac{\partial (r G_r)}{\partial r}= \frac{1}{r}\biggr[ \frac{\partial (r^2 G_r)}{\partial r}\biggr]+\frac{1}{\sin\theta}\frac{B r^2}{2c}\biggr[ \frac{\partial (\sin^2 \theta) }{\partial \theta}\biggr],
 \end{eqnarray}
i.e.,
\begin{eqnarray}
&& G_r+r\frac{\partial ( G_r)}{\partial r}=  2 G_r+r \frac{\partial (G_r)}{\partial r}
+\frac{1}{\sin\theta}\frac{B r^2}{2c} 2 \sin\theta\cos\theta,
 \end{eqnarray}
i.e.,
\begin{eqnarray}\label{eq:Gr-1}
&& G_r=-\frac{B r^2}{c} \cos\theta.
 \end{eqnarray}
However, in Eq. (\ref{eq:Gr-1}) $G_r$ depends on both $r$ and $\theta$, which contradicts with Eq. (\ref{eq:Gr-2}). Thus, the condition (\ref{eq:cond-1}) is not strictly satisfied.

Let us restrict $r$ to the region $\rho < r_0$. The condition (\ref{eq:cond-1}) can be satisfied approximately from viewpoint of physics. Because $r_0$ is the radius of the solenoid, which is very small, i.e., $r_0 \approx 0$. When $\theta$ tends to $\pi/2$ (i.e., $\cos\theta \rightarrow 0$), one has $r \rightarrow r_0$, thus
\begin{eqnarray}
&& G_r=-\frac{B r^2}{c} \cos\theta \approx 0.
\end{eqnarray}
When $\theta \rightarrow 0$ (i.e., $\cos\theta\rightarrow 1$), $r\approx r\cos\theta=z$, by considering the length of the solenoid is finite, i.e., $z$ is finite in practice, then we can approximately have $G_r\approx 0$. However, such an approximation may not seem appropriate. Because the orders of $G_\theta=\frac{B r^2 \sin\theta}{2c}$ and $G_r=-\frac{B r^2}{c} \cos\theta$ are the same, when $G_r\approx 0$, one accordingly have $G_\theta\approx 0$. A more suitable approximation is provided in \emph{Example 4}, where the vector potential and the magnetic field of a linear solenoid can be naturally obtained from those of a toroidal solenoid. $\blacksquare$
\end{remark}

\begin{remark}
\textcolor{blue}{Approximation from the Wu-Yang U(1) Monopole Vector Potential.} Our task is derive the vector potential
\begin{eqnarray}
\vec{A}&=&\dfrac{B \sqrt{x^2+y^2}}{2} \hat{e}_\phi=\dfrac{B r \sin\theta}{2} \hat{e}_\phi = -\frac{B}{2} (\vec{r}\times \hat{z}),
\end{eqnarray}
and the corresponding uniform magnetic field along the $z$-direction
\begin{eqnarray}
\vec{{B}} &=& \vec{\nabla}\times\vec{{A}}=B \hat{e}_z.
\end{eqnarray}
However, the previous remark shows that the strict derivation cannot be achieved. In the following, we would like to derive them approximately from the Wu-Yang U(1) monopole vector potential.

Now, let us study the local approximation of a magnetic monopole near the north pole. Near the north pole of a sphere, the magnetic field and vector potential of a magnetic monopole can be approximated as a uniform magnetic field along the $z$-axis and its corresponding vector potential in symmetric gauge.

\emph{Analysis 1.---\textcolor{blue}{Approximation of the Magnetic Field.}} The magnetic field of a monopole is given by
\begin{eqnarray}
\vec{B}=g\,\frac{\vec{r}}{r^{3}},
\end{eqnarray}
where $g$ is the magnetic charge. On a sphere of radius $R$, the field is radial.
Near the north pole, i.e., the polar angle $\theta\ll 1$, then one has
\begin{eqnarray}
&& \sin \theta \approx \theta,\;\;\;\;\; 1 - \cos \theta \approx \frac{\theta^2}{2},\;\;\;\;\;
\frac{1 - \cos \theta}{\sin \theta} \approx \frac{\theta}{2}.
\end{eqnarray}
We introduce a local Cartesian coordinate system with the $z$-axis pointing radially outward. Then
\begin{eqnarray}
\hat{e}_{r}\approx\hat{e}_z,\qquad r\approx R,
\end{eqnarray}
so that
\begin{eqnarray}
\vec{B}\approx\frac{g}{R^{2}}\,\hat{e}_z.
\end{eqnarray}
This is a uniform magnetic field along the $z$-direction, with magnitude $B=g/R^{2}$.

\emph{Analysis 2.---\textcolor{blue}{Approximation of the Vector Potential.}} The Wu--Yang monopole potential near the north pole is given by
\begin{eqnarray}
\vec{A}_{a}=\frac{g}{r}\,\frac{1-\cos\theta}{\sin\theta}\,\hat{e}_{\phi},
\end{eqnarray}
where $\hat{e}_{\phi}=(-\sin\phi,\cos\phi,0)$ is the azimuthal unit vector. For small $\theta$,
\begin{eqnarray}
\frac{1-\cos\theta}{\sin\theta}\approx\frac{\theta}{2},\qquad r\approx R.
\end{eqnarray}
On the tangent plane, the cylindrical radius is $\rho=R\sin\theta\approx R\theta$, so \(\theta \approx\rho/R\). Substituting,
\begin{eqnarray}
\vec{A}_{a}\approx\frac{g}{R}\cdot\frac{\rho}{2R}\,\hat{e}_{\phi}
=\frac{g\rho}{2R^{2}}\,\hat{e}_{\phi}.
\end{eqnarray}
After expressing $\hat{e}_{\phi}$ in Cartesian coordinates:
\begin{eqnarray}
\hat{e}_{\phi}=\frac{1}{\rho}(-y,x,0),
\end{eqnarray}
we obtain
\begin{eqnarray}
\vec{A}_{a}\approx\frac{g\rho}{2R^{2}}\cdot\frac{1}{\rho}(-y,x,0)
=\frac{g}{2R^{2}}(-y,x,0).
\end{eqnarray}

\emph{Analysis 3.---\textcolor{blue}{Correspondence with the Uniform-Field Vector Potential.}}
After setting
\begin{eqnarray}
B=\frac{g}{R^{2}},
\end{eqnarray}
we have
\begin{eqnarray}
\vec{A}=\frac{B}{2}(-y,x,0),
\end{eqnarray}
which is the symmetric-gauge vector potential for a uniform magnetic field $\vec{B}=B\, \hat{e}_z$. Indeed,
\begin{eqnarray}
\vec{\nabla}\times\vec{A}=B \, \hat{e}_z.
\end{eqnarray}
Thus, near the north pole of the sphere, the monopole fields approximate to
\begin{eqnarray}
\vec{B}\approx\frac{g}{R^{2}}\, \hat{e}_z,\qquad \vec{A}\approx\frac{g}{2R^{2}}(-y,x,0).
\end{eqnarray}
This ends the derivation. $\blacksquare$
\end{remark}

\begin{remark}
The other Wu--Yang potential $\vec{A}_{b}$ in region b is singular at the north pole and is not used in this region. It can be transformed into $\vec{A}_{a}$ by a gauge transformation. This approximation treats the spherical surface locally as a flat plane, neglecting higher-order curvature effects, and is a standard technique in local analysis. $\blacksquare$
\end{remark}

\subsubsection{Example 4}

For a toroidal solenoid, the internal magnetic field in spherical coordinates is given by
\begin{eqnarray}\label{eq:Btoro}
\vec{B} = \frac{\mu_0 NI}{2\pi r \sin\theta} \hat{e}_\phi.
\end{eqnarray}
Here $\mu_0$ is the vacuum permeability, a fundamental physical constant describing the response of vacuum to a magnetic field, with units H/m (henries per meter); $N$ is the total number of turns of the toroidal solenoid, i.e., the total number of windings of the wire; $I$ is the current passing through the solenoid wire, in amperes (A); $r$ is the radial distance (in spherical coordinates) from the origin (the center of the torus) to a point in space; $\theta$ is the polar angle in spherical coordinates, measured from the polar axis (the symmetry axis of the torus). In this formula, $\rho=r\sin\theta$ equals the perpendicular distance from the point to the symmetry axis of the toroidal solenoid (i.e., the radial coordinate
$\rho$ in toroidal coordinates); $\hat{e}_\phi$ is the unit vector in the azimuthal direction in spherical coordinates, perpendicular to both the radial and polar directions, and pointing along the toroidal direction.

The formula (\ref{eq:Btoro}) describes the magnetic field inside a toroidal solenoid. A toroidal solenoid is a coil wound into a ring shape; its internal magnetic field is completely confined within the toroidal volume, directed tangentially along the torus (i.e., the $\hat{e}_\phi$
direction). The magnitude is inversely proportional to the perpendicular distance $r\sin\theta$ from the symmetry axis and proportional to both the current and the number of turns. The factor $2\pi r \sin\theta$ in the denominator is essentially the circumferential length of the magnetic circuit at the given location, reflecting how the magnetic flux density varies with the radial position.

The vector potential of the toroidal solenoid is given by
\begin{eqnarray}
\vec{A} &=& A_r \hat{e}_r+   A_\theta  \hat{e}_\theta+  A_\phi \hat{e}_\phi,
\end{eqnarray}
with
\begin{eqnarray}
A_r=0, \;\;\;\;\; A_\theta=\frac{\mu_0 NI}{2\pi \sin\theta}, \;\;\;\;\;   A_\phi =0.
\end{eqnarray}
One may check that
\begin{eqnarray}
\vec{{B}} &=& \vec{\nabla}\times\vec{{A}}\nonumber\\
&=&  \frac{1}{r \sin \theta} \left[ \frac{\partial}{\partial \theta} (\sin \theta \, A_\phi) - \frac{\partial A_\theta}{\partial \phi} \right] \hat{e}_r
+ \frac{1}{r} \left[ \frac{1}{\sin \theta} \frac{\partial A_r}{\partial \phi} - \frac{\partial}{\partial r} (r A_\phi) \right] \hat{e}_\theta
+ \frac{1}{r} \left[ \frac{\partial}{\partial r} (r A_\theta) - \frac{\partial A_r}{\partial \theta} \right] \hat{e}_\phi \nonumber\\
&=&  \frac{1}{r \sin \theta} \left[ - \frac{\partial A_\theta}{\partial \phi} \right] \hat{e}_r
+ \frac{1}{r} \left[ \frac{\partial}{\partial r} (r A_\theta)\right] \hat{e}_\phi
= \frac{1}{r} \left[ \frac{\partial}{\partial r} (r A_\theta)\right] \hat{e}_\phi \nonumber\\
&=& \frac{1}{r} A_\theta\,  \hat{e}_\phi=\frac{\mu_0 NI}{2\pi r \sin\theta} \hat{e}_\phi,
\end{eqnarray}
which is just Eq. (\ref{eq:Btoro}).

In the following, let us demonstrate that the vector potential can be determined from the \emph{vector potential extraction approach}. To reach this purpose, we only need to prove such an operator $\vec{G}$ exists. Let
\begin{eqnarray}
\vec{G} &=& G_r \hat{e}_r+   G_\theta  \hat{e}_\theta+  G_\phi \hat{e}_\phi,
\end{eqnarray}
with
\begin{eqnarray}
G_r =0.
\end{eqnarray}
Then from
\begin{eqnarray}
\vec{A} &=& c \frac{\vec{r} \times \vec{G}}{r^2} = \frac{c}{r} \left( G_\theta \hat{e}_\phi - G_\phi \hat{e}_\theta \right)
=\frac{\mu_0 NI}{2\pi \sin\theta} \hat{e}_\theta,
\end{eqnarray}
one has
\begin{eqnarray}\label{eq:Gphi}
G_\theta =0, \;\;\;\;\; G_\phi=- \frac{\mu_0 NI r}{2\pi c \sin\theta}.
\end{eqnarray}
Furthermore, we need to check the condition (\ref{M-13}) requiring by the angular momentum, i.e.,
\begin{eqnarray}
&&  \vec{\nabla} (\vec{r}\cdot \vec{G})= \vec{r} \left(\vec{\nabla}\cdot\vec{G}\right).
 \end{eqnarray}
In spherical coordinate system, we can have
\begin{eqnarray}
&&  \vec{\nabla} (r G_r)= \vec{r} \left(\vec{\nabla}\cdot\vec{G}\right).
 \end{eqnarray}
Namely, we need to check
\begin{eqnarray}
&&  \vec{\nabla}\cdot\vec{G}=0,
 \end{eqnarray}
i.e.,
\begin{eqnarray}
&&  \frac{1}{r^2}\biggr[ \frac{\partial (r^2 G_r)}{\partial r}\biggr]+\frac{1}{r\sin\theta}\biggr[ \frac{\partial (\sin\theta G_\theta)}{\partial \theta}\biggr]+\frac{1}{r\sin\theta}\biggr[ \frac{\partial G_\phi}{\partial \phi}\biggr]=0,
 \end{eqnarray}
i.e.,
\begin{eqnarray}
&&  \frac{1}{r\sin\theta}\biggr[ \frac{\partial G_\phi}{\partial \phi}\biggr]=0,
 \end{eqnarray}
i.e.,
\begin{eqnarray}
&&   \frac{\partial G_\phi}{\partial \phi}=0,
 \end{eqnarray}
which is obviously valid due to Eq. (\ref{eq:Gphi}). This ends the demonstration.

\begin{remark} \textcolor{blue}{Transition from a Toroidal Solenoid to an Infinitely Long Straight Solenoid.}
The magnetic field inside a toroidal solenoid is given in spherical coordinates by Eq. (\ref{eq:Btoro}),
where $r\sin\theta = \rho$ is the perpendicular distance from the symmetry axis.
When the toroidal radius \(R\) is very large, the curvature can be neglected locally.
In a small segment near a point on the torus, the toroidal solenoid approximates an infinitely long straight solenoid.
Taking the limit \(R\to\infty\) while keeping the local geometry, the magnetic field becomes
\begin{eqnarray}
\vec{B} \approx \frac{\mu_0 N I}{2\pi R}\,\hat{e}_\phi = B\,\hat{e}_z,
\end{eqnarray}
which is a uniform axial field -- the standard field inside an infinite straight solenoid.
Conversely, a topological argument shows that the infinite solenoid (an infinite cylinder) can be compactified by identifying its two ends at infinity; the resulting quotient space is homeomorphic to a torus, i.e., a toroidal solenoid.
Thus, the two systems are physically equivalent in the sense of a local approximation and also topologically related by compactification. $\blacksquare$
\end{remark}

\begin{remark}\textcolor{blue}{Further Relation between the Infinitely Long Straight Solenoid and the Toroidal Solenoid.}
Physically, the magnetic fields of the infinitely long solenoid and the toroidal solenoid are consistent, differing only in their spatial topology. By gluing the points at infinity together, these two models become perfectly unified.

Here, we give a strict proof. For the infinite solenoid (infinite cylinder), it can be written as
\begin{eqnarray}
C=(\rho\cos\theta,\rho\sin\theta,z)=(\rho\cos\theta,\rho\sin\theta)\times(z), \quad\quad (\rho=\text{const}, \quad\theta\in\mathbb{R}, \quad z\in\mathbb{R}).
\end{eqnarray}
Consider the function
\begin{eqnarray}
f(z) = \frac{2}{\pi}\arctan z.
\end{eqnarray}
It is a homeomorphism from $\mathbb{R}$ to $(-1,1)$. Then we have
\begin{eqnarray}
C \cong (\rho\cos\theta,\rho\sin\theta)\times(f(z)) = (\rho\cos\theta,\rho\sin\theta)\times(z_1),\quad (\rho=\text{const},\; \theta\in\mathbb{R},\; z_1\in(-1,1)).
\end{eqnarray}
After adding the endpoints $\{-1,1\}$, we obtain the compactified cylinder
\begin{eqnarray}
\bar{C} = (\rho\cos\theta,\rho\sin\theta)\times(t),\quad \;\;\; (\rho=\text{const},\; \theta\in\mathbb{R},\; t\in[-1,1]).
\end{eqnarray}
Under this compactification, $z\to -\infty$ corresponds to $t=-1$ and $z\to +\infty$ corresponds to $t=1$.

Now we glue the points at infinity together by defining the equivalence relation
\begin{eqnarray}
\big((\rho\cos\theta,\rho\sin\theta), -1\big) \sim \big((\rho\cos\theta,\rho\sin\theta), 1\big),\qquad \forall\theta\in\mathbb{R}.
\end{eqnarray}
Thus we form the quotient space
\begin{eqnarray}
\tilde{C} = \bar{C}/\sim \; = \big((\rho\cos\theta,\rho\sin\theta)\times[-1,1]\big)/\sim.
\end{eqnarray}
Define a map
\begin{eqnarray}
h\big((\rho\cos\theta,\rho\sin\theta), t\big) = \left( (\rho_1\cos\theta,\rho_1\sin\theta),\; \big(\rho_2\cos(t\pi+\pi),\rho_2\sin(t\pi+\pi)\big) \right),
\end{eqnarray}
where $\rho_1,\rho_2>0$ are constants (e.g., $\rho_1=\rho$, $\rho_2=1$) and $t\in[-1,1]$. Since
\begin{eqnarray}
\cos(t\pi+\pi)\big|{t=-1} &=& \cos(0)=1,\quad \sin(t\pi+\pi)\big|{t=-1} = \sin(0)=0,\nonumber\\
\cos(t\pi+\pi)\big|{t=1} &=& \cos(2\pi)=1,\quad \sin(t\pi+\pi)\big|{t=1} = \sin(2\pi)=0.
\end{eqnarray}
Thus $h$ takes the same value at $t=-1$ and $t=1$, so it descends to a well-defined map on the quotient $\tilde{C}$.

Let $\varphi(t)=(\cos(t\pi+\pi),\sin(t\pi+\pi))$. Then $\varphi:[-1,1]\to S^1$ is surjective and $\varphi(-1)=\varphi(1)$, hence it induces a homeomorphism $[-1,1]/{-1,1}\cong S^1$. Therefore
\begin{eqnarray}
h(\tilde{C}) = \left( (\rho_1\cos\theta,\rho_1\sin\theta), (\rho_2\cos(t\pi+\pi),\rho_2\sin(t\pi+\pi)) \right) \cong S^1\times S^1,
\end{eqnarray}
which is the toroidal solenoid. Thus the infinite solenoid is homeomorphic to a toroidal solenoid after adding the points at infinity and gluing them together. $\blacksquare$
\end{remark}

\subsection{Brief Review of the Vector Potential Extraction Approach for the Non-Abelian Case}

For the non-Abelian case, the approach is similar. The difference is that now the components of $\vec{G}$ are non-abelian operators. For a simple example, let
\begin{eqnarray}
q\vec{G}=\vec{S},
\end{eqnarray}
where $\vec{S}$ is a spin-$s$ angular momentum operator $(s=0, 1/2, 1, 3/2, \cdots)$. In this situation, it is easy to have that the resultant vector
\begin{eqnarray}
\vec{L}&=&\vec{r}\times\vec{p}+q \vec{G}=\vec{r}\times\vec{p}+\vec{S}
\end{eqnarray}
is also an angular momentum operator. Then due to the \emph{vector potential extraction approach}, we can derive the previous spin vector potential
\begin{eqnarray}\label{eq:svp-1}
 \vec{A}&=& c \dfrac{\vec{r}\times\vec{G}}{r^2}=\frac{c}{q} \dfrac{\vec{r}\times\vec{S}}{r^2}.
\end{eqnarray}
Then we explain the form of spin vector potential in Eq. (\ref{eq:Aphi-2}) with
\begin{eqnarray}
\tilde{k}=\frac{c}{q}.
\end{eqnarray}

\subsection{Determining the Most General Spin Vector Potential Based on VPEA}

From previous section, we have known that the spin vector potential takes the following form
\begin{eqnarray}\label{eq:svp-2}
 \vec{A} & \propto&  \dfrac{\vec{r}\times\vec{S}}{r^2},
\end{eqnarray}
which comes form the angular momentum operator
\begin{eqnarray}\label{eq:ls-1}
\vec{L}&=& \vec{\ell}+\vec{S}=\vec{r}\times\vec{p}+\vec{S}
\end{eqnarray}
by the \emph{vector potential extraction approach}. It gives rise to natural question: Is Eq. (\ref{eq:svp-2}) the most general form of the spin vector potential $\vec{A}$? Mathematically, from the viewpoint of VPEA, it is equivalent to ask: Given the orbital angular momentum operator $\vec{\ell}=\vec{r}\times \vec{p}$ and the spin angular momentum operator $\vec{S}$, and we define
\begin{eqnarray}
\vec{L}=F(\vec{\ell}, \vec{S})
\end{eqnarray}
as an operator function of $\vec{\ell}$ and $\vec{S}$, then what is the most general form of $\vec{L}$ if it is still required to be an angular momentum operator? In the next subsection, let us study the general form of $\vec{L}=F(\vec{\ell}, \vec{S})$.

\subsubsection{The General Form of $\vec{L}=F(\vec{\ell}, \vec{S})$}

{Frankly speaking, the expression  of (\ref{eq:ls-1}) is obtained by a simple ``displacement'' (i.e., $\vec{L} = \vec{\ell}+\vec{S}$), which is merely a simple realization of $\vec{L}=F(\vec{\ell}, \vec{S})$. Based on two facts: (i) $\vec{\ell}$ and $\vec{L}$ are angular momentum operator, and (ii) the operator $\vec{L}$ is generated from the operator $\vec{\ell}$, thus the other way to construct $\vec{L}$ is to introduce the unitary transformation $\mathcal{U}$ (i.e., a ``rotation'').} Explicitly, one has
\begin{eqnarray}\label{eq:ls-2}
\vec{L}&=& \mathcal{U} \,\vec{\ell}\,  \mathcal{U}^\dagger.
\end{eqnarray}
Such a unitary transformation acting on $\vec{\ell}$ automatically guarantees that $\vec{L}$ is an angular momentum operator.

Since we expect that the spin vector potential depends only on the position operator $\vec{r}$ and the spin operator $\vec{S}$ (i.e., it does not depend on the momentum operator $\vec{p}$), thus the unitary matrix $\mathcal{U}$ is a function of $\vec{r}$ and $\vec{S}$. In this work, for simplicity, let us restrict to spin-1/2 case, then we may write
\begin{eqnarray}\label{eq:U-1}
\mathcal{U}= {\rm e}^{{\rm i} \theta \vec{\Gamma}\cdot \hat{r}},
\end{eqnarray}
with
\begin{eqnarray}\label{eq:U-1a}
\vec{S}=\frac{\hbar}{2}\, \vec{\Gamma}, \;\;\;\; \hat{r} \equiv \hat{e}_r= \frac{\vec{r}}{r},
\end{eqnarray}
$\vec{\Gamma}\equiv\vec{\sigma}$ is the vector of Pauli matrices, and
\begin{eqnarray}\label{eq:U-1b}
(\vec{\Gamma} \cdot \hat{r})^2=\openone.
\end{eqnarray}
%
%
Because
\begin{eqnarray}
&& \mathcal{U}= {\rm e}^{{\rm i} \theta \vec{\Gamma}\cdot \hat{r}} = \cos\theta \openone+ {\rm i}  \sin\theta (\vec{\Gamma} \cdot \hat{r}), \nonumber\\
&& \mathcal{U}^\dagger= {\rm e}^{-{\rm i} \theta \vec{\Gamma}\cdot \hat{r}}= \cos \theta \openone- {\rm i} \sin \theta (\vec{\Gamma} \cdot \hat{r}),
\end{eqnarray}
we can have
\begin{eqnarray}
&&(\cos\theta \openone+ {\rm i} \vec{\Gamma} \cdot \hat{r} \sin\theta) \, \vec{\ell}\, (\cos\theta \openone - {\rm i} \vec{\Gamma} \cdot \hat{r}\sin\theta)\nonumber\\
&=& \vec{\ell} \cos^2\theta + (\vec{\Gamma} \cdot \hat{r}) \,\vec{\ell} \, (\vec{\Gamma} \cdot \hat{r}) \sin^2\theta + {\rm i} [ \vec{\Gamma} \cdot\hat{r}, \; \vec{\ell} ] \sin\theta \cos\theta.
\end{eqnarray}
Because
\begin{eqnarray}
 [ \ell_z, \; \vec{\Gamma} \cdot \hat{r} ]& =& \left[ \ell_z, \; \Gamma_x \frac{x}{r}+ \Gamma_y \frac{y}{r}+\Gamma_z \frac{z}{r} \right]
= \frac{1}{r}\left[ \ell_z, \; \Gamma_x x+ \Gamma_y y+\Gamma_z z \right]
= \frac{1}{r}\left[ \ell_z, \; \Gamma_x x+ \Gamma_y y \right]  \nonumber\\
&=& \frac{1}{r} \left\{ \Gamma_x \left[ \ell_z, \; x \right]+ \Gamma_y \left[ \ell_z, \; y \right] \right\}
=\frac{1}{r} \left[ ({\rm i}\hbar) \Gamma_x y - ({\rm i}\hbar)  \Gamma_y x \right] \nonumber\\
 &=&{\rm i} \hbar (\vec{\Gamma} \times \hat{r})_z=-{\rm i} \hbar ( \hat{r}\times \vec{\Gamma})_z,
\end{eqnarray}
then we have
\begin{eqnarray}
 [ \vec{\ell}, \; \vec{\Gamma} \cdot \hat{r} ]= -{\rm i} \hbar ( \hat{r}\times \vec{\Gamma}),
\end{eqnarray}
or
\begin{eqnarray}
 [\vec{\Gamma} \cdot \hat{r}, \; \vec{\ell} ]= {\rm i} \hbar ( \hat{r}\times \vec{\Gamma}).
\end{eqnarray}
Because
\begin{eqnarray}
(\vec{\Gamma} \times \hat{r})_z (\vec{\Gamma} \cdot \hat{r}) &=& \frac{1}{r^2}(\Gamma_x y - \Gamma_y x) (\Gamma_x x + \Gamma_y y + \Gamma_z z) = \frac{1}{r^2} \left[\Gamma_x y (\Gamma_x x + \Gamma_y y + \Gamma_z z)-\Gamma_y x (\Gamma_x x + \Gamma_y y + \Gamma_z z)  \right]\nonumber\\
&=& \frac{1}{r^2} \left[(y x + {\rm i}\Gamma_z y^2 -{\rm i} \Gamma_y yz)- (-{\rm i}\Gamma_z x^2 +  xy + {\rm i}\Gamma_x xz)  \right]= \frac{1}{r^2} \left[{\rm i}\Gamma_z (x^2+y^2) -{\rm i} z (\Gamma_y y+\Gamma_x x)  \right]\nonumber\\
&=& \frac{1}{r^2} \left[{\rm i}\Gamma_z (x^2+y^2+z^2) -{\rm i} z (\Gamma_x x+\Gamma_y y+\Gamma_x x)  \right]\nonumber\\
&=& {\rm i} \Gamma_z - {\rm i}  (\vec{\Gamma}\cdot \hat{r}) \hat{e}_z,
\end{eqnarray}
then we have
\begin{eqnarray}
(\vec{\Gamma} \times \hat{r}) (\vec{\Gamma} \cdot \hat{r}) &=&  {\rm i} \vec{\Gamma} - {\rm i}  (\vec{\Gamma}\cdot \hat{r}) \hat{r}.
\end{eqnarray}
Consider that
\begin{eqnarray}
 (\vec{\Gamma} \cdot \hat{r})\; \vec{\ell}\; (\vec{\Gamma} \cdot \hat{r})  &=& \vec{\ell} - [ \vec{\ell}, \; \vec{\Gamma} \cdot \hat{r} ] (\vec{\Gamma} \cdot \hat{r}) = \vec{\ell} - {\rm i} \hbar (\vec{\Gamma} \times \hat{r}) (\vec{\Gamma} \cdot \hat{r})\nonumber\\
 &=& \vec{\ell} + \hbar \left[\vec{\Gamma} - (\vec{\Gamma}\cdot \hat{r}) \hat{r}\right],
\end{eqnarray}
we can have
\begin{eqnarray}\label{eq:L-1}
\vec{L}&=&(\cos\theta + {\rm i} \vec{\Gamma} \cdot \hat{r} \sin\theta) \, \vec{\ell} \, (\cos\theta - {\rm i} \vec{\Gamma} \cdot \hat{r}\sin\theta)\nonumber\\
&=& \vec{\ell} \cos^2\theta + (\vec{\Gamma} \cdot \hat{r}) \,\vec{\ell} \, (\vec{\Gamma} \cdot \hat{r}) \sin^2\theta + {\rm i} [ \vec{\Gamma} \cdot\hat{r}, \; \vec{\ell} ] \sin\theta \cos\theta\nonumber\\
&=& \vec{\ell} \cos^2\theta + \left\{\vec{\ell} + \hbar \left[\vec{\Gamma} - (\vec{\Gamma}\cdot \hat{r}) \hat{r}\right]\right\} \sin^2\theta
-\hbar ( \hat{r}\times \vec{\Gamma}) \sin\theta \cos\theta\nonumber\\
&=& \vec{\ell}- \hbar (  \hat{r}\times\vec{\Gamma})\sin\theta \cos\theta+ \hbar\left[\vec{\Gamma}-(\vec{\Gamma} \cdot \hat{r})\hat{r}\right]\sin^2\theta,
\end{eqnarray}
or in terms of $\vec{S}$ and $\vec{r}$ as
\begin{eqnarray}\label{eq:disp-1}
\vec{L}&=& \vec{\ell}- 2 \sin\theta \cos\theta\frac{\vec{r}\times\vec{S}}{r} + 2 \sin^2\theta \left[\vec{S}-\frac{(\vec{S} \cdot \vec{r})\vec{r}}{r^2}\right].
\end{eqnarray}

\begin{remark} From the viewpoint of displacement, Eq. (\ref{eq:disp-1}) can be written as
\begin{eqnarray}\label{eq:disp-2}
\vec{L}&=& \vec{\ell} +q \vec{G},
\end{eqnarray}
with
\begin{eqnarray}\label{eq:disp-3}
q \vec{G}&=&  - 2 \sin\theta \cos\theta\frac{\vec{r}\times\vec{S}}{r} + 2 \sin^2\theta \left[\vec{S}-\frac{(\vec{S} \cdot \vec{r})\vec{r}}{r^2}\right].
\end{eqnarray}

(i) When $\theta=\pi/4$, one has $2 \sin\theta \cos\theta=1$, $2 \sin^2\theta=1$, and
\begin{eqnarray}
\vec{L}&=& \vec{\ell}- 2 \sin\theta \cos\theta\frac{\vec{r}\times\vec{S}}{r} + 2 \sin^2\theta \left[\vec{S}-\frac{(\vec{S} \cdot \vec{r})\vec{r}}{r^2}\right]\nonumber\\
&=& \vec{\ell}- \frac{\vec{r}\times\vec{S}}{r} + \left[\vec{S}-\frac{(\vec{S} \cdot \vec{r})\vec{r}}{r^2}\right]\nonumber\\
&=& \vec{\ell} + \vec{S}-  \left[\frac{\vec{r}\times\vec{S}}{r}+\frac{(\vec{S} \cdot \vec{r})\vec{r}}{r^2}\right],
\end{eqnarray}
which is different from $\vec{L}= \vec{\ell} + \vec{S}$ by an additional term $-  \left[\frac{\vec{r}\times\vec{S}}{r}+\frac{(\vec{S} \cdot \vec{r})\vec{r}}{r^2}\right]$. This also means that the ``rotation'' $\mathcal{U} \,\vec{\ell}\,  \mathcal{U}^\dagger$ cannot generate the ``displacement'' $\vec{L}= \vec{\ell} + \vec{S}$.

(ii) When $\theta=\pi/2$, one has $2 \sin\theta \cos\theta=0$, $2 \sin^2\theta=2$, and
\begin{eqnarray}
\vec{L}&=& \vec{\ell}- 2 \sin\theta \cos\theta\frac{\vec{r}\times\vec{S}}{r} + 2 \sin^2\theta \left[\vec{S}-\frac{(\vec{S} \cdot \vec{r})\vec{r}}{r^2}\right]\nonumber\\
&=& \vec{\ell}+2\vec{S}-2\frac{(\vec{S} \cdot \vec{r})\vec{r}}{r^2}=\vec{\ell}+2\vec{S}-2(\vec{S} \cdot \hat{r})\hat{r},
\end{eqnarray}
which is also an angular momentum operator, but is different from $\vec{L}= \vec{\ell} + \vec{S}$. $\blacksquare$
\end{remark}

\begin{remark}Based on (i) the simple ``displacement'' $\vec{L}= \vec{\ell} + \vec{S}$, and (ii) the result obtained from the ``rotation''
$\vec{L}= \vec{\ell}- 2 \sin\theta \cos\theta\frac{\vec{r}\times\vec{S}}{r} + 2 \sin^2\theta \left[\vec{S}-\frac{(\vec{S} \cdot \vec{r})\vec{r}}{r^2}\right]$, we can find that the most general displacement is of the following form
\begin{eqnarray}\label{eq:disp-4}
\vec{L}&=& \vec{\ell}+ a_1 \vec{S} + a_2 (\vec{S} \cdot \hat{r})\hat{r} + a_3 (\hat{r}\times\vec{S}),
\end{eqnarray}
i.e.,
\begin{eqnarray}
\vec{L}&=& \vec{\ell} +q \vec{G},
\end{eqnarray}
with
\begin{eqnarray}\label{eq:disp-5}
q \vec{G}&=&  a_1 \vec{S} + a_2 (\vec{S} \cdot \hat{r})\hat{r} + a_3 (\hat{r}\times\vec{S}),
\end{eqnarray}
namely, $q \vec{G}$ is a linear combination of $\vec{S}$, $(\vec{S} \cdot \hat{r})\hat{r}$, and $(\hat{r}\times\vec{S})$, here $a_1$, $a_2$, $a_3$ are {some coefficients}, which will be determined by the following definition of angular momentum operator
\begin{eqnarray}
\vec{L} \times \vec{L}={\rm i} \hbar \vec{L}.
\end{eqnarray}

\emph{The Calculation.---}For simplicity, let us denote
\begin{eqnarray}\label{eq:disp-5a}
\vec{K} &=& q \vec{G}= a_1 \vec{S} + a_2 (\vec{S} \cdot \hat{r})\hat{r} + a_3 (\hat{r}\times\vec{S}).
\end{eqnarray}
Without loss of generality, we check the $z$-component, i.e., and we can have
\begin{eqnarray}
(\vec{L} \times \vec{L})_z=[L_x, L_y] = [\ell_x + K_x, \ell_y + K_y] = [\ell_x, \ell_y] + [\ell_x, K_y] + [K_x, \ell_y] + [K_x, K_y].
\end{eqnarray}

(i) For the first term, we can have
\begin{eqnarray}
 [\ell_x, \ell_y] ={\rm i} \hbar\ell_z.
\end{eqnarray}

(ii) For the second term,we can have
\begin{eqnarray}
[\ell_x, K_y]& =& \left[\ell_x, a_1 S_y + a_2 (\vec{S} \cdot \hat{r})\frac{y}{r} + a_3 (\hat{r}\times\vec{S})_y\right]\nonumber\\
& =& a_2 \left[\ell_x,(\vec{S} \cdot \hat{r})\frac{y}{r}\right]+ a_3 \left[\ell_x, (\hat{r}\times\vec{S})_y\right].
\end{eqnarray}
Because
\begin{eqnarray}
[\ell_x, (\vec{S} \cdot \hat{r})] = \frac{1}{r}[\ell_x, S_x x + S_y y + S_z z] = S_y \frac{{\rm i}\hbar z}{r} + S_z \frac{-{\rm i}\hbar y}{r}
={\rm i}\hbar (\vec{S}\times \hat{r})_{{x}}= -{\rm i}\hbar (\hat{r}\times \vec{S})_{{x}},
\end{eqnarray}
\begin{eqnarray}
 \left[\ell_x,\frac{y}{r}\right] &=& \frac{1}{r}\left[\ell_x, y\right]= {\rm i}\hbar \frac{z}{r},
\end{eqnarray}
then we have
\begin{eqnarray}
\left[\ell_x,(\vec{S} \cdot \hat{r})\frac{y}{r}\right] &=& [\ell_x, (\vec{S} \cdot \hat{r})]\frac{y}{r}+ (\vec{S} \cdot \hat{r}) \left[\ell_x,\frac{y}{r}\right]= \left[-{\rm i}\hbar (\hat{r}\times \vec{S})_{{x}}\right]\frac{y}{r}+ (\vec{S} \cdot \hat{r}) {\rm i}\hbar \frac{z}{r}.
\end{eqnarray}
Because
\begin{eqnarray}
\left[\ell_x, (\hat{r}\times\vec{S})_y\right]&=&
\frac{1}{r}\left[\ell_x, z S_x- x S_z\right]=\frac{1}{r} \left(-{\rm i}\hbar y S_x\right)= -{\rm i}\hbar S_x \frac{y}{r},
\end{eqnarray}
then we have
\begin{eqnarray}
[\ell_x, K_y]
& =& a_2 \left[\ell_x,(\vec{S} \cdot \hat{r})\frac{y}{r}\right]+ a_3 \left[\ell_x, (\hat{r}\times\vec{S})_y\right]\nonumber\\
&=& a_2 \left\{\left[-{\rm i}\hbar (\hat{r}\times \vec{S})_{{x}}\right]\frac{y}{r}+ (\vec{S} \cdot \hat{r}) {\rm i}\hbar \frac{z}{r}\right\}
+a_3 \left[-{\rm i}\hbar S_x \frac{y}{r}\right].
\end{eqnarray}

(iii) For the third term, we can have
\begin{eqnarray}
&& [K_x, \ell_y] = -[\ell_y, K_x],
\end{eqnarray}
and
\begin{eqnarray}
[\ell_y, K_x]&=& \left[\ell_y, a_1 S_x + a_2 (\vec{S} \cdot \hat{r}) \frac{x}{r} + a_3 (\hat{r}\times\vec{S})_x\right]\nonumber\\
&=& a_2\left[\ell_y, (\vec{S} \cdot \hat{r}) \frac{x}{r}\right]
+ a_3 \left[\ell_y, (\hat{r}\times\vec{S})_x\right].
\end{eqnarray}
Because
\begin{eqnarray}
[\ell_y, (\vec{S} \cdot \hat{r})] = \frac{1}{r}[\ell_y, S_x x + S_y y + S_z z] = S_x \frac{-{\rm i}\hbar z}{r} + S_z \frac{{\rm i}\hbar x}{r}
={\rm i}\hbar (\vec{S}\times \hat{r})_y= -{\rm i}\hbar (\hat{r}\times \vec{S})_y,
\end{eqnarray}
\begin{eqnarray}
 \left[\ell_y,\frac{x}{r}\right] &=& \frac{1}{r}\left[\ell_y, x\right]= -{\rm i}\hbar \frac{z}{r},
\end{eqnarray}
then we have
\begin{eqnarray}
\left[\ell_y, (\vec{S} \cdot \hat{r}) \frac{x}{r}\right] &=&
\left[\ell_y, (\vec{S} \cdot \hat{r})\right] \frac{x}{r} + (\vec{S} \cdot \hat{r})  \left[\ell_y, \frac{x}{r}\right]\nonumber\\
&=& \left[-{\rm i}\hbar (\hat{r}\times \vec{S})_y\right] \frac{x}{r} +(\vec{S} \cdot \hat{r}) \left(-{\rm i}\hbar \frac{z}{r}\right).
\end{eqnarray}
Because
\begin{eqnarray}
\left[\ell_y, (\hat{r}\times\vec{S})_x\right]&=&
\frac{1}{r}\left[\ell_y, y S_z- z S_y\right]=\frac{1}{r} \left(-{\rm i}\hbar x S_y\right)= -{\rm i}\hbar S_y \frac{{x}}{r},
\end{eqnarray}
then we have
\begin{eqnarray}
 [K_x, \ell_y] &=& -[\ell_y, K_x]=-a_2\left[\ell_y, (\vec{S} \cdot \hat{r}) \frac{x}{r}\right]
- a_3 \left[\ell_y, (\hat{r}\times\vec{S})_x\right] \nonumber\\
&=&  -a_2 \left\{\left[-{\rm i}\hbar (\hat{r}\times \vec{S})_y\right] \frac{x}{r} +(\vec{S} \cdot \hat{r}) \left(-{\rm i}\hbar \frac{z}{r}\right)\right\} -a_3 \left(-{\rm i}\hbar S_y \frac{{x}}{r}\right)\nonumber\\
&=&  a_2 \left\{\left[{\rm i}\hbar (\hat{r}\times \vec{S})_y\right] \frac{x}{r} +(\vec{S} \cdot \hat{r}) \left({\rm i}\hbar \frac{z}{r}\right)\right\} +a_3 \left({\rm i}\hbar S_y \frac{{x}}{r}\right).
\end{eqnarray}

Based on the results of (ii) and (iii), we have known that
\begin{eqnarray}
[\ell_x, K_y]&=& a_2 \left\{\left[-{\rm i}\hbar (\hat{r}\times \vec{S})_{{x}}\right]\frac{y}{r}+ (\vec{S} \cdot \hat{r}) {\rm i}\hbar \frac{z}{r}\right\}
+a_3 \left[-{\rm i}\hbar S_x \frac{y}{r}\right],
\end{eqnarray}
\begin{eqnarray}
 [K_x, \ell_y] &=&  a_2 \left\{\left[{\rm i}\hbar (\hat{r}\times \vec{S})_y\right] \frac{x}{r} +(\vec{S} \cdot \hat{r}) \left({\rm i}\hbar \frac{z}{r}\right)\right\} +a_3 \left({\rm i}\hbar S_y \frac{{x}}{r}\right),
\end{eqnarray}
which lead to
\begin{eqnarray}
[\ell_x, K_y] + [K_x, \ell_y] &=& a_2 \left\{ \left[-{\rm i}\hbar (\hat{r}\times \vec{S})_x\right]\frac{y}{r}+ (\vec{S} \cdot \hat{r}) {\rm i}\hbar \frac{z}{r} \right\} - a_3 {\rm i}\hbar S_x \frac{y}{r}  + a_2 \left\{ \left[{\rm i}\hbar (\hat{r}\times \vec{S})_y\right]\frac{x}{r}+ (\vec{S} \cdot \hat{r}) {\rm i}\hbar \frac{z}{r} \right\} + a_3 {\rm i}\hbar S_y \frac{x}{r} \nonumber\\
&=& {\rm i}\hbar a_2 \left[ \frac{x}{r}(\hat{r}\times \vec{S})_y - \frac{y}{r}(\hat{r}\times \vec{S})_x + 2(\vec{S} \cdot \hat{r})\frac{z}{r} \right] + {\rm i}\hbar a_3 \left( \frac{x}{r}S_y - \frac{y}{r}S_x \right) \nonumber\\
&=& {\rm i}\hbar a_2 \left[ (\hat{r} \times (\hat{r}\times \vec{S}))_z + 2(\vec{S} \cdot \hat{r})\frac{z}{r} \right] + {\rm i}\hbar a_3 (\hat{r}\times \vec{S})_z \nonumber\\
&=& {\rm i}\hbar a_2 \left[ (\vec{S} \cdot \hat{r})\frac{z}{r} - S_z + 2(\vec{S} \cdot \hat{r})\frac{z}{r} \right] + {\rm i}\hbar a_3 (\hat{r}\times \vec{S})_z \nonumber\\
&=& {\rm i}\hbar \left[ -a_2 S_z + 3a_2 (\vec{S} \cdot \hat{r})\frac{z}{r} + a_3 (\hat{r}\times \vec{S})_z \right].
\end{eqnarray}

(iv) For the fourth term, we can have
\begin{eqnarray}
[K_x, K_y] &=& \left[a_1 S_x + a_2 (\vec{S} \cdot \hat{r}) \frac{x}{r} + a_3 (\hat{r}\times\vec{S})_x, a_1 S_y + a_2 (\vec{S} \cdot \hat{r}) \frac{y}{r} + a_3 (\hat{r}\times\vec{S})_y\right].
\end{eqnarray}
Expanding this commutator gives nine terms. We can calculate each term according to the coefficient.

(iv-1) For the $a_1^2$ term, we can have
\begin{eqnarray}
a_1^2[S_x, S_y] &=& {\rm i}\hbar a_1^2 S_z.
\end{eqnarray}

(iv-2) For the cross terms between $a_1$ and $a_2$, because
\begin{eqnarray}
[S_x, \vec{S} \cdot \hat{r}] = {\rm i}\hbar \frac{y}{r} S_z- {\rm i}\hbar \frac{z}{r}  S_y= {\rm i}\hbar (\hat{r}\times\vec{S})_x,
\end{eqnarray}
\begin{eqnarray}
[\vec{S} \cdot \hat{r},S_y] = {\rm i}\hbar \frac{x}{r} S_z-  {\rm i}\hbar \frac{z}{r} S_x= -{\rm i}\hbar (\hat{r}\times\vec{S})_y,
\end{eqnarray}
we can have
\begin{eqnarray}
 a_1 a_2 \left[S_x, (\vec{S} \cdot \hat{r}) \frac{y}{r}\right] + a_1 a_2 \left[(\vec{S} \cdot \hat{r}) \frac{x}{r}, S_y\right]
&=& a_1 a_2 \left[S_x, \vec{S} \cdot \hat{r}\right] \frac{y}{r} + a_1 a_2 \left[\vec{S} \cdot \hat{r},S_y\right] \frac{x}{r}\nonumber\\
&=& a_1 a_2 \left[{\rm i}\hbar (\hat{r}\times\vec{S})_x\right] \frac{y}{r} - a_1 a_2 \left[{\rm i}\hbar (\hat{r}\times\vec{S})_y\right] \frac{x}{r}\nonumber\\
&=& -{\rm i}\hbar a_1 a_2 \left[ \frac{x}{r}(\hat{r}\times\vec{S})_y - \frac{y}{r}(\hat{r}\times\vec{S})_x \right]\nonumber\\
&=& -{\rm i}\hbar a_1 a_2 \left[ \hat{r} \times (\hat{r}\times\vec{S}) \right]_z\nonumber\\
&=& -{\rm i}\hbar a_1 a_2 \left[ (\vec{S} \cdot \hat{r})\frac{z}{r} - S_z \right] \nonumber\\
&=& {\rm i}\hbar a_1 a_2 \left[ S_z - (\vec{S} \cdot \hat{r})\frac{z}{r} \right].
\end{eqnarray}

(iv-3) For the cross terms between $a_1$ and $a_3$, we have
\begin{eqnarray}
 a_1 a_3 \left[S_x, (\hat{r}\times\vec{S})_y\right] + a_1 a_3 \left[(\hat{r}\times\vec{S})_x, S_y\right]
&=& a_1 a_3 \left[S_x, \frac{z}{r}S_x - \frac{x}{r}S_z\right] + a_1 a_3 \left[\frac{y}{r}S_z - \frac{z}{r}S_y, S_y\right]\nonumber\\
&=& a_1 a_3 \left( -\frac{x}{r} [S_x, S_z] \right) + a_1 a_3 \left( \frac{y}{r} [S_z, S_y] \right)\nonumber\\
&=& a_1 a_3 \left( {\rm i}\hbar S_y \frac{x}{r} \right) + a_1 a_3 \left( -{\rm i}\hbar S_x \frac{y}{r} \right)\nonumber\\
&=& {\rm i}\hbar a_1 a_3 \left( \frac{x}{r}S_y - \frac{y}{r}S_x \right) \nonumber\\
&=& {\rm i}\hbar a_1 a_3 (\hat{r}\times\vec{S})_z.
\end{eqnarray}

(iv-4) For the $a_2^2$ term, we have
\begin{eqnarray}
a_2^2 \left[(\vec{S} \cdot \hat{r}) \frac{x}{r}, (\vec{S} \cdot \hat{r}) \frac{y}{r}\right] &=& 0.
\end{eqnarray}

(iv-5) For the cross terms between $a_2$ and $a_3$, we have
\begin{eqnarray}
&& a_2 a_3 \left[(\vec{S} \cdot \hat{r}) \frac{x}{r}, (\hat{r}\times\vec{S})_y\right] + a_2 a_3 \left[(\hat{r}\times\vec{S})_x, (\vec{S} \cdot \hat{r}) \frac{y}{r}\right]= a_2 a_3 \frac{x}{r} \left[\vec{S} \cdot \hat{r}, (\hat{r}\times\vec{S})_y\right] - a_2 a_3 \frac{y}{r} \left[\vec{S} \cdot \hat{r}, (\hat{r}\times\vec{S})_x\right].
\end{eqnarray}
Because
\begin{eqnarray}
[\vec{S} \cdot \hat{r}, S_x] = -{\rm i}\hbar (\hat{r}\times\vec{S})_x,
\end{eqnarray}
\begin{eqnarray}
[\vec{S} \cdot \hat{r}, S_z] = -{\rm i}\hbar (\hat{r}\times\vec{S})_z,
\end{eqnarray}
 we have
\begin{eqnarray}
\left[\vec{S} \cdot \hat{r}, (\hat{r}\times\vec{S})_y\right] &=& \left[\vec{S} \cdot \hat{r}, \frac{z}{r}S_x - \frac{x}{r}S_z\right] = -{\rm i}\hbar \left[ \frac{z}{r}(\hat{r}\times\vec{S})_x - \frac{x}{r}(\hat{r}\times\vec{S})_z \right],
\end{eqnarray}
and similarly,
\begin{eqnarray}
\left[\vec{S} \cdot \hat{r}, (\hat{r}\times\vec{S})_x\right] &=& -{\rm i}\hbar \left[ \frac{y}{r}(\hat{r}\times\vec{S})_z - \frac{z}{r}(\hat{r}\times\vec{S})_y \right].
\end{eqnarray}
Substituting these back, we can have
\begin{eqnarray}
&& -{\rm i}\hbar a_2 a_3 \left\{ \frac{x}{r} \left[ \frac{z}{r}(\hat{r}\times\vec{S})_x - \frac{x}{r}(\hat{r}\times\vec{S})_z \right] - \frac{y}{r} \left[ \frac{y}{r}(\hat{r}\times\vec{S})_z - \frac{z}{r}(\hat{r}\times\vec{S})_y \right] \right\}\nonumber\\
&=& -{\rm i}\hbar a_2 a_3 \left[ \frac{z}{r} \left( \frac{x}{r}(\hat{r}\times\vec{S})_x + \frac{y}{r}(\hat{r}\times\vec{S})_y \right) - \frac{x^2+y^2}{r^2} (\hat{r}\times\vec{S})_z \right]\nonumber\\
&=&-{\rm i}\hbar a_2 a_3 \left[ -\frac{z^2}{r^2}(\hat{r}\times\vec{S})_z - \frac{x^2+y^2}{r^2} (\hat{r}\times\vec{S})_z \right] \nonumber\\
&=& {\rm i}\hbar a_2 a_3 (\hat{r}\times\vec{S})_z.
\end{eqnarray}

(iv-6) For the $a_3^2$ term, we have
\begin{eqnarray}
a_3^2 \left[(\hat{r}\times\vec{S})_x, (\hat{r}\times\vec{S})_y\right] &=& a_3^2 \left[\frac{y}{r}S_z - \frac{z}{r}S_y, \frac{z}{r}S_x - \frac{x}{r}S_z\right]\nonumber\\
&=& a_3^2 \left( \frac{y z}{r^2}[S_z, S_x] + \frac{z x}{r^2}[S_y, S_z] - \frac{z^2}{r^2}[S_y, S_x] \right)\nonumber\\
&=& {\rm i}\hbar a_3^2 \left( \frac{y z}{r^2}S_y + \frac{z x}{r^2}S_x + \frac{z^2}{r^2}S_z \right) \nonumber\\
&=& {\rm i}\hbar a_3^2 (\vec{S} \cdot \hat{r})\frac{z}{r}.
\end{eqnarray}

After adding all of these terms together, we can have
\begin{eqnarray}
[K_x, K_y] &=& {\rm i}\hbar \left[ a_1^2 S_z + a_1 a_2 \left( S_z - (\vec{S} \cdot \hat{r})\frac{z}{r} \right) + a_1 a_3 (\hat{r}\times\vec{S})_z + a_2 a_3 (\hat{r}\times\vec{S})_z + a_3^2 (\vec{S} \cdot \hat{r})\frac{z}{r} \right]\nonumber\\
&=& {\rm i}\hbar \left[ (a_1^2 + a_1 a_2) S_z + (a_3^2 - a_1 a_2) (\vec{S} \cdot \hat{r})\frac{z}{r} + (a_1 a_3 + a_2 a_3) (\hat{r}\times\vec{S})_z \right].
\end{eqnarray}
Finally, we can calculate $[L_x, L_y]$ as
\begin{eqnarray}[L_x, L_y] &=& [\ell_x, \ell_y] +[\ell_x, K_y] + [K_x, \ell_y] + [K_x, K_y] \nonumber\\
&=& {\rm i}\hbar \ell_z + {\rm i}\hbar \left[ -a_2 S_z + 3a_2 (\vec{S} \cdot \hat{r})\frac{z}{r} + a_3 (\hat{r}\times \vec{S})_z \right] \nonumber\\
&& + {\rm i}\hbar \left[ (a_1^2 + a_1 a_2) S_z + (a_3^2 - a_1 a_2) (\vec{S} \cdot \hat{r})\frac{z}{r} + (a_1 a_3 + a_2 a_3) (\hat{r}\times\vec{S})_z \right] \nonumber\\
&=& {\rm i}\hbar \left\{ \ell_z + (a_1^2 + a_1 a_2 - a_2) S_z + (3a_2 + a_3^2 - a_1 a_2) (\vec{S} \cdot \hat{r})\frac{z}{r} + (a_3 + a_1 a_3 + a_2 a_3) (\hat{r}\times\vec{S})_z \right\}.
\end{eqnarray}
What we need is $[L_x, L_y] = {\rm i}\hbar L_z$, i.e.,
\begin{eqnarray}
[L_x, L_y] &=& {\rm i}\hbar \left[ \ell_z + a_1 S_z + a_2 (\vec{S} \cdot \hat{r})\frac{z}{r} + a_3 (\hat{r}\times\vec{S})_z \right].
\end{eqnarray}
After comparing the coefficients of the corresponding terms, we obtain the following equations for $a_1, a_2, a_3$:
\begin{eqnarray}\label{eq:a123}
a_1^2 + a_1 a_2 - a_2 &=& a_1, \nonumber\\
3a_2 + a_3^2 - a_1 a_2 &=& a_2, \nonumber\\
a_3 + a_1 a_3 + a_2 a_3 &=& a_3.
\end{eqnarray}

\emph{Analysis 1.---}From the third of Eq. (\ref{eq:a123}), we have
\begin{eqnarray}
a_3(a_1 + a_2) = 0.
\end{eqnarray}
If $a_1 + a_2\neq0$, we must have $a_3 = 0$, the remaining equations give $a_1 = 1$ and $a_2 = 0$, which corresponds to the simple ``displacement'' result $\vec{L} = \vec{\ell} + \vec{S}$.

If $a_1 +a_2=0$, the  equations yield $a_2=-a_1$, and
\begin{eqnarray}
-2a_1 + a_3^2 + a_1^2 &=& 0,
\end{eqnarray}
i.e.,
\begin{eqnarray}\label{eq:a2a3}
(a_1 - 1)^2 + a_3^2 = 1.
\end{eqnarray}
If the coefficients $a_2$ and $a_3$ are real numbers, then Eq. (\ref{eq:a2a3})  describes a continuous family of solutions parameterized by $\theta$. One can parameterize them as
\begin{eqnarray}
&& a_1 - 1 = -\cos 2\theta, \nonumber\\
&& a_3 = -\sin 2\theta.
\end{eqnarray}
(Note: since $\theta$ is arbitrary, one can parameterize $a_3$ as either $a_3 = -\sin 2\theta$ or $a_3 = \sin 2\theta$). We find that
\begin{eqnarray}
a_1 &=& 2\sin^2\theta, \nonumber\\
a_2 &=& -2\sin^2\theta, \nonumber\\
a_3 &=& - 2\sin\theta\cos\theta,
\end{eqnarray}
which just comes from the ``rotation'' solution
\begin{eqnarray}
\vec{L}&=& \vec{\ell}- 2 \sin\theta \cos\theta\frac{\vec{r}\times\vec{S}}{r} + 2 \sin^2\theta \left[\vec{S}-\frac{(\vec{S} \cdot \vec{r})\vec{r}}{r^2}\right].
\end{eqnarray}

\emph{Analysis 2.---} Note that the derivation in this \emph{Remark} only depends on the commutation relations of the angular momentum operator (i.e., $\vec{S} \times \vec{S} ={\rm i} \hbar \vec{S}$), without using the special spin-1/2 condition (i.e., $(\vec{\Gamma} \cdot \hat{r})^2=\openone$), thus this result is valid for any spin-$s$ angular momentum operator.
In summary, for arbitrary spin-$s$ angular momentum, the solution of the function $\vec{L}=F(\vec{\ell}, \vec{S})$ is given by
\begin{equation}
    \vec{L}=F(\vec{\ell}, \vec{S}) = \begin{cases}
        \vec{\ell}+\vec{S}, \\
        \vec{\ell}- 2 \sin\theta \cos\theta\dfrac{\vec{r}\times\vec{S}}{r} + 2 \sin^2\theta \left[\vec{S}-\dfrac{(\vec{S} \cdot \vec{r})\vec{r}}{r^2}\right].
    \end{cases}
\end{equation}
However, for simplicity, in this work, we restrict our study on the spin-1/2 case.

\emph{Analysis 3.---}Usually, the angular momentum operator $\vec{L}=(L_x, L_y, L_z)$ is a hermitian operator, i.e.,
 \begin{eqnarray}
\vec{L}=(\vec{L})^\dagger,
\end{eqnarray}
or equivalently
 \begin{eqnarray}
L_j=L_j^\dagger, \;\;\;\;\;\; (j=x, y, z).
\end{eqnarray}
If one relaxes the hermitian condition, but only keeps the following angular-momentum condition
\begin{eqnarray}
\vec{L} \times \vec{L} ={\rm i} \hbar \vec{L},
\end{eqnarray}
then one can have other solutions of the complex $\vec{L}$ with complex coefficients $a_1$, $a_2$, and $a_3$. Due to the following identity of hyperbolic function
\begin{eqnarray}
\cosh^2 \vartheta-\sinh^2 \vartheta=1,
\end{eqnarray}
(i) one can parameterize $a_1$ and $a_3$ as
\begin{eqnarray}
(a_1 - 1)^2 =\cosh^2 \vartheta, \;\;\;\;\ a_3^2 =-\sinh^2 \vartheta,
\end{eqnarray}
i.e.,
\begin{eqnarray}
a_1 - 1 =\pm \cosh \vartheta, \;\;\;\;\ a_3 = {\rm i} \, \sinh \vartheta,
\end{eqnarray}
i.e.,
\begin{eqnarray}
a_1 =1 \pm \cosh \vartheta, \;\;\;\;\ a_3 = {\rm i} \, \sinh \vartheta.
\end{eqnarray}
{(ii) Alternatively, one can parameterize $a_2$ and $a_3$ as
\begin{eqnarray}
 a_3^2=\cosh^2 \vartheta, \;\;\;\;\ (a_1 - 1)^2 =-\sinh^2 \vartheta,
\end{eqnarray}
i.e.,
\begin{eqnarray}
 a_3=\pm \cosh \vartheta, \;\;\;\;\ a_1 =1+ {\rm i}\sinh^2 \vartheta.
\end{eqnarray}
}
$\blacksquare$
\end{remark}

%
%
%

\subsubsection{Extracting General Spin Vector Potential from $\vec{L}=F(\vec{\ell}, \vec{S})$ }

From previous section, we have known that the general solutions of the angular momentum operator $\vec{L}=F(\vec{\ell}, \vec{S})$ are given by
\begin{equation}
    \vec{L}=F(\vec{\ell}, \vec{S}) =  \vec{\ell} + a_1 \vec{S} + a_2 (\vec{S} \cdot \hat{r})\hat{r} + a_3 (\hat{r}\times\vec{S}),
\end{equation}
with \emph{solution 1} as $\{a_1=1, a_2=0, a_3=0\}$ and \emph{solution 2} as $\{a_2=-a_1, (a_1 - 1)^2 + a_3^2 = 1\}$. Let us denote
\begin{eqnarray}
\vec{L}&=& \vec{\ell} +q \vec{G},
\end{eqnarray}
with
\begin{eqnarray}\label{eq:disp-5b}
q \vec{G}&=&  a_1 \vec{S} + a_2 (\vec{S} \cdot \hat{r})\hat{r} + a_3 (\hat{r}\times\vec{S}).
\end{eqnarray}
Then based on the \emph{vector potential extraction approach}, we can extract the spin vector potential as
\begin{eqnarray}
 \vec{A}&=& c \dfrac{\vec{r}\times\vec{G}}{r^2}=\frac{c}{q} \dfrac{\hat{r}\times  \left[a_1 \vec{S} + a_2 (\vec{S} \cdot \hat{r})\hat{r} + a_3 (\hat{r}\times\vec{S})\right]}{r}\nonumber\\
&=&\frac{c}{q} \dfrac{\hat{r}\times  \left[a_1 \vec{S} +  a_3 (\hat{r}\times\vec{S})\right]}{r}= a_1 \frac{c}{q} \dfrac{\hat{r}\times \vec{S}}{r}+
a_3\frac{c}{q} \dfrac{\hat{r}\times (\hat{r}\times\vec{S})}{r}\nonumber\\
&=&  a_1 \frac{c}{q} \dfrac{\hat{r}\times \vec{S}}{r}+
a_3\frac{c}{q} \dfrac{(\vec{S} \cdot \hat{r})\hat{r} - \vec{S}}{r}\nonumber\\
&=& \frac{1}{r} \frac{c}{q}  \left[a_1 (\hat{r}\times \vec{S})-a_3 \vec{S} +
a_3 (\vec{S} \cdot \hat{r})\hat{r}\right].
\end{eqnarray}

\begin{remark}
Specially, for $a_1=1$ and $a_3=0$, we can recover the previous spin vector potential in Eq. (\ref{eq:svp-1}), i.e.,
\begin{eqnarray}
 \vec{A}&=& c \dfrac{\vec{r}\times\vec{G}}{r^2}=\frac{c}{q} \dfrac{\vec{r}\times\vec{S}}{r^2}.
\end{eqnarray}
$\blacksquare$
\end{remark}

\begin{remark}
In the spherical coordinate system, the spin vector potential can be written as
\begin{eqnarray}
\vec{A} &=& A_r \hat{e}_r+   A_\theta  \hat{e}_\theta+  A_\phi \hat{e}_\phi.
\end{eqnarray}
One can check that $A_r=0$. Explicitly, one has
\begin{eqnarray}
 A_r &=& \vec{A}\cdot \hat{e}_r = \frac{1}{r} \frac{c}{q}  \left[a_1 (\hat{r}\times \vec{S})-a_3 \vec{S} +
a_3 (\vec{S} \cdot \hat{r})\hat{r}\right]\cdot \hat{e}_r \nonumber\\
&=& \frac{1}{r} \frac{c}{q}  \left[a_1 (\hat{r}\times \vec{S})-a_3 \vec{S} +
a_3 (\vec{S} \cdot \hat{r})\hat{r}\right]\cdot \hat{r} =
\frac{1}{r} \frac{c}{q}  \left[-a_3 \vec{S} +
a_3 (\vec{S} \cdot \hat{r})\hat{r}\right]\cdot \hat{r}\nonumber\\
& =&
\frac{1}{r} \frac{c}{q} a_3 \left[-\vec{S}\cdot \hat{r} +
 (\vec{S} \cdot \hat{r})\right]=0.
\end{eqnarray}
$\blacksquare$
\end{remark}

%
%

\subsubsection{Discussion}

In the previous subsection, we have studied the spin vector potential from the viewpoint of angular momentum operator. To deepen understanding this topic, in this subsection, let us study the spin vector potential from the viewpoint of linear momentum operator. As before, we restrict our discussion on spin-1/2 case.

\textcolor{blue}{\emph{Discussion 1: The Gauge Transformation of the Momentum.---}}Now, we take the unitary transformation $\mathcal{U}={\rm e}^{{\rm i} \theta \vec{\Gamma}\cdot \hat{r}}$ as the gauge transformation, and we apply it to the linear momentum $\vec{p}$, i.e.,
\begin{eqnarray}
\vec{p} \mapsto \vec{\Pi}=\mathcal{U} \, \vec{p}\,  \mathcal{U}^\dagger,
\end{eqnarray}
{here $\vec{\Pi}$ plays a role of canonical momentum.} Because
\begin{eqnarray}
(\mathcal{U}\, \vec{p} \, \mathcal{U}^\dagger) \psi = \mathcal{U} (-{\rm i} \hbar \vec{\nabla}) (\mathcal{U}^\dagger \psi)
= -{\rm i}\hbar \, \mathcal{U} \left[ (\vec{\nabla} \mathcal{U}^\dagger) \psi + \mathcal{U}^\dagger \vec{\nabla} \psi \right]
= -{\rm i} \hbar \,\mathcal{U} ( \vec{\nabla} \mathcal{U}^\dagger) \psi - {\rm i} \hbar \vec{\nabla} \psi,
\end{eqnarray}
\begin{eqnarray}
\vec{\nabla} (\mathcal{U} \mathcal{U}^\dagger)=(\vec{\nabla} \mathcal{U}) \mathcal{U}^\dagger + \mathcal{U} (\vec{\nabla} \mathcal{U}^\dagger) = 0,
\end{eqnarray}
we can have
\begin{eqnarray}
(\mathcal{U} \, \vec{p}\, \mathcal{U}^\dagger) \psi = {\rm i} \hbar (\vec{\nabla} \mathcal{U}) \mathcal{U}^\dagger \psi - {\rm i}\hbar \vec{\nabla} \psi = \left[ \vec{p} + {\rm i}\hbar (\vec{\nabla} \mathcal{U}) \mathcal{U}^\dagger \right] \psi,
\end{eqnarray}
i.e.,
\begin{eqnarray}
\vec{p} \mapsto \vec{\Pi}= \mathcal{U} \, \vec{p} \, \mathcal{U}^\dagger = \vec{p} + {\rm i} \hbar (\vec{\nabla} \mathcal{U}) \mathcal{U}^\dagger.
\end{eqnarray}

Because
\begin{eqnarray}
&& \mathcal{U}= {\rm e}^{{\rm i} \theta \vec{\Gamma}\cdot \hat{r}} = \cos\theta \openone+ {\rm i}  \sin\theta (\vec{\Gamma} \cdot \hat{r}), \nonumber\\
&& \mathcal{U}^\dagger= {\rm e}^{-{\rm i} \theta \vec{\Gamma}\cdot \hat{r}}= \cos \theta \openone- {\rm i} \sin \theta (\vec{\Gamma} \cdot \hat{r}),
\end{eqnarray}
we have
\begin{eqnarray}
\partial_i \mathcal{U} = {\rm i} \sin \theta \, \partial_i (\vec{\Gamma} \cdot \hat{r}) = \frac{{\rm i} \sin \theta}{r}
\left( \Gamma_i - (\vec{\Gamma} \cdot \hat{r})\hat{r}_i \right),\quad\quad (i=x, y, z),
\end{eqnarray}
therefore
\begin{eqnarray}
(\partial_i \mathcal{U}) \mathcal{U}^\dagger &=&
\frac{{\rm i} \sin \theta}{r} \left( \Gamma_i - (\vec{\Gamma} \cdot \hat{r})\hat{r}_i \right)\left[ \cos \theta - {\rm i} \sin \theta (\vec{\Gamma} \cdot \hat{r})\right]\nonumber\\
&=&\frac{{\rm i} }{r} \left[ \sin \theta \cos \theta \left( \Gamma_i -  (\vec{\Gamma} \cdot \hat{r})\hat{r}_i \right)
 - {\rm i}\, {\sin^2 \theta} \left(  \hat{r}_i {+} {\rm i} (\hat{r}\times\vec{\Gamma}  )_i\right)
 +{\rm i}\, {\sin^2 \theta} \hat{r}_i \right]\nonumber\\
&=&\frac{{\rm i}}{r} \left[ \sin \theta \cos \theta \left( \Gamma_i -  (\vec{\Gamma} \cdot \hat{r})\hat{r}_i \right)
 +\sin^2 \theta (\hat{r} \times \vec{\Gamma})_i \right],
\end{eqnarray}
where we used
\begin{eqnarray}
\Gamma_z  (\vec{\Gamma} \cdot \hat{r}) &=&   \Gamma_z  (\Gamma_x x +\Gamma_y y +\Gamma_z z )\frac{1}{r}
=({\rm i}\Gamma_y x -{\rm i}\Gamma_x y + z )\frac{1}{r}\nonumber\\
&=& \frac{z}{r}+ {\rm i}(x\Gamma_y  -y\Gamma_x)\frac{1}{r} =\frac{z}{r}+ {\rm i}( \hat{r} \times \vec{\Gamma})_z,
\end{eqnarray}
\begin{eqnarray}
\Gamma_i  (\vec{\Gamma} \cdot \hat{r})= \hat{r}_i {+} {\rm i} (\hat{r}\times\vec{\Gamma}  )_i, \;\;\;\;\; (\vec{\Gamma} \cdot \hat{r})^2=\openone.
\end{eqnarray}
Then one has
\begin{eqnarray}
(\vec{\nabla} \mathcal{U}) \mathcal{U}^\dagger
&=&\frac{{\rm i}}{r} \left[ \sin \theta \cos \theta \left( \vec{\Gamma} -  (\vec{\Gamma} \cdot \hat{r}) \hat{r} \right)
 +\sin^2 \theta (\hat{r} \times \vec{\Gamma}) \right]\nonumber\\
&=&\frac{{\rm i}}{r} \left[\sin^2 \theta (\hat{r} \times \vec{\Gamma})+ \sin \theta \cos \theta \left( \vec{\Gamma} -  (\vec{\Gamma} \cdot \hat{r}) \hat{r} \right)
 \right],
\end{eqnarray}
thus
\begin{eqnarray}
\vec{p} \mapsto \vec{\Pi}=\mathcal{U} \, \vec{p}\,  \mathcal{U}^\dagger  &=&  \vec{p} + {\rm i} \hbar (\vec{\nabla} \mathcal{U}) \mathcal{U}^\dagger \nonumber\\
&=&\vec{p} - \frac{\hbar}{r} \left[\sin^2 \theta (\hat{r} \times \vec{\Gamma})+ \sin \theta \cos \theta \left( \vec{\Gamma} -  (\vec{\Gamma} \cdot \hat{r}) \hat{r} \right)
 \right].
\end{eqnarray}


\textcolor{blue}{\emph{Discussion 2: Getting the Vector Potential from the Gauge Transformation.---}}We'd like to verify the unification of linear momentum method and angular momentum method. Notice that
\begin{eqnarray}
\hat{r}\times \left[\sin^2 \theta (\hat{r} \times \vec{\Gamma})+ \sin \theta \cos \theta \left( \vec{\Gamma} -  (\vec{\Gamma} \cdot \hat{r}) \hat{r} \right)
 \right]=\left[ \sin\theta\cos\theta (\hat{r} \times \vec{\Gamma}) + \sin^2\theta \left((\vec{\Gamma}\cdot \hat{r})\hat{r} - \vec{\Gamma}\right) \right].
\end{eqnarray}
We can have
\begin{eqnarray}
\vec{L}=\mathcal{U} \, \vec{\ell}\,  \mathcal{U}^\dagger=\hat{r}\times \mathcal{U} \, \vec{p}\,  \mathcal{U}^\dagger =\hat{r}\times \vec{p} - \frac{\hbar}{r} \left[ \sin\theta\cos\theta (\hat{r} \times \vec{\Gamma}) + \sin^2\theta \left((\vec{\Gamma}\cdot \hat{r})\hat{r} - \vec{\Gamma}\right) \right],
\end{eqnarray}
which is the same as Eq. (\ref{eq:L-1}). Hence, the two methods are unified.

Now, we use minimal coupling to find the potential lying under the gauge transformation. Remember that
\begin{eqnarray}
\vec{\Pi}=\vec{p}-\frac{g}{c}\vec{A}.
\end{eqnarray}
Comparing with
\begin{eqnarray}
\vec{\Pi}=\mathcal{U} \,\vec{p}\, \mathcal{U}^\dagger = \vec{p} + {\rm i} \hbar (\vec{\nabla} \mathcal{U}) \mathcal{U}^\dagger,
\end{eqnarray}
we can have
\begin{eqnarray}
A_i=-{\rm i}\frac{c\hbar}{g}(\partial_i \mathcal{U}) \mathcal{U}^\dagger, \quad\quad (i=1, 2, 3).
\end{eqnarray}
Because
\begin{eqnarray}
(\partial_i \mathcal{U}) \mathcal{U}^\dagger =  \frac{{\rm i}}{r} \left[\sin^2 \theta (\hat{r} \times \vec{\Gamma})_i + \sin \theta \cos \theta \left( \Gamma_i - (\vec{\Gamma} \cdot \hat{r})\hat{r}_i \right) \right],
\end{eqnarray}
we finally find
\begin{eqnarray}
\vec{A}={\rm i}\frac{c}{g}(\vec{\nabla} \mathcal{U}) \mathcal{U}^\dagger=\frac{c}{g} \frac{\hbar}{r} \left[ \sin^2 \theta (\hat{r} \times \vec{\Gamma})+\sin \theta \cos \theta \left( \vec{\Gamma} - (\vec{\Gamma} \cdot \hat{r})\hat{r} \right)  \right].
 \end{eqnarray}

\begin{remark}
We illustrate the central role of gauge transformations of the field. Under a general gauge transformation, we have obtained
\begin{eqnarray}
A_{\mu}' = \mathcal{U} A_{\mu} \mathcal{U}^{-1} + {\rm i} \frac{c\hbar}{g}(\partial_{\mu}\mathcal{U} )\mathcal{U}^{-1}.
\end{eqnarray}
Now, taking $A_{\mu}=0$ to adapt to the case of a completely absent gauge potential, we have
\begin{eqnarray}
A_{\mu}' =  {\rm i} \frac{c\hbar}{g}(\partial_{\mu}\mathcal{U})\mathcal{U}^{-1}, \quad (i=1, 2, 3).
\end{eqnarray}
Note that
\begin{eqnarray}
A_{\mu} = (\varphi, -\vec{A}),
\end{eqnarray}
\begin{eqnarray}
A_{i}' =  -{\rm i} \frac{c\hbar}{g}(\partial_{i}\mathcal{U})\mathcal{U}^{-1},  \;\;\;\;\; (\mu=0, 1, 2, 3;\; i=1, 2, 3).
\end{eqnarray}
This is precisely the result obtained from linear momentum method (In this case, the gauge transformation does not involve time). $\blacksquare$
\end{remark}

\begin{remark}
We find that a symmetric algebra is at work here:
\begin{eqnarray}
\hat{r}\times \left[ \sin^2 \theta (\hat{r} \times \vec{\Gamma})+\sin \theta \cos \theta \left( \vec{\Gamma} - (\vec{\Gamma} \cdot \hat{r})\hat{r} \right)  \right]=
\left[ \sin\theta\cos\theta (\hat{r} \times \vec{\Gamma}) + \sin^2\theta \left((\vec{\Gamma}\cdot \hat{r})\hat{r} - \vec{\Gamma}\right) \right],
\end{eqnarray}
\begin{eqnarray}
\hat{r}\times
\left[ \sin\theta\cos\theta (\hat{r} \times \vec{\Gamma}) + \sin^2\theta \left((\vec{\Gamma}\cdot \hat{r})\hat{r} - \vec{\Gamma}\right) \right]=-\left[ \sin^2 \theta (\hat{r} \times \vec{\Gamma})+\sin \theta \cos \theta \left( \vec{\Gamma} - (\vec{\Gamma} \cdot \hat{r})\hat{r} \right)  \right].
\end{eqnarray}
That is to say, the linear momentum operator and the angular momentum operator exhibit a certain dual structure after the gauge transformation. $\blacksquare$
\end{remark}

\newpage

\section{General Static Solutions for the SU(2) Yang-Mills Theory}\label{sec-6}

In previous section, we have determined the general angular momentum operator $\vec{L}=F(\vec{\ell}, \vec{S})$ as
\begin{equation}
    \vec{L}=F(\vec{\ell}, \vec{S}) =  \vec{\ell} + a_1 \vec{S} + a_2 (\vec{S} \cdot \hat{r})\hat{r} + a_3 (\hat{r}\times\vec{S}),
\end{equation}
where \emph{solution 1} is $\{a_1=1, a_2=0, a_3=0\}$, which comes from the simple ``displacement'', and \emph{solution 2} is $\{a_2=-a_1, (a_1 - 1)^2 + a_3^2 = 1\}$, which comes from the ``rotation'' (i.e., the unitary transformation, or the gauge transformation). Accordingly, based on the \emph{vector potential extraction approach}, one can extract the spin vector potential as
\begin{eqnarray}
 \vec{A}&=& \frac{1}{r} \frac{c}{q}  \left[a_1 (\hat{r}\times \vec{S})-a_3 \vec{S} +
a_3 (\vec{S} \cdot \hat{r})\hat{r}\right].
\end{eqnarray}
Note that the term $a_2 (\vec{S} \cdot \hat{r})\hat{r}$ has no contribution to the vector potential, thus $\{a_1=1, a_3=0\}$ corresponds to the vector potential of \emph{solution 1}, and $\{a_2=-a_1, (a_1 - 1)^2 + a_3^2 = 1\}$ corresponds to vector potential of \emph{the solution 2}.

\begin{remark} The spin vector potential of \emph{solution 1} is given by
\begin{eqnarray}
 \vec{A}^{(1)} \equiv \vec{A}&=& \tilde{k} \frac{\vec{r}\times\vec{S}}{r^2},
\end{eqnarray}
with $\tilde{k} =\frac{c}{g}$ (and we denote $q=g$). In Ref. \cite{Zhou2025}, by setting the vector potential and the scalar potential as
\begin{eqnarray}\label{eq:Coul-1}
     &&  \vec{A}= \tilde{k} \frac{\vec{r} \times \vec{S}}{r^2}, \;\;\;\; \varphi = \tilde{f}_1(r) \left(\vec{r} \cdot \vec{S}\right) +  \tilde{f}_2(r),
\end{eqnarray}
and substitute them into the Yang-Mills equations, one can have the solutions listed in Table \ref{tab:gk}.

\begin{table}[H]
\caption{Summary of the consistent solutions for $\tilde{f}_1(r)$ and $\tilde{f}_2(r)$. Note that for the case $\tilde{g}\neq 0$, $\tilde{k}\neq 0$, solutions exist only when $\tilde{g}\hbar \tilde{k} = 1$ or $2$.}\label{tab:gk}
\centering
\begin{tabular}{|c|c|c|c|}
\hline
(i) $\tilde{g}=0$, $\tilde{k}=0$ & (ii) $\tilde{g}=0$, $\tilde{k}\neq 0$ & (iii) $\tilde{g}\neq 0$, $\tilde{k}=0$ & (iv) $\tilde{g}\neq 0$, $\tilde{k}\neq 0$ \\
\hline
\makecell[l]{$\tilde{f}_1(r) = \dfrac{\kappa_2}{r^3} + \kappa_3$ \\ $\tilde{f}_2(r) = \dfrac{\kappa_1}{r}$} &
\makecell[l]{No solutions} &
\makecell[l]{$\tilde{f}_1(r) = 0$ \\ $\tilde{f}_2(r) = \dfrac{\kappa_1}{r}$} &
\makecell[l]{If $\tilde{g}\hbar \tilde{k} = 2$: $\tilde{f}_1(r) = 0$ \\ If $\tilde{g}\hbar \tilde{k} = 1$: $\tilde{f}_1(r) = \dfrac{\kappa_2}{r} + \dfrac{\kappa_3}{r^2}$ \\ $\tilde{f}_2(r) = \dfrac{\kappa_1}{r}$} \\
\hline
\end{tabular}
\end{table}

Note that in the subsequent calculations, for convenience, we shall use uniformly the notation of vector potential and scalar potential as
\begin{eqnarray}
&& \vec{A} = \frac{1}{r} \left[ k_1 (\hat{r} \times \vec{\Gamma}) \right], \;\;\;\; \varphi = f_1(r) \left( \vec{\Gamma}\cdot \hat{r}\right) +  f_2(r),
\end{eqnarray}
i.e.,
\begin{eqnarray}
&& \{\vec{S}, \vec{r}\} \mapsto \{\vec{\Gamma}, \hat{r}\}, \;\;\;\; \vec{S}=\frac{\hbar}{2}\, \vec{\Gamma}, \;\;\;\; \hat{r} \equiv \hat{e}_r= \frac{\vec{r}}{r},
\end{eqnarray}
thus
\begin{eqnarray}
&& \tilde{k} \frac{\hbar}{2} \mapsto k_1, \;\;\;\; r \tilde{f}_1(r) \frac{\hbar}{2} \mapsto \tilde{f}_1(r), \;\;\;\; \tilde{f}_2(r) \mapsto f_2(r).
\end{eqnarray}
Therefore, in terms of the language of $\{\vec{\Gamma}, \hat{r}\}$, in Table \ref{tab:gk}, the solution of $f_1(r)$ will be multiplied by $r$, and the solution of $f_2(r)$ is unchanged. The merit of using $\{\vec{\Gamma}, \hat{r}\}$ is that $\{\vec{\Gamma}, \hat{r}\}$ are dimensionless.  $\blacksquare$
\end{remark}

\begin{remark} The spin vector potential of \emph{solution 2} is given by
\begin{eqnarray}
\vec{A}^{(2)} \equiv \vec{A}&=& \frac{1}{r} \frac{c}{q}  \left[a_1 (\hat{r}\times \vec{S})-a_3 \vec{S} +
a_3 (\vec{S} \cdot \hat{r})\hat{r}\right].
\end{eqnarray}
When $a_i$'s are real numbers, they can be parameterized as
\begin{eqnarray}
&& a_1 = 2\sin^2\theta, \;\;\;\;\; a_3 = - 2\sin\theta\cos\theta.
\end{eqnarray}
Here $\vec{A}$ is a pure gauge solution for the Yang-Mills equations. A pure gauge solution means that the field strength tensor $F_{\mu\nu}$ vanishes identically, i.e., the gauge potential can be expressed as
$ A_\mu =  -{\rm i} \frac{c\hbar}{g}(\partial_{i}\mathcal{U})\mathcal{U}^{-1} $
for some gauge transformation {$\mathcal{U}$}. In the present case, we consider a static pure gauge with $\varphi = 0$. $\blacksquare$

\end{remark}

\begin{remark} The spin vector potentials of \emph{solution 1} and \emph{solution 2} can be rewritten in the following form
\begin{eqnarray}
\vec{A} = \frac{1}{r} \left[ k_1 (\hat{r} \times \vec{\Gamma}) + k_2 \vec{\Gamma} + k_3 (\vec{\Gamma} \cdot \hat{r})\hat{r} \right],
\end{eqnarray}
with
\begin{eqnarray}
&& \vec{S}=\frac{\hbar}{2}\, \vec{\Gamma}, \;\;\;\; \hat{r} \equiv \hat{e}_r= \frac{\vec{r}}{r},\nonumber\\
&& k_1= {\frac{c\hbar}{2q}} a_1, \;\;\;\; k_2=-\frac{c\hbar}{2q}a_3, \;\;\;\; k_3=\frac{c\hbar}{2q}a_3,
\end{eqnarray}
and the constraint
\begin{eqnarray}
&& k_2+k_3=0.
\end{eqnarray}
In this work, we restrict our study on the spin-1/2 case, then $\vec{\Gamma}$ is the vector of Pauli's matrices. In the following, by setting the scalar potential as
\begin{eqnarray}\label{eq:Coul-1a}
    &&   \varphi = f_1(r) \left( \vec{\Gamma}\cdot \hat{r}\right) +  f_2(r),
\end{eqnarray}
we shall solve the general static solutions for the SU(2) Yang-Mills equations. Here, for convenience of calculation, we have used $\vec{\Gamma}$ to replace $\vec{S}$, and also we shall regard the parameter $q$ as the parameter $g$. $\blacksquare$
\end{remark}

\subsection{Calculation of the Magnetic-like Field and the Electric-like Field}

In this section, let us calculate the magnetic-like field and the electric-like field, i.e.,
\begin{eqnarray}
\vec{{B}} &=& \vec{\nabla}\times\vec{{A}}
                        - \frac{{\rm i} g}{c\hbar}\left(\vec{{A}}\times\vec{{A}}\right),
                     \nonumber\\
\vec{{E}} &=& -\dfrac{1}{c}
                              \dfrac{\partial\,\vec{{A}}}{\partial\,t}
                        -\vec{\nabla}\varphi- \frac{{\rm i} g}{c\hbar} \left[\varphi,\ \vec{{A}}
\right].
\end{eqnarray}

\begin{remark} \textcolor{blue}{Calculation of the Magnetic-like Field.}
Due to symmetry, we only need to calculate the $z$-component of $\vec{B}$, i.e.,
\begin{eqnarray}
B_z=\partial_x A_y-\partial_y A_x-{\rm i}\frac{g}{\hbar c}[A_x,A_y].
\end{eqnarray}
We have
\begin{eqnarray}
A_x=\frac{1}{r}\left[k_1\left(\hat{r}_y\Gamma_z-\hat{r}_z\Gamma_y\right)+k_2\Gamma_x+k_3 \hat{r}_x (\vec{\Gamma} \cdot \hat{r})\right],
\end{eqnarray}
\begin{eqnarray}
A_y=\frac{1}{r}\left[k_1\left(\hat{r}_z\Gamma_x-\hat{r}_x\Gamma_z\right)+k_2\Gamma_y+k_3 \hat{r}_y (\vec{\Gamma} \cdot \hat{r})\right],
\end{eqnarray}
with the simple notations
\begin{eqnarray}
\hat{r}_x=\frac{x}{r}, \;\;\;\;  \hat{r}_y=\frac{y}{r}, \;\;\;\; \hat{r}_z=\frac{z}{r}.
\end{eqnarray}

(i) \emph{The Curl Part.---}Based on
\begin{eqnarray}
&&\frac{\partial \left(\dfrac{1}{r}\right)}{\partial x}
=-\frac{x}{r^3}=-\frac{\hat{r}_x}{r^2}, \;\;\;\; \frac{\partial \left(\dfrac{1}{r}\right)}{\partial y}
=-\frac{y}{r^3}=-\frac{\hat{r}_y}{r^2}, \;\;\;\; \frac{\partial \left(\dfrac{1}{r}\right)}{\partial z}
=-\frac{z}{r^3}=-\frac{\hat{r}_z}{r^2}, \nonumber\\
&&\frac{\partial \hat{r}_x}{\partial x}=\frac{\partial \left(\dfrac{x}{r}\right)}{\partial x}
=\frac{1}{r}- x \frac{x}{r^3}=\frac{1}{r}- \frac{x^2}{r^3}=\frac{1}{r}- \frac{\hat{r}_x^2}{r}=\frac{1-\hat{r}_x^2}{r},
\;\;\;\;\frac{\partial \hat{r}_y}{\partial y}=\frac{1-\hat{r}_y^2}{r}, \;\;\;\;
\frac{\partial \hat{r}_z}{\partial z}=\frac{1-\hat{r}_z^2}{r}, \nonumber\\
&&\frac{\partial \hat{r}_y}{\partial x}=\frac{\partial \left(\dfrac{y}{r}\right)}{\partial x}
= y \frac{\partial \left(\dfrac{1}{r}\right)}{\partial x} =- y \frac{\hat{r}_x}{r^2}=-\frac{\hat{r}_x \hat{r}_y}{r},
\;\;\;\; \frac{\partial \hat{r}_z}{\partial x}=-\frac{\hat{r}_x \hat{r}_z}{r},
\end{eqnarray}

\begin{eqnarray}
\frac{\partial \left( \vec{\Gamma} \cdot \hat{r} \right)}{\partial x}&=&\frac{\partial (\Gamma_x \hat{r}_x+\Gamma_y \hat{r}_y+\Gamma_z \hat{r}_z)}{\partial x}
=\Gamma_x \frac{\partial  \hat{r}_x}{\partial x}+\Gamma_y \frac{\partial  \hat{r}_y}{\partial x}+\Gamma_z \frac{\partial  \hat{r}_z}{\partial x}\nonumber\\
&=&\Gamma_x \frac{1-\hat{r}_x^2}{r} + \Gamma_y \left(-\frac{\hat{r}_x \hat{r}_y}{r}\right)+\Gamma_z \left(-\frac{\hat{r}_x \hat{r}_z}{r}\right)\nonumber\\
&=& \Gamma_x \frac{1-\hat{r}_x^2}{r} - \Gamma_y \left(\frac{\hat{r}_x \hat{r}_y}{r}\right)-\Gamma_z \left(-\frac{\hat{r}_x \hat{r}_z}{r}\right)
=\frac{1}{r} \Gamma_x - \frac{\hat{r}_x}{r} (\vec{\Gamma}\cdot\hat{r}),
\end{eqnarray}
\begin{eqnarray}
\frac{\partial \left( \vec{\Gamma} \cdot \hat{r} \right)}{\partial y}&=&
\frac{1}{r} \Gamma_y - \frac{\hat{r}_y}{r} (\vec{\Gamma}\cdot\hat{r}),
\end{eqnarray}
we have
\begin{eqnarray}
\partial_x A_y &=&\partial_x \left\{ \frac{1}{r} \left[ k_1 \left( \hat{r}_z\Gamma_x-\hat{r}_x\Gamma_z \right) + k_2\Gamma_y + k_3 \hat{r}_y \left( \vec{\Gamma} \cdot \hat{r} \right) \right] \right\} \nonumber\\
&=&-\frac{\hat{r}_x}{r^2} \left[ k_1 \left( \hat{r}_z\Gamma_x-\hat{r}_x\Gamma_z \right) + k_2\Gamma_y + k_3 \hat{r}_y \left( \vec{\Gamma} \cdot \hat{r} \right) \right] \nonumber\\
&&+ \frac{1}{r} \left[ -k_1\frac{\hat{r}_x \hat{r}_z}{r}\Gamma_x - k_1\frac{1 - \hat{r}_x^2}{r}\Gamma_z
+ k_3 \left(-\frac{\hat{r}_x \hat{r}_y}{r}\right) \left( \vec{\Gamma} \cdot \hat{r} \right)
+ k_3 \hat{r}_y \left(\frac{1}{r} \Gamma_x - \frac{\hat{r}_x}{r} (\vec{\Gamma}\cdot\hat{r})\right)\right] \nonumber\\
&=&-\frac{\hat{r}_x}{r^2} \left[ k_1 \left( \hat{r}_z\Gamma_x-\hat{r}_x\Gamma_z \right) + k_2\Gamma_y + k_3 \hat{r}_y \left( \vec{\Gamma} \cdot \hat{r} \right) \right] \nonumber\\
&&+ \frac{1}{r} \left[ -k_1\frac{\hat{r}_x \hat{r}_z}{r}\Gamma_x - k_1\frac{1 - \hat{r}_x^2}{r}\Gamma_z + \frac{k_3}{r} \left( \hat{r}_y \Gamma_x - 2\hat{r}_x \hat{r}_y \left( \vec{\Gamma} \cdot \hat{r} \right) \right) \right] \nonumber\\
&=&-\frac{1}{r^2} \left\{ k_1 \left[ 2\hat{r}_x \hat{r}_z \Gamma_x + \left( 1 - 2\hat{r}_x^2 \right) \Gamma_z \right] + k_2 \hat{r}_x \Gamma_y - k_3 \left[ \hat{r}_y \Gamma_x - 3\hat{r}_x \hat{r}_y \left( \vec{\Gamma} \cdot \hat{r} \right) \right] \right\},
\end{eqnarray}
and
\begin{eqnarray}
\partial_y A_x &=&\partial_y \left\{ \frac{1}{r} \left[ k_1 \left( \hat{r}_y \Gamma_z - \hat{r}_z \Gamma_y \right) + k_2 \Gamma_x + k_3 \hat{r}_x \left( \vec{\Gamma} \cdot \hat{r} \right) \right] \right\} \nonumber\\
&=&-\frac{\hat{r}_y}{r^2} \left[ k_1 \left( \hat{r}_y \Gamma_z - \hat{r}_z \Gamma_y \right) + k_2 \Gamma_x + k_3 \hat{r}_x \left( \vec{\Gamma} \cdot \hat{r} \right) \right] \nonumber\\
&+& \frac{1}{r} \left[ k_1 \frac{1 - \hat{r}_y^2}{r} \Gamma_z + k_1 \frac{\hat{r}_y \hat{r}_z}{r} \Gamma_y + \frac{k_3}{r} \left( \hat{r}_x \Gamma_y - 2\hat{r}_x \hat{r}_y \left( \vec{\Gamma} \cdot \hat{r} \right) \right) \right] \nonumber\\
&=&-\frac{1}{r^2} \left\{ -k_1 \left[ \left( 1 - 2\hat{r}_y^2 \right) \Gamma_z + 2\hat{r}_y \hat{r}_z \Gamma_y \right] + k_2 \hat{r}_y \Gamma_x - k_3 \left[ \hat{r}_x \Gamma_y - 3\hat{r}_x \hat{r}_y \left( \vec{\Gamma} \cdot \hat{r} \right) \right] \right\}.
\end{eqnarray}
So  we can have
\begin{eqnarray}
\partial_x A_y - \partial_y A_x &=& -\frac{1}{r^2} \left\{ k_1 \left[ 2\hat{r}_x \hat{r}_z \Gamma_x + \Gamma_z - 2\hat{r}_x^2 \Gamma_z \right] + k_2 \hat{r}_x \Gamma_y - k_3 \hat{r}_y \Gamma_x + 3k_3 \hat{r}_x \hat{r}_y \left( \vec{\Gamma} \cdot \hat{r} \right) \right\} \nonumber\\
&& + \frac{1}{r^2} \left\{ -k_1 \left[ \Gamma_z - 2\hat{r}_y^2 \Gamma_z + 2\hat{r}_y \hat{r}_z \Gamma_y \right] + k_2 \hat{r}_y \Gamma_x - k_3 \hat{r}_x \Gamma_y + 3k_3 \hat{r}_x \hat{r}_y \left( \vec{\Gamma} \cdot \hat{r} \right) \right\} \nonumber\\
&=& -\frac{1}{r^2} \biggr\{ k_1 \left[ 2\hat{r}_x \hat{r}_z \Gamma_x + 2\hat{r}_y \hat{r}_z \Gamma_y + 2\Gamma_z - 2 \left( \hat{r}_x^2 + \hat{r}_y^2 \right) \Gamma_z \right]  \nonumber\\
&&  + \left( k_2 + k_3 \right) \hat{r}_x \Gamma_y - \left( k_2 + k_3 \right) \hat{r}_y \Gamma_x + \left[ 3k_3 \hat{r}_x \hat{r}_y \left( \vec{\Gamma} \cdot \hat{r} \right) - 3k_3 \hat{r}_x \hat{r}_y \left( \vec{\Gamma} \cdot \hat{r} \right) \right] \biggr\} \nonumber\\
&=& -\frac{1}{r^2} \left\{ 2k_1 \left[ \hat{r}_z \left( \hat{r}_x \Gamma_x + \hat{r}_y \Gamma_y \right) + \Gamma_z \left( 1 - \hat{r}_x^2 - \hat{r}_y^2 \right) \right] + \left( k_2 + k_3 \right) \left( \hat{r}_x \Gamma_y - \hat{r}_y \Gamma_x \right) \right\} \nonumber\\
&=& -\frac{1}{r^2} \left\{ 2k_1 \left[ \hat{r}_z \left( \hat{r}_x \Gamma_x + \hat{r}_y \Gamma_y \right) + \Gamma_z \hat{r}_z^2 \right] + \left( k_2 + k_3 \right) \left( \hat{r} \times \vec{\Gamma} \right)_z \right\} \nonumber\\
&=& -\frac{1}{r^2} \left[ 2k_1 \hat{r}_z \left( \vec{\Gamma} \cdot \hat{r} \right) + \left( k_2 + k_3 \right) \left( \hat{r} \times \vec{\Gamma} \right)_z \right]\nonumber\\
&=& -\frac{1}{r^2} \left[ 2k_1 \left( \vec{\Gamma} \cdot \hat{r} \right) \hat{r}_z + \left( k_2 + k_3 \right) \left( \hat{r} \times \vec{\Gamma} \right)_z \right].
\end{eqnarray}

(ii) \emph{The Commutator Part.---}We have
\begin{eqnarray}
[A_x,A_y]=\frac{1}{r^2}\left[k_1\left(\hat{r}_y\Gamma_z-\hat{r}_z\Gamma_y\right)+k_2\Gamma_x+k_3 \hat{r}_x (\vec{\Gamma} \cdot \hat{r}), \; k_1\left(\hat{r}_z\Gamma_x-\hat{r}_x\Gamma_z\right)+k_2\Gamma_y+k_3 \hat{r}_y (\vec{\Gamma} \cdot \hat{r})\right].
\end{eqnarray}
We calculate the nine commutators as follows.

\begin{eqnarray}
 \left[ \hat{r}_y \Gamma_z - \hat{r}_z \Gamma_y, \hat{r}_z \Gamma_x - \hat{r}_x \Gamma_z \right] &=& \hat{r}_y \hat{r}_z \left[\Gamma_z, \Gamma_x\right] - \hat{r}_y \hat{r}_x \left[\Gamma_z, \Gamma_z\right] - \hat{r}_z^2 \left[\Gamma_y, \Gamma_x\right] + \hat{r}_z \hat{r}_x \left[\Gamma_y, \Gamma_z\right] \nonumber\\
&=& \hat{r}_y \hat{r}_z \left(2 {\rm i} \Gamma_y\right) - 0 - \hat{r}_z^2 \left(-2i \Gamma_z\right) + \hat{r}_z \hat{r}_x \left(2 {\rm i} \Gamma_x\right) \nonumber\\
&=& 2{\rm i} \hat{r}_z \left(\hat{r}_x \Gamma_x + \hat{r}_y \Gamma_y + \hat{r}_z \Gamma_z\right) = 2 {\rm i} \hat{r}_z \left(\vec{\Gamma} \cdot \hat{r}\right)= 2 {\rm i} \left(\vec{\Gamma} \cdot \hat{r}\right)\hat{r}_z,
\end{eqnarray}

\begin{eqnarray}
\left[ \hat{r}_y \Gamma_z - \hat{r}_z \Gamma_y, \Gamma_y \right] = \hat{r}_y \left[\Gamma_z, \Gamma_y\right] - \hat{r}_z \left[\Gamma_y, \Gamma_y\right] = \hat{r}_y \left(-2 {\rm i} \Gamma_x\right) - 0 = -2 {\rm i} \hat{r}_y \Gamma_x,
\end{eqnarray}

\begin{eqnarray}
 \left[ \hat{r}_y \Gamma_z - \hat{r}_z \Gamma_y, \hat{r}_y \left(\vec{\Gamma} \cdot \hat{r}\right) \right]
&=& \hat{r}_y \left\{ \hat{r}_y \left[ \Gamma_z, \hat{r}_x \Gamma_x + \hat{r}_y \Gamma_y + \hat{r}_z \Gamma_z \right] - \hat{r}_z \left[ \Gamma_y, \hat{r}_x \Gamma_x + \hat{r}_y \Gamma_y + \hat{r}_z \Gamma_z \right] \right\} \nonumber\\
&=& \hat{r}_y \left\{ \hat{r}_y \left( \hat{r}_x \left[ 2 {\rm i} \Gamma_y \right] - \hat{r}_y \left[ 2 {\rm i} \Gamma_x \right] \right) - \hat{r}_z \left( \hat{r}_x \left[ -2 {\rm i} \Gamma_z \right] + \hat{r}_z \left[ 2 {\rm i} \Gamma_x \right] \right) \right\} \nonumber\\
&=& 2 {\rm i} \hat{r}_y \left\{ \hat{r}_x \left( \hat{r}_y \Gamma_y + \hat{r}_z \Gamma_z \right) - \Gamma_x \left( \hat{r}_y^2 + \hat{r}_z^2 \right) \right\} \nonumber\\
&=& 2 {\rm i} \hat{r}_y \left\{ \hat{r}_x \left( \hat{r}_y \Gamma_y + \hat{r}_z \Gamma_z + \hat{r}_x \Gamma_x - \hat{r}_x \Gamma_x \right) - \Gamma_x \left( 1 - \hat{r}_x^2 \right) \right\} \nonumber\\
&=& 2 {\rm i} \hat{r}_y \left[ \hat{r}_x \left( \vec{\Gamma} \cdot \hat{r} \right) - \Gamma_x \right],
\end{eqnarray}

\begin{eqnarray}
\left[ \Gamma_x, \hat{r}_z \Gamma_x - \hat{r}_x \Gamma_z \right] = \hat{r}_z \left[\Gamma_x, \Gamma_x\right] - \hat{r}_x \left[\Gamma_x, \Gamma_z\right] = 0 - \hat{r}_x \left(-2{\rm i} \Gamma_y\right) = 2 {\rm i} \hat{r}_x \Gamma_y,
\end{eqnarray}

\begin{eqnarray}
\left[ \Gamma_x, \Gamma_y \right] = 2{\rm i}  \Gamma_z,
\end{eqnarray}

\begin{eqnarray}
\left[ \Gamma_x, \hat{r}_y \left(\vec{\Gamma} \cdot \hat{r}\right) \right] = \hat{r}_y \left[\Gamma_x, \left(\vec{\Gamma} \cdot \hat{r}\right)\right] = \hat{r}_y \left[ 2 {\rm i} \left(\hat{r}_y \Gamma_z - \hat{r}_z \Gamma_y\right) \right] = 2 {\rm i} \hat{r}_y \left(\hat{r}_y \Gamma_z - \hat{r}_z \Gamma_y\right),
\end{eqnarray}

\begin{eqnarray}
 \left[ \hat{r}_x \left(\vec{\Gamma} \cdot \hat{r}\right), \hat{r}_z \Gamma_x - \hat{r}_x \Gamma_z \right]
&=& \hat{r}_x \left\{ \hat{r}_z \left[ \hat{r}_x \Gamma_x + \hat{r}_y \Gamma_y + \hat{r}_z \Gamma_z, \Gamma_x\right] - \hat{r}_x \left[ \hat{r}_x \Gamma_x + \hat{r}_y \Gamma_y + \hat{r}_z \Gamma_z, \Gamma_z\right] \right\} \nonumber\\
&=& \hat{r}_x \left\{ \hat{r}_z \left( \hat{r}_y \left[-2 {\rm i} \Gamma_z\right] + \hat{r}_z \left[2 {\rm i} \Gamma_y\right] \right) - \hat{r}_x \left( \hat{r}_x \left[-2 {\rm i} \Gamma_y\right] + \hat{r}_y \left[2 {\rm i} \Gamma_x\right] \right) \right\} \nonumber\\
&=& 2{\rm i}  \hat{r}_x \left\{ \hat{r}_z^2 \Gamma_y - \hat{r}_z \hat{r}_y \Gamma_z + \hat{r}_x^2 \Gamma_y - \hat{r}_x \hat{r}_y \Gamma_x \right\} \nonumber\\
&=& 2{\rm i} \hat{r}_x \left\{ \left(\hat{r}_x^2 + \hat{r}_z^2\right) \Gamma_y - \hat{r}_y \left( \hat{r}_x \Gamma_x + \hat{r}_z \Gamma_z \right) \right\} \nonumber\\
&=& 2{\rm i} \hat{r}_x \left\{ \left(1 - \hat{r}_y^2\right) \Gamma_y - \hat{r}_y \left( \hat{r}_x \Gamma_x + \hat{r}_z \Gamma_z \right) \right\} \nonumber\\
&=& 2{\rm i} \hat{r}_x \left[ \Gamma_y - \hat{r}_y \left(\vec{\Gamma} \cdot \hat{r}\right) \right],
\end{eqnarray}
\begin{eqnarray}
\left[ \hat{r}_x \left(\vec{\Gamma} \cdot \hat{r}\right), \Gamma_y \right] = \hat{r}_x \left[\left(\vec{\Gamma} \cdot \hat{r}\right), \Gamma_y\right] = \hat{r}_x \left[ -2{\rm i} \left(\hat{r}_z \Gamma_x - \hat{r}_x \Gamma_z\right) \right] = 2{\rm i} \hat{r}_x \left(\hat{r}_x \Gamma_z - \hat{r}_z \Gamma_x\right),
\end{eqnarray}

\begin{eqnarray}
\left[ \hat{r}_x \left(\vec{\Gamma} \cdot \hat{r}\right), \hat{r}_y \left(\vec{\Gamma} \cdot \hat{r}\right) \right] = \hat{r}_x \hat{r}_y \left[ \left(\vec{\Gamma} \cdot \hat{r}\right), \left(\vec{\Gamma} \cdot \hat{r}\right) \right] = 0.
\end{eqnarray}
By combining these nine terms, we can obtain
\begin{eqnarray}
\left[ A_x, A_y \right] &=& \frac{2{\rm i}}{r^2} \biggr\{  k_1^2 \hat{r}_z \left( \vec{\Gamma} \cdot \hat{r} \right) - k_1k_2\hat{r}_y \Gamma_x + k_1k_3\hat{r}_y \left[ \hat{r}_x \left( \vec{\Gamma} \cdot \hat{r} \right) - \Gamma_x \right] \nonumber\\
&& + k_2k_1 \hat{r}_x \Gamma_y + k_2^2\Gamma_z + k_2k_3\hat{r}_y \left( \hat{r}_y \Gamma_z - \hat{r}_z \Gamma_y \right) \nonumber\\
&& + k_3k_1 \hat{r}_x \left[ \Gamma_y - \hat{r}_y \left(\vec{\Gamma} \cdot \hat{r}\right) \right]+ k_3k_2\hat{r}_x \left( \hat{r}_x \Gamma_z - \hat{r}_z \Gamma_x \right)  \biggr\}\nonumber\\
&=& \frac{2{\rm i}}{r^2} \biggr\{ k_1^2 \hat{r}_z \left( \vec{\Gamma} \cdot \hat{r} \right) + k_2^2 \Gamma_z + k_1 k_2 \left( \hat{r}_x \Gamma_y - \hat{r}_y \Gamma_x \right)  \nonumber\\
&& + k_1 k_3 \left( \hat{r}_y \hat{r}_x \left( \vec{\Gamma} \cdot \hat{r} \right) - \hat{r}_y \Gamma_x + \hat{r}_x \Gamma_y - \hat{r}_x \hat{r}_y \left( \vec{\Gamma} \cdot \hat{r} \right) \right) \nonumber\\
&&  + k_2 k_3 \left[ \left( \hat{r}_x^2 + \hat{r}_y^2 \right) \Gamma_z - \hat{r}_z \left( \hat{r}_x \Gamma_x + \hat{r}_y \Gamma_y \right) \right] \biggr\} \nonumber\\
&=& \frac{2{\rm i}}{r^2} \left\{ k_1^2 \hat{r}_z \left( \vec{\Gamma} \cdot \hat{r} \right) + k_2^2 \Gamma_z + k_1 \left( k_2 + k_3 \right) \left( \hat{r}_x \Gamma_y - \hat{r}_y \Gamma_x \right) + k_2 k_3 \left[ \Gamma_z - \hat{r}_z \left( \vec{\Gamma} \cdot \hat{r} \right) \right] \right\} \nonumber\\
&=& \frac{2{\rm i}}{r^2} \left[ \left( k_1^2 - k_2 k_3 \right) \hat{r}_z \left( \vec{\Gamma} \cdot \hat{r} \right) + k_2 \left( k_2 + k_3 \right) \Gamma_z + k_1 \left( k_2 + k_3 \right) \left( \hat{r}_x \Gamma_y - \hat{r}_y \Gamma_x \right) \right]\nonumber\\
&=&\frac{2{\rm i}}{r^2} \left[ \left( k_1^2 - k_2 k_3 \right) \hat{r}_z \left( \vec{\Gamma} \cdot \hat{r} \right) + k_2 \left( k_2 + k_3 \right) \Gamma_z + k_1 \left( k_2 + k_3 \right) \left(\hat{r}\times\vec{\Gamma}\right)_z \right].
\end{eqnarray}
So we can have
\begin{eqnarray}
B_z&=&\partial_x A_y-\partial_y A_x-{\rm i}\frac{g}{\hbar c}\left[A_x,A_y\right]\nonumber\\
&=&-\frac{1}{r^2} \left[ 2k_1 \hat{r}_z \left( \vec{\Gamma} \cdot \hat{r} \right) + \left( k_2 + k_3 \right) \left( \hat{r} \times \vec{\Gamma} \right)_z \right] \nonumber\\
&&+\frac{2g}{\hbar c}\frac{1}{r^2} \left[ \left( k_1^2 - k_2 k_3 \right) \hat{r}_z \left( \vec{\Gamma} \cdot \hat{r} \right) + k_2 \left( k_2 + k_3 \right) \Gamma_z + k_1 \left( k_2 + k_3 \right) \left(\hat{r}\times\vec{\Gamma}\right)_z \right]\nonumber\\
&=& \frac{1}{r^2} \left\{ \left[ 2k_1 \left( \frac{g}{\hbar c} k_1 - 1 \right) - \frac{2g}{\hbar c} k_2 k_3 \right] \left( \vec{\Gamma} \cdot \hat{r} \right) \hat{r}_z + \left( k_2 + k_3 \right) \left( \frac{2g}{\hbar c} k_1 - 1 \right) \left( \hat{r} \times \vec{\Gamma} \right)_z + \frac{2g}{\hbar c} k_2 \left( k_2 + k_3 \right) \Gamma_z \right\}.
\end{eqnarray}
By symmetry, we finally have
\begin{eqnarray}\label{eq:B-1}
\vec{B}=\frac{1}{r^2} \left\{ \left[ 2k_1 \left( \frac{g}{\hbar c} k_1 - 1 \right) - \frac{2g}{\hbar c} k_2 k_3 \right] \left( \vec{\Gamma} \cdot \hat{r} \right) \hat{r}+ \left( k_2 + k_3 \right) \left( \frac{2g}{\hbar c} k_1 - 1 \right) \left( \hat{r} \times \vec{\Gamma} \right) + \frac{2g}{\hbar c} k_2 \left( k_2 + k_3 \right) \vec{\Gamma} \right\}.
\end{eqnarray}
If one considers the constraint $k_2+k_3=0$, then he simply has
\begin{eqnarray}
\vec{B}&=&\frac{1}{r^2} \left\{ \left[ 2k_1 \left( \frac{g}{\hbar c} k_1 - 1 \right) - \frac{2g}{\hbar c} k_2 k_3 \right] \left( \vec{\Gamma} \cdot \hat{r} \right) \hat{r}\right\}\nonumber\\
&=& \left[ 2k_1 \left( \frac{g}{\hbar c} k_1 - 1 \right) + \frac{2g}{\hbar c} k_3^2 \right] \left( \vec{\Gamma} \cdot \hat{r} \right)  \frac{\vec{r}}{r^3}.
\end{eqnarray}
By considering
\begin{eqnarray}
&& k_1= {\frac{c\hbar}{2q}} a_1, \;\;\;\; k_3=\frac{c\hbar}{2q}a_3,\;\;\;\; q=g,
\end{eqnarray}
one has
\begin{eqnarray}
\vec{B}
&=& \left[ 2k_1 \left( \frac{g}{\hbar c} k_1 - 1 \right) + \frac{2g}{\hbar c} k_3^2 \right] \left( \vec{\Gamma} \cdot \hat{r} \right)  \frac{\vec{r}}{r^3}\nonumber\\
&=& \left[ 2\frac{c\hbar}{2g}a_1 \left( \frac{a_1}{2} - 1 \right) + \frac{c\hbar}{2g}a_3^2 \right] \left( \vec{\Gamma} \cdot \hat{r} \right)  \frac{\vec{r}}{r^3}\nonumber\\
&=& \left[ \frac{c\hbar}{2g}a_1 \left( a_1 - 2 \right) + \frac{c\hbar}{2g}a_3^2 \right] \left( \vec{\Gamma} \cdot \hat{r} \right)  \frac{\vec{r}}{r^3}\nonumber\\
&=& \left[ a_1 \left( a_1 - 2 \right) + a_3^2 \right] \frac{c\hbar}{2g}\left( \vec{\Gamma} \cdot \hat{r} \right)  \frac{\vec{r}}{r^3}\nonumber\\
&=& \left[ (a_1-1)^2+ a_3^2-1 \right] \frac{c\hbar}{2g}\left( \vec{\Gamma} \cdot \hat{r} \right)  \frac{\vec{r}}{r^3}.
\end{eqnarray}
Due to the constraint $(a_1-1)^2+ a_3^2=1$, actually we have
\begin{eqnarray}
\vec{B} &=& 0.
\end{eqnarray}
In other words, if we do not impose the constraint condition on $k_i$'s $(i=1, 2, 3)$, then the expression of the ``magnetic'' field is given in
Eq. (\ref{eq:B-1}), while if we consider the constraint condition obtained by the VPEA, i.e.,
\begin{eqnarray}
&& k_1= {\frac{c\hbar}{2q}} a_1, \;\;\;\; k_2=-\frac{c\hbar}{2q}a_3, \;\;\;\; k_3=\frac{c\hbar}{2q}a_3,\nonumber\\
&&(a_1-1)^2+ a_3^2=1,
\end{eqnarray}
then the ``magnetic'' field is zero. $\blacksquare$
\end{remark}

\begin{remark} \textcolor{blue}{Calculation of the Electric-like Field.} We can have that
\begin{eqnarray}
 \vec{E} = -\dfrac{1}{c}\dfrac{\partial\,\vec{A}}{\partial\,t}   -\vec{\nabla}\varphi-{\rm i}\,\frac{g}{\hbar c}\left[\varphi,\ \vec{A}\right]=-\vec{\nabla}\varphi-{\rm i}\,\frac{g}{\hbar c}\left[\varphi,\ \vec{A}\right]
\end{eqnarray}
By symmetry, it is sufficient to calculate the $x$-direction, which is
\begin{eqnarray}
E_x=-\partial_x\varphi-{\rm i}\,\frac{g}{\hbar c}\left[\varphi,A_x\right]
\end{eqnarray}

(i) \emph{The Gradient Part.---}Because
\begin{eqnarray}
\varphi=f_1(r)(\vec{\Gamma}\cdot\hat{r})+f_2(r),
\end{eqnarray}
and
\begin{eqnarray}
\frac{\partial \left( \vec{\Gamma} \cdot \hat{r} \right)}{\partial x}&=&\frac{1}{r} \Gamma_x - \frac{\hat{r}_x}{r} (\vec{\Gamma}\cdot\hat{r})
={\frac{1}{r} \Gamma_x - \frac{x}{r^2} (\vec{\Gamma}\cdot\hat{r})},
\end{eqnarray}
we can have
\begin{eqnarray}
\partial_x \varphi &=& \partial_x \left[ f_1(r)\left(\vec{\Gamma}\cdot\hat{r}\right)+f_2(r)  \right] \nonumber\\
&=&\left[\partial_x f_1(r)\right] \left(\vec{\Gamma}\cdot\hat{r}\right) +f_1(r)\left[\partial_x \left(\vec{\Gamma}\cdot\hat{r}\right)  \right]+\partial_x f_2(r)  \nonumber\\
&=&\frac{x}{r} \frac{\partial f_1(r)}{\partial r} +f_1(r) \left[ {\frac{1}{r} \Gamma_x - \frac{x}{r^2} (\vec{\Gamma}\cdot\hat{r})} \right]
+\frac{x}{r} \frac{\partial f_2(r)}{\partial r}  \nonumber\\
&=&\frac{\Gamma_x f_1(r)}{r} + \frac{x}{r} \left[\frac{\partial f_1(r)}{\partial r} - \frac{f_1(r)}{r}\right] (\vec{\Gamma} \cdot \hat{r})+\frac{x}{r} \frac{\partial f_2(r)}{\partial r} \nonumber\\
&=&\frac{\Gamma_x f_1(r)}{r} + \hat{r}_x \left[\frac{\partial f_1(r)}{\partial r} - \frac{f_1(r)}{r}\right] (\vec{\Gamma} \cdot \hat{r})
+\hat{r}_x \frac{\partial f_2(r)}{\partial r}.
\end{eqnarray}

(ii) \emph{The Commutator Part.---}We have
\begin{eqnarray}
\left[\varphi,A_x\right]=\left[f_1(r)(\vec{\Gamma}\cdot\hat{r})+f_2(r),\, A_x\right]=f_1(r)\left[(\vec{\Gamma}\cdot\hat{r}),\, A_x\right].
\end{eqnarray}
Consider that
\begin{eqnarray}
A_x=\frac{1}{r}\left[k_1\left(\hat{r}_y\Gamma_z-\hat{r}_z\Gamma_y\right)+k_2\Gamma_x+k_3 \hat{r}_x (\vec{\Gamma} \cdot \hat{r})\right].
\end{eqnarray}
Note that the $k_3 $ term commutes with $A_x$, so we only need to calculate the following three commutators
\begin{eqnarray}
\left[\vec{\Gamma}\cdot\hat{r},\; \Gamma_z\right]&=&\hat{r}_x\left[\Gamma_x,\Gamma_z\right]+\hat{r}_y\left[\Gamma_y,\Gamma_z\right] = -2 {\rm i} \left(\hat{r}_x \Gamma_y - \hat{r}_y \Gamma_x\right),\nonumber\\
\left[\vec{\Gamma}\cdot\hat{r},\; \Gamma_y\right] &=& -2{\rm i}  \left(\hat{r}_z \Gamma_x - \hat{r}_x \Gamma_z\right), \nonumber\\
\left[\vec{\Gamma}\cdot\hat{r}, \;\Gamma_x\right] &=& -2{\rm i}  \left(\hat{r}_y \Gamma_z - \hat{r}_z \Gamma_y\right).
\end{eqnarray}
By combining them, we get
\begin{eqnarray}
\left[\vec{\Gamma}\cdot\hat{r}, \; A_x\right]&=&-2{\rm i}\frac{1}{r}\left\{k_1\left[\hat{r}_y\left(\hat{r}_x \Gamma_y - \hat{r}_y \Gamma_x\right)-\hat{r}_z\left(\hat{r}_z \Gamma_x - \hat{r}_x \Gamma_z\right)\right]+k_2\left(\hat{r}_y \Gamma_z - \hat{r}_z \Gamma_y\right)\right\}\nonumber\\
&=&-2 {\rm i} \frac{1}{r}  \left\{  k_1\left[\hat{r}_x \left(\hat{r} \cdot \vec{\Gamma}\right) - \Gamma_x\right]+k_2\left(\hat{r} \times \vec{\Gamma}\right)_x   \right\}.
\end{eqnarray}
Based on above calculation, we can have
\begin{eqnarray}
E_x&=&-\partial_x\varphi-{\rm i}\,\frac{g}{\hbar c}\left[\varphi,\; A_x\right]\nonumber\\
&=&-\left[\frac{\Gamma_x f_1(r)}{r} + \hat{r}_x \left[\frac{\partial f_1(r)}{\partial r} - \frac{f_1(r)}{r}\right] (\vec{\Gamma} \cdot \hat{r})
+\hat{r}_x \frac{\partial f_2(r)}{\partial r}\right]-\frac{2g}{\hbar c} \frac{f_1(r)}{r}  \left\{  k_1\left(\hat{r}_x \left(\hat{r} \cdot \vec{\Gamma}\right) - \Gamma_x\right)+k_2\left(\hat{r} \times \vec{\Gamma}\right)_x   \right\}\nonumber\\
&=& - \hat{r}_x \frac{\partial f_2(r)}{\partial r} - \frac{f_1(r)}{r} \left( 1 - \frac{2g k_1}{\hbar c} \right) \Gamma_x - \hat{r}_x (\vec{\Gamma} \cdot \hat{r}) \left[ \frac{\partial f_1(r)}{\partial r} - \frac{f_1(r)}{r} \left( 1 - \frac{2gk_1}{\hbar c} \right) \right] - \frac{2gk_2  }{\hbar c } \frac{f_1(r)}{r}\left(\hat{r} \times \vec{\Gamma}\right)_x. \nonumber\\
\end{eqnarray}
By symmetry, we have the electric-like field as
\begin{eqnarray}
\vec{E}=-  \frac{\partial f_2(r)}{\partial r} \hat{r}- \frac{f_1(r)}{r} \left( 1 - \frac{2g k_1}{\hbar c} \right) \vec{\Gamma} - \hat{r} (\vec{\Gamma} \cdot \hat{r}) \left[ \frac{\partial f_1(r)}{\partial r} - \frac{f_1(r)}{r} \left( 1 - \frac{2gk_1}{\hbar c} \right) \right] - \frac{2gk_2  }{\hbar c } \frac{f_1(r)}{r}\left(\hat{r} \times \vec{\Gamma}\right),
\end{eqnarray}
which does not depend on $k_3$.
$\blacksquare$
\end{remark}

\subsection{The Constraint Equations Obtained from Solving the Yang-Mills Equations}


We now come to solve the Yang-Mills equations and we shall obtain some constraint equations. We just need to solve
      \begin{subequations}
            \begin{eqnarray}
                  && \vec{\nabla}\cdot\vec{E}\ {-}\boxed{{\rm i}\,\frac{g}{\hbar c}\Bigl(
                        \vec{A}\cdot\vec{E}
                        -\vec{E}\cdot\vec{A}\Bigr)}=0, \label{eq:DivEYM} \\
                  && \vec{\nabla}\times\vec{B}  {-}\boxed{{\rm i}\,\frac{g}{\hbar c}\biggl(\Bigl[\varphi,\
                                    \vec{E}\Bigr]
                              +\vec{A}\times\vec{B}
                              +\vec{B}\times\vec{A}\biggr)}=0, \label{eq:CurlBYM}
            \end{eqnarray}
      \end{subequations}
with the following quantities
\begin{eqnarray}
\vec{A} = \frac{1}{r} \left[ k_1 (\hat{r} \times \vec{\Gamma}) + k_2 \vec{\Gamma} + k_3 \hat{r} (\vec{\Gamma} \cdot \hat{r}) \right],
 \end{eqnarray}
\begin{eqnarray}
\varphi=f_1(r)(\vec{\Gamma}\cdot\hat{r})+f_2(r),
 \end{eqnarray}
\begin{eqnarray}\label{eq:B-2}
\vec{B}=\frac{1}{r^2} \left\{ \left[ 2k_1 \left( \frac{g}{\hbar c} k_1 - 1 \right) - \frac{2g}{\hbar c} k_2 k_3 \right] \left( \vec{\Gamma} \cdot \hat{r} \right) \hat{r}+ \left( k_2 + k_3 \right) \left( \frac{2g}{\hbar c} k_1 - 1 \right) \left( \hat{r} \times \vec{\Gamma} \right) + \frac{2g}{\hbar c} k_2 \left( k_2 + k_3 \right) \vec{\Gamma} \right\},
\end{eqnarray}
\begin{eqnarray}
\vec{E}=-  \frac{\partial f_2(r)}{\partial r} \hat{r}- \frac{f_1(r)}{r} \left( 1 - \frac{2g k_1}{\hbar c} \right) \vec{\Gamma} - \hat{r} (\vec{\Gamma} \cdot \hat{r}) \left[ \frac{\partial f_1(r)}{\partial r} - \frac{f_1(r)}{r} \left( 1 - \frac{2gk_1}{\hbar c} \right) \right] - \frac{2gk_2  }{\hbar c } \frac{f_1(r)}{r}\left(\hat{r} \times \vec{\Gamma}\right).
\end{eqnarray}

\subsubsection{Checking the Equation of the Magnetic-like Field}

Let us first check (\ref{eq:CurlBYM}). We need
\begin{eqnarray}
 \vec{\nabla}\times\vec{B}  {-}\boxed{{\rm i}\,\frac{g}{\hbar c}\biggl(\Bigl[\varphi,\vec{E}\Bigr]+\vec{A}\times\vec{B}+\vec{B}\times\vec{A}\biggr)}=0
\end{eqnarray}
For the vector identity, due to symmetry, we only need to verify the $z$-component. The individual terms are calculated as follows:

(i) Consider the magnetic field commutator term:
\begin{eqnarray}
\left(\vec{A}\times\vec{B}+\vec{B}\times\vec{A}\right)_z=\left[A_x, B_y\right] - \left[A_y, B_x\right].
\end{eqnarray}
Because
\begin{eqnarray}
A_x=\frac{1}{r}\left[k_1\left(\hat{r}_y\Gamma_z-\hat{r}_z\Gamma_y\right)+k_2\Gamma_x+k_3 \hat{r}_x (\vec{\Gamma} \cdot \hat{r})\right],
\end{eqnarray}
\begin{eqnarray}
B_y= \frac{1}{r^2} \left\{ \left[ 2k_1 \left( \frac{g}{\hbar c} k_1 - 1 \right) - \frac{2g}{\hbar c} k_2 k_3 \right] \left( \vec{\Gamma} \cdot \hat{r} \right) \hat{r}_y + \left( k_2 + k_3 \right) \left( \frac{2g}{\hbar c} k_1 - 1 \right) \left( \hat{r}_z\Gamma_x - \hat{r}_x\Gamma_z \right) + \frac{2g}{\hbar c} k_2 \left( k_2 + k_3 \right) \Gamma_y \right\},
\end{eqnarray}
we can calculate the following nine commutators:
\begin{eqnarray}
\left[ \hat{r}_y\Gamma_z - \hat{r}_z\Gamma_y, \left( \vec{\Gamma} \cdot \hat{r} \right) \hat{r}_y \right] &=& 2 {\rm i}\hat{r}_y \left[ \hat{r}_x \left( \hat{r}_y\Gamma_y + \hat{r}_z\Gamma_z \right) - \left( \hat{r}_y^2 + \hat{r}_z^2 \right) \Gamma_x \right] \nonumber \\
&=& 2 {\rm i}\hat{r}_y \left[ \hat{r}_x \left( \vec{\Gamma} \cdot \hat{r} - \hat{r}_x\Gamma_x \right) - \left( 1 - \hat{r}_x^2 \right) \Gamma_x \right] \nonumber \\
&=& 2{\rm i}\hat{r}_y \left[ \hat{r}_x \left( \vec{\Gamma} \cdot \hat{r} \right) - \Gamma_x \right],
\end{eqnarray}
\begin{eqnarray}
\left[ \hat{r}_y\Gamma_z - \hat{r}_z\Gamma_y, \hat{r}_z\Gamma_x - \hat{r}_x\Gamma_z \right] &=& \hat{r}_y\hat{r}_z \left[ \Gamma_z, \Gamma_x \right] - \hat{r}_z^2 \left[ \Gamma_y, \Gamma_x \right] + \hat{r}_x\hat{r}_z \left[ \Gamma_y, \Gamma_z \right] \nonumber \\
&=& 2 {\rm i}\hat{r}_y\hat{r}_z \Gamma_y + 2{\rm i}\hat{r}_z^2 \Gamma_z + 2{\rm i}\hat{r}_x\hat{r}_z \Gamma_x \nonumber \\
&=& 2{\rm i}\hat{r}_z \left( \vec{\Gamma} \cdot \hat{r} \right),
\end{eqnarray}
\begin{eqnarray}
\left[ \hat{r}_y\Gamma_z - \hat{r}_z\Gamma_y, \Gamma_y \right] &=& \hat{r}_y \left[ \Gamma_z, \Gamma_y \right] - \hat{r}_z \left[ \Gamma_y, \Gamma_y \right] \nonumber \\
&=& \hat{r}_y \left( -2{\rm i}\Gamma_x \right) - 0 \nonumber \\
&=& -2{\rm i} \hat{r}_y \Gamma_x,
\end{eqnarray}
\begin{eqnarray}
\left[ \Gamma_x, \left( \vec{\Gamma} \cdot \hat{r} \right) \hat{r}_y \right] &=& \hat{r}_y \left[ \hat{r}_x \left[ \Gamma_x, \Gamma_x \right] + \hat{r}_y \left[ \Gamma_x, \Gamma_y \right] + \hat{r}_z \left[ \Gamma_x, \Gamma_z \right] \right] \nonumber \\
&=& \hat{r}_y \left[ 0 + \hat{r}_y \left( 2{\rm i}\Gamma_z \right) + \hat{r}_z \left( -2{\rm i}\Gamma_y \right) \right] \nonumber \\
&=& 2{\rm i}\hat{r}_y \left( \hat{r}_y\Gamma_z - \hat{r}_z\Gamma_y \right),
\end{eqnarray}
\begin{eqnarray}
\left[ \Gamma_x, \hat{r}_z\Gamma_x - \hat{r}_x\Gamma_z \right] &=& \hat{r}_z \left[ \Gamma_x, \Gamma_x \right] - \hat{r}_x \left[ \Gamma_x, \Gamma_z \right] \nonumber \\
&=& \hat{r}_z \cdot 0 - \hat{r}_x \left( -2{\rm i}\Gamma_y \right) \nonumber \\
&=& 2{\rm i}\hat{r}_x \Gamma_y,
\end{eqnarray}
\begin{eqnarray}
\left[ \Gamma_x, \Gamma_y \right] = 2{\rm i}\Gamma_z,
\end{eqnarray}
\begin{eqnarray}
\left[ \hat{r}_x \left( \vec{\Gamma} \cdot \hat{r} \right), \left( \vec{\Gamma} \cdot \hat{r} \right) \hat{r}_y \right] &=& \hat{r}_x\hat{r}_y \left[ \vec{\Gamma} \cdot \hat{r}, \vec{\Gamma} \cdot \hat{r} \right] = 0,
\end{eqnarray}
\begin{eqnarray}
\left[ \hat{r}_x \left( \vec{\Gamma} \cdot \hat{r} \right), \hat{r}_z\Gamma_x - \hat{r}_x\Gamma_z \right] &=& \hat{r}_x\hat{r}_z \left[ \vec{\Gamma} \cdot \hat{r}, \Gamma_x \right] - \hat{r}_x^2 \left[ \vec{\Gamma} \cdot \hat{r}, \Gamma_z \right] \nonumber \\
&=& \hat{r}_x\hat{r}_z \left( 2 {\rm i}\hat{r}_z\Gamma_y - 2{\rm i}\hat{r}_y\Gamma_z \right) - \hat{r}_x^2 \left( 2{\rm i}\hat{r}_y\Gamma_x - 2{\rm i}\hat{r}_x\Gamma_y \right) \nonumber \\
&=& 2{\rm i}\hat{r}_x \left[ \left( \hat{r}_z^2 + \hat{r}_x^2 \right) \Gamma_y - \hat{r}_y \left( \hat{r}_x\Gamma_x + \hat{r}_z\Gamma_z \right) \right] \nonumber \\
&=& 2{\rm i}\hat{r}_x \left[ \left( 1 - \hat{r}_y^2 \right) \Gamma_y - \hat{r}_y \left( \vec{\Gamma} \cdot \hat{r} - \hat{r}_y\Gamma_y \right) \right] \nonumber \\
&=& 2{\rm i}\hat{r}_x \left[ \Gamma_y - \hat{r}_y \left( \vec{\Gamma} \cdot \hat{r} \right) \right],
\end{eqnarray}
\begin{eqnarray}
\left[ \hat{r}_x \left( \vec{\Gamma} \cdot \hat{r} \right), \Gamma_y \right] &=& \hat{r}_x \left[ \vec{\Gamma} \cdot \hat{r}, \Gamma_y \right] \nonumber \\
&=& \hat{r}_x \left( \hat{r}_x \left[ \Gamma_x, \Gamma_y \right] + \hat{r}_z \left[ \Gamma_z, \Gamma_y \right] \right) \nonumber \\
&=& 2{\rm i}\hat{r}_x \left( \hat{r}_x\Gamma_z - \hat{r}_z\Gamma_x \right).
\end{eqnarray}
By combining them according to their coefficients, we have
\begin{eqnarray}
\left[ A_x, B_y \right] &=& \frac{2{\rm i}}{r^3} \left\{ \left[ 2k_1 \left( \frac{g}{\hbar c} k_1 - 1 \right) - \frac{2g}{\hbar c} k_2 k_3 \right] \hat{r}_y \left[ k_1 \left( \hat{r}_x \left( \vec{\Gamma} \cdot \hat{r} \right) - \Gamma_x \right) + k_2 \left( \hat{r}_y \Gamma_z - \hat{r}_z \Gamma_y \right) \right] \right. \nonumber \\
&& + \left[ \left( k_2 + k_3 \right) \left( \frac{2g}{\hbar c} k_1 - 1 \right) \right] \left[ k_1 \hat{r}_z \left( \vec{\Gamma} \cdot \hat{r} \right) + k_2 \hat{r}_x \Gamma_y + k_3 \hat{r}_x \left( \Gamma_y - \hat{r}_y \left( \vec{\Gamma} \cdot \hat{r} \right) \right) \right] \nonumber \\
&& \left. + \left[ \frac{2g}{\hbar c} k_2 \left( k_2 + k_3 \right) \right] \left[ -k_1 \hat{r}_y \Gamma_x + k_2 \Gamma_z + k_3 \hat{r}_x \left( \hat{r}_x \Gamma_z - \hat{r}_z \Gamma_x \right) \right] \right\}.
\end{eqnarray}

Similarly, since
\begin{eqnarray}
A_y=\frac{1}{r}\left[k_1\left(\hat{r}_z\Gamma_x-\hat{r}_x\Gamma_z\right)+k_2\Gamma_y+k_3 \hat{r}_y \left(\vec{\Gamma} \cdot \hat{r}\right)\right],
\end{eqnarray}
\begin{eqnarray}
B_x= \frac{1}{r^2} \left\{ \left[ 2k_1 \left( \frac{g}{\hbar c} k_1 - 1 \right) - \frac{2g}{\hbar c} k_2 k_3 \right] \left( \vec{\Gamma} \cdot \hat{r} \right) \hat{r}_x + \left( k_2 + k_3 \right) \left( \frac{2g}{\hbar c} k_1 - 1 \right) \left( \hat{r}_y\Gamma_z - \hat{r}_z\Gamma_y \right) + \frac{2g}{\hbar c} k_2 \left( k_2 + k_3 \right) \Gamma_x \right\},
\end{eqnarray}
we can calculate the nine fundamental commutators:
\begin{eqnarray}
\left[ \hat{r}_z\Gamma_x - \hat{r}_x\Gamma_z, \left( \vec{\Gamma} \cdot \hat{r} \right) \hat{r}_x \right] &=& 2{\rm i}\hat{r}_x \left[ \hat{r}_y \left( \hat{r}_x\Gamma_x + \hat{r}_z\Gamma_z \right) - \left( \hat{r}_x^2 + \hat{r}_z^2 \right) \Gamma_y \right] \nonumber \\
&=& 2{\rm i}\hat{r}_x \left[ \hat{r}_y \left( \vec{\Gamma} \cdot \hat{r} - \hat{r}_y\Gamma_y \right) - \left( 1 - \hat{r}_y^2 \right) \Gamma_y \right] \nonumber \\
&=& 2{\rm i}\hat{r}_x \left[ \hat{r}_y \left( \vec{\Gamma} \cdot \hat{r} \right) - \Gamma_y \right],
\end{eqnarray}
\begin{eqnarray}
\left[ \hat{r}_z\Gamma_x - \hat{r}_x\Gamma_z, \hat{r}_y\Gamma_z - \hat{r}_z\Gamma_y \right] &=& \hat{r}_z\hat{r}_y \left[ \Gamma_x, \Gamma_z \right] - \hat{r}_z^2 \left[ \Gamma_x, \Gamma_y \right] + \hat{r}_x\hat{r}_z \left[ \Gamma_z, \Gamma_y \right] \nonumber \\
&=& -2{\rm i}\hat{r}_z\hat{r}_y \Gamma_y - 2{\rm i}\hat{r}_z^2 \Gamma_z - 2{\rm i}\hat{r}_x\hat{r}_z \Gamma_x \nonumber \\
&=& -2{\rm i}\hat{r}_z \left( \vec{\Gamma} \cdot \hat{r} \right),
\end{eqnarray}

\begin{eqnarray}
\left[ \hat{r}_z\Gamma_x - \hat{r}_x\Gamma_z, \Gamma_x \right] &=& -\hat{r}_x \left[ \Gamma_z, \Gamma_x \right] = -2{\rm i} \hat{r}_x \Gamma_y,
\end{eqnarray}

\begin{eqnarray}
\left[ \Gamma_y, \left( \vec{\Gamma} \cdot \hat{r} \right) \hat{r}_x \right] &=& \hat{r}_x \left[ \hat{r}_x \left[ \Gamma_y, \Gamma_x \right] + \hat{r}_y \left[ \Gamma_y, \Gamma_y \right] + \hat{r}_z \left[ \Gamma_y, \Gamma_z \right] \right] \nonumber \\
&=& \hat{r}_x \left[ \hat{r}_x \left( -2{\rm i}\Gamma_z \right) + 0 + \hat{r}_z \left( 2{\rm i}\Gamma_x \right) \right] \nonumber \\
&=& 2{\rm i}\hat{r}_x \left( \hat{r}_z\Gamma_x - \hat{r}_x\Gamma_z \right),
\end{eqnarray}

\begin{eqnarray}
\left[ \Gamma_y, \hat{r}_y\Gamma_z - \hat{r}_z\Gamma_y \right] &=& \hat{r}_y \left[ \Gamma_y, \Gamma_z \right] = 2{\rm i}\hat{r}_y \Gamma_x,
\end{eqnarray}

\begin{eqnarray}
\left[ \Gamma_y, \Gamma_x \right] = -2{\rm i} \Gamma_z,
\end{eqnarray}

\begin{eqnarray}
\left[ \hat{r}_y \left( \vec{\Gamma} \cdot \hat{r} \right), \left( \vec{\Gamma} \cdot \hat{r} \right) \hat{r}_x \right] = 0,
\end{eqnarray}

\begin{eqnarray}
\left[ \hat{r}_y \left( \vec{\Gamma} \cdot \hat{r} \right), \hat{r}_y\Gamma_z - \hat{r}_z\Gamma_y \right] &=& \hat{r}_y^2 \left[ \vec{\Gamma} \cdot \hat{r}, \Gamma_z \right] - \hat{r}_y\hat{r}_z \left[ \vec{\Gamma} \cdot \hat{r}, \Gamma_y \right] \nonumber \\
&=& 2{\rm i}\hat{r}_y \left[ \left( \hat{r}_y^2 + \hat{r}_z^2 \right) \Gamma_x - \hat{r}_x \left( \hat{r}_y\Gamma_y + \hat{r}_z\Gamma_z \right) \right] \nonumber \\
&=& 2{\rm i}\hat{r}_y \left[ \left( 1 - \hat{r}_x^2 \right) \Gamma_x - \hat{r}_x \left( \vec{\Gamma} \cdot \hat{r} - \hat{r}_x\Gamma_x \right) \right] \nonumber \\
&=& 2{\rm i}\hat{r}_y \left[ \Gamma_x - \hat{r}_x \left( \vec{\Gamma} \cdot \hat{r} \right) \right],
\end{eqnarray}

\begin{eqnarray}
\left[ \hat{r}_y \left( \vec{\Gamma} \cdot \hat{r} \right), \Gamma_x \right] &=& \hat{r}_y \left[ \vec{\Gamma} \cdot \hat{r}, \Gamma_x \right] = 2{\rm i}\hat{r}_y \left( \hat{r}_z\Gamma_y - \hat{r}_y\Gamma_z \right).
\end{eqnarray}
By combining them, we get
\begin{eqnarray}
\left[ A_y, B_x \right] &=& \frac{2{\rm i}}{r^3} \left\{ \left[ 2k_1 \left( \frac{g}{\hbar c} k_1 - 1 \right) - \frac{2g}{\hbar c} k_2 k_3 \right] \hat{r}_x \left[ k_1 \left( \hat{r}_y \left( \vec{\Gamma} \cdot \hat{r} \right) - \Gamma_y \right) + k_2 \left( \hat{r}_z \Gamma_x - \hat{r}_x \Gamma_z \right) \right] \right. \nonumber \\
&& + \left[ \left( k_2 + k_3 \right) \left( \frac{2g}{\hbar c} k_1 - 1 \right) \right] \left[ -k_1 \hat{r}_z \left( \vec{\Gamma} \cdot \hat{r} \right) + k_2 \hat{r}_y \Gamma_x + k_3 \hat{r}_y \left( \Gamma_x - \hat{r}_x \left( \vec{\Gamma} \cdot \hat{r} \right) \right) \right] \nonumber \\
&& \left. + \left[ \frac{2g}{\hbar c} k_2 \left( k_2 + k_3 \right) \right] \left[ -k_1 \hat{r}_x \Gamma_y - k_2 \Gamma_z + k_3 \hat{r}_y \left( \hat{r}_z \Gamma_y - \hat{r}_y \Gamma_z \right) \right] \right\}.
\end{eqnarray}
As a result, we can have
\begin{eqnarray}\label{eq:AB}
&&\left[ A_x, B_y \right] - \left[ A_y, B_x \right] \nonumber\\
&=& \frac{2{\rm i}}{r^3} \left\{ \left[ 2k_1 \left( \frac{g}{\hbar c} k_1 - 1 \right) - \frac{2g}{\hbar c} k_2 k_3 \right] \left[ k_1 \left( \hat{r}_x \Gamma_y - \hat{r}_y \Gamma_x \right) + k_2 \left( \Gamma_z - \hat{r}_z \left( \vec{\Gamma} \cdot \hat{r} \right) \right) \right] \right. \nonumber \\
&& + \left[ \left( k_2 + k_3 \right) \left( \frac{2g}{\hbar c} k_1 - 1 \right) \right] \left[ 2k_1 \hat{r}_z \left( \vec{\Gamma} \cdot \hat{r} \right) + \left( k_2 + k_3 \right) \left( \hat{r}_x \Gamma_y - \hat{r}_y \Gamma_x \right) \right] \nonumber \\
&& \left. + \left[ \frac{2g}{\hbar c} k_2 \left( k_2 + k_3 \right) \right] \left[ k_1 \left( \hat{r}_x \Gamma_y - \hat{r}_y \Gamma_x \right) + \left( 2k_2 + k_3 \right) \Gamma_z - k_3 \hat{r}_z \left( \vec{\Gamma} \cdot \hat{r} \right) \right] \right\}\nonumber\\
&=& \frac{2{\rm i}}{r^3} \left\{ \left[ k_1 \left( 2k_1 \left( \frac{g}{\hbar c} k_1 - 1 \right) - \frac{2g}{\hbar c} k_2 k_3 \right) + \left( k_2 + k_3 \right)^2 \left( \frac{2g}{\hbar c} k_1 - 1 \right) + \frac{2g}{\hbar c} k_1 k_2 \left( k_2 + k_3 \right) \right] \left( \hat{r}_x \Gamma_y - \hat{r}_y \Gamma_x \right) \right. \nonumber\\
&& + \left[ k_2 \left( 2k_1 \left( \frac{g}{\hbar c} k_1 - 1 \right) - \frac{2g}{\hbar c} k_2 k_3 \right) + \frac{2g}{\hbar c} k_2 \left( k_2 + k_3 \right) \left( 2k_2 + k_3 \right) \right] \Gamma_z \nonumber\\
&& \left. + \left[ -k_2 \left( 2k_1 \left( \frac{g}{\hbar c} k_1 - 1 \right) - \frac{2g}{\hbar c} k_2 k_3 \right) + 2k_1 \left( k_2 + k_3 \right) \left( \frac{2g}{\hbar c} k_1 - 1 \right) - \frac{2g}{\hbar c} k_2 k_3 \left( k_2 + k_3 \right) \right] \hat{r}_z \left( \vec{\Gamma} \cdot \hat{r} \right) \right\}.
\end{eqnarray}

\begin{remark}
When we consider the following constraint
\begin{eqnarray}
&& k_1= {\frac{c\hbar}{2q}} a_1, \;\;\;\; k_2=-\frac{c\hbar}{2q}a_3, \;\;\;\; k_3=\frac{c\hbar}{2q}a_3, \;\;\; q=g\nonumber\\
&&(a_1-1)^2+ a_3^2=1,
\end{eqnarray}
we can have
\begin{eqnarray}
&& k_2 + k_3=0, \;\;\;\;  2k_1 \left( \frac{g}{\hbar c} k_1 - 1 \right) - \frac{2g}{\hbar c} k_2 k_3 =0,
\end{eqnarray}
which leads to $\vec{B}=0$, thus Eq. (\ref{eq:AB}) is valid. $\blacksquare$
\end{remark}

(ii) Consider the curl term. Since
\begin{eqnarray}
B_x= \frac{1}{r^2} \left\{ \left[ 2k_1 \left( \frac{g}{\hbar c} k_1 - 1 \right) - \frac{2g}{\hbar c} k_2 k_3 \right] \left( \vec{\Gamma} \cdot \hat{r} \right) \hat{r}_x + \left( k_2 + k_3 \right) \left( \frac{2g}{\hbar c} k_1 - 1 \right) \left( \hat{r}_y\Gamma_z - \hat{r}_z\Gamma_y \right) + \frac{2g}{\hbar c} k_2 \left( k_2 + k_3 \right) \Gamma_y \right\},
\end{eqnarray}
\begin{eqnarray}
B_y= \frac{1}{r^2} \left\{ \left[ 2k_1 \left( \frac{g}{\hbar c} k_1 - 1 \right) - \frac{2g}{\hbar c} k_2 k_3 \right] \left( \vec{\Gamma} \cdot \hat{r} \right) \hat{r}_y + \left( k_2 + k_3 \right) \left( \frac{2g}{\hbar c} k_1 - 1 \right) \left( \hat{r}_z\Gamma_x - \hat{r}_x\Gamma_z \right) + \frac{2g}{\hbar c} k_2 \left( k_2 + k_3 \right) \Gamma_y \right\},
\end{eqnarray}
We can have
\begin{eqnarray}
\partial_x B_y &=& \frac{\partial}{\partial x} \left[ \frac{k_2+k_3}{r^2} \left( \frac{2g}{\hbar c} k_1 - 1 \right) \left( \hat{r}_z\Gamma_x - \hat{r}_x\Gamma_z \right) +
\left[\frac{2k_1}{r^2} \left( \frac{gk_1}{\hbar c} - 1 \right)
- \frac{2g}{\hbar c} \frac{k_2 k_3}{r^2} \right]\left( \vec{\Gamma} \cdot \hat{r} \right) \hat{r}_y
+ \frac{2g}{\hbar c} \frac{k_2 (k_2+k_3)}{r^2} \Gamma_y \right] \nonumber \\
&=& (k_2+k_3) \left( \frac{2g}{\hbar c} k_1 - 1 \right) \left[ -\frac{2\hat{r}_x}{r^3} \left( \hat{r}_z\Gamma_x - \hat{r}_x\Gamma_z \right) + \frac{1}{r^2} \left( \frac{-\hat{r}_x\hat{r}_z}{r}\Gamma_x - \frac{1 - \hat{r}_x^2}{r}\Gamma_z \right) \right] \nonumber \\
& & + \left[ 2k_1 \left( \frac{gk_1}{\hbar c} - 1 \right) - \frac{2g}{\hbar c} k_2 k_3 \right] \left[ -\frac{2\hat{r}_x}{r^3} \left( \vec{\Gamma} \cdot \hat{r} \right) \hat{r}_y + \frac{1}{r^2} \left( \frac{\Gamma_x - \hat{r}_x \left( \vec{\Gamma} \cdot \hat{r} \right)}{r} \hat{r}_y + \left( \vec{\Gamma} \cdot \hat{r} \right) \frac{-\hat{r}_x\hat{r}_y}{r} \right) \right] \nonumber \\
& & + \frac{2g}{\hbar c} k_2 (k_2+k_3) \left( -\frac{2\hat{r}_x}{r^3} \Gamma_y \right) \nonumber \\
&=& \frac{1}{r^3} \left\{ (k_2+k_3) \left( \frac{2g}{\hbar c} k_1 - 1 \right) \left[ -3\hat{r}_x\hat{r}_z\Gamma_x + (3\hat{r}_x^2 - 1) \Gamma_z \right] \right. \nonumber \\
& & \left. + \left[ 2k_1 \left( \frac{gk_1}{\hbar c} - 1 \right) - \frac{2g}{\hbar c} k_2 k_3 \right] \left[ \hat{r}_y\Gamma_x - 4\hat{r}_x\hat{r}_y \left( \vec{\Gamma} \cdot \hat{r} \right) \right] - \frac{4g}{\hbar c} k_2 (k_2+k_3) \hat{r}_x \Gamma_y \right\},
\end{eqnarray}
\begin{eqnarray}
\partial_y B_x &=& \frac{\partial}{\partial y} \left[ \frac{k_2+k_3}{r^2} \left( \frac{2g}{\hbar c} k_1 - 1 \right) \left( \hat{r}_y\Gamma_z - \hat{r}_z\Gamma_y \right)
+ \left[\frac{2k_1}{r^2} \left( \frac{gk_1}{\hbar c} - 1 \right)
- \frac{2g}{\hbar c} \frac{k_2 k_3}{r^2} \right] \left( \vec{\Gamma} \cdot \hat{r} \right) \hat{r}_x
+ \frac{2g}{\hbar c} \frac{k_2 (k_2+k_3)}{r^2} \Gamma_x \right] \nonumber \\
&=& (k_2+k_3) \left( \frac{2g}{\hbar c} k_1 - 1 \right) \left[ -\frac{2\hat{r}_y}{r^3} \left( \hat{r}_y\Gamma_z - \hat{r}_z\Gamma_y \right) + \frac{1}{r^2} \left( \frac{1 - \hat{r}_y^2}{r}\Gamma_z - \frac{-\hat{r}_y\hat{r}_z}{r}\Gamma_y \right) \right] \nonumber \\
& & + \left[ 2k_1 \left( \frac{gk_1}{\hbar c} - 1 \right) - \frac{2g}{\hbar c} k_2 k_3 \right] \left[ -\frac{2\hat{r}_y}{r^3} \left( \vec{\Gamma} \cdot \hat{r} \right) \hat{r}_x + \frac{1}{r^2} \left( \frac{\Gamma_y - \hat{r}_y \left( \vec{\Gamma} \cdot \hat{r} \right)}{r} \hat{r}_x + \left( \vec{\Gamma} \cdot \hat{r} \right) \frac{-\hat{r}_y\hat{r}_x}{r} \right) \right] \nonumber \\
& & + \frac{2g}{\hbar c} k_2 (k_2+k_3) \left( -\frac{2\hat{r}_y}{r^3} \Gamma_x \right) \nonumber \\
&=& \frac{1}{r^3} \left\{ (k_2+k_3) \left( \frac{2g}{\hbar c} k_1 - 1 \right) \left[ (1 - 3\hat{r}_y^2) \Gamma_z + 3\hat{r}_y\hat{r}_z\Gamma_y \right] \right. \nonumber \\
& & \left. + \left[ 2k_1 \left( \frac{gk_1}{\hbar c} - 1 \right) - \frac{2g}{\hbar c} k_2 k_3 \right] \left[ \hat{r}_x\Gamma_y - 4\hat{r}_x\hat{r}_y \left( \vec{\Gamma} \cdot \hat{r} \right) \right] - \frac{4g}{\hbar c} k_2 (k_2+k_3) \hat{r}_y \Gamma_x \right\}.
\end{eqnarray}
So we can have
\begin{eqnarray}
\partial_x B_y - \partial_y B_x &=& \frac{k_2+k_3}{r^3} \left( \frac{2gk_1}{\hbar c} - 1 \right) \left[ \Gamma_z - 3\hat{r}_z \left( \vec{\Gamma} \cdot \hat{r} \right) \right] \nonumber \\
& & - \frac{1}{r^3} \left[ 2k_1 \left( \frac{gk_1}{\hbar c} - 1 \right) - \frac{2g}{\hbar c} k_2 k_3 + \frac{4g}{\hbar c} k_2 (k_2+k_3) \right] \left( \hat{r}_x\Gamma_y - \hat{r}_y\Gamma_x \right) \nonumber \\
&=& \frac{k_2+k_3}{r^3} \left( \frac{2gk_1}{\hbar c} - 1 \right) \left[ \Gamma_z - 3\hat{r}_z \left( \vec{\Gamma} \cdot \hat{r} \right) \right] - \frac{2}{r^3} \left[ \frac{g(k_1^2 - k_2 k_3 + 2k_2^2 + 2k_2 k_3)}{\hbar c} - k_1 \right] \left( \hat{r}_x \Gamma_y - \hat{r}_y \Gamma_x \right) \nonumber \\
&=& \frac{k_2+k_3}{r^3} \left( \frac{2gk_1}{\hbar c} - 1 \right) \left[ \Gamma_z - 3\hat{r}_z \left( \vec{\Gamma} \cdot \hat{r} \right) \right] - \frac{2}{r^3} \left[ \frac{g \left( {k_1^2 + 2k_2^2 + k_2 k_3} \right)}{\hbar c} - k_1 \right] \left( \hat{r}_x \Gamma_y - \hat{r}_y \Gamma_x \right).
\end{eqnarray}

(iii) Consider the electric-like field commutator term. We can have
\begin{eqnarray}\label{eq:phiE}
\left[\varphi, E_z\right] = f_1(r) \left[\left(\vec{\Gamma}\cdot\hat{r}\right), E_z\right].
\end{eqnarray}
Because
\begin{eqnarray}
{E}_z=-  \frac{\partial f_2(r)}{\partial r} \hat{r}_z- \frac{f_1(r)}{r} \left( 1 - \frac{2g k_1}{\hbar c} \right) {\Gamma}_z - \hat{r}_z (\vec{\Gamma} \cdot \hat{r}) \left[ \frac{\partial f_1(r)}{\partial r} - \frac{f_1(r)}{r} \left( 1 - \frac{2gk_1}{\hbar c} \right) \right] - \frac{2gk_2  }{\hbar c } \frac{f_1(r)}{r}\left(\hat{r} \times \vec{\Gamma}\right)_z,
\end{eqnarray}
and consider that the $\hat{r}_x$ term commutes with the potential, we only need to calculate the following two commutators:
\begin{eqnarray}
[\vec{\Gamma}\cdot\hat{r}, \; \Gamma_z] = -2{\rm i}(\hat{r}_x \Gamma_y - \hat{r}_y \Gamma_x) = -2{\rm i}(\hat{r} \times \vec{\Gamma})_z,
\end{eqnarray}
\begin{eqnarray}
[\vec{\Gamma}\cdot\hat{r},\;  \hat{r}_x \Gamma_y - \hat{r}_y \Gamma_x] &=& \hat{r}_x [(\vec{\Gamma}\cdot\hat{r}), \Gamma_y] - \hat{r}_y [(\vec{\Gamma}\cdot\hat{r}), \Gamma_x] \nonumber\\
&=& \hat{r}_x [-2{\rm i}(\hat{r}_z \Gamma_x - \hat{r}_x \Gamma_z)] - \hat{r}_y [-2i(\hat{r}_y \Gamma_z - \hat{r}_z \Gamma_y)] \nonumber\\
&=& -2{\rm i} [ \hat{r}_z (\hat{r}_x \Gamma_x + \hat{r}_y \Gamma_y) - (\hat{r}_x^2 + \hat{r}_y^2) \Gamma_z ] \nonumber\\
&=& -2{\rm i} [ \hat{r}_z (\vec{\Gamma}\cdot\hat{r} - \hat{r}_z \Gamma_z) - (1 - \hat{r}_z^2) \Gamma_z ] \nonumber\\
&=& -2{\rm i} [ \hat{r}_z (\vec{\Gamma}\cdot\hat{r}) - \Gamma_z ].
\end{eqnarray}
Substituting these results into Eq. (\ref{eq:phiE}), we can have
\begin{eqnarray}
[\varphi, E_z] &=& 2{\rm i}f_1(r) \biggl\{\frac{f_1(r)}{r} \left( 1 - \frac{2g k_1}{\hbar c} \right)(\hat{r} \times \vec{\Gamma})_z+\frac{2gk_2  }{\hbar c } \frac{f_1(r)}{r} [ \hat{r}_z (\vec{\Gamma}\cdot\hat{r}) - \Gamma_z ] \biggr\}.
\end{eqnarray}

\begin{remark}In summary, from above we have known that
\begin{eqnarray}
&&\left[ A_x, B_y \right] - \left[ A_y, B_x \right] \nonumber\\
&=& \frac{2i}{r^3} \left\{ \left[ k_1 \left( 2k_1 \left( \frac{g}{\hbar c} k_1 - 1 \right) - \frac{2g}{\hbar c} k_2 k_3 \right) + \left( k_2 + k_3 \right)^2 \left( \frac{2g}{\hbar c} k_1 - 1 \right) + \frac{2g}{\hbar c} k_1 k_2 \left( k_2 + k_3 \right) \right] \left( \hat{r}_x \Gamma_y - \hat{r}_y \Gamma_x \right) \right. \nonumber\\
&& + \left[ k_2 \left( 2k_1 \left( \frac{g}{\hbar c} k_1 - 1 \right) - \frac{2g}{\hbar c} k_2 k_3 \right) + \frac{2g}{\hbar c} k_2 \left( k_2 + k_3 \right) \left( 2k_2 + k_3 \right) \right] \Gamma_z \nonumber\\
&& \left. + \left[ -k_2 \left( 2k_1 \left( \frac{g}{\hbar c} k_1 - 1 \right) - \frac{2g}{\hbar c} k_2 k_3 \right) + 2k_1 \left( k_2 + k_3 \right) \left( \frac{2g}{\hbar c} k_1 - 1 \right) - \frac{2g}{\hbar c} k_2 k_3 \left( k_2 + k_3 \right) \right] \hat{r}_z \left( \vec{\Gamma} \cdot \hat{r} \right) \right\},
\end{eqnarray}

\begin{eqnarray}
\partial_x B_y - \partial_y B_x &=& \frac{k_2 + k_3}{r^3} \left( 2\kappa k_1 - 1 \right) \left[ \Gamma_z - 3\hat{r}_z \left( \vec{\Gamma} \cdot \hat{r} \right) \right]  - \frac{2}{r^3} \left[ \kappa \left( {k_1^2 + 2k_2^2 + k_2 k_3} \right) - k_1 \right] \left( \hat{r}_x \Gamma_y - \hat{r}_y \Gamma_x \right),
\end{eqnarray}

\begin{eqnarray}
[\varphi, E_z] &=&= 2if_1(r) \biggl\{\frac{f_1(r)}{r} \left( 1 - \frac{2g k_1}{\hbar c} \right)(\hat{r} \times \vec{\Gamma})_z+\frac{2gk_2 f_1(r) }{\hbar c r}  [ \hat{r}_z (\vec{\Gamma}\cdot\hat{r}) - \Gamma_z ] \biggr\}.
\end{eqnarray}
After substituting them into
\begin{eqnarray}
 \partial_x B_y - \partial_y B_x  {-}\boxed{{\rm i}\,\frac{g}{\hbar c}\biggl(\Bigl[\varphi,E_z\Bigr]+[A_x, B_y] - [A_y, B_x]\biggr)}=0,
\end{eqnarray}
we can get three equations from the coefficients matching of $(\hat{r} \times \vec{\Gamma})_z $, $\Gamma_z$ and $\hat{r}_z (\vec{\Gamma}\cdot\hat{r})$, respectively. Explicitly, we have

(a) The coefficient of $(\hat{r} \times \vec{\Gamma})_z$:
\begin{eqnarray}
&& -\frac{2}{r^3} \left( \frac{g \left( {k_1^2 + 2k_2^2 + k_2 k_3} \right)}{\hbar c} - k_1 \right) + \frac{2g f_1^2 \left( r \right)}{\hbar c r} \left( 1 - \frac{2g k_1}{\hbar c} \right) \nonumber \\
&& + \frac{2g}{\hbar c r^3} \left[ k_1 \left( 2k_1 \left( \frac{g k_1}{\hbar c} - 1 \right) - \frac{2g k_2 k_3}{\hbar c} \right) + \left( k_2 + k_3 \right)^2 \left( \frac{2g k_1}{\hbar c} - 1 \right) + \frac{2g k_1 k_2 \left( k_2 + k_3 \right)}{\hbar c} \right] = 0.
\end{eqnarray}

(b) The coefficient of $\Gamma_z$:
\begin{eqnarray}
&& \frac{k_2 + k_3}{r^3} \left( \frac{2g k_1}{\hbar c} - 1 \right) - \frac{4g^2 k_2 f_1^2 \left( r \right)}{\left( \hbar c \right)^2 r} \nonumber \\
&& + \frac{2g}{\hbar c r^3} \left[ k_2 \left( 2k_1 \left( \frac{g k_1}{\hbar c} - 1 \right) - \frac{2g k_2 k_3}{\hbar c} \right) + \frac{2g k_2 \left( k_2 + k_3 \right) \left( 2k_2 + k_3 \right)}{\hbar c} \right] = 0.
\end{eqnarray}

(c) The coefficient of $\hat{r}_z (\vec{\Gamma} \cdot \hat{r})$:
\begin{eqnarray}
&& -\frac{3 \left( k_2 + k_3 \right)}{r^3} \left( \frac{2g k_1}{\hbar c} - 1 \right) + \frac{4g^2 k_2 f_1^2 \left( r \right)}{\left( \hbar c \right)^2 r} \nonumber \\
&& + \frac{2g}{\hbar c r^3} \left[ -k_2 \left( 2k_1 \left( \frac{g k_1}{\hbar c} - 1 \right) - \frac{2g k_2 k_3}{\hbar c} \right) + 2k_1 \left( k_2 + k_3 \right) \left( \frac{2g k_1}{\hbar c} - 1 \right) - \frac{2g k_2 k_3 \left( k_2 + k_3 \right)}{\hbar c} \right] = 0.
\end{eqnarray}
$\blacksquare$
\end{remark}

\subsubsection{Checking the Equation of the Electric-like Field}

We now verify Eq. (\ref{eq:DivEYM}).

(i) The divergence term:
\begin{eqnarray}
\vec{E}=- \hat{r} f_2'(r) - \frac{f_1(r)}{r} \left( 1 - \frac{2g k_1}{\hbar c} \right)\vec{ \Gamma} -\left[ f_1'(r) - \frac{f_1(r)}{r} \left( 1 - \frac{2gk_1}{\hbar c} \right) \right](\vec{\Gamma} \cdot \hat{r}) \hat{r}   - \frac{2gk_2 f_1(r) }{\hbar c r} \left(\hat{r} \times \vec{\Gamma}\right),
\end{eqnarray}
where for convenience, we have denote $f_j'(r)=\dfrac{\partial f_j(r)}{\partial r}$, $(j=1, 2)$. We can have
\begin{eqnarray}
\nabla \cdot \vec{E} & = & \nabla \cdot \left\{ - \hat{r} f_2'\left(r\right) - \frac{f_1\left(r\right)}{r} \left( 1 - \frac{2g k_1}{\hbar c} \right)\vec{ \Gamma} - \left[ f_1'\left(r\right) - \frac{f_1\left(r\right)}{r} \left( 1 - \frac{2gk_1}{\hbar c} \right) \right]\left(\vec{\Gamma} \cdot \hat{r}\right) \hat{r} - \frac{2gk_2 f_1\left(r\right) }{\hbar c r} \left(\hat{r} \times \vec{\Gamma}\right) \right\} \nonumber \\
& = & \nabla \cdot \left[ - f_2'\left(r\right) \hat{r} \right] - \left( 1 - \frac{2g k_1}{\hbar c} \right) \vec{\Gamma} \cdot \nabla \left[ \frac{f_1\left(r\right)}{r} \right] - \nabla \cdot \left\{ \left[ f_1'\left(r\right) - \frac{f_1\left(r\right)}{r} \left( 1 - \frac{2gk_1}{\hbar c} \right) \right] \left(\vec{\Gamma} \cdot \hat{r}\right) \hat{r} \right\} \nonumber \\
& = & - \frac{1}{r^2} \frac{d}{dr} \left[ r^2 f_2'\left(r\right) \right] - \left( 1 - \frac{2g k_1}{\hbar c} \right) \vec{\Gamma} \cdot \left\{ \hat{r} \frac{d}{dr} \left[ \frac{f_1\left(r\right)}{r} \right] \right\} - \left(\vec{\Gamma} \cdot \hat{r}\right) \frac{1}{r^2} \frac{d}{dr} \left\{ r^2 \left[ f_1'\left(r\right) - \frac{f_1\left(r\right)}{r} \left( 1 - \frac{2gk_1}{\hbar c} \right) \right] \right\} \nonumber \\
& = & - \left[ f_2''\left(r\right) + \frac{2}{r} f_2'\left(r\right) \right] - \left( 1 - \frac{2g k_1}{\hbar c} \right) \left(\vec{\Gamma} \cdot \hat{r}\right) \left[ \frac{f_1'\left(r\right)}{r} - \frac{f_1\left(r\right)}{r^2} \right] \nonumber \\
& & - \left(\vec{\Gamma} \cdot \hat{r}\right) \left\{ \frac{2}{r} \left[ f_1'\left(r\right) - \frac{f_1\left(r\right)}{r} \left( 1 - \frac{2gk_1}{\hbar c} \right) \right] + \left[ f_1''\left(r\right) - \left( 1 - \frac{2gk_1}{\hbar c} \right) \left( \frac{f_1'\left(r\right)}{r} - \frac{f_1\left(r\right)}{r^2} \right) \right] \right\} \nonumber \\
& = & - \left[ f_2''\left(r\right) + \frac{2}{r} f_2'\left(r\right) \right] - \left(\vec{\Gamma} \cdot \hat{r}\right) \left\{ \left( 1 - \frac{2g k_1}{\hbar c} \right) \left[ \frac{f_1'\left(r\right)}{r} - \frac{f_1\left(r\right)}{r^2} \right] + \frac{2 f_1'\left(r\right)}{r} \right. \nonumber \\
& & \left. - \frac{2 f_1\left(r\right)}{r^2} \left( 1 - \frac{2gk_1}{\hbar c} \right) + f_1''\left(r\right) - \left( 1 - \frac{2gk_1}{\hbar c} \right) \left[ \frac{f_1'\left(r\right)}{r} - \frac{f_1\left(r\right)}{r^2} \right] \right\} \nonumber \\
& = & - \left[ f_2''\left(r\right) + \frac{2}{r} f_2'\left(r\right) \right] - \left(\vec{\Gamma} \cdot \hat{r}\right) \left[ f_1''\left(r\right) + \frac{2}{r} f_1'\left(r\right) - \frac{2 f_1\left(r\right)}{r^2} \left( 1 - \frac{2gk_1}{\hbar c} \right) \right].
\end{eqnarray}

(ii) The commutator term. We just need to calculate
\begin{eqnarray}
\left[A_x,E_x\right].
\end{eqnarray}
Because
\begin{eqnarray}
A_x=\frac{1}{r}\left[k_1\left(\hat{r}_y\Gamma_z-\hat{r}_z\Gamma_y\right)+k_2\Gamma_x+k_3 \hat{r}_x \left(\vec{\Gamma} \cdot \hat{r}\right)\right],
\end{eqnarray}
\begin{eqnarray}
E_x=- \hat{r}_x f_2'(r) - \frac{f_1(r)}{r} \left( 1 - \frac{2g k_1}{\hbar c} \right) \Gamma_x - \hat{r}_x (\vec{\Gamma} \cdot \hat{r}) \left[ f_1'(r) - \frac{f_1(r)}{r} \left( 1 - \frac{2gk_1}{\hbar c} \right) \right] - \frac{2gk_2 f_1(r) }{\hbar c r} \left(\hat{r} \times \vec{\Gamma}\right)_x.
\end{eqnarray}
We need to calculate the following nine commutators:
\begin{eqnarray}
\left[\left(\hat{r}_y\Gamma_z-\hat{r}_z\Gamma_y\right), \Gamma_x\right] & = & \hat{r}_y \left[\Gamma_z, \Gamma_x\right] - \hat{r}_z \left[\Gamma_y, \Gamma_x\right] \nonumber\\
& = & \hat{r}_y \left(2{\rm i}\Gamma_y\right) - \hat{r}_z \left(-2{\rm i}\Gamma_z\right) \nonumber\\
& = & 2{\rm i}\left(\hat{r}_y \Gamma_y + \hat{r}_z \Gamma_z\right),
\end{eqnarray}
\begin{eqnarray}
\left[\left(\hat{r}_y\Gamma_z-\hat{r}_z\Gamma_y\right), \left(\vec{\Gamma} \cdot \hat{r}\right) \hat{r}_x\right] & = & \hat{r}_x \left[\hat{r}_y \Gamma_z - \hat{r}_z \Gamma_y, \hat{r}_x \Gamma_x + \hat{r}_y \Gamma_y + \hat{r}_z \Gamma_z\right] \nonumber\\
& = & \hat{r}_x \left\{ \hat{r}_y \left[\Gamma_z, \hat{r}_x \Gamma_x + \hat{r}_y \Gamma_y\right] - \hat{r}_z \left[\Gamma_y, \hat{r}_x \Gamma_x + \hat{r}_z \Gamma_z\right] \right\} \nonumber\\
& = & \hat{r}_x \left\{ \hat{r}_y \left( 2{\rm i} \hat{r}_x \Gamma_y - 2{\rm i} \hat{r}_y \Gamma_x \right) - \hat{r}_z \left( -2i \hat{r}_x \Gamma_z + 2i \hat{r}_z \Gamma_x \right) \right\} \nonumber\\
& = & 2{\rm i} \hat{r}_x \left[ \hat{r}_x \hat{r}_y \Gamma_y - \hat{r}_y^2 \Gamma_x + \hat{r}_x \hat{r}_z \Gamma_z - \hat{r}_z^2 \Gamma_x \right] \nonumber\\
& = & 2{\rm i} \hat{r}_x \left[ \hat{r}_x \left(\hat{r}_y \Gamma_y + \hat{r}_z \Gamma_z\right) - \left(\hat{r}_y^2 + \hat{r}_z^2\right) \Gamma_x \right] \nonumber\\
& = & 2{\rm i} \hat{r}_x \left[ \hat{r}_x \left(\vec{\Gamma} \cdot \hat{r} - \hat{r}_x \Gamma_x\right) - \left(1 - \hat{r}_x^2\right) \Gamma_x \right] \nonumber\\
& = & 2{\rm i} \left[ \hat{r}_x^2 \left(\vec{\Gamma} \cdot \hat{r}\right) - \hat{r}_x^3 \Gamma_x - \hat{r}_x \Gamma_x + \hat{r}_x^3 \Gamma_x \right] \nonumber\\
& = & 2{\rm i} \left[ \hat{r}_x^2 \left(\vec{\Gamma} \cdot \hat{r}\right) - \hat{r}_x \Gamma_x \right],
\end{eqnarray}
\begin{eqnarray}
\left[\left(\hat{r}_y\Gamma_z-\hat{r}_z\Gamma_y\right), \left(\hat{r}_y \Gamma_z - \hat{r}_z \Gamma_y\right)\right] = 0,
\end{eqnarray}
\begin{eqnarray}
\left[\Gamma_x, \Gamma_x\right] = 0,
\end{eqnarray}
\begin{eqnarray}
\left[\Gamma_x, \left(\vec{\Gamma} \cdot \hat{r}\right) \hat{r}_x\right] & = & \hat{r}_x \left[\Gamma_x, \hat{r}_x \Gamma_x + \hat{r}_y \Gamma_y + \hat{r}_z \Gamma_z\right] \nonumber\\
& = & \hat{r}_x \left\{ \hat{r}_x \left[\Gamma_x, \Gamma_x\right] + \hat{r}_y \left[\Gamma_x, \Gamma_y\right] + \hat{r}_z \left[\Gamma_x, \Gamma_z\right] \right\} \nonumber\\
& = & \hat{r}_x \left[ 0 + \hat{r}_y \left(2{\rm i}\Gamma_z\right) + \hat{r}_z \left(-2{\rm i}\Gamma_y\right) \right] \nonumber\\
& = & 2{\rm i} \hat{r}_x \left(\hat{r}_y \Gamma_z - \hat{r}_z \Gamma_y\right),
\end{eqnarray}
\begin{eqnarray}
\left[\Gamma_x, \left(\hat{r}_y \Gamma_z - \hat{r}_z \Gamma_y\right)\right] & = & \hat{r}_y \left[\Gamma_x, \Gamma_z\right] - \hat{r}_z \left[\Gamma_x, \Gamma_y\right] \nonumber\\
& = & \hat{r}_y \left(-2{\rm i}\Gamma_y\right) - \hat{r}_z \left(2{\rm i}\Gamma_z\right) = -2{\rm i}\left(\hat{r}_y \Gamma_y + \hat{r}_z \Gamma_z\right),
\end{eqnarray}
\begin{eqnarray}
\left[ \hat{r}_x (\vec{\Gamma} \cdot \hat{r}), \Gamma_x \right] &=& \hat{r}_x \left[ \hat{r}_y \Gamma_y + \hat{r}_z \Gamma_z, \Gamma_x \right] \nonumber \\
&=& \hat{r}_x \left( \hat{r}_y [-2{\rm i}\Gamma_z] + \hat{r}_z [2{\rm i}\Gamma_y] \right) = -2{\rm i} \hat{r}_x (\hat{r}_y \Gamma_z - \hat{r}_z \Gamma_y),
\end{eqnarray}
\begin{eqnarray}
\left[ \hat{r}_x (\vec{\Gamma} \cdot \hat{r}), (\hat{r}_y \Gamma_z - \hat{r}_z \Gamma_y) \right] &=& - \left[ (\hat{r}_y \Gamma_z - \hat{r}_z \Gamma_y), \hat{r}_x (\vec{\Gamma} \cdot \hat{r}) \right] = -2{\rm i} \left[ \hat{r}_x^2 (\vec{\Gamma} \cdot \hat{r}) - \hat{r}_x \Gamma_x \right].
\end{eqnarray}
By combining them, we obtain
\begin{eqnarray}
[A_x, E_x] &=& \frac{2{\rm i}}{r} \Biggl\{
\left[ \frac{2g k_2^2 f_1(r)}{\hbar c r} - \frac{k_1 f_1(r)}{r} \left( 1 - \frac{2g k_1}{\hbar c} \right) \right] (\hat{r}_y \Gamma_y + \hat{r}_z \Gamma_z) \nonumber \\
&& - \left[ k_2 \left( f_1'(r) - \frac{f_1(r)}{r} \left( 1 - \frac{2g k_1}{\hbar c} \right) \right) - \frac{2g k_1 k_3 f_1(r)}{\hbar c r} - \frac{k_3 f_1(r)}{r} \left( 1 - \frac{2g k_1}{\hbar c} \right) \right] \hat{r}_x (\hat{r}_y \Gamma_z - \hat{r}_z \Gamma_y) \nonumber \\
&& - \left[ k_1 \left( f_1'(r) - \frac{f_1(r)}{r} \left( 1 - \frac{2g k_1}{\hbar c} \right) \right) - \frac{2g k_2 k_3 f_1(r)}{\hbar c r} \right] \left[ \hat{r}_x^2 (\vec{\Gamma} \cdot \hat{r}) - \hat{r}_x \Gamma_x \right]
\Biggr\}.
\end{eqnarray}
From symmetry, it is obvious that the second and third terms vanish after summing over \(x, y, z\). Thus, we have:
\begin{eqnarray}
[A_x, E_x] +[A_y, E_y]+[A_z, E_z]=-4i\frac{1}{r^2}\left(k_1-\frac{2g k_1^2}{\hbar c}-\frac{2g k_2^2}{\hbar c}\right)f_1(r)(\vec{\Gamma} \cdot \hat{r}).
\end{eqnarray}
From Eq. (\ref{eq:DivEYM}) we have
\begin{eqnarray}
 \vec{\nabla}\cdot\vec{E}\ {-}\boxed{{\rm i}\,\frac{g}{\hbar c}\Bigl([A_x, E_x] +[A_y, E_y]+[A_z, E_z]\Bigr)}=0,
\end{eqnarray}
we can get two equations from matching coefficients. Explicitly, we have

(a) The coefficient of constant term:
\begin{eqnarray}
 f_2''(r) + \frac{2}{r} f_2'(r)=0
\end{eqnarray}

(b) The coefficient of $\vec{\Gamma} \cdot \hat{r}$:
\begin{eqnarray}
4\frac{1}{r^2}\frac{g}{\hbar c}\left(k_1-\frac{2g k_1^2}{\hbar c}-\frac{2g k_2^2}{\hbar c}\right)f_1(r) +  \left[f_1''(r) + \frac{2}{r} f_1'(r) - \frac{2 f_1(r)}{r^2} \left(1 - \frac{2gk_1}{\hbar c}\right)\right]=0
\end{eqnarray}
In fact, the \(k_3\) term does not affect the divergence equation of the electric field.

\subsection{Solving the Constraint Equations for $\vec{A}^{(1)}$ and $\vec{A}^{(2)}$}

Now, we obtain a set of constraint equations for two functions $f_1(r)$, $f_2(r)$, the parameters $g$ and $k_j$'s $(j=1, 2, 3)$,  which are as follows
\begin{eqnarray}\label{eq:f2}
 f_2''(r) + \frac{2}{r} f_2'(r)=0,
\end{eqnarray}

\begin{subequations}
\begin{align}
&4\frac{1}{r^2}\frac{g}{\hbar c}\left(k_1-\frac{2g k_1^2}{\hbar c}-\frac{2g k_2^2}{\hbar c}\right)f_1(r) +  \left[f_1''(r) + \frac{2}{r} f_1'(r) - \frac{2 f_1(r)}{r^2} \left(1 - \frac{2gk_1}{\hbar c}\right)\right]=0, \label{eq:f1-a}\\
& -\frac{2}{r^3} \left( \frac{g {\left( k_1^2 + 2k_2^2 + k_2 k_3 \right)}}{\hbar c} - k_1 \right) + \frac{2g f_1^2 \left( r \right)}{\hbar c r} \left( 1 - \frac{2g k_1}{\hbar c} \right) \nonumber \\
& + \frac{2g}{\hbar c r^3} \left[ k_1 \left( 2k_1 \left( \frac{g k_1}{\hbar c} - 1 \right) - \frac{2g k_2 k_3}{\hbar c} \right) + \left( k_2 + k_3 \right)^2 \left( \frac{2g k_1}{\hbar c} - 1 \right) + \frac{2g k_1 k_2 \left( k_2 + k_3 \right)}{\hbar c} \right] = 0,\label{eq:f1-b}\\
& \frac{k_2 + k_3}{r^3} \left( \frac{2g k_1}{\hbar c} - 1 \right) - \frac{4g^2 k_2 f_1^2 \left( r \right)}{\left( \hbar c \right)^2 r} \nonumber \\
& + \frac{2g}{\hbar c r^3} \left[ k_2 \left( 2k_1 \left( \frac{g k_1}{\hbar c} - 1 \right) - \frac{2g k_2 k_3}{\hbar c} \right) + \frac{2g k_2 \left( k_2 + k_3 \right) \left( 2k_2 + k_3 \right)}{\hbar c} \right] = 0,\label{eq:f1-c}\\
& -\frac{3 \left( k_2 + k_3 \right)}{r^3} \left( \frac{2g k_1}{\hbar c} - 1 \right) + \frac{4g^2 k_2 f_1^2 \left( r \right)}{\left( \hbar c \right)^2 r} \nonumber \\
& + \frac{2g}{\hbar c r^3} \left[ -k_2 \left( 2k_1 \left( \frac{g k_1}{\hbar c} - 1 \right) - \frac{2g k_2 k_3}{\hbar c} \right) + 2k_1 \left( k_2 + k_3 \right) \left( \frac{2g k_1}{\hbar c} - 1 \right) - \frac{2g k_2 k_3 \left( k_2 + k_3 \right)}{\hbar c} \right] = 0.\label{eq:f1-d}
  \end{align}
      \end{subequations}
Eq. (\ref{eq:f2}) only involves the function $f_2(r)$, whose solution is
\begin{eqnarray}
f_2(r)=\frac{C_1}{r} + C_0,
\end{eqnarray}
where $C_0$ is trivial, one may set $C_0=0$. Eq. (\ref{eq:f1-a})-Eq. (\ref{eq:f1-d}) involves the function $f_1(r)$ and the parameters $g$ and $k_j$'s. In this subsection, we shall determine them by considering the spin vector potential of \emph{solution 1} (i.e., $\vec{A}^{(1)}$) and the spin vector potential of \emph{solution 2} (i.e., $\vec{A}^{(2)}$).

\subsubsection{The Case of $\vec{A}^{(1)}$}

In this subsection, let us consider the spin vector potential of \emph{solution 1}, for which
\begin{eqnarray}
\{a_1=1, a_2=0, a_3=0\}.
\end{eqnarray}
Here we only consider the non-trivial case of $g\neq 0$ (for the special case of $g=0$, we shall make a uniformly treatment in behind). Then in terms $\{\vec{\Gamma}, \hat{r}\}$ the spin vector potential can be written as
\begin{eqnarray}
 \vec{A}^{(1)} \equiv \vec{A}&=& \frac{1}{r} \left[ k_1 (\hat{r} \times \vec{\Gamma})\right].
\end{eqnarray}
This means that
\begin{eqnarray}
&& k_1= {\frac{c\hbar}{2g}} a_1 = {\frac{c\hbar}{2g}}, \;\;\;\; k_2=-\frac{c\hbar}{2g}a_3=0, \;\;\;\; k_3=\frac{c\hbar}{2g}a_3=0,
\end{eqnarray}
i.e., we have the constraint condition
\begin{eqnarray}\label{eq:con-a}
&& \frac{2gk_1}{c\hbar}=1,
\end{eqnarray}
and
\begin{eqnarray}
&& k_2=k_3=0, \;\;\; \frac{gk_1}{c\hbar}-1=-\frac{1}{2},
\end{eqnarray}
then Eq. (\ref{eq:f1-a})-Eq. (\ref{eq:f1-d}) turn to
\begin{eqnarray}
& f_1''(r) + \dfrac{2}{r} f_1'(r) =0, 
\end{eqnarray}
which leads to
\begin{eqnarray}
f_1(r)=\frac{C_3}{r}+C_2.
\end{eqnarray}

\begin{remark}
Due to
\begin{eqnarray}
\tilde{g}=\frac{g}{c\hbar}, \; \quad \tilde{k}=k_1 \frac{2}{\hbar},
\end{eqnarray}
The condition (\ref{eq:con-a}) is equivalent to
\begin{eqnarray}\label{eq:con-b}
&& \frac{2gk_1}{c\hbar}=\frac{2}{c\hbar}gk_1 =\frac{2}{c\hbar} (c\hbar \tilde{g}) \left(\frac{\hbar}{2}\tilde{k}\right)=\tilde{g} \hbar \tilde{k}=1.
\end{eqnarray}
Thus we recover the result in Table \ref{tab:gk} with the constraint $\tilde{g} \hbar \tilde{k}=1$. $\blacksquare$
\end{remark}

\subsubsection{The Case of $\vec{A}^{(2)}$}

In this subsection, let us consider the spin vector potential of \emph{solution 2}, for which
\begin{eqnarray}\label{eq:a1a2a3-1}
\{a_2=-a_1, (a_1 - 1)^2 + a_3^2 = 1\}.
\end{eqnarray}
Similarly, here we only consider the non-trivial case of $g\neq 0$.
There are two special cases for Eq. (\ref{eq:a1a2a3-1}). The first one
\begin{eqnarray}\label{eq:a1a2a3-2}
\{a_1=2, a_2=-2, a_3=0\}.
\end{eqnarray}
The second one is
\begin{eqnarray}
\{a_1=0, a_2=0, a_3=0\}.
\end{eqnarray}

(i) The case of $\{a_1=2, a_2=-2, a_3=0\}$. In this case, in terms $\{\vec{\Gamma}, \hat{r}\}$ the spin vector potential can be written as
\begin{eqnarray}
 \vec{A}^{(1)} \equiv \vec{A}&=& \frac{1}{r} \left[ k_1 (\hat{r} \times \vec{\Gamma})\right],
\end{eqnarray}
This means that
\begin{eqnarray}
&& k_1= {\frac{c\hbar}{2g}} a_1 = {\frac{c\hbar}{g}}, \;\;\;\; k_2=-\frac{c\hbar}{2g}a_3=0, \;\;\;\; k_3=\frac{c\hbar}{2g}a_3=0,
\end{eqnarray}
i.e., we have the constraint condition
\begin{eqnarray}\label{eq:con-aaa}
&& \frac{gk_1}{c\hbar}=1,
\end{eqnarray}
and
\begin{eqnarray}
&& k_2=k_3=0, \;\;\; \frac{2gk_1}{c\hbar}-1=1.
\end{eqnarray}
Based on $k_2=k_3=0$ and $g\neq 0$, from Eq. (\ref{eq:f1-a})-Eq. (\ref{eq:f1-d}) we have
\begin{subequations}
\begin{align}
&4\frac{1}{r^2}\frac{gk_1}{\hbar c}\left(1-\frac{2g k_1}{\hbar c}\right)f_1(r) +  \left[f_1''(r) + \frac{2}{r} f_1'(r) - \frac{2 f_1(r)}{r^2} \left(1 - \frac{2gk_1}{\hbar c}\right)\right]=0, \label{eq:f1-aa}\\
& -\frac{2k_1}{r^3} \left( \frac{g k_1}{\hbar c} - 1 \right) + \frac{2g f_1^2 \left( r \right)}{\hbar c r} \left( 1 - \frac{2g k_1}{\hbar c} \right) + \frac{2g}{\hbar c r^3} \left[ k_1 \left( 2k_1 \left( \frac{g k_1}{\hbar c} - 1 \right)\right)  \right] = 0.\label{eq:f1-bb}
  \end{align}
\end{subequations}
Furthermore, by considering $\dfrac{gk_1}{c\hbar}=1$, the above two equations become
\begin{eqnarray}
\label{eq:f1-aaa}
&&f_1''(r) + \frac{2}{r} f_1'(r) - \frac{2 f_1(r)}{r^2}=0\nonumber\\
&&\frac{gf_1^2(r)}{c\hbar r}=0,
\end{eqnarray}
because $g\neq 0$, one has
\begin{eqnarray}
f_1(r)=0.
\end{eqnarray}
Thus we recover the result in Table \ref{tab:gk} with the constraint $\tilde{g} \hbar \tilde{k}=2$.

(ii) The case of $\{a_1=0, a_2=0, a_3=0\}$. In this case, in terms $\{\vec{\Gamma}, \hat{r}\}$ the spin vector potential can still be written as
\begin{eqnarray}
 \vec{A}^{(1)} \equiv \vec{A}&=& \frac{1}{r} \left[ k_1 (\hat{r} \times \vec{\Gamma})\right],
\end{eqnarray}
but with $k_1=0$. After substituting $k_1=0$ into Eqs. (\ref{eq:f1-aa}) and (\ref{eq:f1-bb}), we still have Eq. (\ref{eq:f1-aaa}). Then we have
$f_1(r)=0$. Thus we recover the result in Table \ref{tab:gk} with $\tilde{g}\neq 0$ and $\tilde{k}=0$.

(iii) The general case $\{a_2=-a_1, (a_1 - 1)^2 + a_3^2 = 1\}$ with $a_3\neq 0$. In this case, in terms $\{\vec{\Gamma}, \hat{r}\}$ the spin vector potential can be written as
\begin{eqnarray}
\vec{A} = \frac{1}{r} \left[ k_1 (\hat{r} \times \vec{\Gamma}) + k_2 \vec{\Gamma} + k_3 \hat{r} (\vec{\Gamma} \cdot \hat{r}) \right].
 \end{eqnarray}
This means that
\begin{eqnarray}
&& k_1= {\frac{c\hbar}{2g}} a_1, \;\;\;\; k_2=-\frac{c\hbar}{2g}a_3, \;\;\;\; k_3=\frac{c\hbar}{2g}a_3.
\end{eqnarray}
Based on $k_2+k_3=0$ (or $k_2=-k_3$) and $g\neq 0$, from Eq. (\ref{eq:f1-a})-Eq. (\ref{eq:f1-d}) we have
\begin{subequations}
\begin{align}
&4\frac{1}{r^2}\frac{g}{\hbar c}\left(k_1-\frac{2g k_1^2}{\hbar c}-\frac{2g k_3^2}{\hbar c}\right)f_1(r)
+  \left[f_1''(r) + \frac{2}{r} f_1'(r) - \frac{2 f_1(r)}{r^2} \left(1 - \frac{2gk_1}{\hbar c}\right)\right]=0, \label{eq:f1-aaa1}\\
& -\frac{2}{r^3} \left( \frac{g
\left( {k_1^2 + k_3^2} \right)}{\hbar c}
- k_1 \right) + \frac{2g f_1^2 \left( r \right)}{\hbar c r} \left( 1 - \frac{2g k_1}{\hbar c} \right)  + \frac{2g}{\hbar c r^3} \left[ k_1 \left( 2k_1 \left( \frac{g k_1}{\hbar c} - 1 \right) + \frac{2g k_3^2}{\hbar c} \right)
 \right] = 0,\label{eq:f1-bbb}\\
&  \frac{4g^2 k_3 f_1^2 \left( r \right)}{\left( \hbar c \right)^2 r}
+ \frac{2g}{\hbar c r^3} \left[ -k_3 \left( 2k_1 \left( \frac{g k_1}{\hbar c} - 1 \right) + \frac{2g k_3^2}{\hbar c} \right)\right] = 0,\label{eq:f1-ccc}\\
& - \frac{4g^2 k_3 f_1^2 \left( r \right)}{\left( \hbar c \right)^2 r}
+ \frac{2g}{\hbar c r^3} \left[ k_3 \left( 2k_1 \left( \frac{g k_1}{\hbar c} - 1 \right) + \frac{2g k_3^2}{\hbar c} \right)
 \right] = 0. \label{eq:f1-ddd}
  \end{align}
 \end{subequations}
i.e.,
\begin{subequations}
\begin{align}
&4\frac{1}{r^2}\frac{g}{\hbar c}\left(k_1-\frac{2g k_1^2}{\hbar c}-\frac{2g k_3^2}{\hbar c}\right)f_1(r)
+  \left[f_1''(r) + \frac{2}{r} f_1'(r) - \frac{2 f_1(r)}{r^2} \left(1 - \frac{2gk_1}{\hbar c}\right)\right]=0, \label{eq:f1-aaaa}\\
& -\frac{2}{r^3} \left( \frac{g
\left( {k_1^2 + k_3^2 } \right)}{\hbar c}
- k_1 \right) + \frac{2g f_1^2 \left( r \right)}{\hbar c r} \left( 1 - \frac{2g k_1}{\hbar c} \right)  + \frac{2g}{\hbar c r^3} \left[ k_1 \left( 2k_1 \left( \frac{g k_1}{\hbar c} - 1 \right) + \frac{2g k_3^2}{\hbar c} \right)
 \right] = 0,\label{eq:f1-bbbb}\\
&  \frac{4g^2 f_1^2 \left( r \right)}{\left( \hbar c \right)^2 r}
 -\frac{2g}{\hbar c r^3} \left[  2k_1 \left( \frac{g k_1}{\hbar c} - 1 \right) + \frac{2g k_3^2}{\hbar c} \right] = 0.\label{eq:f1-cccc}
  \end{align}
 \end{subequations}
Because
\begin{eqnarray}
\frac{g}{\hbar c}\left(k_1-\frac{2g k_1^2}{\hbar c}-\frac{2g k_3^2}{\hbar c}\right)
&=&\frac{g k_1}{\hbar c}- 2 \left(\frac{g k_1}{\hbar c}\right)^2-  2 \left(\frac{g k_3}{\hbar c}\right)^2
=\frac{a_1}{2}-2 \left(\frac{a_1}{2}\right)^2-2 \left(\frac{a_3}{2}\right)^2\nonumber\\
&=&\frac{1}{2} \left(a_1-a_1^2-a_3^2\right)=\frac{1}{2} \left[a_1-a_1^2+\left[(a_1-1)^2-1\right]\right]\nonumber\\
&=&\frac{1}{2} \left[a_1-a_1^2+a_1^2-2a_1+1-1\right]=-\frac{a_1}{2},
\end{eqnarray}
\begin{eqnarray}
\frac{g}{\hbar c} \left[  2k_1 \left( \frac{g k_1}{\hbar c} - 1 \right) + \frac{2g k_3^2}{\hbar c} \right]
&=& 2 \left( \frac{g k_1}{\hbar c}\right)^2 -2 \left( \frac{g k_1}{\hbar c}\right)+2 \left( \frac{g k_3}{\hbar c}\right)^2\nonumber\\
&=& - \left( \frac{g k_1}{\hbar c}\right)-\left[\left( \frac{g k_1}{\hbar c}\right) - 2 \left( \frac{g k_1}{\hbar c}\right)^2-2 \left( \frac{g k_3}{\hbar c}\right)^2\right]\nonumber\\
&=& - \left( \frac{g k_1}{\hbar c}\right) +\frac{a_1}{2}=-\frac{a_1}{2}+\frac{a_1}{2}=0,
\end{eqnarray}
\begin{eqnarray}
\left( \frac{g \left( {k_1^2 + k_3^2 } \right)}{\hbar c}- k_1 \right)
=\frac{1}{2} \left(2\frac{g \left( k_1^2 + k_3^2 \right)}{\hbar c} -2k_1\right)=\frac{1}{2}\left[  2k_1 \left( \frac{g k_1}{\hbar c} - 1 \right) + \frac{2g k_3^2}{\hbar c} \right]=0.
\end{eqnarray}
Then we have
\begin{subequations}
\begin{align}
&-\frac{2a_1}{r^2}f_1(r)
+  \left[f_1''(r) + \frac{2}{r} f_1'(r) - \frac{2 f_1(r)}{r^2} \left(1 - \frac{2gk_1}{\hbar c}\right)\right]=0, \\
&  \frac{g^2 f_1^2 \left( r \right)}{\left( \hbar c \right)^2 r}=0,  \end{align}
 \end{subequations}
i.e.,
\begin{subequations}
\begin{align}
&f_1''(r) + \frac{2}{r} f_1'(r) - \frac{2 f_1(r)}{r^2}=0, \\
&  g^2 f_1^2 (r)=0.  \end{align}
 \end{subequations}
because $g\neq 0$, one has
\begin{eqnarray}
f_1(r)=0.
\end{eqnarray}

\subsection{Solving the Constraint Equations for the General Spin Vector Potential $\vec{A}$}


The ``general'' spin vector potential is given by
\begin{eqnarray}
\vec{A} = \frac{1}{r} \left[ k_1 (\hat{r} \times \vec{\Gamma}) + k_2 \vec{\Gamma} + k_3 \hat{r} (\vec{\Gamma} \cdot \hat{r}) \right].
 \end{eqnarray}
 For the ``general'' case, we means that we do not impose the constraint conditions on the parameters $k_j$'s $(j=1, 2, 3)$ in the beginning.
From Eq. (\ref{eq:f2}), we have determined the function $f_2(r)$ as $f_2(r)=\frac{C_1}{r}$. Our task is to solve the function $f_(r)$
for Eq. (\ref{eq:f1-a})-Eq. (\ref{eq:f1-d}), but without imposing the constraint conditions on the parameters $k_j$'s in the beginning.

For convenience, let us denote $\tilde{g}=\dfrac{g}{\hbar c}$, then Eq. (\ref{eq:f1-a})-Eq. (\ref{eq:f1-d}) become
\begin{eqnarray}\label{eq:e1}
4\frac{\tilde{g}}{r^2} \left( k_1 - 2\tilde{g} k_1^2 - 2\tilde{g} k_2^2 \right) f_1(r) + f_1''(r) + \frac{2}{r} f_1'(r) - \frac{2f_1(r)}{r^2} \left( 1 - 2\tilde{g} k_1 \right) = 0,
\end{eqnarray}
\begin{eqnarray}\label{eq:e2}
& & -\frac{2}{r^3} \left[ \tilde{g} \left( k_1^2 + 2k_2^2 + k_2 k_3 \right) - k_1 \right] + \frac{2\tilde{g} f_1^2 \left( r \right)}{r} \left( 1 - 2\tilde{g} k_1 \right) \nonumber \\
& & + \frac{2\tilde{g}}{r^3} \left[ k_1 \left( 2k_1 \left( \tilde{g} k_1 - 1 \right) - 2\tilde{g} k_2 k_3 \right) + \left( k_2 + k_3 \right)^2 \left( 2\tilde{g} k_1 - 1 \right) + 2\tilde{g} k_1 k_2 \left( k_2 + k_3 \right) \right] = 0,
\end{eqnarray}
\begin{eqnarray}\label{eq:e3}
& & \frac{k_2 + k_3}{r^3} \left( 2\tilde{g} k_1 - 1 \right) - \frac{4\tilde{g}^2 k_2 f_1^2 \left( r \right)}{r} \nonumber \\
& & + \frac{2\tilde{g}}{r^3} \left[ k_2 \left( 2k_1 \left( \tilde{g} k_1 - 1 \right) - 2\tilde{g} k_2 k_3 \right) + 2\tilde{g} k_2 \left( k_2 + k_3 \right) \left( 2k_2 + k_3 \right) \right] = 0,
\end{eqnarray}
\begin{eqnarray}\label{eq:e4}
& & -\frac{3 \left( k_2 + k_3 \right)}{r^3} \left( 2\tilde{g} k_1 - 1 \right) + \frac{4\tilde{g}^2 k_2 f_1^2 \left( r \right)}{r} \nonumber \\
& & + \frac{2\tilde{g}}{r^3} \left[ -k_2 \left( 2k_1 \left( \tilde{g} k_1 - 1 \right) - 2\tilde{g} k_2 k_3 \right) + 2k_1 \left( k_2 + k_3 \right) \left( 2\tilde{g} k_1 - 1 \right) - 2\tilde{g} k_2 k_3 \left( k_2 + k_3 \right) \right] = 0.
\end{eqnarray}
We shall divide the study into four cases: (1) $g=0$ and $k_1=0$;  (1) $g=0$ and $k_1\neq 0$; (1) $g\neq 0$ and $k_1=0$; (1) $g\neq 0$ and $k_1\neq 0$
%
%

\subsubsection{The Case of $\{g=0, k_1=0\}$}

In this case, from Eqs. (\ref{eq:e1})-(\ref{eq:e4}) we have
\begin{eqnarray}\label{eq:e1-A}
 f_1''(r) + \frac{2}{r} f_1'(r) - \frac{2f_1(r)}{r^2}= 0,
\end{eqnarray}
\begin{eqnarray}
 k_2+k_3=0.
\end{eqnarray}
Then we have
\begin{eqnarray}
 f_1(r)=\frac{C_2}{r^2}+C_3 r,
\end{eqnarray}
with $k_2=-k_3$. Here $C_2$ and $C_3$ can be arbitrary real and complex numbers.

\begin{remark}
{For convenience to \textcolor{blue}{build} Tables \ref{Tab:real} and \ref{Tab:complex} in Sec. \ref{sec-6}, let us list the real and complex static solutions as follows.

\emph{Real Static Solution.---} We have the real static solution as
\begin{eqnarray}
f_2(r)=\frac{C_1}{r}, \;\;\;\; f_1(r)=\frac{C_2}{r^2}+C_3 r, \;\;\;\; k_1=0, \;\;\; k_2 \in \mathbb{R}, \;\;\; k_3=-k_2,
\end{eqnarray}
where $\{C_1, C_2, C_3, k_1, k_2, k_3\}$ are all real numbers.

\emph{Complex Static Solution.---} We have the complex static solution as
\begin{eqnarray}
f_2(r)=\frac{C_1}{r}, \;\;\;\; f_1(r)=\frac{C_2}{r^2}+C_3 r, \;\;\;\; k_1=0, \;\;\; k_2 \in \mathbb{R}, \;\;\; k_3=-k_2,
\end{eqnarray}
where at least one of $\{C_1, C_2, C_3, k_1, k_2, k_3\}$ is a complex number. $\blacksquare$
}
\end{remark}

\subsubsection{The Case of $\{g=0, k_1\neq 0\}$}

In this case, from Eqs. (\ref{eq:e1})-(\ref{eq:e4}) we have
\begin{eqnarray}
 f_1''(r) + \frac{2}{r} f_1'(r) - \frac{2f_1(r)}{r^2}= 0,
\end{eqnarray}
\begin{eqnarray}\label{eq:e2-A}
& & \frac{2}{r^3} k_1=0,
\end{eqnarray}
\begin{eqnarray}
 k_2+k_3=0.
\end{eqnarray}
However, since $k_1\neq 0$, then Eq. (\ref{eq:e2-A}) cannot be valid. Therefore, no solution for the case of $\{g=0, k_1\neq 0\}$.

\subsubsection{The Case of $\{g\neq 0, k_1=0\}$}

In this case, from Eqs. (\ref{eq:e1})-(\ref{eq:e4}) we have
\begin{eqnarray}\label{eq:e1-a}
4\frac{\tilde{g}}{r^2} \left( - 2\tilde{g} k_2^2 \right) f_1(r) + f_1''(r) + \frac{2}{r} f_1'(r) - \frac{2f_1(r)}{r^2} = 0,
\end{eqnarray}

\begin{eqnarray}\label{eq:e2-a}
& & -\frac{2}{r^3} \left[ \tilde{g} \left(2k_2^2 + k_2 k_3 \right) \right] + \frac{2\tilde{g} f_1^2 \left( r \right)}{r} + \frac{2\tilde{g}}{r^3} \left[- \left( k_2 + k_3 \right)^2 \right] = 0,
\end{eqnarray}

\begin{eqnarray}\label{eq:e3-a}
& & -\frac{k_2 + k_3}{r^3} - \frac{4\tilde{g}^2 k_2 f_1^2 \left( r \right)}{r}  + \frac{2\tilde{g}}{r^3} \left[ k_2 \left(- 2\tilde{g} k_2 k_3 \right) + 2\tilde{g} k_2 \left( k_2 + k_3 \right) \left( 2k_2 + k_3 \right) \right] = 0,
\end{eqnarray}

\begin{eqnarray}\label{eq:e4-a}
& & \frac{3 \left( k_2 + k_3 \right)}{r^3} + \frac{4\tilde{g}^2 k_2 f_1^2 \left( r \right)}{r} + \frac{2\tilde{g}}{r^3} \left[ -k_2 \left( - 2\tilde{g} k_2 k_3 \right)  - 2\tilde{g} k_2 k_3 \left( k_2 + k_3 \right) \right] = 0.
\end{eqnarray}
By adding Eq. (\ref{eq:e3-a}) and Eq. (\ref{eq:e4-a}), one has
\begin{eqnarray}
& & \frac{2 \left( k_2 + k_3 \right)}{r^3}  + \frac{2\tilde{g}}{r^3} \left[ 2\tilde{g} k_2 \left( k_2 + k_3 \right) \left( 2k_2 + k_3 \right) - 2\tilde{g} k_2 k_3 \left( k_2 + k_3 \right) \right] = 0,
\end{eqnarray}
\begin{eqnarray}
& & \frac{ \left( k_2 + k_3 \right)}{r^3}  + \frac{\left( k_2 + k_3 \right) \tilde{g}}{r^3} \left[ 2\tilde{g} k_2  \left( 2k_2 + k_3 \right) - 2\tilde{g} k_2 k_3 \right] = 0,
\end{eqnarray}
\begin{eqnarray}
& & \frac{ \left( k_2 + k_3 \right)}{r^3}  + \frac{\left( k_2 + k_3 \right) \tilde{g}}{r^3} \left[ 4\tilde{g} k_2^2  \right] = 0,
\end{eqnarray}
i.e.,
\begin{eqnarray}
& & \left( k_2 + k_3 \right) \frac{1+{4}\tilde{g}^2 k_2^2}{r^3} = 0,
\end{eqnarray}
{which leads to two possibilities:
(i) $k_2 + k_3= 0$;
(ii) $k_2 + k_3\neq  0$ and  $1+4 \tilde{g}^2 k_2^2=0$;
}

\vspace{4mm}

\centerline{\emph{3.1. The Case of $k_2 + k_3= 0$}}

\vspace{4mm}

In this case, one has
\begin{eqnarray}\label{eq:k2k3}
& & k_3=-k_2.
\end{eqnarray}
Based on Eq. (\ref{eq:k2k3}), after adding Eq. (\ref{eq:e3-a}) and Eq. (\ref{eq:e4-a}) we have
\begin{eqnarray}
& & \frac{4\tilde{g}^2 k_2 f_1^2 \left( r \right)}{r} + \frac{2\tilde{g}}{r^3}k_2 \left(2\tilde{g} k_2 k_3 \right) = 0,
\end{eqnarray}
i.e.,
\begin{eqnarray}
& & 4\tilde{g}^2 k_2 f_1^2 (r) + \frac{4\tilde{g}^2}{r^2}  k_2^2  k_3 = 0,
\end{eqnarray}
i.e.,
\begin{eqnarray}
& &  k_2 f_1^2 (r) + \frac{1}{r^2}  k_2^2  k_3 = 0,
\end{eqnarray}
i.e.,
\begin{eqnarray}
& &  -k_3 f_1^2 (r) + \frac{1}{r^2}  k_3^3 = 0,
\end{eqnarray}
i.e.,
\begin{eqnarray}
& &  k_3 f_1^2 (r)= \frac{1}{r^2}  k_3^3.
\end{eqnarray}
There are two possibilities: (i) $k_3=0$; (ii) $k_3\neq 0$ and $f_1^2(r)=\dfrac{k_3^2}{r^2}$.

\emph{ (i) The Case of $k_3=0$.---} In this case, one has $k_2=k_3=0$. Then Eq. (\ref{eq:e1-a}) and Eq. (\ref{eq:e2-a}) become
\begin{eqnarray}\label{eq:e1-aa}
 f_1''(r) + \frac{2}{r} f_1'(r) - \frac{2f_1(r)}{r^2} = 0,
\end{eqnarray}
\begin{eqnarray}\label{eq:e2-aa}
& & \frac{2\tilde{g} f_1^2 \left( r \right)}{r} = 0,
\end{eqnarray}
which yield
\begin{eqnarray}
 f_1(r) = 0.
\end{eqnarray}

\emph{(ii) The Case of $k_3\neq 0$ and $f_1^2(r)=\dfrac{k_3^2}{r^2}$.---} In this case, one has $k_2=-k_3$. Then Eq. (\ref{eq:e2-a}) becomes
\begin{eqnarray}
& & -\frac{2}{r^3} \left[ \tilde{g} \left(k_3^2 \right) \right] + \frac{2\tilde{g} f_1^2 \left( r \right)}{r} = 0,
\end{eqnarray}
i.e.,
\begin{eqnarray}
& &  f_1^2(r)= \frac{k_3^2}{r^2},
\end{eqnarray}
which is automatically valid. Similarly, Eq. (\ref{eq:e1-a}) becomes
\begin{eqnarray}
-\frac{8 \tilde{g}^2 k_3^2}{r^2} f_1(r) + f_1''(r) + \frac{2}{r} f_1'(r) - \frac{2f_1(r)}{r^2} = 0.
\end{eqnarray}
Let $f_1(r)=\dfrac{C}{r}$, with $C=\pm \sqrt{k_3^2}$, one has
\begin{eqnarray}
-\frac{8 \tilde{g}^2 k_3^2}{r^2} \left(\dfrac{C}{r}\right) + \left(\dfrac{C}{r}\right)'' + \frac{2}{r} \left(\dfrac{C}{r}\right)' - \frac{2}{r^2}\left(\dfrac{C}{r}\right) = 0,
\end{eqnarray}
i.e.,
\begin{eqnarray}
-\frac{8 \tilde{g}^2 k_3^2}{r^2} \left(\dfrac{1}{r}\right) + \left(\dfrac{1}{r}\right)'' + \frac{2}{r} \left(\dfrac{1}{r}\right)' - \frac{2}{r^2}\left(\dfrac{1}{r}\right) = 0,
\end{eqnarray}
i.e.,
\begin{eqnarray}
-\frac{8 \tilde{g}^2 k_3^2}{r^3}  + \dfrac{2}{r^3} + \frac{2}{r} \times \dfrac{-1}{r^2} - \frac{2}{r^3}= 0,
\end{eqnarray}
i.e.,
\begin{eqnarray}
8 \tilde{g}^2 k_3^2+2= 0,
\end{eqnarray}
i.e.,
\begin{eqnarray}
4 \tilde{g}^2 k_3^2+1= 0,
\end{eqnarray}
i.e.,
\begin{eqnarray}
4 \tilde{g}^2 k_2^2+1= 0.
\end{eqnarray}
This implies that when $k_2 + k_3=  0$ and  $1+4 \tilde{g}^2 k_2^2=0$, the solution is $f_1(r)=\dfrac{C}{r}$, with $C=\pm \sqrt{k_3^2}$, and
\begin{eqnarray}
k_3=-k_2=\mp \frac{{\rm i}}{2|\tilde{g}|},
\end{eqnarray}
which is a pure imaginary number.

\begin{remark}
{For convenience to \textcolor{blue}{build} Tables \ref{Tab:real} and \ref{Tab:complex} in Sec. \ref{sec-6}, let us list the real and complex static solutions as follows.

\emph{Real Static Solution.---} We have the real static solution as
\begin{eqnarray}
f_2(r)=\frac{C_1}{r}, \;\;\;\; f_1(r)=0, \;\;\;\; k_1=0, \;\;\; k_2=0, \;\;\; k_3=-k_2=0,
\end{eqnarray}
where $C_1$ is a real number.

\emph{Complex Static Solution.---} We have the complex static solution as

(a)
\begin{eqnarray}
f_2(r)=\frac{C_1}{r}, \;\;\;\; f_1(r)=0, \;\;\;\; k_1=0, \;\;\; k_2=0, \;\;\; k_3=-k_2=0,
\end{eqnarray}
with $C_1$ is a complex number.

(b)
\begin{eqnarray}
f_2(r)=\frac{C_1}{r}, \;\;\;\; f_1(r)=\frac{C}{r}=\frac{\pm \frac{{\rm i}}{2|\tilde{g}|}}{r}, \;\;\;\; k_1=0, \;\;\; k_2 = \pm\frac{{\rm i}}{2|\tilde{g}|}, \;\;\; k_3=-k_2=\mp \frac{{\rm i}}{2|\tilde{g}|},
\end{eqnarray}
where at least one of $\{C_1, C, k_2, k_3\}$ is a complex number. $\blacksquare$
}
\end{remark}

\vspace{4mm}

\centerline{\emph{3.2. The Case of $k_2 + k_3\neq 0$ and $1+4\tilde{g}^2 k_2^2=0$}}

\vspace{4mm}

{
In this case, we have \begin{eqnarray}
4\tilde{g}^2 k_2^2=-1,
\end{eqnarray}
i.e.,
\begin{eqnarray}
k_2^2=-\frac{1}{4\tilde{g}^2},
\end{eqnarray}
i.e.,
\begin{eqnarray}
k_2=\pm \frac{{\rm i}}{2|\tilde{g}|},
\end{eqnarray}
namely, $k_2\neq 0$. From Eq. (\ref{eq:e3-a}) we have
\begin{eqnarray}
& & \frac{4\tilde{g}^2 k_2 f_1^2 \left( r \right)}{r}= -\frac{k_2 + k_3}{r^3}   + \frac{2\tilde{g}}{r^3} \left[ k_2 \left(- 2\tilde{g} k_2 k_3 \right) + 2\tilde{g} k_2 \left( k_2 + k_3 \right) \left( 2k_2 + k_3 \right) \right],
\end{eqnarray}
i.e.,
\begin{eqnarray}
& & 4\tilde{g}^2 k_2 f_1^2(r)= -\frac{k_2 + k_3}{r^2}   + \frac{2\tilde{g}}{r^2} \left[ k_2 \left(- 2\tilde{g} k_2 k_3 \right) + 2\tilde{g} k_2 \left( k_2 + k_3 \right) \left( 2k_2 + k_3 \right) \right],
\end{eqnarray}
i.e.,
\begin{eqnarray}
& &  f_1^2(r)= k_2 \left\{\frac{k_2 + k_3}{r^2} - \frac{2\tilde{g}}{r^2} \left[ k_2 \left(- 2\tilde{g} k_2 k_3 \right) + 2\tilde{g} k_2 \left( k_2 + k_3 \right) \left( 2k_2 + k_3 \right) \right]\right\},
\end{eqnarray}
i.e.,
\begin{eqnarray}
& &  f_1^2(r)= k_2 \left\{\frac{k_2 + k_3}{r^2} - \frac{2\tilde{g}}{r^2} \left[ -2 \tilde{g} k_2^2 k_3
+ \left( 2\tilde{g} k_2^2 + 2\tilde{g} k_2  k_3 \right) \left( 2k_2 + k_3 \right) \right]\right\},
\end{eqnarray}
i.e.,
\begin{eqnarray}
& &  f_1^2(r)= k_2 \left\{\frac{k_2 + k_3}{r^2} - \frac{1}{r^2} \left[ -4 \tilde{g}^2 k_2^2 k_3
+ \left( 4\tilde{g}^2 k_2^2 + 4\tilde{g}^2 k_2  k_3 \right) \left( 2k_2 + k_3 \right) \right]\right\},
\end{eqnarray}
i.e.,
\begin{eqnarray}
& &  f_1^2(r)= k_2 \left\{\frac{k_2 + k_3}{r^2} - \frac{1}{r^2} \left[  k_3
+ \left(-1 + 4\tilde{g}^2 k_2  k_3 \right) \left( 2k_2 + k_3 \right) \right]\right\},
\end{eqnarray}
i.e.,
\begin{eqnarray}
& &  f_1^2(r)= k_2 \left\{\frac{k_2}{r^2} +\frac{1}{r^2} \left[ \left(1 -4\tilde{g}^2 k_2  k_3 \right) \left( 2k_2 + k_3 \right) \right]\right\},
\end{eqnarray}
i.e.,
\begin{eqnarray}
& &  f_1^2(r)= k_2 \left\{\frac{k_2}{r^2} +\frac{1}{r^2} \left[ \left(1 + \frac{k_3}{k_2} \right) \left( 2k_2 + k_3 \right) \right]\right\},
\end{eqnarray}
i.e.,
\begin{eqnarray}
& &  f_1^2(r)= \left\{\frac{k_2^2}{r^2} +\frac{1}{r^2} \left[ \left(k_2 + k_3 \right) \left( 2k_2 + k_3 \right) \right]\right\},
\end{eqnarray}
i.e.,
\begin{eqnarray}
& &  f_1^2(r)= \frac{k_2^2+ \left(k_2 + k_3 \right) \left( 2k_2 + k_3 \right)}{r^2},
\end{eqnarray}
i.e.,
\begin{eqnarray}
& &  f_1^2(r)= \frac{3 k_2^2+ 3k_2 k_3 +k_3^2}{r^2}.
\end{eqnarray}
Hence the function $f_1(r)$ has the structure of
\begin{eqnarray}
& &  f_1(r)= \frac{C}{r},
\end{eqnarray}
with
\begin{eqnarray}
& &  C^2=3 k_2^2+ 3k_2 k_3 +k_3^2.
\end{eqnarray}

(i) If $C=0$, then we have
\begin{eqnarray}
& &  f_1(r)=0.
\end{eqnarray}
with
\begin{eqnarray}
3 k_2^2+ 3k_2 k_3 +k_3^2=0,
\end{eqnarray}
i.e.,
\begin{eqnarray}
k_3= \frac{-3\pm {\rm i} \sqrt{3}}{2} k_2.
\end{eqnarray}

(ii) If $C\neq 0$, then Eq. (\ref{eq:e1-a}) becomes
\begin{eqnarray}
4\frac{\tilde{g}}{r^2} \left( - 2\tilde{g} k_2^2 \right) \left(\dfrac{C}{r}\right) + \left(\dfrac{C}{r}\right)'' + \frac{2}{r} \left(\dfrac{C}{r}\right)' - \frac{2}{r^2} \left(\dfrac{C}{r}\right)= 0,
\end{eqnarray}
i.e.,
\begin{eqnarray}
4\frac{\tilde{g}}{r^2} \left( - 2\tilde{g} k_2^2 \right) \left(\dfrac{C}{r}\right) - \frac{2}{r^2} \left(\dfrac{C}{r}\right) = 0,
\end{eqnarray}
i.e.,
\begin{eqnarray}
4\tilde{g}^2  k_2^2=-1,
\end{eqnarray}
which is satisfied automatically.
}

\begin{remark}
{For convenience to {build} Tables \ref{Tab:real} and \ref{Tab:complex} in Sec. \ref{sec-6}, let us list the real and complex static solutions as follows.

\emph{Real Static Solution.---} We do not have the real static solution because $k_2$ is a complex number.

\emph{Complex Static Solution.---} We have the complex static solution as

(a)
\begin{eqnarray}
f_2(r)=\frac{C_1}{r}, \;\;\;\; f_1(r)=0, \;\;\;\; k_1=0, \;\;\; k_2=\pm \frac{{\rm i}}{2|\tilde{g}|},  \;\;\; k_3= \frac{-3\pm {\rm i} \sqrt{3}}{2} k_2,
\end{eqnarray}
with $C_1$ is a real or complex number.

(b)
\begin{eqnarray}
f_2(r)=\frac{C_1}{r}, \;\;\;\; f_1(r)=\frac{C}{r}=\frac{\pm \sqrt{3 k_2^2+ 3k_2 k_3 +k_3^2}}{r}, \;\;\;\; k_1=0, \;\;\; k_2=\pm \frac{{\rm i}}{2|\tilde{g}|},, \;\;\; k_3\neq -k_2,
\end{eqnarray}
with $C_1$ is a real or complex number. $\blacksquare$
}
\end{remark}

\subsubsection{The Case of $\{g\neq 0, k_1\neq 0\}$}

In this case, from Eqs. (\ref{eq:e1})-(\ref{eq:e4}) we have
\begin{eqnarray}\label{eq:e1-b}
4\frac{\tilde{g}}{r^2} \left( k_1 - 2\tilde{g} k_1^2 - 2\tilde{g} k_2^2 \right) f_1(r) + f_1''(r) + \frac{2}{r} f_1'(r) - \frac{2f_1(r)}{r^2} \left( 1 - 2\tilde{g} k_1 \right) = 0,
\end{eqnarray}
\begin{eqnarray}\label{eq:e2-b}
& & -\frac{2}{r^3} \left[ \tilde{g} \left( k_1^2 + 2k_2^2 + k_2 k_3 \right) - k_1 \right] + \frac{2\tilde{g} f_1^2 \left( r \right)}{r} \left( 1 - 2\tilde{g} k_1 \right) \nonumber \\
& & + \frac{2\tilde{g}}{r^3} \left[ k_1 \left( 2k_1 \left( \tilde{g} k_1 - 1 \right) - 2\tilde{g} k_2 k_3 \right) + \left( k_2 + k_3 \right)^2 \left( 2\tilde{g} k_1 - 1 \right) + 2\tilde{g} k_1 k_2 \left( k_2 + k_3 \right) \right] = 0,
\end{eqnarray}
\begin{eqnarray}\label{eq:e3-b}
& & \frac{k_2 + k_3}{r^3} \left( 2\tilde{g} k_1 - 1 \right) - \frac{4\tilde{g}^2 k_2 f_1^2 \left( r \right)}{r} \nonumber \\
& & + \frac{2\tilde{g}}{r^3} \left[ k_2 \left( 2k_1 \left( \tilde{g} k_1 - 1 \right) - 2\tilde{g} k_2 k_3 \right)
+ 2\tilde{g} k_2 \left( k_2 + k_3 \right) \left( 2k_2 + k_3 \right) \right] = 0,
\end{eqnarray}
\begin{eqnarray}\label{eq:e4-b}
& & -\frac{3 \left( k_2 + k_3 \right)}{r^3} \left( 2\tilde{g} k_1 - 1 \right) + \frac{4\tilde{g}^2 k_2 f_1^2 \left( r \right)}{r} \nonumber \\
& & + \frac{2\tilde{g}}{r^3} \left[ -k_2 \left( 2k_1 \left( \tilde{g} k_1 - 1 \right) - 2\tilde{g} k_2 k_3 \right) + 2k_1 \left( k_2 + k_3 \right) \left( 2\tilde{g} k_1 - 1 \right) - 2\tilde{g} k_2 k_3 \left( k_2 + k_3 \right) \right] = 0.
\end{eqnarray}
After adding Eq. (\ref{eq:e3-b}) and Eq. (\ref{eq:e4-b}), we have
\begin{eqnarray}
  -\frac{2 \left( k_2 + k_3 \right)}{r^3} \left( 2\tilde{g} k_1 - 1 \right)  + \frac{2\tilde{g}}{r^3} \left[  2\tilde{g} k_2 \left( k_2 + k_3 \right) \left( 2k_2 + k_3 \right)+ 2k_1 \left( k_2 + k_3 \right) \left( 2\tilde{g} k_1 - 1 \right) - 2\tilde{g} k_2 k_3 \left( k_2 + k_3 \right) \right] = 0,
\end{eqnarray}
i.e.,
\begin{eqnarray}
  -\left( k_2 + k_3 \right) \left( 2\tilde{g} k_1 - 1 \right)  + \tilde{g} \left[  2\tilde{g} k_2 \left( k_2 + k_3 \right) \left( 2k_2 + k_3 \right)+ 2k_1 \left( k_2 + k_3 \right) \left( 2\tilde{g} k_1 - 1 \right) - 2\tilde{g} k_2 k_3 \left( k_2 + k_3 \right) \right] = 0,
\end{eqnarray}
i.e.,
\begin{eqnarray}
  -\left( k_2 + k_3 \right) \left( 2\tilde{g} k_1 - 1 \right)
  + 2\tilde{g} \left( k_2 + k_3 \right)\left[  \tilde{g} k_2  \left( 2k_2 + k_3 \right)+ k_1  \left( 2\tilde{g} k_1 - 1 \right)
   - \tilde{g} k_2 k_3  \right] = 0,
\end{eqnarray}
i.e.,
\begin{eqnarray}
  -\left( k_2 + k_3 \right) \left( 2\tilde{g} k_1 - 1 \right)
  + 2\tilde{g} \left( k_2 + k_3 \right)\left[  2 \tilde{g} k_2^2 + 2 \tilde{g} k_1^2- k_1\right] = 0,
\end{eqnarray}
i.e.,
\begin{eqnarray}
  \left( k_2 + k_3 \right) \left\{ (1-2\tilde{g} k_1)
  + 2\tilde{g} \left[  2 \tilde{g} k_2^2 + 2 \tilde{g} k_1^2- k_1\right]\right\} = 0,
\end{eqnarray}
i.e.,
\begin{eqnarray}
  \left( k_2 + k_3 \right) \left[ 1
  + 4\tilde{g} \left(   \tilde{g} k_2^2 + \tilde{g} k_1^2- k_1\right)\right] = 0.
\end{eqnarray}
Then one has {three} possibilities: (i) $k_2=k_3=0$; (ii) $k_2 =-k_3$ and $k_3\neq 0$; (iii) $k_2 + k_3\neq 0$ and $\left[ 1  + 4\tilde{g} \left(   \tilde{g} k_2^2 + \tilde{g} k_1^2- k_1\right)\right]=0$.

\vspace{4mm}

\centerline{\emph{4.1. The Case of $k_2=k_3=0$}}

\vspace{4mm}

In this case, Eq. (\ref{eq:e3-b}) and Eq. (\ref{eq:e4-b}) are automatically satisfied. Eq. (\ref{eq:e2-b}) become
\begin{eqnarray}
& & -\frac{2}{r^3} \left[ \tilde{g}  k_1^2  - k_1 \right] + \frac{2\tilde{g} f_1^2 \left( r \right)}{r} \left( 1 - 2\tilde{g} k_1 \right)+ \frac{2\tilde{g}}{r^3} \left[ 2 k_1^2  \left( \tilde{g} k_1 - 1 \right)  \right] = 0,
\end{eqnarray}
i.e.,
\begin{eqnarray}
& & -\frac{k_1}{r^2} \left( \tilde{g}  k_1  - 1 \right) + \tilde{g} f_1^2(r)  \left( 1 - 2\tilde{g} k_1 \right)+ \frac{\tilde{g}}{r^2} \left[ 2 k_1^2  \left( \tilde{g} k_1 - 1 \right)  \right] = 0,
\end{eqnarray}
i.e.,
\begin{eqnarray}
& & -\frac{k_1\left( \tilde{g}  k_1  - 1 \right)}{r^2}  + \tilde{g} f_1^2(r)  \left( 1 - 2\tilde{g} k_1 \right)+ \frac{k_1\left( \tilde{g}  k_1  - 1 \right)}{r^2} \left[ 2 \tilde{g} k_1 \right] = 0,
\end{eqnarray}
i.e.,
\begin{eqnarray}
& & \tilde{g} f_1^2(r)  \left( 1 - 2\tilde{g} k_1 \right)=\frac{k_1\left( \tilde{g}  k_1  - 1 \right)}{r^2} \left(1- 2 \tilde{g} k_1 \right),
\end{eqnarray}
which leads to two possibilities: (i) $2\tilde{g} k_1=1$; (ii) $2\tilde{g} k_1-1\neq 0$ and $f_1^2(r)=\dfrac{k_1\left( \tilde{g}  k_1  - 1 \right)/\tilde{g}}{r^2}$.

(i) The case of $2\tilde{g} k_1=1$. {This situation also means $\left[ 1  + 4\tilde{g} \left(   \tilde{g} k_2^2 + \tilde{g} k_1^2- k_1\right)\right]=0$ with $k_2=0$}. In this case, Eq. (\ref{eq:e1-b}) becomes
\begin{eqnarray}
 f_1''(r) + \frac{2}{r} f_1'(r)= 0,
\end{eqnarray}
which yields
\begin{eqnarray}
 f_1(r)=\frac{C_2}{r} + C_3.
\end{eqnarray}

(ii) The case of $2\tilde{g} k_1-1\neq 0$ and $f_1^2(r)=\dfrac{k_1\left( \tilde{g}  k_1  - 1 \right)/\tilde{g}}{r^2}$. For simplicity, we let
{\begin{eqnarray}\label{eq:kappa-1}
 f_1(r)=\frac{C}{r}, \;\;\;\;\; C= \pm \sqrt{k_1\left( \tilde{g}  k_1  - 1 \right)/\tilde{g}}.
\end{eqnarray}
}

(ii-1) If $C=0$, then we have
\begin{eqnarray}
 \tilde{g}  k_1  = 1,
\end{eqnarray}
thus $f_1(r)=0$. In this situation, one may check that Eq. (\ref{eq:e1-b}) is automatically satisfied.

(ii-2)  If $C\neq 0$ or $\tilde{g}  k_1  \neq  1$, after substituting Eq. (\ref{eq:kappa-1}) into Eq. (\ref{eq:e1-b}), we have
\begin{eqnarray}
4\frac{\tilde{g}}{r^2} \left( k_1 - 2\tilde{g} k_1^2 \right) \left(\dfrac{C}{r}\right) + \left(\dfrac{C}{r}\right)'' + \frac{2}{r} \left(\dfrac{C}{r}\right)' - \frac{2}{r^2} \left( 1 - 2\tilde{g} k_1 \right) \left(\dfrac{C}{r}\right)= 0,
\end{eqnarray}
i.e.,
\begin{eqnarray}
4\frac{\tilde{g}}{r^2} \left( k_1 - 2\tilde{g} k_1^2 \right) \left(\dfrac{1}{r}\right) + \left(\dfrac{1}{r}\right)'' + \frac{2}{r} \left(\dfrac{1}{r}\right)' - \frac{2}{r^2} \left( 1 - 2\tilde{g} k_1 \right) \left(\dfrac{1}{r}\right)= 0,
\end{eqnarray}
i.e.,
\begin{eqnarray}
4\frac{\tilde{g}}{r^2} \left( k_1 - 2\tilde{g} k_1^2  \right) \left(\dfrac{1}{r}\right) +
\dfrac{2}{r^3} + \frac{2}{r} \times \dfrac{-1}{r^2}- \frac{2}{r^2} \left( 1 - 2\tilde{g} k_1 \right) \left(\dfrac{1}{r}\right)= 0,
\end{eqnarray}
i.e.,
\begin{eqnarray}
4\frac{\tilde{g}}{r^2} \left( k_1 - 2\tilde{g} k_1^2  \right) \left(\dfrac{1}{r}\right)
- \frac{2}{r^2} \left( 1 - 2\tilde{g} k_1 \right) \left(\dfrac{1}{r}\right)= 0,
\end{eqnarray}
i.e.,
\begin{eqnarray}
2\tilde{g} k_1\left( 1 - 2\tilde{g} k_1 \right) - \left( 1 - 2\tilde{g} k_1 \right) = 0,
\end{eqnarray}
i.e.,
\begin{eqnarray}
-(2\tilde{g} k_1 -1)^2= 0,
\end{eqnarray}
which cannot be valid for $2\tilde{g} k_1-1\neq 0$. Thus there is no solution for this case.

\begin{remark}
{For convenience to \textcolor{blue}{build} Tables \ref{Tab:real} and \ref{Tab:complex} in Sec. \ref{sec-6}, let us list the real and complex static solutions as follows.

\emph{Real Static Solution.---} We have the real static solution as

(a)
\begin{eqnarray}
f_2(r)=\frac{C_1}{r}, \;\;\;\; f_1(r)=\frac{C_2}{r} + C_3, \;\;\;\; k_1=\frac{1}{2\tilde{g}}, \;\;\; k_2=0,  \;\;\; k_3=0,
\end{eqnarray}
with $\{C_1, C_2, C_3\}$ are real numbers.

(b)
\begin{eqnarray}
f_2(r)=\frac{C_1}{r}, \;\;\;\; f_1(r)=0, \;\;\;\; k_1=\frac{1}{\tilde{g}}, \;\;\; k_2=0, \;\;\; k_3=0,
\end{eqnarray}
with $C_1$ is a real number.

\emph{Complex Static Solution.---} We have the complex static solution as

(a)
\begin{eqnarray}
f_2(r)=\frac{C_1}{r}, \;\;\;\; f_1(r)= \frac{C_2}{r} + C_3, \;\;\;\; k_1=\frac{1}{2\tilde{g}}, \;\;\; k_2=0,  \;\;\; k_3=0,
\end{eqnarray}
with at least one of $\{C_1, C_2, C_3\}$ is complex number.

(b)
\begin{eqnarray}
f_2(r)=\frac{C_1}{r}, \;\;\;\; f_1(r)=0, \;\;\;\; k_1=\frac{1}{\tilde{g}}, \;\;\; k_2=0, \;\;\; k_3=0,
\end{eqnarray}
with $C_1$ is a complex number. $\blacksquare$
}
\end{remark}

%
%

\vspace{4mm}

\centerline{\emph{4.2. $k_2 =-k_3$ and $k_3\neq 0$}}

\vspace{4mm}

In this case, Eq. (\ref{eq:e3-b}) and Eq. (\ref{eq:e4-b}) reduces to
\begin{eqnarray}
& & \frac{4\tilde{g}^2 k_2 f_1^2 \left( r \right)}{r}  + \frac{2\tilde{g}}{r^3} \left[ -k_2 \left( 2k_1 \left( \tilde{g} k_1 - 1 \right) - 2\tilde{g} k_2 k_3 \right)\right] = 0,
\end{eqnarray}
i.e.,
\begin{eqnarray}
& & 4\tilde{g}^2 k_2 f_1^2(r) = \frac{4\tilde{g} k_2}{r^2} \left[ k_1 \left( \tilde{g} k_1 - 1 \right) - \tilde{g} k_2 k_3 \right],
\end{eqnarray}
i.e.,
\begin{eqnarray}
& & \tilde{g} f_1^2(r) = \frac{1}{r^2} \left[ k_1 \left( \tilde{g} k_1 - 1 \right) +\tilde{g} k_3^2 \right],
\end{eqnarray}
i.e.,
\begin{eqnarray}\label{eq:ff-1}
& & f_1^2(r) = \frac{\left( \tilde{g} k_1^2+\tilde{g} k_3^2 -k_1\right)/\tilde{g}}{r^2}.
\end{eqnarray}
After substituting Eq. (\ref{eq:ff-1}) into Eq. (\ref{eq:e2-b}), we have
\begin{eqnarray}
& & -\frac{2}{r^3} \left[ \tilde{g} \left( k_1^2 + 2k_2^2 + k_2 k_3 \right) - k_1 \right] + \frac{2\tilde{g} f_1^2 \left( r \right)}{r} \left( 1 - 2\tilde{g} k_1 \right)  + \frac{2\tilde{g}}{r^3} \left[ k_1 \left( 2k_1 \left( \tilde{g} k_1 - 1 \right) - 2\tilde{g} k_2 k_3 \right)
  \right] = 0,
\end{eqnarray}
i.e.,
\begin{eqnarray}
& & -\frac{1}{r^3} \left[ \tilde{g} \left( k_1^2 + k_3^2 \right) - k_1 \right] + \frac{\tilde{g} f_1^2 \left( r \right)}{r} \left( 1 - 2\tilde{g} k_1 \right)  + \frac{\tilde{g}}{r^3} \left[ k_1 \left( 2k_1 \left( \tilde{g} k_1 - 1 \right) + 2\tilde{g} k_3^2 \right)
  \right] = 0,
\end{eqnarray}
i.e.,
\begin{eqnarray}
& & -\left[ \tilde{g} \left( k_1^2 + k_3^2 \right) - k_1 \right] + \left( \tilde{g} k_1^2+\tilde{g} k_3^2 -k_1\right) \left( 1 - 2\tilde{g} k_1 \right)
+ \tilde{g} k_1 \left[ 2k_1 \left( \tilde{g} k_1 - 1 \right) + 2\tilde{g} k_3^2 \right] = 0,
\end{eqnarray}
i.e.,
{
\begin{eqnarray}
& & -\left[ \tilde{g}  k_1^2 + \tilde{g} k_3^2  - k_1 \right] + \left( \tilde{g} k_1^2+\tilde{g} k_3^2 -k_1\right) \left( 1 - 2\tilde{g} k_1 \right)
+ 2 \tilde{g} k_1 \left[ \tilde{g}  k_1^2 + \tilde{g} k_3^2  - k_1  \right] = 0,
\end{eqnarray}
}
which is satisfied automatically.

Now, the function $f_1(r)$ has the structure
\begin{eqnarray}
 f_1(r)=\frac{C}{r}, \;\;\;\;\; C= \pm \sqrt{\left( \tilde{g} k_1^2+\tilde{g} k_3^2 -k_1\right)/\tilde{g}}.
\end{eqnarray}

(i) If $C=0$, we have $f_1(r)=0$, one may check that Eq. (\ref{eq:e1-b}) is automatically satisfied.
In this situation, we have
\begin{eqnarray}\label{eq:ka}
 \tilde{g} k_1^2+\tilde{g} k_3^2 -k_1=0.
\end{eqnarray}
Note: By considering
\begin{eqnarray}
&& k_1= {\frac{c\hbar}{2g}} a_1, \;\;\;\; k_2=-\frac{c\hbar}{2g}a_3, \;\;\;\; k_3=\frac{c\hbar}{2g}a_3, \;\;\;\;\tilde{g}=\frac{g}{c\hbar},
\end{eqnarray}
Eq. (\ref{eq:ka}) becomes
\begin{eqnarray}
 \tilde{g}^2 k_1^2+\tilde{g}^2 k_3^2 - \tilde{g} k_1=0,
\end{eqnarray}
i.e.,
\begin{eqnarray}
 4\tilde{g}^2 k_1^2+4\tilde{g}^2 k_3^2 - 4\tilde{g} k_1=0,
\end{eqnarray}
i.e.,
\begin{eqnarray}
 a_1^2+a_3^2 - 2 a_1=0,
\end{eqnarray}
which is just the previous constraint
\begin{eqnarray}
(a_1 - 1)^2 + a_3^2 = 1.
\end{eqnarray}

For the real static solution, one can have the following parameterization
\begin{eqnarray}
&& a_1 = 2\sin^2\theta, \;\;\;\; a_3 = - 2\sin\theta\cos\theta,
\end{eqnarray}
thus
\begin{eqnarray}
&& k_1= \frac{c\hbar}{2g} a_1 = \frac{c\hbar}{g} \sin^2\theta = \frac{1}{\tilde{g}} \sin^2\theta , \;\;\;\; k_3=-\frac{1}{\tilde{g}}\sin\theta\cos\theta,
\end{eqnarray}
with $k_3 \neq 0$, i.e., $\theta \neq \pi/2, 3\pi/2$.

For the complex static solution, one can have the following parameterizations:
\begin{eqnarray}
a_1 =1 \pm \cosh \vartheta, \;\;\;\;\ a_3 = {\rm i} \, \sinh \vartheta.
\end{eqnarray}
\begin{eqnarray}
&& k_1=  \frac{1}{2\tilde{g}} (1 \pm \cosh \vartheta), \;\;\;\; k_3= \frac{{\rm i}}{2\tilde{g}} \sinh \vartheta,
\end{eqnarray}
with $\vartheta \neq 0$, or
\begin{eqnarray}
 a_3=\pm \cosh \vartheta, \;\;\;\;\ a_1 =1+ {\rm i}\sinh^2 \vartheta.
\end{eqnarray}
\begin{eqnarray}
&& k_1=  \frac{1}{2\tilde{g}} 1+ {\rm i}\sinh^2 \vartheta, \;\;\;\; k_3= \pm \cosh \vartheta.
\end{eqnarray}
Here $\theta$ and $\vartheta$ are real numbers.

(ii) If $C\neq 0$, i.e.,
\begin{eqnarray}
 \tilde{g} k_1^2+\tilde{g} k_3^2 -k_1 \neq 0,
\end{eqnarray}
then from Eq. (\ref{eq:e1-b}) we have
\begin{eqnarray}
4\frac{\tilde{g}}{r^2} \left( k_1 - 2\tilde{g} k_1^2 - 2\tilde{g} k_2^2 \right) \left(\dfrac{C}{r}\right) + \left(\dfrac{C}{r}\right)'' + \frac{2}{r} \left(\dfrac{C}{r}\right)' - \frac{2}{r^2} \left( 1 - 2\tilde{g} k_1 \right) \left(\dfrac{C}{r}\right)= 0,
\end{eqnarray}
i.e.,
\begin{eqnarray}
4\frac{\tilde{g}}{r^2} \left( k_1 - 2\tilde{g} k_1^2 - 2\tilde{g} k_2^2 \right) \left(\dfrac{C}{r}\right)
  - \frac{2}{r^2} \left( 1 - 2\tilde{g} k_1 \right) \left(\dfrac{C}{r}\right)= 0,
\end{eqnarray}
i.e.,
\begin{eqnarray}
4\tilde{g}\left( k_1 - 2\tilde{g} k_1^2 - 2\tilde{g} k_2^2 \right)
  - 2 \left( 1 - 2\tilde{g} k_1 \right) = 0,
\end{eqnarray}
i.e.,
\begin{eqnarray}
2\tilde{g}\left( k_1 - 2\tilde{g} k_1^2 - 2\tilde{g} k_2^2 \right)
  - \left( 1 - 2\tilde{g} k_1 \right) = 0,
\end{eqnarray}
i.e.,
\begin{eqnarray}
2\tilde{g}\left(2 k_1 - 2\tilde{g} k_1^2 - 2\tilde{g} k_2^2 \right)
  -1 = 0,
\end{eqnarray}
i.e.,
\begin{eqnarray}
4\tilde{g}\left(k_1 - \tilde{g} k_1^2 - \tilde{g} k_3^2 \right)
  -1 = 0,
\end{eqnarray}
i.e.,
\begin{eqnarray}
4\tilde{g}\left(\tilde{g} k_1^2 + \tilde{g} k_3^2 -k_1 \right)
  +1 = 0,
\end{eqnarray}
i.e.,
\begin{eqnarray}
\tilde{g} k_1^2 + \tilde{g} k_3^2 -k_1 =-\frac{1}{4\tilde{g}}.
\end{eqnarray}
Thus
\begin{eqnarray}
 C= \pm \sqrt{\left( \tilde{g} k_1^2+\tilde{g} k_3^2 -k_1\right)/\tilde{g}}=\pm \sqrt{-\frac{1}{4\tilde{g}^2}}=\pm \frac{{\rm i}}{2|\tilde{g}|},
\end{eqnarray}
 which is a complex number.

\begin{remark}
{For convenience to \textcolor{blue}{build} Tables \ref{Tab:real} and \ref{Tab:complex} in Sec. \ref{sec-6}, let us list the real and complex static solutions as follows.

\emph{Real Static Solution.---} We have the real static solution as
\begin{eqnarray}
f_2(r)=\frac{C_1}{r}, \;\;\;\; f_1(r)= 0, \;\;\;\; k_1= \frac{1}{\tilde{g}} \sin^2\theta , \;\;\;\; k_3=-\frac{1}{\tilde{g}}\sin\theta\cos\theta, \;\;\; k_2=-k_3,  \;\;\; k_3\neq 0,
\end{eqnarray}
with $C_1$ is a real number. Here $k_1$ and $k_3$ satisfy the constraint $\tilde{g} k_1^2+\tilde{g} k_3^2 -k_1=0$, i.e.,
$k_3= \pm \sqrt{-\tilde{g} k_1^2+(k_1/\tilde{g})}$.

\emph{Complex Static Solution.---} We have the complex static solution as

(a)
\begin{eqnarray}
f_2(r)=\frac{C_1}{r}, \;\;\;\; f_1(r)= 0, \;\;\;\; k_1= \frac{1}{\tilde{g}} \sin^2\theta , \;\;\;\; k_3=-\frac{1}{\tilde{g}}\sin\theta\cos\theta, \;\;\; k_2=-k_3,  \;\;\; k_3\neq 0,
\end{eqnarray}
with $C_1$ is a complex number. Here $k_1$ and $k_3$ satisfy the constraint $\tilde{g} k_1^2+\tilde{g} k_3^2 -k_1=0$.

(b)
\begin{eqnarray}
f_2(r)=\frac{C_1}{r}, \;\;\;\; f_1(r)= 0, \;\;\;\; k_1=  \frac{1}{2\tilde{g}} (1 \pm \cosh \vartheta), \;\;\;\; k_3= \frac{{\rm i}}{2\tilde{g}} \sinh \vartheta, \;\;\; k_2=-k_3,  \;\;\; k_3\neq 0,
\end{eqnarray}
with $C_1$ is a real or complex number. Here $k_1$ and $k_3$ satisfy the constraint $\tilde{g} k_1^2+\tilde{g} k_3^2 -k_1=0$.

(c)
\begin{eqnarray}
f_2(r)=\frac{C_1}{r}, \;\;\;\; f_1(r)= 0, \;\;\;\; k_1=  \frac{1}{2\tilde{g}} 1+ {\rm i}\sinh^2 \vartheta, \;\;\;\; k_3= \pm \cosh \vartheta \;\;\; k_2=-k_3,  \;\;\; k_3\neq 0,
\end{eqnarray}
with $C_1$ is a real or complex number. Here $k_1$ and $k_3$ satisfy the constraint $\tilde{g} k_1^2+\tilde{g} k_3^2 -k_1=0$.

(d)
\begin{eqnarray}
f_2(r)=\frac{C_1}{r}, \;\;\;\; f_1(r)= \frac{\pm \frac{{\rm i}}{2|\tilde{g}|}}{r}, \;\;\;\; \tilde{g} k_1^2 + \tilde{g} k_3^2 -k_1 =-\frac{1}{4\tilde{g}}, \;\;\; k_2=-k_3,  \;\;\; k_3\neq 0,
\end{eqnarray}
with $C_1$ is a real or complex number. Here $k_1$ and $k_3$ satisfy the constraint $\tilde{g} k_1^2+\tilde{g} k_3^2 -k_1=-\frac{1}{4\tilde{g}}$, i.e.,
$k_3= \pm \sqrt{-\tilde{g} k_1^2+\left(k_1-\frac{1}{4\tilde{g}}\right)/\tilde{g}}$. $\blacksquare$
}
\end{remark}

%
%
%
%
%

%
%

\vspace{4mm}

\centerline{\emph{4.3. The Case of $k_2 + k_3\neq 0$ and $\left[ 1  + 4\tilde{g} \left(   \tilde{g} k_2^2 + \tilde{g} k_1^2- k_1\right)\right]=0$}}

\vspace{4mm}

In this case, we have
\begin{eqnarray}
1  + 4\tilde{g} \left(   \tilde{g} k_2^2 + \tilde{g} k_1^2- k_1\right) =0.
\end{eqnarray}

(i) If $k_2=0$ and $k_3\neq 0$, then we have
\begin{eqnarray}
1  + 4\tilde{g} \left(   \tilde{g} k_1^2- k_1\right) =0,
\end{eqnarray}
i.e.,
\begin{eqnarray}
4\tilde{g}^2 k_1^2 -4\tilde{g} k_1 +1=0,
\end{eqnarray}
i.e.,
\begin{eqnarray}
(2\tilde{g} k_1 -1)^2=0,
\end{eqnarray}
i.e.,
\begin{eqnarray}
2\tilde{g} k_1=1,
\end{eqnarray}
or
\begin{eqnarray}
k_1=\frac{1}{2\tilde{g}}.
\end{eqnarray}
From Eq. (\ref{eq:e3-b}) we have
\begin{eqnarray}
& & \frac{k_3}{r^3} \left( 2\tilde{g} k_1 - 1 \right)= 0,
\end{eqnarray}
which is automatically satisfied.

From Eq. (\ref{eq:e2-b}) we have
\begin{eqnarray}
& & -\frac{2}{r^3} \left[ \tilde{g} \left( k_1^2 + 2k_2^2 + k_2 k_3 \right) - k_1 \right]
  + \frac{2\tilde{g}}{r^3} \left[ k_1 \left( 2k_1 \left( \tilde{g} k_1 - 1 \right) - 2\tilde{g} k_2 k_3 \right)
+ 2\tilde{g} k_1 k_2 \left( k_2 + k_3 \right) \right] = 0,
\end{eqnarray}
i.e.,
\begin{eqnarray}
& & -\frac{2}{r^3} \left[ \tilde{g}  k_1^2  - k_1 \right]
  + \frac{2\tilde{g}}{r^3} \left[ 2k_1^2  \left( \tilde{g} k_1 - 1 \right) \right] = 0,
\end{eqnarray}
i.e.,
\begin{eqnarray}
& & -2k_1 \left[ \tilde{g}  k_1  - 1 \right]
  + 4\tilde{g} k_1^2  \left( \tilde{g} k_1 - 1 \right)  = 0,
\end{eqnarray}
i.e.,
\begin{eqnarray}
& & -1  + 2\tilde{g} k_1 = 0,
\end{eqnarray}
which is valid.

From Eq. (\ref{eq:e1-b}) we have
\begin{eqnarray}
f_1''(r) + \frac{2}{r} f_1'(r) = 0,
\end{eqnarray}
which yields
\begin{eqnarray}
 f_1(r)=\frac{C_2}{r} + C_3.
\end{eqnarray}

(ii) If $k_2\neq 0$ and $k_3\neq -k_2$, then we have
\begin{eqnarray}
(2\tilde{g} k_2)^2  + (2\tilde{g} k_1 -1)^2 =0,
\end{eqnarray}
i.e.,
\begin{eqnarray}
k_2^2=- \left(\frac{2\tilde{g} k_1 -1}{2\tilde{g}}\right)^2,
\end{eqnarray}
i.e.,
\begin{eqnarray}
k_2= \pm \sqrt{- \left(\frac{2\tilde{g} k_1 -1}{2\tilde{g}}\right)^2}=\pm {\rm i} \left|\frac{2\tilde{g} k_1 -1}{2\tilde{g}}\right|.
\end{eqnarray}
Note that $k_2\neq 0$ implies that $2\tilde{g} k_1 -1\neq 0$, and here $k_2$ is generally a complex number.

From Eq. (\ref{eq:e3-b}) we have
\begin{eqnarray}
& & \frac{k_2 + k_3}{r^3} \left( 2\tilde{g} k_1 - 1 \right) - \frac{4\tilde{g}^2 k_2 f_1^2 \left( r \right)}{r} \nonumber \\
& & + \frac{2\tilde{g}}{r^3} \left[ k_2 \left( 2k_1 \left( \tilde{g} k_1 - 1 \right) - 2\tilde{g} k_2 k_3 \right)
+ 2\tilde{g} k_2 \left( k_2 + k_3 \right) \left( 2k_2 + k_3 \right) \right] = 0,
\end{eqnarray}
i.e.,
\begin{eqnarray}
& & \frac{4\tilde{g}^2 k_2 f_1^2 \left( r \right)}{r} = \frac{k_2 + k_3}{r^3} \left( 2\tilde{g} k_1 - 1 \right) + \frac{2\tilde{g}}{r^3} \left[ k_2 \left( 2k_1 \left( \tilde{g} k_1 - 1 \right) - 2\tilde{g} k_2 k_3 \right)
+ 2\tilde{g} k_2 \left( k_2 + k_3 \right) \left( 2k_2 + k_3 \right) \right],
\end{eqnarray}
i.e.,
\begin{eqnarray}
& & 4\tilde{g}^2 k_2 f_1^2 (r)= \frac{k_2 + k_3}{r^2} \left( 2\tilde{g} k_1 - 1 \right) + \frac{2\tilde{g}}{r^2} \left[ k_2 \left( 2k_1 \left( \tilde{g} k_1 - 1 \right) - 2\tilde{g} k_2 k_3 \right)
+ 2\tilde{g} k_2 \left( k_2 + k_3 \right) \left( 2k_2 + k_3 \right) \right],
\end{eqnarray}
i.e.,
\begin{eqnarray}
& & 4\tilde{g}^2 k_2 f_1^2 (r)= \frac{k_2 + k_3}{r^2} \left( 2\tilde{g} k_1 - 1 \right)
+ \frac{4\tilde{g} k_2}{r^2} \left[ \left( k_1 \left( \tilde{g} k_1 - 1 \right) -\tilde{g} k_2 k_3 \right)
+ \tilde{g} \left( k_2 + k_3 \right) \left( 2k_2 + k_3 \right) \right],
\end{eqnarray}
i.e.,
\begin{eqnarray}
& & 4\tilde{g}^2 k_2 f_1^2 (r)= \frac{k_2 + k_3}{r^2} \left( 2\tilde{g} k_1 - 1 \right)
+ \frac{4\tilde{g} k_2}{r^2} \left[  \tilde{g} k_1^2 -k_1 -\tilde{g} k_2 k_3 +2 \tilde{g} k_2^2 +\tilde{g} k_2 k_3
+ \tilde{g} k_3 \left( 2k_2 + k_3 \right) \right],
\end{eqnarray}
i.e.,
\begin{eqnarray}\label{eq:k3-1}
& & 4\tilde{g}^2 k_2 f_1^2 (r)= \frac{k_2 + k_3}{r^2} \left( 2\tilde{g} k_1 - 1 \right)
+ \frac{4\tilde{g} k_2}{r^2} \left[  \tilde{g} k_1^2 -k_1  +2 \tilde{g} k_2^2
+ \tilde{g} k_3 \left( 2k_2 + k_3 \right) \right].
\end{eqnarray}
Obviously, the function $f_1(r)$ has the following structure
\begin{eqnarray}\label{eq:kappa-2}
 f_1(r)=\frac{C}{r},
\end{eqnarray}
and we shall {analyze} and simplify the number $C$ later.

(ii-1) If $C=0$, then we have $f_1(r)=0$. In this situation, we have
\begin{eqnarray}
& &  C^2= \frac{(k_2 + k_3)\left( 2\tilde{g} k_1 - 1 \right)
+ 4\tilde{g} k_2 \left[  \tilde{g} k_1^2 -k_1  +2 \tilde{g} k_2^2
+ \tilde{g} k_3 \left( 2k_2 + k_3 \right) \right]}{4\tilde{g}^2 k_2}=0,
\end{eqnarray}
i.e.,
\begin{eqnarray}
& &  (k_2 + k_3)\left( 2\tilde{g} k_1 - 1 \right)
+ k_2 \left[ 4\tilde{g} (\tilde{g} k_1^2 -k_1  +2 \tilde{g} k_2^2)
+ 4\tilde{g}^2 k_3 \left( 2k_2 + k_3 \right) \right]=0,
\end{eqnarray}
i.e.,
\begin{eqnarray}
& &  (k_2 + k_3)\left( 2\tilde{g} k_1 - 1 \right)
+ k_2 \left[ 4\tilde{g} (\tilde{g} k_1^2 -k_1  + \tilde{g} k_2^2)+4\tilde{g}^2 k_2^2
+ 4\tilde{g}^2 k_3 \left( 2k_2 + k_3 \right) \right]=0,
\end{eqnarray}
i.e.,
\begin{eqnarray}
& &  (k_2 + k_3)\left( 2\tilde{g} k_1 - 1 \right)
+ k_2 \left[ -1+4\tilde{g}^2(k_2+k_3)^2\right]=0,
\end{eqnarray}
which is the constraint among $k_1$, $k_2$ and $k_3$, with $k_2\neq 0$, $k_2+k_3\neq 0$ and $\left[ 1  + 4\tilde{g} \left(   \tilde{g} k_2^2 + \tilde{g} k_1^2- k_1\right)\right]=0$.

(ii-2) If $C\neq 0$, after substituting Eq. (\ref{eq:kappa-2}) into Eq. (\ref{eq:e1-b}), one has
\begin{eqnarray}
4\frac{\tilde{g}}{r^2} \left( k_1 - 2\tilde{g} k_1^2 - 2\tilde{g} k_2^2 \right) f_1(r) + f_1''(r) + \frac{2}{r} f_1'(r) - \frac{2f_1(r)}{r^2} \left( 1 - 2\tilde{g} k_1 \right) = 0,
\end{eqnarray}
i.e.,
\begin{eqnarray}
4\frac{\tilde{g}}{r^2} \left( k_1 - 2\tilde{g} k_1^2 - 2\tilde{g} k_2^2 \right) \left(\dfrac{C}{r}\right) + \left(\dfrac{C}{r}\right)'' + \frac{2}{r} \left(\dfrac{C}{r}\right)' - \frac{2}{r^2} \left( 1 - 2\tilde{g} k_1 \right) \left(\dfrac{C}{r}\right)= 0,
\end{eqnarray}
i.e.,
\begin{eqnarray}
4\frac{\tilde{g}}{r^2} \left( k_1 - 2\tilde{g} k_1^2 - 2\tilde{g} k_2^2 \right) \left(\dfrac{1}{r}\right) + \left(\dfrac{1}{r}\right)'' + \frac{2}{r} \left(\dfrac{1}{r}\right)' - \frac{2}{r^2} \left( 1 - 2\tilde{g} k_1 \right) \left(\dfrac{1}{r}\right)= 0,
\end{eqnarray}
i.e.,
\begin{eqnarray}
4\frac{\tilde{g}}{r^2} \left( k_1 - 2\tilde{g} k_1^2 - 2\tilde{g} k_2^2 \right) \left(\dfrac{1}{r}\right) +
\dfrac{2}{r^3} + \frac{2}{r} \times \dfrac{-1}{r^2}- \frac{2}{r^2} \left( 1 - 2\tilde{g} k_1 \right) \left(\dfrac{1}{r}\right)= 0,
\end{eqnarray}
i.e.,
\begin{eqnarray}
4\frac{\tilde{g}}{r^2} \left( k_1 - 2\tilde{g} k_1^2 - 2\tilde{g} k_2^2 \right) \left(\dfrac{1}{r}\right) - \frac{2}{r^2} \left( 1 - 2\tilde{g} k_1 \right) \left(\dfrac{1}{r}\right)= 0,
\end{eqnarray}
i.e.,
\begin{eqnarray}
2\tilde{g} \left( k_1 - 2\tilde{g} k_1^2 - 2\tilde{g} k_2^2 \right) -  \left( 1 - 2\tilde{g} k_1 \right) = 0,
\end{eqnarray}
i.e.,
\begin{eqnarray}
-2\tilde{g} \left( k_1 - 2\tilde{g} k_1^2 - 2\tilde{g} k_2^2 \right) + \left( 1 - 2\tilde{g} k_1 \right) = 0,
\end{eqnarray}
i.e.,
\begin{eqnarray}
(2\tilde{g} k_2)^2  + (2\tilde{g} k_1 -1)^2 =0,
\end{eqnarray}
which is valid automatically.

Now let us come to see what is the number $C$.

\emph{Analysis 1.---}If $k_3=0$, then Eq. (\ref{eq:k3-1}) becomes
\begin{eqnarray}
& & 4\tilde{g}^2 k_2 f_1^2 (r)= \frac{k_2}{r^2} \left( 2\tilde{g} k_1 - 1 \right)
+ \frac{4\tilde{g} k_2}{r^2} \left[  \tilde{g} k_1^2 -k_1  +2 \tilde{g} k_2^2
\right],
\end{eqnarray}
i.e.,
\begin{eqnarray}
& & 4\tilde{g}^2  f_1^2 (r)= \frac{1}{r^2} \left( 2\tilde{g} k_1 - 1 \right)
+ \frac{4\tilde{g} }{r^2} \left[  \tilde{g} k_1^2 -k_1  +2 \tilde{g} k_2^2
\right],
\end{eqnarray}
i.e.,
\begin{eqnarray}
& & 4\tilde{g}^2  f_1^2 (r)= \frac{1}{r^2} \left[ 2\tilde{g} k_1 - 1
+ 4\tilde{g}  \left(  \tilde{g} k_1^2 -k_1  +2 \tilde{g} k_2^2
\right)\right].
\end{eqnarray}
Due to
\begin{eqnarray}
1  + 4\tilde{g} \left(   \tilde{g} k_2^2 + \tilde{g} k_1^2- k_1\right) =0,
\end{eqnarray}
one has
i.e.,
\begin{eqnarray}
& & 4\tilde{g}^2  f_1^2 (r)= \frac{1}{r^2} \left[ 2\tilde{g} k_1 - 1
+4\tilde{g}  \left(  \tilde{g} k_1^2 -k_1  + \tilde{g} k_2^2
\right)+4\tilde{g}^2k_2^2 \right],
\end{eqnarray}
i.e.,
\begin{eqnarray}
& & 4\tilde{g}^2  f_1^2 (r)= \frac{1}{r^2} \left[ 2\tilde{g} k_1 - 2+4\tilde{g}^2k_2^2 \right],
\end{eqnarray}
i.e.,
\begin{eqnarray}
& & 2\tilde{g}^2  f_1^2 (r)= \frac{1}{r^2} \left[ \tilde{g} k_1 - 1+2\tilde{g}^2k_2^2 \right].
\end{eqnarray}
Thus $C$ satisfies
\begin{eqnarray}\label{eq:c2}
& & C^2 =\frac{\tilde{g} k_1 - 1+2\tilde{g}^2k_2^2}{2\tilde{g}^2}.
\end{eqnarray}

\emph{Analysis 2.---}If $k_3\neq 0$, then from Eq. (\ref{eq:k3-1}) we can have
\begin{eqnarray}
& &  C^2= \frac{(k_2 + k_3)\left( 2\tilde{g} k_1 - 1 \right)
+ 4\tilde{g} k_2 \left[  \tilde{g} k_1^2 -k_1  +2 \tilde{g} k_2^2
+ \tilde{g} k_3 \left( 2k_2 + k_3 \right) \right]}{4\tilde{g}^2 k_2},
\end{eqnarray}
i.e.,
\begin{eqnarray}
& &  C^2= \frac{(k_2 + k_3)\left( 2\tilde{g} k_1 - 1 \right)
+ k_2 \left[ 4\tilde{g} (\tilde{g} k_1^2 -k_1  +2 \tilde{g} k_2^2)
+ 4\tilde{g}^2 k_3 \left( 2k_2 + k_3 \right) \right]}{4\tilde{g}^2 k_2},
\end{eqnarray}
i.e.,
\begin{eqnarray}
& &  C^2= \frac{(k_2 + k_3)\left( 2\tilde{g} k_1 - 1 \right)
+ k_2 \left[ 4\tilde{g} (\tilde{g} k_1^2 -k_1  + \tilde{g} k_2^2)+4\tilde{g}^2 k_2^2
+ 4\tilde{g}^2 k_3 \left( 2k_2 + k_3 \right) \right]}{4\tilde{g}^2 k_2},
\end{eqnarray}
i.e.,
\begin{eqnarray}
& &  C^2= \frac{(k_2 + k_3)\left( 2\tilde{g} k_1 - 1 \right)
+ k_2 \left[ -1+4\tilde{g}^2(k_2+k_3)^2\right]}{4\tilde{g}^2 k_2},
\end{eqnarray}
which reduces to Eq. (\ref{eq:c2}) when $k_3$ becomes 0.

\begin{remark}
{For convenience to  {build} Tables \ref{Tab:real} and \ref{Tab:complex} in Sec. \ref{sec-6}, let us list the real and complex static solutions as follows.

\emph{Real Static Solution.---} We have the real static solution as
\begin{eqnarray}
f_2(r)=\frac{C_1}{r}, \;\;\;\;  f_1(r)=\frac{C_2}{r} + C_3, \;\;\;\; k_1= \frac{1}{2 \tilde{g}}, \;\;\;\; k_2=0, \;\;\; k_3\neq 0,
\end{eqnarray}
with $\{C_1, C_2, C_3\}$ are real numbers. Here the constraint is $2\tilde{g} k_1=1$.

\emph{Complex Static Solution.---} We have the complex static solution as

(a)
\begin{eqnarray}
f_2(r)=\frac{C_1}{r}, \;\;\;\;  f_1(r)=\frac{C_2}{r} + C_3, \;\;\;\; k_1= \frac{1}{2 \tilde{g}}, \;\;\;\; k_2=0, \;\;\; k_3\neq 0,
\end{eqnarray}
with at least one of $\{C_1, C_2, C_3\}$ is complex number. Here the constraint is $2\tilde{g} k_1=1$.

(b)
\begin{eqnarray}
&& f_2(r)=\frac{C_1}{r}, \;\;\;\; f_1(r)= 0, \;\;\;\; k_1\neq  \frac{1}{2\tilde{g}}, \;\;\;\;k_2= \pm {\rm i} \left|\frac{2\tilde{g} k_1 -1}{2\tilde{g}}\right|,\nonumber\\
&&(k_2 + k_3)\left( 2\tilde{g} k_1 - 1 \right)+ k_2 \left[ -1+4\tilde{g}^2(k_2+k_3)^2\right]=0,
\end{eqnarray}
with $C_1$ is a real or complex number. Here $k_1$, $k_2$and $k_3$ satisfy the constraint $(k_2 + k_3)\left( 2\tilde{g} k_1 - 1 \right)
+ k_2 \left[ -1+4\tilde{g}^2(k_2+k_3)^2\right]=0$.

(c)
\begin{eqnarray}
f_2(r)=\frac{C_1}{r}, \;\;\;\; f_1(r)= \frac{C}{r}, \;\;\;\; k_1\neq  \frac{1}{2\tilde{g}}, \;\;\;\;k_2= \pm {\rm i} \left|\frac{2\tilde{g} k_1 -1}{2\tilde{g}}\right|, \;\;\;\;\; k_3 \; is\; arbitrary,
\end{eqnarray}
with $C_1$ and $k_3$ are real or complex numbers, $C^2= \dfrac{(k_2 + k_3)\left( 2\tilde{g} k_1 - 1 \right)
+ k_2 \left[ -1+4\tilde{g}^2(k_2+k_3)^2\right]}{4\tilde{g}^2 k_2}$.  $\blacksquare$
}
\end{remark}

\subsection{Summary: Static Solutions of the Yang-Mills Equations}

\subsubsection{The Real Static Solutions}

Based on the results in previous sections, we would like to list all the real static solutions of the Yang-Mills equations in
Table \ref{Tab:real}.

\begin{table}[H]
\centering
\caption{All the real static solutions of the Yang-Mills equations. The spin vector potential is given by $\vec{A} = \frac{1}{r} \left[ k_1 (\hat{r} \times \vec{\Gamma}) + k_2 \vec{\Gamma} + k_3 (\vec{\Gamma} \cdot \hat{r})\hat{r} \right]$, and the scalar potential is $ \varphi = f_1(r) ( \vec{\Gamma}\cdot \hat{r}) +  f_2(r)$. Here ``static'' means the solutions $\{\vec{A}, \varphi\}$ are time-independent, ``real'' means three parameters $k_1$, $k_2$, $k_3$ and two functions $f_1(r)$, $f_2(r)$ are all real. The parameter $\tilde{g}=g/c\hbar$, with $\hbar$ being Planck's constant, $c$ is the speed of light in vacuum, $g$ is the gauge coupling parameter, and they are all real numbers. $g$ can be greater than, equal to, or less than 0. The solution of $f_2(r)$ is given by $f_2(r)=\frac{C_1}{r}$. $C_1$ is real. }
\begin{tabular}{|c|c|c|c|c|}
\hline
\(f_1(r)\) & \(k_1\) & \(k_2\) & \(k_3\) & Remarks \\
\hline
\multicolumn{5}{|c|}{\textbf{Case 1: $\{g=0, k_1=0\}$}} \\
\hline
\(\dfrac{C_2}{r^2}+C_3 r\) & \(0\) & $k_2 \in \mathbb{R}$ & \(-k_2\) & $C_2$ and $C_3$ are real \\
\hline
\multicolumn{5}{|c|}{\textbf{Case 2: $\{g=0, k_1\neq 0\}$}} \\
\hline
\(--\) & \(--\) & \(--\) & \(--\) & No Solutions\\
\hline
\multicolumn{5}{|c|}{\textbf{Case 3: $\{g\neq 0, k_1= 0\}$}} \\
\hline
\(0\) & \(0\) & \(0\) & \(0\) & $\vec{A}=0$, and $\varphi$ is Coulomb-type potential  \\
\hline
\multicolumn{5}{|c|}{\textbf{Case 4: $\{g\neq 0, k_1\neq  0\}$}} \\
\hline
\(\dfrac{C_2}{r} + C_3\) & \(\dfrac{1}{2\tilde{g}}\) & \(0\) & \(0\) & The constraint condition $2\tilde{g} k_1=1$ \\
\hline
\(0\) & \(\dfrac{1}{\tilde{g}}\) & \(0\) & \(0\) & The constraint condition $\tilde{g} k_1=1$ \\
\hline
\(0\) & \(\dfrac{1}{\tilde{g}} \sin^2\theta\) & \(-k_3\) & \(-\dfrac{1}{\tilde{g}}\sin\theta\cos\theta \neq 0\) & Constraint $\tilde{g} k_1^2+\tilde{g} k_3^2 -k_1=0$ \\
\hline
\(\dfrac{C_2}{r} + C_3\) & \(\dfrac{1}{2\tilde{g}}\) & \(0\) & \(k_3\neq 0\) & The constraint condition $2\tilde{g} k_1=1$ \\
\hline
\end{tabular}\label{Tab:real}
\end{table}

\subsubsection{The Complex Static Solutions}

Based on the results in previous sections, we would like to list the complex static solutions of the Yang-Mills equation in
Table \ref{Tab:complex}.

\begin{table}[H]
\centering
\footnotesize
\caption{All the complex static solutions of the Yang-Mills equations. The spin vector potential is given by $\vec{A} = \frac{1}{r} \left[ k_1 (\hat{r} \times \vec{\Gamma}) + k_2 \vec{\Gamma} + k_3 (\vec{\Gamma} \cdot \hat{r})\hat{r} \right]$, and the scalar potential is $ \varphi = f_1(r) ( \vec{\Gamma}\cdot \hat{r}) +  f_2(r)$. Here ``static'' means the solutions $\{\vec{A}, \varphi\}$ are time-independent, ``complex'' means at least one of three parameters $k_1$, $k_2$, $k_3$ or two functions $f_1(r)$, $f_2(r)$ is complex. The parameter $\tilde{g}=g/c\hbar$, with $\hbar$ being Planck's constant, $c$ is the speed of light in vacuum, $g$ is the gauge coupling parameter, and they are all real numbers. $g$ can be greater than, equal to, or less than 0. The solution of $f_2(r)$ is given by $f_2(r)=\frac{C_1}{r}$. }
\begin{tabular}{|c|c|c|c|c|}
\hline
\(f_1(r)\) & \(k_1\) & \(k_2\) & \(k_3\) & Remarks \\
\hline
\multicolumn{5}{|c|}{\textbf{Case 1: $\{g=0, k_1=0\}$}} \\
\hline
\(\dfrac{C_2}{r^2}+C_3 r\) & \(0\) & $k_2 \in \mathbb{C}$ & \(-k_2\) & At least one of $\{C_1, C_2, C_3, k_2\}$ is complex \\
\hline
\multicolumn{5}{|c|}{\textbf{Case 2: $\{g=0, k_1\neq 0\}$}} \\
\hline
\(--\) & \(--\) & \(--\) & \(--\) & No Solutions\\
\hline
\multicolumn{5}{|c|}{\textbf{Case 3: $\{g\neq 0, k_1= 0\}$}} \\
\hline
\(0\) & \(0\) & \(0\) & \(0\) & $C_1$ is complex  \\
\hline
\(\dfrac{\pm \frac{{\rm i}}{2|\tilde{g}|}}{r}\) & \(0\) & \(\pm\frac{{\rm i}}{2|\tilde{g}|}\) & \(\mp \frac{{\rm i}}{2|\tilde{g}|}\) & At least one of $\{C_1, C, k_2, k_3\}$ is complex \\
\hline
\(0\) & \(0\) & \(\pm \frac{{\rm i}}{2|\tilde{g}|}\) & \(\frac{-3\pm {\rm i} \sqrt{3}}{2} k_2\) & $C_1$ is a real or complex number \\
\hline
\(\dfrac{\pm \sqrt{3 k_2^2+ 3k_2 k_3 +k_3^2}}{r}\) & \(0\) & \(\pm \frac{{\rm i}}{2|\tilde{g}|}\) & \(k_3\neq -k_2\) & $C_1$ is a real or complex number \\
\hline
\multicolumn{5}{|c|}{\textbf{Case 4: $\{g\neq 0, k_1\neq  0\}$}} \\
\hline
\(\dfrac{C_2}{r} + C_3\) & \(\dfrac{1}{2\tilde{g}}\) & \(0\) & \(0\) & At least one of $\{C_1, C_2, C_3\}$ is complex \\
\hline
\(0\) & \(\dfrac{1}{\tilde{g}}\) & \(0\) & \(0\) & $C_1$ is a complex number \\
\hline
\(0\) & \(\dfrac{1}{\tilde{g}} \sin^2\theta\) & \(-k_3\) & \(-\dfrac{1}{\tilde{g}}\sin\theta\cos\theta \neq 0\) & $C_1$ complex, constraint $\tilde{g} k_1^2+\tilde{g} k_3^2 -k_1=0$ \\
\hline
\(0\) & \(\dfrac{1}{2\tilde{g}} (1 \pm \cosh \vartheta)\) & \(-k_3\) & \(\dfrac{{\rm i}}{2\tilde{g}} \sinh \vartheta \neq 0\) & Constraint $\tilde{g} k_1^2+\tilde{g} k_3^2 -k_1=0$ \\
\hline
\(0\) & \(\dfrac{1}{2\tilde{g}} (1+ {\rm i}\sinh^2 \vartheta)\) & \(-k_3\) & \(\pm \cosh \vartheta \neq 0\) & Constraint $\tilde{g} k_1^2+\tilde{g} k_3^2 -k_1=0$ \\
\hline
\(\dfrac{\pm \frac{{\rm i}}{2|\tilde{g}|}}{r}\) & Satisfies constraint & \(-k_3\) & \(k_3\neq 0\) & Constraint $\tilde{g} k_1^2+\tilde{g} k_3^2 -k_1=-\frac{1}{4\tilde{g}}$ \\
\hline
\(\dfrac{C_2}{r} + C_3\) & \(\dfrac{1}{2\tilde{g}}\) & \(0\) & \(k_3\neq 0\) & Constraint $2\tilde{g} k_1=1$, at least one of $\{C_1, C_2, C_3\}$ is complex \\
\hline
\(0\) & \(k_1\neq  \frac{1}{2\tilde{g}}\) & \(\pm {\rm i} \left|\frac{2\tilde{g} k_1 -1}{2\tilde{g}}\right|\) & Satisfies constraint & $(k_2 + k_3)( 2\tilde{g} k_1 - 1 ) + k_2[ -1+4\tilde{g}^2(k_2+k_3)^2]=0$ \\
\hline
\(\dfrac{C}{r}\) & \(k_1\neq  \frac{1}{2\tilde{g}}\) & \(\pm {\rm i} \left|\frac{2\tilde{g} k_1 -1}{2\tilde{g}}\right|\) & Arbitrary & $C^2= \dfrac{(k_2 + k_3)( 2\tilde{g} k_1 - 1 ) + k_2[ -1+4\tilde{g}^2(k_2+k_3)^2]}{4\tilde{g}^2 k_2}$ \\
\hline
\end{tabular}\label{Tab:complex}
\end{table}
